\documentclass[%
reprint,
amsmath,amssymb,
aps,
pra,
onecolumn
]{revtex4-1}

\usepackage{graphicx}
\usepackage{dcolumn}
\usepackage{bm}
\usepackage{graphics,dcolumn,bm,epic, eepic,float} 
\usepackage{amssymb,amsmath,amsfonts,multirow,rotate,color,array}
\usepackage{MnSymbol}
\usepackage{soul}
\usepackage{inputenc}
\usepackage{wasysym}
\usepackage{dcolumn}
\usepackage{subfigure}
\usepackage{xcolor}

\usepackage[colorlinks=true,linkcolor=blue,urlcolor=blue,citecolor=blue,anchorcolor=blue]{hyperref}

\newcommand*{\bd}{\boldsymbol}
\newcommand{\vect}[1]{\bm{\mathrm{#1}}}
\usepackage{booktabs}
\usepackage{array}
\usepackage{physics}
\usepackage[symbol*]{footmisc}
\DefineFNsymbolsTM{myfnsymbols}{
	\textasteriskcentered *
	\textdagger    \dagger
	\textdaggerdbl \ddagger
	\textsection   \mathsection
	\textbardbl    \|%
	\textparagraph \mathparagraph
}%
\setfnsymbol{myfnsymbols}

\begin{document}

\title{Networks beyond pairwise interactions: structure and dynamics}

\author{Federico Battiston}\email{battistonf@ceu.edu}
\affiliation{Department of Network and Data Science, Central European University, Budapest 1051, Hungary}

\author{Giulia Cencetti}
\affiliation{Mobs Lab, Fondazione Bruno Kessler, Via Sommarive 18, 38123, Povo, TN, Italy}

\author{Iacopo Iacopini}
\affiliation{School of Mathematical Sciences, Queen Mary University of London, London E1 4NS, United Kingdom}
\affiliation{Centre for Advanced Spatial Analysis, University College London, London, W1T 4TJ, United Kingdom}

\author{Vito Latora}\email{v.latora@qmul.ac.uk}
\affiliation{School of Mathematical Sciences, Queen Mary University of London, London E1 4NS, United Kingdom}
\affiliation{Dipartimento di Fisica ed Astronomia, Universit\`a di Catania and INFN, I-95123 Catania, Italy}
\affiliation{The Alan Turing Institute, The British Library, London NW1 2DB, United Kingdom}
\affiliation{Complexity Science Hub Vienna (CSHV), Vienna, Austria}

\author{Maxime Lucas}
\affiliation{Aix Marseille Univ, CNRS, CPT, Turing Center for Living Systems, Marseille, France}
\affiliation{Aix Marseille Univ, CNRS, IBDM, Turing Center for Living Systems, Marseille, France}
\affiliation{Aix Marseille Univ, CNRS, Centrale Marseille, I2M, Turing Center for Living Systems, Marseille, France}

\author{Alice Patania}
\affiliation{Network Science Institute, Indiana University, Bloomington, IN, USA}

\author{Jean-Gabriel Young}
\affiliation{Center for the Study of Complex Systems, University of Michigan, Ann Arbor, MI, USA, 48109}

\author{Giovanni Petri}\email{giovanni.petri@isi.it}
\affiliation{ISI Foundation, via Chisola 5, 10126 Turin, Italy}
\affiliation{ISI Global Science Foundation, 33 W 42nd St, 10036 New York NY, USA}

\date{\today}

\begin{abstract}

The complexity of many biological, social and technological systems
stems from the richness of the interactions among their units. Over
the past decades, a great variety of complex systems has been
successfully described as networks whose interacting pairs of nodes
are connected by links. Yet, in face-to-face human communication,
chemical reactions and ecological systems, interactions can occur in
groups of three or more nodes and cannot be simply described just in
terms of simple dyads.  Until recently, little attention has been
devoted to the higher-order architecture of real complex systems.
However, a mounting body of evidence is showing that taking the
higher-order structure of these systems into account can greatly enhance
our modeling capacities and help us to understand and predict their
emerging dynamical behaviors. Here, we present a complete overview of
the emerging field of networks beyond pairwise interactions. We first 
discuss the methods to represent higher-order interactions and give a unified
presentation of the different frameworks used to describe higher-order systems, 
highlighting the links between the existing concepts and
representations. We review both the measures designed to characterize
the structure of these systems, and the models proposed in the
literature to generate synthetic structures, such as random and growing
simplicial complexes, bipartite graphs and hypergraphs.
We then introduce and discuss the rapidly growing research on
higher-order dynamical systems and on dynamical 
topology. We focus on novel emergent phenomena characterizing landmark
dynamical processes, such as diffusion, spreading, synchronization and
games, when extended beyond pairwise interactions. We elucidate the
relations between higher-order topology and dynamical properties, and
conclude with a summary of empirical applications, providing an
outlook on current modeling and conceptual frontiers.
\end{abstract}

\maketitle

\tableofcontents


\section{Introduction}
\label{sec:introduction}

Any significant understanding of a complex system must rely on system
level descriptions. Consider the following exercise: take an
ecosystem, and break it into its pieces. No matter how good or
accurate our knowledge at the level of the individual species is,
chances are that our understanding of population dynamics (e.g. how
the abundances of the different species change in time) will be slim
at best. The same holds true when we attempt to explain epileptic
seizures starting from the individual neurons of the human brain; or
viral rumors spreading across societies from individual human
psychology. All these approaches fail because they are missing
a fundamental ingredient of any complex system, 
that is the rich pattern of nonlinear interactions between
the system components. After many years of reductionism, science
has abandoned the idea that the collective behaviors of a complex
system can be simply understood and predicted by considering the
units of the system in isolation~\cite{anderson1972more}, and now more
than ever is embracing the idea of complexity as one of the principles 
governing the world we live in.

Within this paradigm, networks have emerged as a reference modeling
tool for complex systems \cite{barabasi2011network}.  Networks are the
maps that define the physical or virtual space where interactions take
place. Add competitive and cooperative relationships to an ecosystem, 
synaptic connections to the human brain, and human interactions to
rumor spreading, and readily the self-organizing patterns and the collective behavior 
we observe in nature begin to unravel and look less obscure.
Building on earlier work in mathematics, social network analysis and
ecology, a handful of breakthrough papers at the turn of the millennium
has attracted the interest of the scientific community, triggering 
thousands of contributions over the last twenty years and leading 
to the formation of the new multidisciplinary field of 
{\em Network Science}. This new research community has developed an unusual
mixture of graph theory and statistical mechanics into a flourishing
discipline, with applications spanning 
the full range of science, from fundamental physics all the way to the
social sciences.
The boundaries and potential applications are yet to be fully realized
\cite{barabasi2011network}.  Still, the richness in scope and tools
has already made the field of networks an independent discipline,
often referred to as a science of its own. The growth of network
science was also strengthened by the progressively wider availability of
large datasets with detailed information on social, technological and
biological interactions, which provided the raw material for the
empirical validation of network models and predictions.
We refer the reader interested in a first approach to network science
to the several early review papers~\cite{albert2002statistical,
  dorogovtsev2002evolution, newman2003structure,
  boccaletti2006complex} and textbooks~\cite{newman2010networks,
  estrada2011structure, barabasi2016network, latora2017complex} on the
subject.
\\

As exploration of real-world systems deepens, network scientists are
realizing the need to further characterize and enrich the
relationships captured by a network description.  This, however,
creates problem: networks have originally been understood as a
collection of nodes, representing the elementary units of the system,
and edges, describing the existence of interactions between pairs of
such units. Applications to real-world systems, however, require the
possibility to describe more details of an interaction
\cite{butts2009revisiting}, like for example:
directed edges to describe the origin and destination of a message;
edge weights, to highlight the intensity of an interaction; and even
signs on the edges to distinguish whether a link encodes a productive or
detrimental interaction among two units.  In more recent years, a
large effort has been devoted to formalize and develop the
mathematical tools to analyze {\em temporal networks}, where interactions
are not static but unfold in the temporal dimension
\citep{holme2012temporal}. Similarly, many works have recently
considered the case of interacting systems where units can be
connected by links of different nature, and which can be effectively
represented in terms of {\em multiplex networks} or {\em multilayer networks} 
\citep{boccaletti2014structure}.\\
All these aspects have contributed in many cases to a better network representation, but are networks themselves enough to provide a complete 
description of a complex system? \\

The fundamental limit of networks is that they capture pairwise
interactions only, while many systems display group interactions.
Indeed, in social systems, ecology and biology among other examples,
many connections and relationships do not take place between pairs of
nodes, but rather are collective actions at the level of groups of
nodes.  For instance, three or more species routinely compete for food
and territory in complex ecosystems \citep{levine2017beyond}. In other
cases, the presence of a third species influences the interaction
between other two, affecting directly the interaction (the link)
rather than the species involved (the nodes).  Similarly, social
mechanisms, such as peer-pressure, inherently go beyond the idea of
dyadic connections.
Collective interactions are not an entirely new idea, and to some
degree have appeared in early research on networks.  Think for
instance to the majority-rule model for the dynamics of opinion
formation, or the public goods game in evolutionary game theory.
In addition to these examples, one
of the most successful streams of research in network science in
recent years, complex contagion, naturally accounts for multiple
simultaneous interactions \citep{centola2010spread}.  However, in all
cases these applications tried to leverage the language of pairwise
networks to describe interactions of higher order, for example by
using bipartite graphs \cite{newman2001random}.  Can we instead find
mathematical frameworks that can explicitly and naturally describes group
interactions?\\

Simplicial complexes and hypergraphs are the natural candidates to provide such 
descriptions. 
And indeed, over the last few years, a wave of enthusiasm for these representations has revolutionized our vision of
and ability to tackle real-world systems characterized by more than
simple dyadic connections.  The importance of high-order interactions
had been recognized already a long time ago
\cite{atkin1972cohomology,berge1973graphs,atkin1974mathematical}, but
this rejuvenated interest has brought a new, and much deeper
understanding of higher-order representations.  
There are no doubts now that moving beyond dyadic interactions is fundamental to explain and predict collective behaviors that could
not be described before.  
\\

The aim of this report is to provide a review of the state-of-the-art
on the structure and dynamics of complex networks beyond pairwise
interactions, as well as a reference and perspective on crucial open
questions in the field. Together with this introduction, the report is
organized as follows:
\begin{itemize}

\item The first part (Sections \ref{sec:representations} to
  \ref{sec:models}) focuses on the structure of systems with higher-order
  interactions.  In particular, Section
  \ref{sec:representations} provides an introduction to the
  mathematical frameworks underlying higher-order
  representations. Section \ref{sec:measures} describes the most
  common measures and properties currently used to describe the
  structure of systems with many-body interactions. Finally, Section \ref{sec:models}
  reviews random models of higher-order systems and how they are used
  to make statistical inferences.

\item The second part (Sections \ref{sec:diffusion} to
  \ref{sec:games}) focuses on the dynamics of systems with higher-order
  interactions.  In more detail, Section \ref{sec:diffusion} discusses
  models of higher-order diffusion. Section
  \ref{sec:synchronization} describes the generalization of oscillator
  models and synchronization.  Section \ref{sec:social} introduces
  recent models of spreading in social systems with group structure.
  Section \ref{sec:games} reports on models of competition and
  cooperation among multiple agents.

\item Finally, Section \ref{sec:applications} is an overview of real-world
  applications to systems with higher-order interactions.
Our final conclusions and outlook are presented in Section \ref{sec:conclusions}.
\end{itemize}

We conclude this preamble with a final remark. The idea to include
higher-order interactions in network analysis is very simple. To this
end, going beyond pairwise interactions might not look harder than
attaching weights or signs to the edges of a graph. Yet, in practice,
moving from pairs to a more complicated interaction structure is a
difficult issue, and it requires a great deal of sophistication and
novel mathematical tools.  This explains why the analysis of this
aspect of complex system has been heavily delayed compared
to its weighted and signed, and even temporal and multilayer
counterparts. After all, classical physics already knew this: while a
closed-form solution is available for the two-body problem, solving for 
the trajectories of $n$ interacting bodies given their positions and
momenta is still an open problem!


\section{Higher-order representations of networks}
\label{sec:representations}

\subsection{Elementary representations of higher-order interactions}

\subsubsection{Low- versus high-order representations}
\label{subsubsec:low-high-reps}
We begin by first defining more precisely what we consider as
interactions and, as a consequence, as higher-order interactions.  We
define an \emph{interaction} as a set $I = [p_0, p_1, \ldots,
  p_{k-1}]$ containing an arbitrary number $k$ of basic elements 
of the system under study, which we indicate as \emph{nodes} or
\emph{vertices}.  
Such interactions can then describe different situations in real systems,
e.g. the coauthors of a scientific paper, 
a set of genes required to perform a certain function,
the coactivation of a group neurons during a specific task, etc.
In a slightly counterintuive way, we will denote the \emph{order} (or dimension)
of an interaction involving $k$ nodes to be $k-1$: a node interacting
with itself only is a 0-order interaction, an interaction between two
nodes has order 1, one among three nodes has order 2, and so on.
Furthermore, we consider \emph{higher-order interactions} to be
$k$-interactions with $k \geq 2$. Conversely, low order interactions
are those characterized by $k\leq1$. 
In plain terms, low order systems are those in which only self- or pair-wise interactions take place
(like edges in a graph), while \emph{higher-order systems} (HOrSs, from now on) display
interactions in groups of more than two elements. \\

The distinction between low- and high-order interactions is needed for
two reasons.  
First, it highlights the differences between the
graph-theoretic descriptions, that shaped the study of complex systems
in recent decades, and the more recently (re)proposed descriptions
based on genuine group interactions.  
Secondly, it allows us to clearly
frame the connections between such descriptions, their various overlaps and
reciprocal mappings.  Finally, our definition explicitly leaves out
other types of higher-order dependencies between the
components of a system, as for example those defined
by multiple link types in multilayer networks \citep{boccaletti2014structure,kivela2014multilayer,battiston2017new,bianconi2018multilayer,aleta2019multilayer}, or by non-Markovian paths
in time-stamped interaction data  
\citep{holme2012temporal,lambiotte2019networks}.
While these are out of the scope of this review, the interested
reader can find an extensive discussion of these topics in 
the references mentioned above.  \\

\begin{figure*}
\centering
\includegraphics[width=\textwidth]{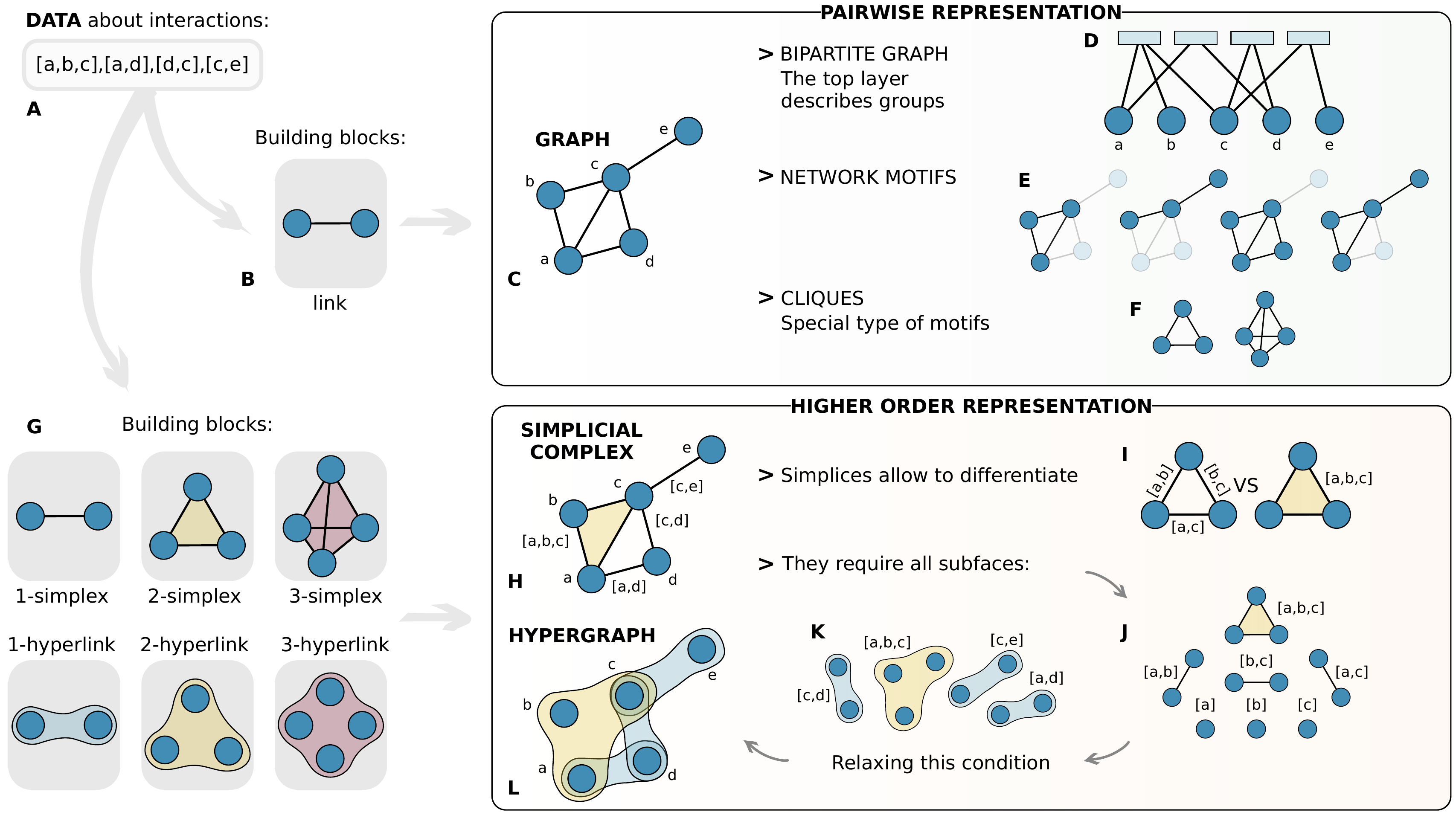}
\caption{\textbf{Representations of higher-order interactions.}
A set of interactions of heterogeneous order (A) can be represented using only pairwise interactions (B). 
Using only low order blocks, the set of interactions can be described in th simplest way by using a graph (C). Alternatively, interactions can be encoded as nodes in one layer of a bipartite graph, where the other layer contains the interaction vertices (D). 
Other examples of high-order coordinated patterns can be encoded using motifs, small subgraphs with specific connectivity structures (E). 
Among motifs,  cliques are especially popular as they represent the densest subgraphs, aking to higher order bricks (F). 
All these representations discard information that was present in the original interaction data (A). 
A solution is to consider explicitly higher order building blocks, in the form of simplices and hyperedges (G). 
Collection of simplices form simplicial complexes (H), which allow to discriminate between genuine higher order interactions and -even complex- sums of low order ones (I).  Unfortunately, simplicial complexes, given a simplex, require the presence of all possible subsimplices (J), which can be too strong an assumption in some systems. Relaxing this condition effectively implies moving from  simplices to hyperedges (K), which are the most general---and less constrained---representation of higher-order interactions (L). 
}
\label{fig:hors-representations}
\end{figure*}

We define an interacting system $(V,\mathcal{I})$ as the family of interactions $\mathcal{I} = \{I_0, \ldots I_n\}$ taking place on a node set $V$.
To aid the intuition, let us make a specific example. 
Consider the node set $V=[a,b,c,d,e]$ and the set of interactions $\mathcal{I} = \{[a,b,c], [a,d], [d,c], [c,e]\}$ (Fig.~\ref{fig:hors-representations}A). 
$\mathcal{I}$ contains three  1-interactions and one 2-interaction. 
While the complete information about the systems is included in the list above, the study of most interesting properties of the system requires the choice of a representation.
For example, measuring the collective effects of the interactions on a specific node requires the capacity to map interactions of different orders in a way that makes them comparable to each other; asking how dense the system is, or whether one node is reachable from another, again requires being able to compare in a controlled way interactions of different order and composition.
\\

\subsubsection{Graph-based representations}
\label{subsubsec:graphs}
Graphs are the most common way to represent families of
interactions (Fig.~\ref{fig:hors-representations}C).  
A graph $G=(V,E)$ is defined by a nodeset $V$ with $n$
elements, and an edgeset $E$ whose $m$ elements are pairs of nodes. 
A graph is then a collection of edges connecting pairs of
nodes. 
In other words, the building blocks of graph representations
are 1-interactions, i.e. interactions of the type $I = [i,j]$.
The most natural choice is then to unfold each higher-order interaction in $\mathcal{I}$ in terms of 1-interactions built from pairs of nodes in $I$. 
Under this assumption, our example $\mathcal{I} = \{[a,b,c], [a,d], [d,c], [c,e]\}$ maps to $\mathcal{I}_{\text{G}} = \{[a,b], [b,c], [c,a],
[a,d], [d,c], [c,e] \}$ (Fig.~\ref{fig:hors-representations}B).
This mapping makes systems amenable to be studied 
using tools developed in both graph theory 
\cite{bondy1976graph} and network science \cite{newman2010networks}.
Indeed, graph representations enabled the growth, depth and breadth of
results on real-world complex networks in the last two decades
\cite{estrada2011structure,barabasi2016network,latora2017complex},
with applications spanning biology
\cite{alon2003biological,kashtan2005spontaneous}, ecology
\cite{grilli2017higher,montoya2006ecological}, social science
\cite{borgatti2009network,mcpherson2001birds}, engineering
\cite{gao2012networks,buldyrev2010catastrophic}, neuroscience
\cite{bullmore2009complex,bassett2017network,medaglia2015cognitive},
all the way to cosmology \cite{boguna2014cosmological}.

Despite the power of graph representations to capture many properties of complex interacting systems, their limits are easily identified: it is impossible to explicitly describe group interactions, or in other terms there is no direct relationship between $\mathcal{I}$ and $\mathcal{I}_G$ nor any way to recover the former from the latter. 
For example, going back to our toy example, at the description level provided by $\mathcal{I}_G$, it is impossible to tell (and hence to describe) whether the original interaction set contained $[a,c,d]$ or not. 
Naturally, in some cases networks can provide information on higher-order interactions, but these are always inferences based on the low order interactions, obtained for example by looking for very dense subsets of nodes using community \cite{fortunato2010community}, clique \cite{palla2005uncovering} or block detection \cite{karrer2011stochastic} techniques. However, such reconstructions are often incomplete and rife with problems \cite{lancichinetti2011limits,abbe2015community,lancichinetti2008benchmark}.   
\\

Bipartite graph representations effectively describe group
interactions.  Solidly within the realms of low-order interactions,
bipartite graphs are graphs defined by two nodesets $(U,W)$ and
edgeset $E$ containing only edges $(u,w)$ such that $u\in U$ and $ w
\in W$.  To represent higher-order interactions, one chooses $U$ to
coincide with the original nodeset $V$, i.e. $U=V$, and $W$ to
coincide with the set of interactions $\mathcal{I}$
\cite{guillaume2004bipartite,guillaume2006bipartite}.  The links in
the bipartite graph connect a node (in $V$) to the interactions (of
arbitrary order) in which it takes part 
(Fig.~\ref{fig:hors-representations}D).  This representation emerges
naturally in many fields: it is used for example in social sciences,
where it provides a way to encode the membership of individuals to
groups of different dimensions \cite{wasserman1994social,newman2002random};
or to describe
the collaboration of actors (nodes) in movies (interactions)
\cite{guimera2007module}; it is also used in ecological bipartite
graphs, where species linked to a common prey represent competition
for resources among more than two species
\cite{montoya2006ecological}; in recommendation systems
\cite{zhou2007bipartite} they describe the relations between customers
and purchased products, and so on.

It is easy to see that the entire information in our toy model is
preserved when interactions are described as a bipartite graph.
In fact, this representation is very general and can indeed well mimic most interaction structures. 
However, at difference with other multilayer graph formulations \cite{boccaletti2014structure,kivela2014multilayer}, in a bipartite graph the nodes of the original system do not interact directly with each other anymore. 
Rather, their relation is always mediated by the interaction layer, which is of a different nature from the node layer itself. 
This implies that any measure or dynamic process define on the bipartite representation needs to take into account this additional complexity. The usual workaround to this problem is to consider the unipartite networks obtained by projecting the bipartite on one of the two layers. Each interaction becomes then a fully connected subgraph among the nodes belonging to the interaction, losing the group structure in the same way as in the simple graph case. In addition, it is usually impossible to translate the information contained in the standard graph operators (e.g. Laplacian) defined on a bipartite graph into the ones corresponding to the unipartite projections \cite{zweig2011systematic,schaub2018flow}\\

Motifs allows to extract additional information on the properties of
an interaction.  They are the---usually small---recurrent subgraphs of
a given network, or of a class of networks of similar origin
\cite{milo2002network}.  Motifs are defined as specific patterns of
edges (1-interactions) between vertices that appear to be statistically
significant in the network (Fig.~\ref{fig:hors-representations}E). They are considered structural
signatures of the function of a network. That is,
different motifs can correspond and reflect different functions or different
optimization solutions to the same function \cite{alon2007network}.
Typically, the statistically validated frequencies (z-scores) 
with which the various motifs are observed 
in a network are collected into a motif profile, which can then be
used for example to discriminate between different networks
\cite{benson2016higher}, e.g. between brain functional networks in
different states \cite{morgan2018low,avena2015network} or between
differently evolved biological networks
\cite{shen2002network,alon2007network,kashtan2005spontaneous}.  Motifs
also found widespread application in the study of social
\cite{fowler2009model} and temporal
\cite{paranjape2017motifs,kovanen2011temporal} systems.
\\ Motifs 
constitute a refinement of the bipartite representation of a system, since, in
addition to a division in groups akin to that of bipartite graphs, they allow 
to specify the interaction pattern in which a node is involved.
The drawback to this is that the number of possible motifs to
investigate grows exponentially with the number of nodes
involved.  This unfortunately makes them quite unwieldy as a
descriptive tool for large graphs and/or motifs. As an example,
generative models aimed to quantify randomness in networks via
motif-based constraints \cite{mahadevan2006systematic} were shown to
become very hard to manage, or even sample, for interactions above
order 2 \cite{orsini2015quantifying}. \\

Because of the exponential growth in the number of motifs, a large
part of the work on analyzing subgraphs focuses on a special type of
motifs: cliques (Fig.~\ref{fig:hors-representations}F).  
A clique of size $k$ is defined as a fully connected subgraph of $k$ nodes. 
Here, we use size for cliques to avoid confusion, since a $k$-clique usually encodes an interaction of order $k-1$. 
The interest in cliques is justified also by the fact that they represent the most obvious definition of
group from a network point of view, because they are the densest and
most uniform motif \cite{derenyi2005clique}.  Also, they directly
encode the idea that every member of the clique interacts with every
other \cite{dunbar1995social,provan1998networks}.  Due to these
properties, cliques are privileged building blocks of a network and its 
communities \cite{palla2005uncovering}. However, we can incur into
problems if we want to use cliques to characterize higher-order
interactions. In fact, going back
to our toy example, we see that both sets $a,b,c$ and $a,d,c$ form
3-cliques. Conversely, in $\mathcal{I}_G$ we only had a true
2-interaction, namely $[a,b,c]$, while the fictitious interaction 
$[a,d,c]$ is emerging as a byproduct of the union of the 1-interactions
$[a,d]$ and $[d,c]$ with the $[a,c]$ edge induced by $[a,b,c]$.  We
can see then that, by considering all cliques present at the graph level,
we would ``fill'' a 2-interaction that was not included in the
original interaction set.  This is somewhat opposite to what happened
when we considered the edges alone. In that case, we lost completely
the notion of group. In the case of cliques instead, we risk
``filling'' too much and thus creating high-order interactions that
were not there to begin with.\\

\subsubsection{Explicit higher-order representations}
\label{subsubsec:high-reps}
To properly describe higher-order interactions, we need to encode them 
explicitly. Why not encoding interactions exactly as they are in fact?
Simplices are the simplest mathematical objects to accomplishes this.
The formal definition of simplices mimicks very closely
the one we gave of higher-order interactions. 
In fact, a \emph{$k$-simplex} 
$\sigma$ is, in its most general form, just a set of $k+1$ nodes $\sigma
=[p_0, p_1, \ldots, p_{k}]$ (Fig.~\ref{fig:hors-representations}G). 
This notation is the standard one borrowed from the literature in algebraic topology \citep{hatcher2002algebraic}, where nodes are often points in a topological space. 
In applications where the interactions are purely combinatorial, one might want to draw attention to the interactions rather than to the underlying space. 
Thus, in these cases, nodes are often denominated as $v_0, v_1, \ldots$ to highlight that they are vertices of interactions, without any reference to an underlying space.
The definition of dimension of a simplex coincides then with the
definition of order of an interaction we gave earlier. 
Based on this parallel between the definitions of
interactions and simplices, it is easy to see that we do not incur anymore in the problems described above.  
However, it is not clear how we can handle interactions of potentially different dimensions together and which are the advantages of such representations.  
Just like graphs are collections of edges, \emph{simplicial complexes} are collections of simplices (Fig.~\ref{fig:hors-representations}H).  
At difference with graphs, they require further properties to be considered valid complexes: a collection of $n$ simplices $K =
\{\sigma_0, \sigma_1 \ldots \sigma_n \}$ is a valid simplicial complex
if, for every $k$-simplex  $\sigma = [p_0, p_1, \ldots, p_{k}] \in K$, 
all its subfaces of any dimensions belong to $K$ too.  
For example, if the triangle $[a,b,c] \in K$, then we also require
$[a],[b],[c], [a,b], [a,c], [b,c]$ to belong to $K$. 
Note that if we were to extract cliques from a graph and consider them as simplices (which is the operative definition of a \emph{clique complex}), it would be impossible to distinguish the two cases in which respectively the triangle was present or not. 
Using a simplicial formalism instead this distinction is immediate, as we only need to check whether the 2-interaction $[a,b,c]$ is included in $K$ (Fig.~\ref{fig:hors-representations}I).   
Simplicial descriptions are very powerful because they come equipped with many nice mathematical gadgets. 
It is in fact straight-forward to define Laplacian operators for any dimension on simplicial complexes \citep{horak2013spectra,muhammad2006control}, they can approximate both regular manifolds and highly irregular structures \cite{costa2016random,bianconi2016network}, and they come naturally equipped with boundary operators stringing together simplices with different dimensions.  
Crucially, these operators describe the topology and shape of simplicial complexes in terms of their cycles, cavities and higher-order topological holes \cite{ghrist2014elementary} and are naturally related to the combinatorial Laplacians \citep{muhammad2006control}.  
In the following sections, we will describe many of these properties in
greater detail, because they represent some the most powerful tools
currently available and are the foundation of recent advances in topological data analysis \citep{carlsson2009topology,patania2017topological,expert2019topological}.
\\
Although simplicial complexes overcome some of the problems 
encountered by other lower dimensional representations, they are still
quite limited by the requirement on the existence of all 
subfaces. 
In some cases, this constraint is too restrictive. 
For instance, when studying social systems, it is important to
be able to describe interactions in groups. 
In this case we can use simplicial complexes as it is rather safe to assume that a group interaction also implies the underlying pairwise
interactions (Fig.~\ref{fig:hors-representations}J). 
The relative importance of pairwise versus group interactions can then be encoded in weights over the simplices.

However, in other cases, the inclusion constraint can be less easily justified: suppose for example that we are studying collaborations in scientific papers, and we observe a paper by three authors and none by the corresponding pairs of authors; or gene pathways were exactly three genes are needed to perform a function, but the subgroups are not responsible for any function on their own. Clearly, it would be useful to be able to describe also these situations (Fig.~\ref{fig:hors-representations}K). \\

Hypergraphs provide the most general and unconstrained description of
higher-order interactions.  
Formally, a \emph{hypergraph} is defined by a nodeset $V$ and a set of hyper-edges $H$ that specify which nodes participate in which way within an interaction. Each hyper-edge is a non-empty subset of $V$. 
It is easy to see that hypergraphs are the most appropriate description of interacting systems $(V,\mathcal{I})$ that we gave at the beginning of
the section (Fig.~\ref{fig:hors-representations}L). 
Notice that a hypergraph can include the 2-interaction $[a,b,c]$ without any requirement on the existence of 1-interactions $[a,b], [a,c]$ and $[b,c]$. 
In fact, hypergraphs are so unconstrained that it is also possible to define hyperedges that include other hyperedges, e.g. given $v,w,z \in V$ and $\gamma = [v,w] \in H$, it is possible to define a new hyperedge $\gamma' = [z,w,v ; \gamma] \in H$.  
Such extreme flexibility comes, as expected, with an additional complexity in treating them.  
For example, while many graph-theoretic concepts can be extended to the case of hypergraphs, such an endevour is often fraught with complications \citep{higuchi1999higher}, to the point
that a proper definition of Laplacian operators
\citep{louis2015hypergraph} on hypergraphs and of their properties,
e.g. the spectral diameter, has only emerged in the few last years
\citep{chan2018spectral,chan2019generalizing} and---to the best of our
knowledge---has found few real world applications,
e.g. degree-generating models \citep{ghoshal2009random} and hypergraph
modularity \citep{kumar2018hypergraph,chodrow2020annotated}.

\subsection{Relations and links between representations}
\label{subsec:rep-links}

\begin{figure*}
\centering
\includegraphics[width=\textwidth]{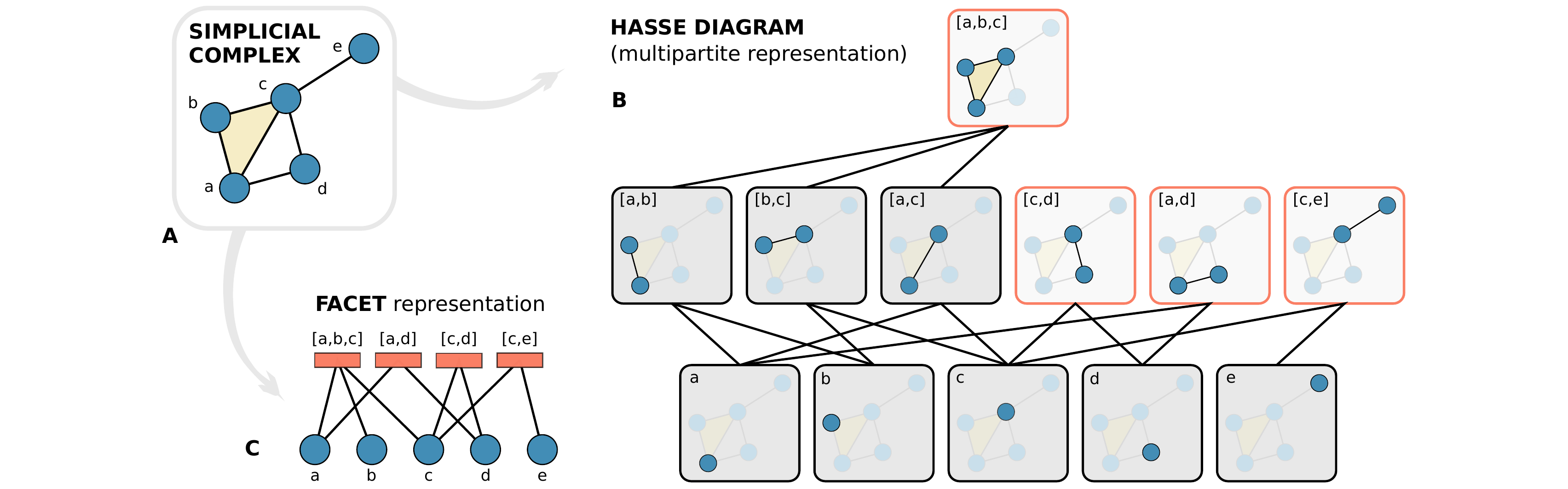}
\caption{\textbf{Relations among representations.}
A simplicial complex (A) is defined by the list of simplices that compose it. The structure of the natural inclusions between simplices can be described as a graph (B), where nodes correspond to simplices and edges the inclusions (in the figure, when two simplices are linked the top one contains the bottom one).  Following the chain of inclusions upward, one reaches the maximal simplices, \emph{facets}, that are not included in any larger simplex. These facets can be used to define a bipartite (or hypergraph) representation of the simplicial complex, identifying the facets with the hyperedges (C). 
}
\label{fig:hors-relations}
\end{figure*}

Naturally, many questions emerge  when discussing different
representations of the same interacting system: how much overlap is
there among two different representations?  
Is it possible to map one onto another in a canonical way? 
What kind of information is preserved (and lost) when moving between representations? 
For example, it seems obvious that a simplicial complex composed only by 1-dimensional simplices (edges) should be the same thing as a graph, right? \\ 
Well, it depends. 
Let us illustrate the links between representations starting from simplicial complexes (Fig.~\ref{fig:hors-relations}A).  
There are many ways of writing down a simplicial complex, but we focus here only on two descriptions, that are equally valid yet carry very different meanings: the Hasse diagram and the facet representation.\\

The Hasse diagram of a simplicial complex $K$ is the directed acyclic graph $HD(K) = (V_{HD}, E_{HD})$, whose nodeset $V_{HD}$ contains a node for each simplex in $K$ ($V_{HD}= \{ \sigma \} \forall \sigma \in K $), while the edgeset $E_{HD}$ contains an edge for each inclusion between simplices that differ in dimension by 1 . 
In other terms, for two simplices $\sigma, \tau \in K$ there exist an edge $(\sigma, \tau)\in E_{HD}$ iff $\sigma \subset \tau$ and $\dim(\tau)= \dim(\sigma)+1$. In Fig.~\ref{fig:hors-relations}B we provide an example of a Hasse Diagram for a toy simplicial complex (Fig.~\ref{fig:hors-relations}A) . 
It is easy to see that the Hasse Diagram unfolds all the structure in the simplicial complex, by making explicit the hierarchy of simplices in the complex via its multipartite structure (one layer per dimension), and thus providing information about its internal organization. 
Importantly, it also gives an explicit way to walk on a simplicial complex: starting from a node (simplex), a walker can follow the links in the Hasse Diagram and explore the whole complex. 
It turns out that the structure of the Hasse Diagram directly relates to the generalization of the graph Laplacian to simplicial complexes and to random walks on complexes. 
We will describe this in detail in Sections \ref{sec:measures} and \ref{sec:diffusion}, but, even without the full theory, it is already possible to understand some of the peculiarities of diffusion on simplicial complexes. 
The operators that link simplices that differ by $\pm1$ in dimension, akin to the links in the Hasse Diagram, are (co)boundary operators. 
For example, given a triangle (2-simplex), the boundary returns a combination of the three edges that form the perimeter of the triangle. Taking its boundary again, however, gives zero, because a boundary has no boundary itself (just like in standard differential geometry). 
It is easy to see now that any operator built on top of such boundary operators, like the combinatorial Laplacian, will only be able to describe diffusion between adjacent simplices with co-dimension 1. 
Similarly, it is easy to imagine that operators defined on different representations are not necessarily equivalent, as for example shown recently by \citet{schaub2018flow}, that found that the Laplacian built on a graph, on its line graph and on the corresponding 1-dimensional complex are not mutually exchangeable.

On the other extreme, the facet representation of a simplicial complex is the most parsimonious in terms of number of stored simplices. 
A \textit{facet} for complex $K$ is a simplex that is not contained in any other simplex in $K$. 
In the Hasse Diagram, facets correspond to nodes that are not included in any other. In Fig.~\ref{fig:hors-relations}B they are indicated as the simplices with an orange contour. 
In this sense, facets are akin to maximal cliques in graphs, and, just like maximal cliques, the list of facets of a complex uniquely identifies it. 
It is also a compressed description because it implies the existence of all the subsimplices without explicitly listing them. 
In truth, the facet representation is a directed bipartite graph under disguise, where facets constitute one layer, vertices the other, and directed edges represent inclusion of a node in a facet. 
It can also be recovered easily from the Hasse diagram, by keeping only the vertex layer and the simplices that have zero outdegree (i.e. without anything above themselves) . 
This bipartite graph associated to the facet representation can then also be studied as a hypergraph, where facet membership defines the hyperedges (Fig.~\ref{fig:hors-relations}C).
Note that the converse is generally false: a bipartite graph (or hypergraph) gives rise to a simplicial complex in the facet representation only if no set of vertex nodes linked to a facet node (hyperedge) is a subset another set of nodes linked to another facet  node (hyperedge); in short, the incident node sets of facets need to respect the non-inclusion properties of facets, or equivalently, no hyperedge can be included in another hyperedge.

\section{Measures}
\label{sec:measures}

In the previous section we have discussed the various ways and levels
at which high-order interactions can be described and represented.
In this section we will focus on observables and
measures that can be used to characterize and quantify the structural
properties of high-order interacting systems, at each level of their
description.
In particular, in the case of cliques,
hyperedges, sets, or simplices, many common notions developed for
ordinary graphs have been generalized to their higher-order
counterparts. We will start by discussing how to
describe interactions in terms of matrices or tensors.
We will then show how standard graph-based measures have been
generalized and what are the insights that can be extracted
using them.

\subsection{Matrix representations of higher-order systems}
\label{sec:measures:incidence}

\subsubsection{Incidence matrix}

In mathematics, the incidence matrix is the classical way to describe
the relationships between two classes of objects. First introduced by
Kirchkoff in 1847 for applications to electrical circuits, the
{\em incidence matrix} of a graph $G=(V,E)$ is a $n\times m$ matrix $I= \{
I_{i \alpha} \} $, where $n$ is the number of nodes and $m$ is the
number of edges. The entry $I_{i\alpha}$ in row $i$ and column
$\alpha$ is $1$ if node $i$ and edge $\alpha$ are incident, and zero
otherwise.
The definition can be easily extended to the case of higher-order interactions, in which case $\alpha$ labels the most general
type of interaction HOrS (Fig.~\ref{fig:matrix-tensorial-representations}A).
For example, in the case of hypergraphs, $n$ is the total
number of nodes while $m$ is the number of hyperedges
\citep{berge1973graphs,estrada2005complex}.
In particular, for hypergraphs allowing for a node to be represented more than once in each
hyperedge, it can be useful to weight the entries of the incidence
matrix.
In this case then the nonzero entries of the incidence matrix
would represent the number of times the vertex $i$ is present in the
relative hyperedge \citep{kaminski2019clustering}.
Notice that the incidence matrix can also be seen as the adjacency matrix of a
bipartite graph with two node sets one of size $n$ and one of size $m$
(see Sec.~\ref{sec:representations}).
In the case of simplicial complexes, the incidence matrix between nodes and simplices can be defined in the same way (Fig.~\ref{fig:matrix-tensorial-representations}B).
In the following, we will use the language of hypergraphs whenever the definitions apply to simplicial complexes as well.

Incidence matrices come in handy when it comes to characterize the various properties of HOrSs.
For instance, the \emph{degree} of a node $i$ in either a graph or a HOrS can be defined as the sum of the elements of the $i$th-row of the incidence matrix.
In a (simple) graph the column of an incidence matrix always sums to 2 as the relationships described are always between two nodes of the graph.
In a hypergraph (simplicial complex), however, the rows of the matrix can have more than two non-zero elements as each hyperedge (simplex) can describe interactions among more than two vertices.
The sum of the elements of the columns of the incidence matrix define the size sequence of the hyperedges (simplices) of the system.
These two local measures, the degree of the nodes and the size of the
hyperedges, are the first measures one can use to study the properties of HOrSs.

\subsubsection{Adjacency matrix}
\label{subsec:adjacency}
\begin{figure*}
\centering
\includegraphics[width=\textwidth]{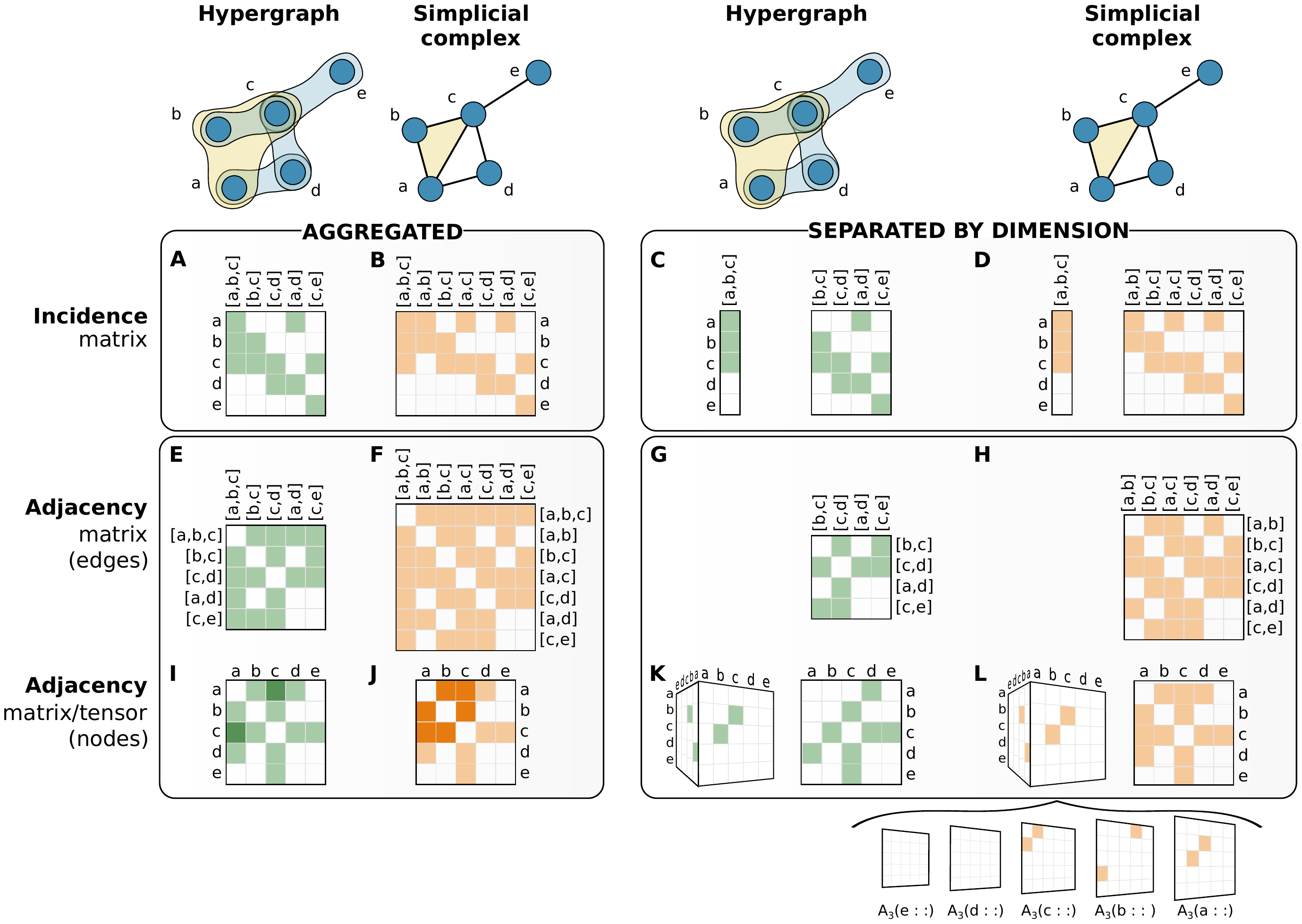}
\caption{\textbf{Matricial and tensorial descriptions of HOrSs}:
  Visualization of incidence matrices and adjacency matrices
  that can be used to represent the structure of HOrSs.
  There are
  three types of matrices: (A-D) incidence matrices relating nodes
  and edges, (E-H) adjacency matrices representing the
  connectivity of edges to edges via the nodes they share in common, and (I-L) adjacency matrices relating nodes to nodes
  via edges.
   Furthermore, one can consider edges aggregated by dimensions (left panels) or only subsets of edges of the same
  dimension, obtaining a collection of matrices, one for each of the
  different sizes of hyperedges present in the HOrS (right panels).
}
\label{fig:matrix-tensorial-representations}
\end{figure*}

From the incidence matrix of a graph we can also construct another
matrix that fully encodes the connectivity of the graph, the
\emph{adjacency matrix} $A$.
Since the matrix product $I\cdot I^T$ is a $n\times n$ matrix whose $i,j$ element is the number of columns of the incidence matrix $I$ that contain both vertices $i$ and $j$, while $i,i$ gives the degree of node $i$, the adjacency matrix of a simple graph can be defined as:
\begin{equation}
  A=II^T-D
  \label{aid}
\end{equation}
where $D$ is the diagonal matrix whose diagonal entries are the nodes
degrees.
The adjacency matrix $A$ is $0$ along the diagonal, while for $i\neq j$ the entry $a_{ij}=1$ iff nodes $i$ and $j$ are adjacent, that is, there
exists an edge connecting them.
We can generalize the notion of adjacency matrix to the case of HOrSs by using the same expression in
Eq.~\ref{aid} and considering as $D$ the diagonal matrix whose diagonal entries are the number of hyperedges a vertex belongs to.
While for simple graphs there can be at most one edge connecting a pair of nodes $i$ and $j$, for HOrSs there can be more than one hyperedge $\alpha$ containing the two nodes.
The adjacency matrix of a HOrS is then a $n\times n$ matrix whose elements $a_{ij}$ are the number of hyperedges that contain both $i$ and $j$ (Figs.~\ref{fig:matrix-tensorial-representations}I,J for hypergraphs and simplicial complexes respectively).
When the hyperedges are weighted, the adjacency matrix of a
hypergraph can be written as $A = IWI^T - D$, where $I$ is the
incidence matrix, $W$ is the diagonal matrix with the weights of the
hyperedges along the diagonal, and $D$ is a diagonal matrix with the
degrees of the nodes along the diagonal \citep{zhou2007learning}.

From the incidence matrix, one can also define the {\em intersection profile} of a HOrS as
\begin{equation}
P = I^T I,
\end{equation}
which is an $m\times m$ matrix, whose elements $P_{\alpha\beta}$ count the number of vertices in common between two hyperedges $\alpha$ and $\beta$ and $m$ is the number of hyperedges (Fig.~\ref{fig:matrix-tensorial-representations}E).
The intersection profile is useful in the statistical study of edge intersections in hypergraphs \citep{chodrow2019configuration}.
The same construction also applies to simplicial complexes (Fig.~\ref{fig:matrix-tensorial-representations}F).\\

The adjacency between two vertices can be defined directly, without any dependence on the definition of an incidence matrix.
This approach is often used when the higher-order structures cannot be
uniquely identified only by the set of nodes involved, or when there
is a theoretical need for a more restrictive notion of adjacency between two hyperedges than just that they intersect in at least two vertices.
For example, when studying motifs in a network, for each motif $M$ one can construct a $n\times n$ adjacency matrix $A_M$, where $n$ is again the number of nodes, and whose entries $a_{ij}$ are the number of times $i$ and $j$ both belong to an instance of motif $M$\citep{benson2016higher}.
Such adjacency matrix can also be seen as that of a weighted network built only of the instances of $M$.
It can be useful to notice that when $M$ is a $d$-clique, then $A_M= A_d$ is
the adjacency matrix built from the incidence matrix containing only
$d$-dimensional hyperedges. This same approach can be used to build incidence matrices representing the relationship between the nodes and HOrS (Figs.~\ref{fig:matrix-tensorial-representations}C,D).

The adjacency matrices $\{A_2, A_3,\cdots, A_d, \cdots\}$ for each $d$-dimensional hyperedge in the HOrS represent the weighted networks underlying the HOrS, and can be collected in a natural way in an adjacency \emph{tensor} of dimension $d$, indexed by the node labels (Figs.~\ref{fig:matrix-tensorial-representations}G,H).
Further insights into the structure of the hypergraphs themselves
\citep{rodriguez2003laplacian} and into the processes taking place
over them \citep{bellaachia2013random,avin2010radio} can be obtained from studying the Laplacians of these networks and their spectra, which will be introduced in Sec.~\ref{subsec:ho-laplacian}.

Another reason for building an adjacency tensor without relying on the incidence matrix is practicality.
For example, simplicial complexes require to explicitly list all $2^k$ simplices included in each $k$-simplex and this can become very impractical.
This is due to the constraint on the existence of all subsimplices of any given simplex, which in turn is fundamental for the correct construction of the useful algebraic structures that come with a simplicial complex (e.g. walks and homology, see sections
\ref{sec:measures:centrality} and \ref{subsec:measures:homology}).
To avoid listing all the simplices in a simplicial complex, one could
only list the maximal simplices \citep{young2017construction}, as mentioned in section \ref{sec:representations}.
However, while this method effectively compresses the global structure of a simplicial complex, it does not encode the relationships between the $k$-simplices in the complex and the $k+1$ and $k-1$-simplices, which are exactly the crucial ones to make the simplicial complex representation  so useful and unique among the other HOrSs.

In order to avoid this problem, one can define two $m_k\times m_k$
adjacency matrices for each dimension $k$ describing respectively an upper adjacency $A_U$ and a lower adjacency $A_L$ for all
$k$-simplices.
Here, $m_k$ is the number of $k$-simplices.
Following standard notation
\citep{goldberg2002combinatorial, maletic2008simplicial,
  duval2002shifted, serrano2019higher}, two $k$-simplices are lower
adjacent if they intersect in a $k-1$-simplex, they are upper adjacent
if they are both faces of the same $k+1$-simplex. Then
$(A_L^k)_{\alpha\beta} = 1$ only if the $k$-simplices $\alpha$ and
$\beta$ are lower adjacent, while $(A_U^k)_{\alpha\beta} =
1$ only if the $k$-simplices $\alpha$ and $\beta$ are upper
adjacent \citep{estrada2018centralities}.
Another way of defining adjacency is to construct a single adjacency
matrix $A^k$ that isolates lower adjacent interactions that are not
involved in upper ones, that is, $A^k_{\alpha\beta}=1$ only if the
$k$-simplices $\alpha$ and $\beta$ are lower adjacent but not upper
adjacent \citep{estrada2018centralities,muhammad2006control}.
Both these definitions of adjacency will be instrumental in the
study of node shortest path centrality defined on paths on $k$-dimensional simplices (Sec.~\ref{sec:measures:centrality}).

It is also possible to define an adjacency matrix which generalizes the standard one used in simple graphs, that is that an element $a_{ij}=1$ when the edge $\{i,j\}$ is present in the graph.
To generalize this idea to higher-order interactions, one needs to consider a combinatorial object $A$ indexed by all possible permutations of $\alpha$.
Then, for each order $d$, one defines an
$\overbrace{n\times n \times \dots \times n}^d$ adjacency tensor
$\mathbf{A}_d $ so that an entry $a_{i_1,\dots,i_d}=a_\alpha$
represents $d$-dimensional set of nodes participating in the
higher-order interaction $\alpha = \{i_1, \dots,i_d\}$.
This means that $a_\alpha = 1$ if the set $\alpha$ is present, while $a_\alpha = 0$ otherwise.
This definition was originally introduced in \citep{courtney2016generalized}
in the context of ensembles of simplicial complexes (Fig.~\ref{fig:matrix-tensorial-representations}L).
However, it can be easily extended to hypergraphs and other set-based HOrS (Fig.~\ref{fig:matrix-tensorial-representations}K).

\subsection{Walks, paths and centrality measures}
\label{sec:measures:centrality}
Network centralities are node-related measures that quantify how ``central'' a node is in a network.
There are many ways in which a node can be considered so: for example, it can be central if it is connected to many other nodes (degree centrality), or relatively to its connectivity to the rest of the network (path based centralities, eigenvector centrality).
In the following, we review some of the most common centralities and their possible generalizations to higher orders.

\subsubsection{Degree centralities}
\label{sec:measures:degree_centrality}
The simplest centrality measure is the degree of a vertex, which counts how many other vertices are incident to it. The degree can easily be defined from any of the adjacency matrices defined in Sec.~\ref{sec:measures:centrality} as
\begin{equation}
\deg(i) = \sum_{j=1}^n a_{ij}.
\end{equation}
Via the adjacency tensor introduced in \citep{courtney2016generalized}, one can define a comprehensive generalized degree which incorporates not only the dependecies of nodes to their higher-order counterparts, but also for any intermediate $\delta < d$-dimension.
In terms of the adjacency tensor, the generalized degree is defined as
\begin{equation}
  k_{d,\delta}(\alpha) = \sum_\alpha a_\alpha'
\end{equation}
and indicates the number of $d$-dimensional simplices $\alpha'$
that are incident on the $\delta$-dimensional simplex $\alpha$.
 For $\delta = 0$ the generalized degree reduces to the standard node degree centrality relative to $d$-simplices.
Finally, a coarser way to quantify node degree centrality in simplicial complexes is to just count the number of maximal simplices (or \emph{facets}) incident on a vertex \citep{patania2017shape}.

When working with weighted hypergraphs, it is slightly trickier to properly define a degree.
For example, \citet{kapoor2013weighted} illustrate how a node's degree centrality can be defined it terms of either its incident hyperedges or its adjacent nodes, where two nodes are considered adjacent if they belong to the same hyperedge.
The degree centrality of a node in a hypergraph becomes then defined as the number of nodes adjacent to it, and the weighted degree centrality as the sum of weights of the ties of the node with the other nodes in the hypergraph.

How to define the weight of a tie is going to be important to identify the meaning of centrality. The weight can be in the tie between two nodes as the number of hyperedges they both belong to, or on the hyperedge itself attached as a function of its multiplicity.
\citet{kapoor2013weighted} compare degree centralities relative to five different definition of hyperedge weight: constant, frequency based, Newman's strength for collaboration networks, Gao's weights for the email dataset, the probability of a contact between the two nodes over $\ell$ interactions in a group of size $k$.

Degrees in HOrSs can also be defined on hyperedges or simplices. In fact, for each hyperedge or simplex $\alpha$ we can define the $\deg(\alpha)$ as the number of hyperedges that are adjacent to $\alpha$ in some of the ways introduced earlier.
In particular, we can define a lower and upper adjacency degree in simplicial complexes \citep{jiang2007spatial,estrada2018centralities} as the number of simplices that are either lower or upper adjacent to $\alpha$, or the number of simplicies that are lower adjacent but not upper adjacent to $\alpha$.
Moreover, these adjacency definitions can be combined as done by Serrano and G\'omez in \citep{serrano2019higher}, where they introduce the maximal simplicial degree $\deg(\alpha) = \deg_A(\alpha)+\deg_U(\alpha)$, that counts the number of upper adjacent simplices to $\alpha$ and the number of lower adjacent simplices that are not upper adjacent.
Degree based centralities can be defined building on any of the above definitions and generalizing on the graph based formulas.
The interested reader can find an exhaustive review of centrality measures for HOrSs in \citep{estrada2018centralities, serrano2019centrality}.

\subsubsection{Paths and path-based centralities}
To define a centrality relative to the entire network, we need to define an acceptable way in which one can traverse a HOrS by defining walks along its connections.
A walk in a simple graph is a sequence of vertices $[v_1, v_2, \cdots, v_\ell]$ such that two consecutive vertices are $v_i, v_{i+1}$ are adjacent to each other.
A walk where a vertex is present only once is called a path.

After having introduced the concept of walks on HOrSs, then centralities can be defined using the classical definitions as either the number of paths that go through node $i$ (betweeness centrality), or the average length of shortest path between a vertex and all vertices in the graph (closeness centrality), the number of closed walks of different lengths starting and ending at the same vertex (subgraph centrality).
In the case of HOrSs, when defining paths, it is easier to consider walks connecting two hyperedges (simplices) than walks connecting two vertices.
This is because any pair of nodes present in the two extremal hyperedges (simplices) of the path will be connected by exactly the same walk.

The easiest way to define a walk in a hypergraph is as a sequence of hyperedges with at least 1 vertex  in common \citep{zhou2007learning,lu2011high}.
This definition follows from the notion of adjacency induced by the incidence matrix.
Then, the sub-hypergraph centrality of vertex $v$ is the number of closed walks of different lengths in the network starting and ending at vertex $v$
\citep{estrada2005complex, estrada2006subgraph}, which can be expressed as
\[
C_sh(v_i) = \sum_{v_j} u_{ij}^2 e_{\lambda_j}
\]
where $u_{ij}$ is the $i$th component of the $j$th eigenvector of the adjacency matrix.

\begin{figure*}
\centering
\includegraphics[width=0.95\textwidth, keepaspectratio = true]{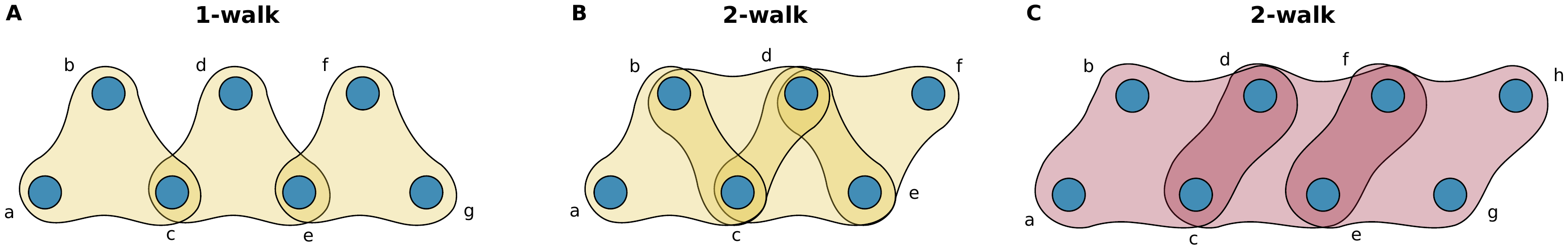}
\caption{\textbf{Example of $k$-walks on hyperedges.}
The simplest walk, a 1-walk, is the one where hyperedges share only one vertex (A) similarly to how walks are defined on graphs. Such walks can be generalised to larger intersections (k-walks), for example 2-walks (B and C). Note that the size of the intersection poses no upper bound on the size of hyperedges along the walk, for example B and C and composed hyperedges of different size while still being 2-walks.
Figures adapted from Ref.~\cite{lu2011high}.
}
\label{fig:hyperwalks}
\end{figure*}

This definition can be generalized to $k$-walks between hyperedges as a sequence of hyperedges such that each pair of successive hyperedges are adjacent and they intersect in at least $k$ vertices~\citep{maletic2008simplicial,aksoy2019hypernetwork}.
This in turn requires that all hyperedges in the walk have dimension $s$ at least $s=k+1$, but poses no constraints on their maximum dimension $s$.
Examples of two simple walks for regular hypergraphs, one 1-walk and two different 2-walks, are shown in Fig.~\ref{fig:hyperwalks}.
The corresponding closeness centrality is then the reciprocal of the average length of the shortest path between the node and all other nodes in the HOrS \citep{aksoy2019hypernetwork}.

The hypergraph definition of a $k$-walk also applies to simplicial complexes \citep{serrano2019centrality} and can be used to define other measures of betweeness and closeness centrality.
However, in simplicial complexes, it is more appropriate to define a $k$-walk only comprised of $k$-simplices that are lower adjacent i.e. have in common $k$ vertices (remember, a $k$ simplex contains $(k+1)$vertices) \citep{estrada2018centralities}.
Just as before, the simplicial closeness of a $k$-simplex is then the reciprocal of the sum of its $k$-shortest path distance to all other $k$-simplices.
The simplicial harmonic closeness centrality of a $k$-simplex is instead the sum of reciprocal $s$-shortest path distance to all other $k$-simplices.

\begin{figure*}
\centering
\includegraphics[width=.8\textwidth]{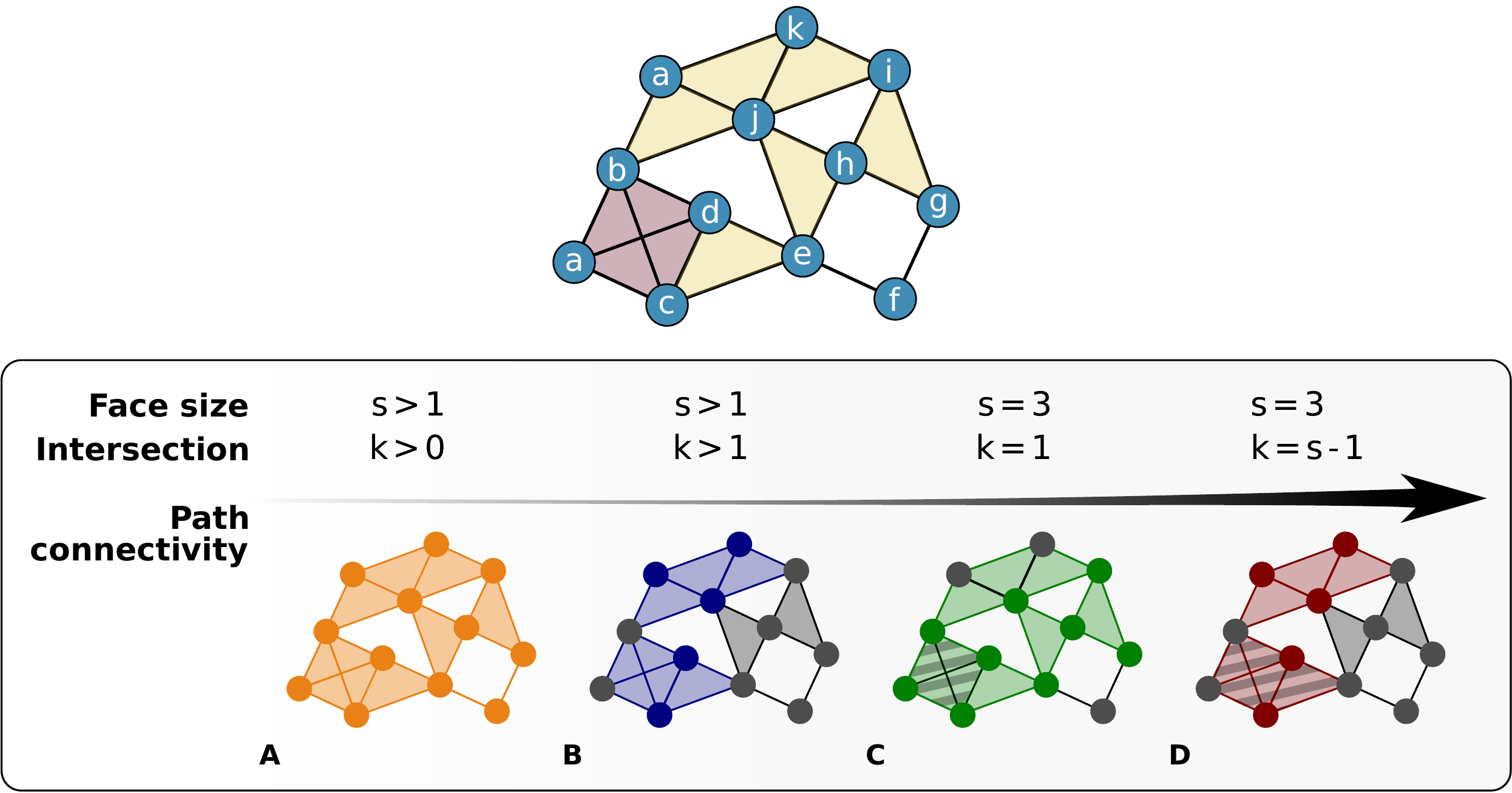}
\caption{\textbf{Walks in HOrSs.} Visualization of the different definitions of walks on an toy example of higher-order network. From left to right, the definition of the walk gets more restrictive.
For each restriction on the defining variable of a walk (what elements are allowed to be considered for a walk and how much do they need to overlap in order to be adjacent)--we color what parts of the HOrS are reachable by at least a walk of length greater than one, and in gray what parts do not allow any walk to pass through them.
In (C,D), some parts of the HOrS are connected only if the HOrS is a simplicial complex, and we visualize those with a striped pattern.
We can see how the less restrictive walk (A),  which can pass through two edges that have at least one vertex in common, yields the same connectivity as its underlying graph.
Restricting the intersection between two adjacent edges to have at least 2 vertices in common, example (B), already highlights different mesoscale connectivity patterns in the HOrS, which can be further studied introducing a further restriction on the size of the edges, examples (C) and (D).
\label{fig:walks}
}
\end{figure*}

In Fig.~\ref{fig:walks} we provide examples of how different configurations on $k$ and $s$ can yield different connectivity structures for the same HOrS.
If we allow any hyperedge or simplex dimension ($s>1$) and any size of the intersection between them ($k>0$), then we find that all paths are valid and the whole toy HOrS is connected.
In fact, we recover the simple graph connectedness (Fig.~\ref{fig:walks}A).

If we instead require that interactions share at least two vertices ($k>1$), some paths are not allowed. For example, the triangle $[2,10,12]$ is not connected to any of the triangles in the tetrahedron $[1,2,3,4]$, because the only intersection is $2$.

Note also at this point that whether the HOrS is a hypergraph or a simplicial complex can make a difference.
In Fig.~\ref{fig:walks}C, if we consider the HOrS to be a simplicial complex, the presence of the tetrahedron $[1,2,3,4]$ implies also the presence of all the subfaces.
Combining this with the requirement $k=1$, this also implies that some paths will not be walkable, e.g. all the triangles inside the tetrahedron share an edge with each other.
Hence one cannot walk from one triangle to the other (shown as the black edges).
However, there exist other paths that make the HOrS connected, for example, $[1,2,4]$ is connected to $[3,4,5]$.
Similar considerations also apply to Fig.~\ref{fig:walks}D, which shows an example of the simplicial $k$-walk described earlier in this section, that is, a walk limited to jumps between lower adjacent simplices.

\subsubsection{Eigenvector centralities}
In some applications, it is important to quantify the influence of a node on the entire network, rather than its centrality relative to possible paths.
First introduced in a sociological context by Bonacich \citep{bonacich1972factoring}, the eigenvector centrality tries to capture this effect using an iterative definition.
In fact, the eigenvector centrality a node depends on the centrality of its neighbours \citep{newman2006modularity}.
In the graph case, it can be written as
\begin{equation}
x_v = \frac{1}{\lambda} \sum_{t \sim v} x_t = \frac{1}{\lambda} \sum_{t\in G} a_{vt} x_t
\end{equation}
where $x_v$ is the eigen-centrality of node $v$ and $a_{vt}$ the adjacency matrix of the graph.
Note that again the generalization to higher order interactions relies only on the definition of connectivity and paths.
This measure has become widely used in a variety of situations ranging from Google's PageRank \citep{langville2004deeper} to neuron's firing rate \citep{fletcher2018structure}.
Interestingly, Bonacich \citep{bonacich1972factoring} also showed that, if association is defined in terms of walks, a family of centralities can be defined based on the length of walk considered.
Degree centrality counts walks of length one, while eigenvalue centrality counts walks of length infinity.
Alternative definitions of association are also reasonable.
Alpha centrality allows vertices to have an external source of influence, while Estrada's subgraph centrality proposes only counting closed paths (triangles, squares, etc) \citep{,estrada2006subgraph,estrada2018centralities}.

\citet{bonacich1991simultaneous} generalized eigenvector centrality to the case of bipartite graphs using their adjacency matrix.
A feature of this definition is that one can compute centrality score for the same eigenvalue for both node sets.
Using the same technique, one can compute eigenvalue centrality scores from incidence matrices for both hypergraphs, which will give an eigencentrality score for both vertices and hyperedges.
A two-mode analysis of an incidence matrix then enables to identify central hyperedges in addition to nodes \citep{bonacich2004hyper}.

For motifs, it is possible to use the spectral features of the weighted motif adjacency matrix $A_M$ defined in \ref{sec:measures:incidence}. For example, the clique motif eigenvector centrality score of node $i$ is given by the $i$th component of the largest real eigenvector of $W$ \citep{benson2019three}.

To incorporate non-linearities, we can make it so that the contribution of the centralities of two nodes in a 3-node hyperedge is multiplicative for the third.
To do so, one can define the centrality using the eigenvector of the tensor $\mathbf{A}_k$, where $k$ labels the considered dimensions.
There are several other types of tensor eigenvectors \citep{qi2017tensor}, and for this reason \citet{benson2019three} use $Z$- and $H$-eigenvectors which are arguably the most well-understood and commonly used tensor eigenvectors.
The $Z$-eigenvector centrality vector is then defined as any positive vector $c$ satisfying
\begin{equation}
Tc^{m - 1}=\lambda c^m \quad\| c\|_1= 1
\end{equation} for some eigenvalue $\lambda  >0$ of the adjacency tensor $T= A_k$, and respectively the $H$-eigenvector centrality vector as the positive real vector $c$ satisfying
\begin{equation}
Tc^{m - 1}=\lambda c^m
\end{equation}.

As we have seen before, for hypergraphs and simplicial complexes, adjacency can be defined not only at the vertex level, but also between hyperedges.
It is possible to introduce notions of centralities for simplices and hyperedges through the components of the principal eigenvector of $A_k$ \citep{estrada2018centralities}.
The simplicial eigenvector centrality of the $k$-simplex $\alpha$ is given by the $\alpha$ component of the principal eigenvector of $A_k$, and its simplicial Katz centrality as $K_{k,\alpha} = [\sum_p=0^\infty x^m A^m_k]_\alpha$ where $0<x<\frac 1 {\lambda_1(A_k)}$.

\subsection{Triadic closure and clustering coefficient}
\label{sec:measures:clustering}
A key concept in network analysis for going beyond node-related measures is triadic closure. It is a concept that comes from sociology \citep{granovetter1977strength}, which argues that a strong social tie between two persons can only occur if it is part of a triangle.
In other terms, my closest friends are the ones I share friends with.
In a graph structure, triadic closure is represented as 2-paths of length 2 that are closed by an third edge.
The fraction of pairs of neighbouring nodes that are themselves linked by an edge defines the node's clustering coefficient.
The clustering coefficient is an important network measure, which informs on the density of a node's neighborhood.
This coefficient can also be computed globally as the total percentage of 2-paths that are closed by and edge, i.e. are part of triangles.

This concept does not generalize well to bipartite graphs, because triangles - as any other odd cycle - do not exist in bipartite graphs.
The global clustering coefficient can however be defined through its one-mode projections, as the number of 4-paths in the bipartite graph that are part of a 6-cycle \citep{opsahl2013triadic, borgatti1997network}.

Other attempts to generalize the concept of clustering coefficient beyond pairwise relations focused on keeping its close relation to the notion of triadic closure.
One possibility is to define a local clustering coefficient from the new definitions of neighborhood that a node can have in a HOrS \citep{kartun2019beyond}.
For example, in a simplicial complex, a neighborhood can also be defined at the maximal simplex level, and can also be defined for higher order simplices, not only for nodes \citep{serrano2019centrality}.

Another possibility is to redefine the notion of paths \citep{estrada2005complex} as sequence of vertices $(v_1, \dots, v_\ell)$ such that two adjacent vertices $v_i,v_{i+1}$ both belong to the same hyperedge $e_i$.
Then, the clustering coefficient follows from its pairwise definition as the ratio of 2-paths that are closed by an edge.
In a HOrS, paths can also be defined as sequence of $k$-cliques, or $k-1$-simplices $(e_1, \dots, e_\ell)$ such that two adjacent cliques $e_i,e_{i+1}$ have $k-1$ nodes in common, which we will call $k$-paths as they are formed of $k$-cliques.
In this case too, the clustering coefficient is then defined as the fraction of $k$-paths of length $k$ that are part of a $k+1$-clique \citep{yin2017local}.

In simplicial complexes, we can distinguish between a closed $k$-path of length $k+1$ and a $k$-simplex.
Hence, the clustering coefficient can be defined as the ratio of closed $k$-paths of length $k+1$ that are closed by a $k$-simplex.
In particular, when considering triangles, this definition of the clustering coefficient can be used to verify the sociological intuition behind the diadic triadic closure idea \citep{patania2017shape}, that is, it is possible to count how many actual ``full'' triangles (2-simplices) among the possible ``empty'' triangles constructed from three edges (a closed path of three 1-simplices).
Finally, this higher-order clustering coefficient can be further generalized to motifs in weighted or growing HOrS \citep{benson2018simplicial}.

\subsection{Simplicial homology}
\label{sec:measures:homology}
One of the main reasons to use simplicial complexes as representations for higher-order datasets is a new algebraic toolset that studies the topology of the HOrS in a unique way: simplicial homology.
Homology is an algebraic topological concept that enables us to study the structure of a simplicial complex at different dimensional scales.\\
Before we can introduce homology, we need to define an algebraic structure on our simplicial complex.
This requires imposing an orientation for each simplex in the complex, formalized as the ordering of the vertices. The orientation can be arbitrarily chosen, just like the choice of node labels in a network, and it is only needed in order to coherently perform the computations.
An orientations is an equivalence class on the vertex orderings, where two orderings are equivalent if they differ by an even permutation~\citep{edelsbrunner2014short, hatcher2002algebraic}.
The orientation issue does not exist in a 0-simplex, since the nodes are not oriented, and only arises when we deal with higher order graphs.
For simplicity, and with no loss of generality, we choose the orientation induced by the ordering of the vertex labels.

\subsubsection{Boundary operators and homology groups}
\label{subsec:measures:homology}
We can combine these oriented simplices in {\it $k-$dimensional chains} $c = r_1\sigma_1 + r_2\sigma_2 + \cdots$ where $\sigma_i$ are $k$-dimensional simplices and $r_i\in \mathbb{F}$ are coefficients in a field $\mathbb{F}$. The collection of all possible $k$-chains in $X$ is the vector space $C_k=\{ r_1\sigma_1 + r_2\sigma_2 + ...| r_i\in \mathbb{F}, \sigma_i\in X_k \}$ where $X_k$ is the set of $k$-simplices in $X$.

We can also define the dual space of $C_k$, denoted as the {\it co-chain space}, $C^k$ as the linear space of all alternating functions $f: C_k \rightarrow \mathbb{R}$.
Chain and co-chain space are just two sides of the same coin, as they encode the same information.
For instance, $C^1$ can be interpreted as the space of edge-flow vectors and each of its elements $f$ assigns a scalar to an edge, representing the intensity of flow along that edge with a sign which represent the agreement or not with the chosen orientation of the edge.

We can relate the $k$-chain space $C_k$ to the $k-1$ using the boundary operator, which maps each $k$-simplex to its $k-1$-dimensional faces $\partial_k: C_k \rightarrow C_{k-1}$.
When applied on a simplex $\alpha = [v_0,...,v_{k}]$, it gives:
\begin{equation}
\partial_k([v_0,...,v_{k}]) = \sum_{i=0}^{k} (-1)^i [v_0,...v_{i-1},v_{i+1},...,v_{k}].
\end{equation}
Basically, in each term of the linear combination, we remove a vertex from the original simplex. In this way, we obtain its boundary as an alternate sum of the $k-1$-order simplices.
In a triangle $[v_0,v_1, v_2]$, for instance, we get the alternate sum of the three edges ($[v_1,v_2]-[v_0,v_2]+[v_0,v_1]$).
The image of the boundary map, $im(\partial_k)$, coincides with the space of $(k-1)$-boundaries. The kernel $\ker(\partial_k)$ is instead the space of $k$-cycles, as it is easy to prove that for every cyclic chain $c$ whose starting point coincides with the ending point $\partial_k c=0$.
Moreover, $\partial_k \circ \partial_{k+1}=0$, which implies that $im(\partial_{k+1}) \subseteq \ker(\partial_k)$. The elements of $\ker(\partial_k)$ which are not included in $im(\partial_{k+1})$ can be denoted with the quotient space
\begin{equation}
\mathcal{H}_k \equiv \frac{\ker(\partial_k)}{im(\partial_{k+1})}
\end{equation}
which takes the name of $k$-th {\it homology group}.
The elements of $\mathcal{H}_k$ correspond to the $k$-cycles that are not induced by a $k$-boundary, namely the $k$-dimensional {\it holes} of our complex~\citep{hatcher2002algebraic, edelsbrunner2014short}.\\
The dimension of the homology group $H_k$ is called the $k$-th Betti number and it represents a way to classify the $k$-dimensional topology of a HOrS.
Specifically, the $0$th Betti number represents the number of connected component in the simplicial complex, the $1st$ Betti number is the number of cycles, the $2nd$ the number of voids enclosed by $2$-dimensional simplices, $3rd$ the number of $4$-dimensional voids etc.\\

\subsubsection{Evolving simplicial complexes}
Homology is an century old concept in algebra and is one the key tools for the study and classification of shapes in mathematics \citep{edelsbrunner2014short}.
Recently the concept has been extended to weighted and growing simplicial complexes \citep{ghrist2008barcodes}.
Inspired by 90s shape theory \citep{verri1993use}, in the early 2000s persistent homology was invented in different research groups around the globe \citep{cagliari2001size,carlsson2009topology,edelsbrunner2000topological} giving birth to the field of Topological Data Analysis \citep{patania2017topological}.
\emph{Persistent homology} is a way of computing the homology of a growing simplicial complex and to follow how its homological features evolve along the filtration \citep{zomorodian2005computing}.
The filtration is a sequence of simplicial complexes that provide progressively finer approximations of the data space under investigation.
The persistence of certain homological features through the multiple scales explored in the filtration is related to their relevance for the data space, with the typical assumption that more persistent features are more important, although the exact interpretation of the persistence of a homological feature depends crucially on how the filtration is constructed (see for example \citep{feng2019persistent}).
In the last 20 years the field has been vastly developed, and new methods for tracking homological features have been introduced for cases when simplicial complexes can also lose simplices along the filtration, \emph{zig-zag homology}~\citep{carlsson2010zigzag}, or for when the growth of the simplicial complex can be described through more than one parameter \emph{multi-persistent homology}\citep{carlsson2009theory}.
The interested reader can find a good introduction to the theory and practice of persistent homology in \citep{edelsbrunner2017persistent} and \citep{otter2017roadmap}.
A large fraction of the work no HOrSs in real datasets requires some version of persistent homology. In section \ref{sec:applications} we provide some examples of its use

\subsubsection{Other measures of shape in simplicial complexes}
\label{sec:measures:other-homology}
Homology in all its variants (persistence, zigzag, multiparameter) is a powerful tool to classify structure according to key mesoscale features.
However, it is important to notice that it depends on the choice of coefficient field $\mathbb F$ used to compute the homology.
Moreover, homological invariants are not exhaustive in general, as they pertain to homological equivalence classes, thus they compress some information away. This is rooted in the invariance of topological properteis to deformations, the classical example being the topological equivalence between a mug and a donut.
Indeed, there are no complete topological classification known, and one can find examples of simplicial complexes homologically indistinguishable from a 3-sphere (a sphere in 4-dimensions) that are not spheres at all~\citep{muldoon1993topology}.
Nonetheless, the homological invariants give unique insights in the dynamics that can exist in data spaces and as mentioned above have found widespread application (see Sec.~\ref{sec:applications} for relevant examples).

In addition to the full homological description, other invariants have been used in applications \citep{adler2017estimating,pranav2019topology}.
Two commonly used ones are the Euler characteristic and the Laplacian spectral entropy.

For any simplicial complex $\Sigma$, the Euler characteristic is defined as the alternating sum $\chi = \sum_{k=0}^D (-1)^k f_k$, where $f_k$ is the number of simplices of dimension $k$ present in the simplicial complex, and $D$ is the maximal dimension of a simplex in $\Sigma$~\citep{muldoon1993topology}.

The spectral entropy, first introduced by \citep{maletic2012combinatorial} for simplicial complexes, provides a measure of the degree of the overlap between simplices in the complex via the study of the eigenvalues of the $L_k$ combinatorial laplacian.
The spectral entropy is then
\begin{equation}
H_k = - \frac 1 {\log(f_k)} \sum_{i=1}^{f_k} p(\lambda_k^i) \log(P(\lambda_k^i))
\end{equation}
where $p(\lambda_k^i) = \frac {\lambda_k^i} {\sum_i \lambda_k^i}$ is the contribution of the eigenvalue $\lambda_k^i$ to the eigenspectrum of the $k$th combinatorial laplacian $L_k$, and $f_k$ the number of simplices of dimension $k$ present in the simplicial complex.
The most general framework for not $k$-uniform hypergraphs requires a deeper analysis and the problem of generalizing a Laplacian for these structures has been addressed by many scientists ~\cite{lim2015hodge,goldberg2002combinatorial,muhammad2006control,lim2015hodge,parzanchevski2017simplicial,schaub2020random}.

\subsection{Higher-order Laplacian operators}
\label{subsec:ho-laplacian}
The Laplacian is an operator that plays a key role in information
processing of relational data, and has analogies with the Laplacian in
differential geometry. Similarly to the adjacency matrix, there is no
unique way to generalize the Laplacian to HOrSs.
However, as networks can be thought of as a special subset of the larger
family of HOrSs, the graph Laplacian is a special case belonging to the
more general family of Hodge Laplacians. The intuition here in the
construction of higher order Laplacians, either for hypergraphs or
for simplicial complexes, is that the role played by the nodes in the
graph Laplacian is, at higher orders, played by links, triangles, tetrahedra and higher dimension analogues.

In analogy with the standard construction in graphs, a straighforward way to introduce a higher order Laplacian is to define a Laplacian matrix $L$ from one of the adjacency
matrices introduced in Sec.~\ref{sec:measures:incidence}.
We can thus write:
\begin{equation}
L = D-A
\end{equation}
where $A$ is the chosen adjacency matrix, and $D$ is the diagonal
matrix with the degree sequence of the nodes along the
diagonal \citep{rodriguez2009laplacian,chung1993laplacian}.
However, this approach yields a matrix $L$ that is equivalent to the Laplacian of
the weighted graph associated to the adjacency matrix $A$ \citep{saito2018hypergraph}.

The graph Laplacian can be interpreted as a particular case of the
discrete Laplace operator representing the flux density of the
gradient flow of a function defined on the vertices of a graph.
For hypergraphs and simplicial complexes, because of their richer structure, there are
multiple ways one can define Laplacians that are compatible with the corresponding differential geometry operator.
In particular, for $k$-uniform hypergraphs and $k$-regular simplicial complexes we can define a
unique adjacency tensor $\mathbf A_k$ that represents the HOrS.
Then, one introduces a Laplacian tensor which is the discretization of the
higher order Laplace-Beltrami operator in differential geometry
\citep{cooper2012spectra,hu2015laplacian}.

This theoretical connection to the continuous operator opens the
possibility to use the spectrum of this tensor to study the diffusion
properties of the HOrS.
For example, in recent years, many authors have defined a Laplacian operator on hypergraphs for specific
diffusion processes
\citep{louis2015hypergraph,chan2018spectral,chan2019generalizing,li2018submodular}.
Another higher-order Laplacian,
designed in the context of synchronization in systems
with higher-order interactions between oscillators places on the nodes,
and introduced in Ref.~\cite{lucas2020multiorder},
is discussed in Sec.~\ref{subsec:phase_oscllators}.
In the following subsections we provide explicit definitions for
Laplacian operators on hypergraphs and simplicial complexes.
In Sec.~\ref{sec:diffusion} instead we discuss in details their
mathematical properties and their link to diffusion.

\subsubsection{Hypergraph Laplacians}
\label{subsubsec:hypergraph-laplacians}

Historically, the first attempt to generalize the Laplacian operator
to hypergraphs along these lines is due to
Chung~\cite{chung1993laplacian}, who considered a simplified type
of hypergraphs, the $s$-uniform hypergraphs, where all the hyperedges
have the same size $s$.
Given an $s$-uniform hypergraph $H$ with $N$ nodes and edge set $E$,
for each $(s-1)$-subset of nodes, $x$, we can define the degree
$d(x)$ as the number of edges involving vertices in $x$, and
the diagonal matrix $D$ such that $D(x,x)= d(x)$.
The adjacency matrix $A$ used in this case is a binary matrix such that $A(x,y)=1$ if subsets $x$ and $y$ are connected $(s-1)$-subsets that share $s-2$ nodes.
Formally, that corresponds to $A(x,y)=1$ if $x = [x_1,x_2,...,x_{s-1}]$ and $y = [y_1,x_2,...,x_{s-1}]$ and $x \cup y \in E$, and 0 otherwise.
The Laplacian can then be defined as:
\begin{equation}
\label{eq:graph-laplacian}
  L = D - A + \rho(K+(s-1)I)
\end{equation}
where $\rho = d/N$, $d$ is the average degree, and $K$ is the matrix
of the complete graph, such that $K(x,y)=1$ if $x =
[x_1,x_2,...,x_{s-1}]$ and $y =[y_1,x_2,...,x_{s-1}]$, and $0$
otherwise.

Another possibility to define a hypergraph Laplacian is to derive
the Laplacian from the transition matrix of a random walk. For instance,
\citet{lu2011high} considered {\it random $k$-walks}
($k<s$) on  $s$-uniform hypergraphs.
These are $k$-walks generated as follows (Fig.~\ref{fig:hyperwalks}).
The walker starts from the sequence of $k$ visited vertices at the initial step $x_0$ edge. At each time step, let $S$ be the set of last $s$ vertices in the sequence of visited vertices in the hypergraph $H = (V,E)$.
A random $(s-k)$-set $T$ is chosen from the neighborhood $\Gamma(S)$ of $S$ uniformly; here $\Gamma(S)$ is given by $\{T |S \cap T = \emptyset \quad \text{and} \quad S \cup T \in E(H)\}$; the vertices in $T$ are added into the sequence one by one in an arbitrary order.
The definition of Laplacian is split in two cases:
\begin{itemize}
\item For $1 \leq k \leq s/2$ the $k$-th Laplacian is defined as the Laplacian of a weighted undirected graph $G^{(k)}$ built such that a random $k$-walk on $H$ is essentially a random walk on $G^{(k)}$.
\item For $s/2 \leq k \leq s-1$ the $k$-th Laplacian is defined as the Laplacian of an Eulerian directed graph $D^{(k)}$ and the random $k$-walk on $H$ is in one-to-one correspondence to the random walk on $D^{(k)}$.
\end{itemize}
Lu and Peng \cite{lu2011high} also introduced $\alpha$-lazy random
$k$-walks, with $0\leq\alpha\leq1$, which are modified random $k$-walks
where with probability $\alpha$ the walker stays at the current edge
and with probability $1-\alpha$ it moves by appending $s-k$ vertices
to the sequence.

\subsubsection{Combinatorial Laplacians}
\label{subsubsec:combinatorial-laplacians}
For general simplicial complexes, a higher-order Laplacian can be defined
for each dimension $k$ via two matrices that encode respectively the roles of upper and lower adjacencies in dimension $k$:
\begin{equation}
 L_k = L^k_U + L^k_L
\end{equation}
where $L^k_U$ and $L^k_L$ are called the upper and lower adjacency Laplacians.
The full Laplacian $L_k$ is usually referred to as combinatorial, and each $k$-order Laplacian encodes the relationships of $k$-simplices with their adjacent $(k+1)-$ (upper adjacency) and $(k-1)$-simplices (lower adjacency).

The link between $L_k$ and the simplices of dimension $k$ with those of dimension $k+1$ and $k-1$ can be readily understood from the definition of $L^k_{U/L}$.
The linear boundary operators $\partial_k$ described in Sec.~\ref{subsec:measures:homology} can be represented as a matrix $B_k$,
whose columns represent all the $k$-dimensional simplices in the complex, and the rows the $(k-1)$-dimensional simplices.
Element $\beta ,\alpha$ of $B_k$, $(B_k)_{\beta \alpha}$, is non-zero if the $k-1$-simplex $\beta$ is a face of $k$-simplex $\alpha$, and can be $+1$ or $-1$ according to the orientation induced on $\beta$ by $\alpha$, that is the coefficient that $\beta$ has in the alternating sum $\partial_k(\alpha)$.
For each boundary operator, there exists also a {\it co-boundary operator} $\partial_k^*: C_k \rightarrow C_{k+1}$, that is, the adjoint of the boundary operator.
In matricial form, this can be represented by the transpose conjugate matrix of $B_k$, $B_k^T$. These boundary matrices define
$L^k_U$ and $L^k_L$ as:
\begin{equation}
L^k_U = B_{k+1}*B_{k+1}^T  \qquad
L^k_L = B_k^T*B_k
\end{equation}
The higher-order combinatorial Laplacian $L_k$ becomes then:
\begin{equation}
\label{eq:combinatorial-laplacian}
L_k = B_k^T*B_k + B_{k+1}*B_{k+1}^T
\end{equation}

\begin{figure}
\centering
\includegraphics[width=0.8\textwidth]{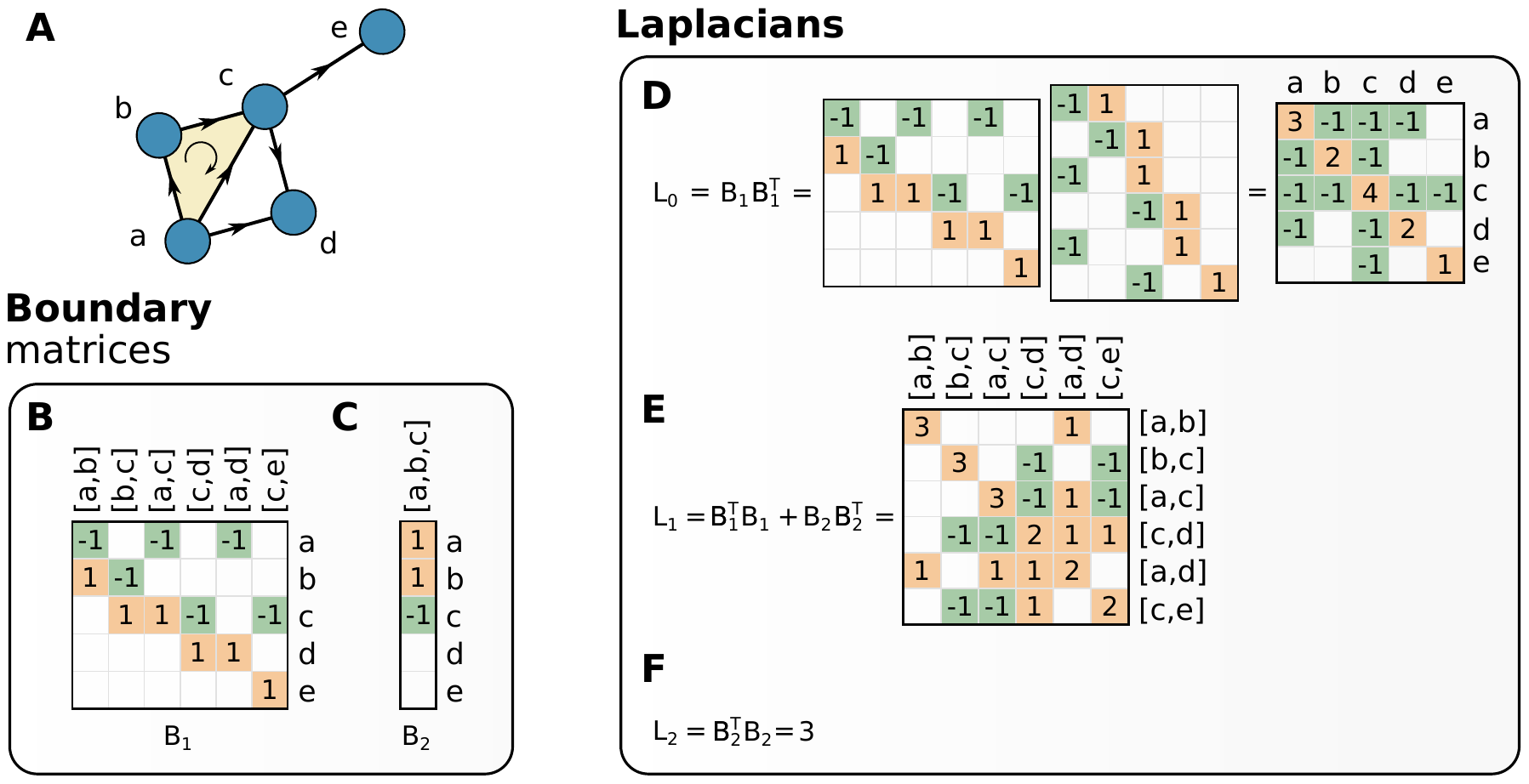}
\caption{\textbf{Construction of combinatorial Laplacian.}
The first step required to define boundary operators is to endow a simplicial complex with an orientation. Here we choose to orient our toy simplicial complex with a simple lexicographic orientation (A).
Once the orientation is fixed, it is possible to define boundary matrices; since there are only simplices with order $\leq2$, we have two non trivial boundary matrices $B_1$ (B) and $B_2$ (C).
Following Eq.~\eqref{eq:combinatorial-laplacian}, we can build the combinatorial Laplacians corresponding to the three different dimensions of simplices in the simplicial complex: $L_0$ defined on nodes, and identical to the standard graph Laplacian (D); $L_1$ defined on the edges (E); and $L_2$ which is a scalar in this case because the simplicial complex contains only one 2-simplex (F). }
\label{fig:combinatorial-laplacian}
\end{figure}

Figure \ref{fig:combinatorial-laplacian} shows the explicit construction of the combinatorial Laplacian for the simplicial complex in Fig.~\ref{fig:matrix-tensorial-representations}.
We first need to choose an orientation for the simplices in the simplicial complex.
Here for simplicity we choose the orientation to follow the lexicographic ordering on the nodes' labels (Fig.~\ref{fig:combinatorial-laplacian}A).
Once fixed this, it is possible to compute the two boundary matrices that we need in this case $B_1$ and $B_2$, corresponding respectively to the the boundary matrix mapping chains of edges into chains of nodes (Fig.~\ref{fig:combinatorial-laplacian}B), and to the boundary matrix mapping chains of triangles into chains of edges  (Fig.~\ref{fig:combinatorial-laplacian}C).
We can then easily compute the corresponding combinatorial Laplacians from Eq.~\eqref{eq:combinatorial-laplacian}, obtaining the matrices in Figs.~\ref{fig:combinatorial-laplacian}D-F.
Note that for $L_0$ and $L_2$ there is only one contribution to the sum, because the simplicial complex does not contain $k$-simplices of order $k$ respectively less than 0 and larger than 1, hence one of the two terms in the sum vanishes.

Since $L^k_{U/L}$ describe the relatedness between upper and lower adjacent simplices, they should be related to the upper and lower adjacency matrices $A^k_{U/L}$ described in section \ref{subsec:adjacency}.
It is possible to rewrite $L_k$ as \citep{muhammad2006control}:
\begin{equation}
L_k = D^k_U - A^k_U +(k+1)I_{n_k} + A^k_L
\end{equation}
where $D^k_U$ is a diagonal matrix with on the upper degree of simplices computed from $A^k_U$, and $n_k$ is the number of $k$-simplices.
Setting $k=0$, one recovers standard graph Laplacian of Eq. \ref{eq:graph-laplacian}.
In terms of boundary operators, it is easy to see also how the combinatorial Laplacian is related to the Hasse Diagram (see Sec.~\ref{sec:representations}) of the Simplicial complex.
The first term on the rhs of Eq.~\eqref{eq:combinatorial-laplacian} corresponds to moving from k-simplices down to $(k-1)$-simplices and then back up to $k$-simplices, while the second term does the opposite, first up to $(k+1)$-simplices and then back down to $k$-simplices.

The Laplacian is a crucial operator in the definition of diffusion
processes on simplicial complexes \citep{maletic2012combinatorial,
  muhammad2006control} which will be discussed thoroughly in Sec.~\ref{sec:diffusion}.
In particular, following from the Combinatorial Hodge theorem \citep{hatcher2002algebraic}, the combinatorial Laplacian decomposes the $k$-chain space $C_k$ into three subspaces: $C_k =  im(B_d)\oplus\ker(L_k)\oplus im(B_{d+1}^T)$,.
These three subspaces represent the globally acyclic, the cyclic and locally acyclic components of the flow defined on the combinatorial structure \citep{maletic2012combinatorial}, which can be used to decompose and study the evolution of dynamical processes on simplicial complexes \citep{schaub2018flow}.
Another interesting consequence comes from the fact that the kernel of the Laplacian $\ker(L_k)=\ker(B_d^T)\cap\ker(B_{d+1}) $ is isomorphic to the $k$-homology group $H_k$.
This in turn implies that the number of (homological) $k$-holes is the same as the dimension of the kernel of $L_k$, establishing a direct link between the homological and spectral representations of a complex's topology.
We leave to the interested reader to check that $L_1$ in Fig.~\ref{fig:combinatorial-laplacian}E has indeed a one-dimensional kernel corresponding to the one-dimensional hole in the simplicial complex in Fig.~\ref{fig:combinatorial-laplacian}A.
Finally, we list a few additional spectral properties of $L_k$, which can be easily checked considering subcomplexes of the toy complex in Fig.~\ref{fig:combinatorial-laplacian}A:
 \begin{itemize}
\item If the simplicial complex consists of disconnected simplicial complexes, then the spectrum of $L_k$is equal to the union of spectra of each component's $k$-th Laplacian.
\item If the simplicial complex is formed by gluing two simplicial complexes along a $k$-face, then the spectrum is the union of the two spectra.
\item If the simplicial complex only consists of one simplex of dimension $q$, then the Laplacian spectrum only has one eigenvalue, $\mu = q$ with multiplicity $q!/[(q-1-k)!(k+1)!]$~\cite{maletic2012combinatorial}.
\end{itemize}


\section{Models}
\label{sec:models} 
%
Models of HOrSs aim to reproduce, explain, and predict the structure of systems best described with interactions that involve two or more elements of the systems.
To allow for variability in their outputs, these models are often specified as collections of random rules, i.e., as stochastic processes.
Hence, they define implicit or explicit distributions over sets of HOrSs.
In what follows, we review a great many models of models for such HOrSs.

To better delineate the similarities and differences between models, we have organized them in two broad categories based on the type of stochastic process used.
In the first subsection (Sec.~\ref{subsec:equilibrium_models}), we review \emph{equilibrium models} defined as static distributions over HOrSs. 
In the second subsection (Sec.~\ref{subsec:outofequilibrium_models}), we review \emph{out-of-equilibrium models}, given as sequences of distributions over HOrSs.
Although the separation is not formally perfect (sequences sometimes converge to equilibrium distributions), these models tend to be quite different in practice, which makes this classification a natural one.
Our choice is motivated also by the fact that the philosophy underlying these models are somewhat at odd \cite{coolen2017generating}.
On the one hand, \emph{equilibrium models} are usually simple and straightforward to analyze; they typically make use of independence assumptions, which leads to distributions over HOrSs that can be written down analytically.
They are therefore well suited to making statistical inference and to acting as substrate for dynamical processes taking place on HOrSs. 
On the other hand, \emph{out-of-equilibrium models} typically lead to complicated outcomes despite their simple specifications; this makes them ideal for explaining how the qualitative properties of real systems can emerge from simple rules.
We note, however, that this classification can sometimes be superficial, especially when there are formal correspondence between equilibrium and growing formulation of models \citep{krioukov2013duality}.

Under these two principal headings, we have also organized models by the representation in which they are expressed (see Sec.~\ref{sec:representations} for an overview).
We have found this separation useful because different threads of the literature have historically favored a single representation, such that models given in the same representation tend to both rest on similar assumptions, and have similar modeling objectives.
At the same time, we recognize that from a strictly mathematical point of view, the choice of representation is again superficial due to the formal correspondence between HOrS representations \cite{aksoy2019hypernetwork}.
Hence, we highlight formal equivalence whenever they exist.

On a final introductory note, we have limited the scope of this review by focusing on models where higher-order interactions appear as ``first-class citizen.''  
As a result, we have excluded network models that happen to generate  higher-order interactions as a byproduct, such as  models of non-trivial clustering \cite{bianconi2003number,bianconi2005loops,serrano2005tuning} or cliques   \cite{bollobas1976cliques,bianconi2006emergence} in networks.

\subsection{Equilibrium models}
\label{subsec:equilibrium_models} 

\subsubsection{Bipartite models}
\label{subsec:models:equilibrium:bipartite}


We begin our review of equilibrium models with the bipartite configuration model (bipartite CM).
It is perhaps the best-known example of an equilibrium model described in the bipartite network representation.
The bipartite CM generates HOrSs where both the size of the interactions and the number of interactions per element can be controlled, allowing one to investigate the impact of these quantities on the higher-order structure.

There are a number of variations on the theme of the bipartite CM~\cite{fosdick2018configuring}.
In general, it is defined as some form of maximally random distribution over all bipartite networks that have a fixed degree sequence or distribution (either on average or exactly).
For example, in one the version of the model, the degrees are fixed exactly: one provides two sequences of degrees $\bm{k}^{(A)}=(k_1^{(A)},...k_m^{(A)})$ and $\bm{k}^{(B)}=(k_1^{(B)},...k_m^{(B)})$, one for the $m$ nodes  in set $A$ and one for the $n$ nodes in set $B$, but all the other properties are randomized.
The probability of a bipartite graph $G$, according to this model, is thus
\begin{equation}
    \label{eq:models:microcanonical_biCM}
    P(G) = \frac{1}{|\Omega(\bm{k}^{(A)}, \bm{k}^{(B)})|}
\end{equation}
for all bipartite networks $G$ in the set $\Omega(\bm{k}^{(A)}, \bm{k}^{(B)})$ of bipartite networks with these degree sequences, and it is 0 for networks outside of this set.
This version is sometimes called the microcanonical bipartite CM, or bipartite CM with hard constraints \cite{coolen2017generating}.
The canonical (or soft constraint) version of the model only fixes the \emph{expected} degree of nodes \cite{newman2001random}; see \citet{fosdick2018configuring} for a rigorous discussion of the various versions of the model.

Early references to the bipartite configuration model first appeared in ecology \cite{diamond1975assembly,connor1979assembly}.
Mathematically equivalent objects---contingency table with fixed row and column sums---were also studied in statistics around the same time \cite{gail1977counting,verbeek1985survey}.
Network science and physics innovations in this area have included thus far: the use of probability generating function to calculate properties of the model \cite{newman2001random}; the introduction of statistical mechanics tools (grand-canonical ensembles) to analyze the model \cite{saracco2015randomizing,payrato2019breaking}; hidden variable formalisms \cite{kitsak2011hidden}; and the addition of fixed degree-degree correlations to the model \cite{boroojeni2017generating}.
Direct generalizations of the bipartite configuration model have also been proposed, focusing on networks that have more than two ``parts.''
Although it is typically not their explicit goal, these generalizations allow one to model $k-1$ simultaneous \emph{types} of  higher-order interactions, where $k$ is the number of parts \cite{soderberg2002general,allard2009heterogeneous} (see also \citet{boccaletti2014structure} for a review).
They collapse back to the bipartite case when $k=2$.

The bipartite configuration model offers a paradigmatic example of how models of HOrSs are used to test hypotheses, via a technique known as ``null modeling.''
The general idea behind null modeling is to generate maximally random HOrSs with a few fixed properties matching that of an observed empirical system. 
If one observes that some other unrelated properties are systematically reproduced in the randomized ensemble, then these properties are, in a way, ``explained'' by the fixed properties.
Hence, null modeling can help identify connections between the property of HOrSs in empirical studies.

In the context of the bipartite CM, the fixed property is the degree sequence of the two node sets.
By applying this technique to bipartite networks of species co-occurrences, for example, it has been shown that the ``number of sites per species'' and the ``number of species per sites'' determine the structure, so much so that one cannot conclude on whether natural assembly rules drive the network formation \cite{diamond1975assembly,connor1979assembly}.
Similar conclusions have been reached with the same method for bipartite networks of plant and pollinators species \cite{payrato2019breaking}.
A variation on the bipartite configuration model has also been shown to reproduce most of the properties of a network of questions and tags built from the \emph{stackoverflow} knowledge database, again showing that degrees can ``explain'' many network properties \cite{fu2019modeling}. 
We note, however, that slightly changing the model can alter the conclusions of network significance analyses a great deal \cite{chodrow2019configuration,fosdick2018configuring}---one should carefully consider the modeling assumptions. \\

The bipartite CM is a special case of a much more general set of random  models, known as exponential random graphs model (ERGM), or logit models \cite{wasserman1996logit}.
These models aim to generate networks in which one controls the relative frequency of arbitrary small subgraphs (motifs).
As we have seen in Sec.~\ref{sec:representations}, these motifs can be used to encode higher-order interactions explicitly---and indeed we discuss versions of the ERGM that use motifs for this explicit purpose in Sec.~\ref{subsec:models:equilibrium:motifs} below.
For now, however, we only focus on those versions of the ERGM that add motifs \emph{within} the framework of bipartite networks.
In other words, we focus on ERGMs that encode higher-order interactions with two node sets, but that also use bipartite motifs to generate a richer distribution over HOrSs.

Formally, one defines a bipartite ERGM by choosing $Q_{\mu}(G)$, the number of times that motif $\mu=1,...,K$  occurs in $G$.
In the context of bipartite graphs, $Q_1$ could refer to the number of isolated pairs of nodes, $Q_2$ to the number of paths of length four, and so on.
The choice of motif set is, to a large extent, arbitrary, although it is usually influenced by sampling and identifiability considerations \cite{snijders2006new}.
An ERGM then assigns the probability
\begin{equation}
    \label{eq:model:ergm}
    P(G|\bm{Q},\bm{\lambda}) = \frac{1}{Z(\bm{\lambda})} e^{\sum_{\mu} \lambda_{\mu}Q_{\mu}(G)} \qquad Z(\bm{\lambda}) = \sum_{G} e^{\sum_{\mu} \lambda_{\mu}Q_{\mu}(G)}
\end{equation}
to the bipartite network $G$, where $\bm{\lambda}=(\lambda_{1},...,\lambda_{K})$ is a set of parameters that control the expected number of motifs of each type in $G$.

It is in the \emph{unipartite} network context that exponential random graphs first appeared  \cite{frank1986markov,holland1981exponential} (see our discussion below in Sec.~\ref{subsec:models:equilibrium:motifs}).
That said, there have since been quite a few explicit treatments of the bipartite case, coming mainly from social network literature where bipartite networks are known as ``two-mode data'' \cite{latapy2008basic}.
A few early models made use of small collections of local motifs, including a model that fixes the number of connected pairs (with one node of each type at the ends) and the number of ``two-stars'' (number of times the pattern $v_1$--$w_1$--$v_2$ is found, where $v_i$ is a node in part $A$ and $w_j$ is a node in part $B$) \cite{iacobucci1990social,skvoretz1999logit}.
More complicated models controlling the distribution of short paths \cite{robins2004small} or that add annotations \cite{agneessens2004choices} soon followed.

It was eventually realized that sampling from ERGMs (both in the bipartite and the unipartite network context) could be extremely challenging due to a \emph{degeneracy problem} \cite{strauss1986general,handcock2003statistical}, where the model place a high weight on empty and fully connected networks only.
This realization prompted the development of subtle physics-inspired sampling methods for specific ERGMs specifications \cite{fischer2015sampling}, and of entirely new specifications for the unipartite model \cite{snijders2006new}.
Analogs to the new specifications were also proposed \cite{wang2009exponential,wang2013exponential} for the bipartite model, leading to improved inference.

Notwithstanding these problems, ERGMs have found many inferential applications.
One popular use-case is generating typical network instances that match a set of motif counts $\bm{Q}$ obtained via empirical surveys \cite{robins2007introduction,smith2012macrostructure}.
Bipartite networks generated in this way have been used to quantify the significance of specific patterns of interactions found in political networks in the US \cite{jasny2012baseline}, and in Russia during Brezhnev’s era \cite{faust2002scaling}.
Another popular use-case of ERGMs is to ``invert the direction'' of inference and consider $P(G|\bm{Q},\bm{\lambda})$ as a \emph{likelihood} that can be used to deduce $\bm{\lambda}$ from a fully observed network $G$ with counts $\bm{Q}$ \cite{wang2009exponential}.
These parameters can then inform us about the presence and strength of structural effects in the network data \cite{robins2007introduction}.
It should be noted, however, that the inference is not necessarily robust since the problems facing ERGMs, like degeneracy, can also affect inference  \cite{snijders2002markov}.
Additionally, it has been recently argued that if we consider---as we probably ought to---that network data we have are usually sub-samples of a larger underlying network, then ERGMs can lead to misleading inferences  \cite{shalizi2013consistency,crane2018probabilistic}. 
\\

The two sets of  models above---bipartite CMs and ERGMs---reproduce \emph{local} features like the number of the neighbor of a node or the number of short loops.
A different set of equilibrium models that instead focus on reproducing their  \emph{mesoscale} features  \cite{young2018universality}.
These models control the frequency and organization of large patterns of connections involving many of, but not all of, the nodes in a network \cite{newman2012communities}.
Examples include disjoint communities of nodes, disassortatively mixing groups \cite{newman2003mixing}, and separations in a core and a periphery \cite{borgatti2000models}.

These models are most often formulated as extensions of the well-known stochastic block model (SBM) \cite{holland1983stochastic} to the bipartite case \cite{doreian2004generalized,rohe2016co}.
The general idea is to split the node into two sets of $K_A$ and $K_B$ ``blocks''---one set for each part of the network---and to then connect the nodes randomly, with probabilities $\bm{\omega}$ that depend on their respective blocks.

One simple instantiation of this idea goes as follows \cite{larremore2014efficiently}.
We denote by  $\bm{g}^{(A)}=(g_1^{(A)},..., g_m^{(A)})$ and $\bm{g}^{(A)}=(g_1^{(B)},..., g_n^{(B)})$ the blocks of the nodes in each part (where $g_i^{(A)}=\ell$ means that node $i$ of part $A$ belongs to block $\ell\in\{1,...,K_A\}$).
We then write the probability of a particular  graph $G$ with incidence matrix $\bm{B}=[b_{ij}]$ as
\begin{equation}
    \label{eq:models:canonical_biSBM}
    P(G|\bm{\omega}, \bm{g}^{(A)},\bm{g}^{(B)}) = \prod_{i\in A}\prod_{j\in B} (1 - \omega_{g_i^{(A)}g_j^{(B)}})^{1-b_{ij}}(\omega_{g_i^{(A)}g_j^{(B)}})^{b_{ij}},
\end{equation}
meaning that an edge is placed between $i$ and $j$ with probability $\omega_{g_i^{(A)}g_j^{(B)}}$.
By varying the connection probably matrix $\bm{\omega}$ and the blocks, one can generate networks with arbitrary mesoscopic structure, ultimately approximating all systems to arbitrary accuracy \cite{olhede2014network}.

\begin{figure*}
    \centering
	\includegraphics[width=\textwidth, keepaspectratio = true]{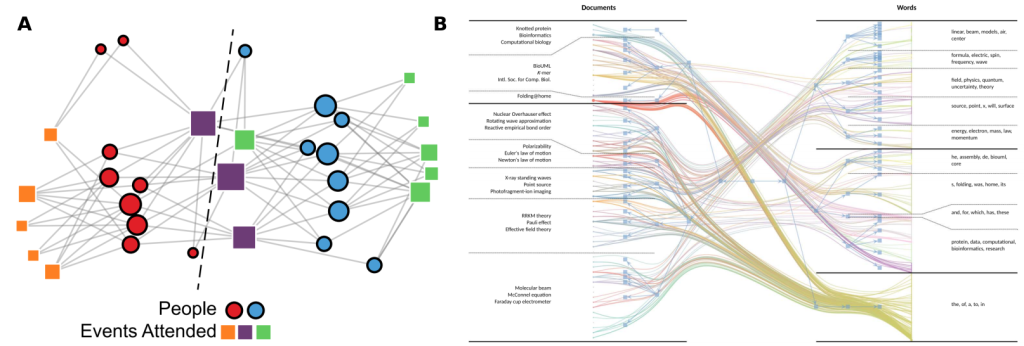}
    \caption{\textbf{Inference with the bipartite stochastic block model.} 
    (A) The data represents people (as circle) interacting through events (as squares).
    A variation on Eq.~\eqref{eq:models:canonical_biSBM} assigns a likelihood to every joint partitions of the people and events.  A high likelihood partition is shown here using colors.  Simpler methods incorrectly split the network along the black line. Figure reproduced from Ref.~\cite{larremore2014efficiently}. (B) An  elaborate hierarchical version of the bipartite SBM, applied to collections of words interacting via texts.  This time the blocks (indicated by vertical dotted lines) are themselves regrouped in high-level blocks, hierarchically. Figure reproduced from Ref.~\cite{gerlach2018network}.}
    \label{fig:models:bisbm}
\end{figure*}

Equation~\eqref{eq:models:canonical_biSBM} refers to a canonical or soft constraint bipartite SBM since the number of neighbors of a node is fixed only on average by $\bm{\omega}$ and the blocks.
Like the bipartite CM, the bipartite SBM, too, comes in many variants.
One can define a microcanonical version along the same lines \cite{peixoto2012entropy}, for example.
Another variant considers that edges have distinguishable types, e.g., in a user--movie bipartite network where edges represent ratings on a fixed scale \cite{guimera2012predicting}.
Yet another variant jointly models the degree sequence \emph{and} the mesoscopic structure, leading to a so-called degree-corrected bipartite SBM \cite{ball2011efficient} that, unlike the classical degree-corrected SBM, makes explicit the assumptions that there are two node types \cite{rohe2016co,larremore2014efficiently,hric2016network} (Fig.~\ref{fig:models:bisbm}A).
Finally, it is also possible to introduce a hierarchy, with blocks that are themselves grouped into larger blocks, and so on \cite{gerlach2018network} (Fig.~\ref{fig:models:bisbm}B).

The many versions of the bipartite SBM are used for what is another prototypical example of a HOrSs inference problem: latent parameter inference.
Indeed, different from the configuration models, the structure of a system does not trivially determine the SBM's parameters. 
In a configuration model, one calculates the degree of all the nodes of a given bipartite network to determine what is the associated randomized ensemble of networks.
In contrast, the SBM assigns nodes to blocks via $(\bm{g}^{(A)}, \bm{g}^{(B)})$, a piece of information that is typically not given---unless metadata is also available \cite{hric2016network}. 
This situation leads to a family of inference problems where the goal is to determine an assignment of nodes to the blocks, from a network’s structure alone (see Fig.~\ref{fig:models:bisbm} A).
This goal has been the main driving force behind the development of sophisticated model families \cite{peixoto2012entropy,larremore2014efficiently,hric2016network}, with formal equivalence to inference problems in other fields, like topic modeling \cite{ball2011efficient,blei2003latent,gerlach2018network} or data biclustering \cite{sheng2003biclustering}.\\

The bipartite SBM assigns nodes to latent \emph{discrete} categories (blocks), and adds connections to the network at random based on these categories.
A somewhat related class of models instead  assign  continuous latent \emph{positions} to nodes and creates connections with probabilities parametrized by the distance between nodes in the latent space.
Much like the case of the bipartite CM and SBM, these models take the form of natural generalizations of simple network models to the bipartite case.

The AB  random geometric graph model \cite{iyer2012percolation}, for instance, is a direct extension of the well-known random geometric graph model \cite{penrose2003random}.
In this model, two sets of nodes are first randomly embedded in a low-dimension Euclidean space.
Nodes \emph{in different sets} are then connected by edges, with a probability that depends on the distance $d_{ij}$ separating them.
The AB random geometric graph model makes use of the simplest possible connection rule: two nodes are connected if and only if  $d_{ij}<r$.
But we note that  some unipartite models consider more complicated functional form for the connection probabilities \cite{waxman1988routing} (see also the model that follows).

Another approach to latent space models builds on results coming from the network geometry literature \cite{serrano2008self}.
In these models, one embed nodes in an \emph{abstract} space of preferences (rather than a \emph{literal} space, like what is done in the AB model).
It is then possible to define general classes of models by specifying different connection rules and embedding spaces \cite{kitsak2017latent}.
The so-called  $\mathbb{S}^1\times  \mathbb{S}^1$ specification is perhaps the one that  has been analyzed the most thus far \cite{serrano2012uncovering,kitsak2017latent}.
In this version of the model, two sets of nodes are embedded uniformly at random on a circle (i.e., on the 1-sphere denoted $\mathbb{S}^1$).
The nodes are then assigned random hidden variables drawn at random from some distribution \cite{kitsak2011hidden}---denoted $\phi_i$ if node $i$ is in set $A$ and $\psi_i$ if it is in set $B$.
Every pairs of nodes $(i,j)$ in different sets are finally connected with a probability that depends on their distance $d_{ij}$ in $\mathbb{S}^1$, as well as the value of their hidden variables $\phi_i$ and $\psi_j$. 
The specific functional form analyzed in  \cite{serrano2012uncovering,kitsak2017latent}  is
\begin{equation}
    p_{ij}=\sigma\left(\frac{d_{ij}}{\mu\ \phi_i \psi_j}\right)  
\end{equation}
where  $\mu>0$ and  $\sigma$ is any integrable function with image in $[0,1]$.
The likelihood of the whole network is therefore $P(G|\bm{d},\mu,\phi,\psi) = \prod_{ij} p_{ij}^{b_{ij}}(1 - p_{ij})^{1-b_{ij}}$ where $b_{ij}$ is an entry of the incidence matrix.
The role of the hidden variable is to allow for variations in the degrees \cite{kitsak2011hidden}, while the embedding helps control the level of clustering  \cite{krioukov2016clustering}.

The main use-case for latent space models is, again, inference.
With a latent space model, one can take a real bipartite network as input, and find the embedding in a latent space that best matches the network, using the likelihood $P(G|\bm{d},\mu,\phi,\psi)$ to guide the search.
This inference technique has been used to, for instance, infer the latent geometry of bipartite networks of metabolites and of the reactions they intervene in \cite{serrano2012uncovering}.

\subsubsection{Motifs models}
\label{subsec:models:equilibrium:motifs}
Motifs-based models are formulated as assembly rules for  arbitrary collections of small graphs, like triangles, short-loops, or cliques.
They can be viewed as high-order models because they build systems from relationships that are not strictly pairwise, even though these models are ultimately defined as distributions over classical networks.
Motifs-based models have a rich history going back to the early days of network science \cite{newman2003properties}, preceded by work in sociology \cite{davis1967structure,holland1976local} and statistics \cite{frank1986markov}.\\

The first motifs-based models of networks appeared in sociometry, motivated by the need for survey methods that would categorize and quantify patterns of interaction among small subsets of individuals of a larger directed social network \cite{davis1967structure,holland1976local}.
As we have mentioned above in Sec.~\ref{subsec:models:equilibrium:bipartite}, Markov random graphs \cite{holland1981exponential,frank1986markov} and their exponential random graph model (ERGM) generalization \cite{wasserman1996logit}  have been proposed early on to fulfill this role.
In a nutshell, these model describe maximally random distributions over networks $G$ whose properties $\bm{Q}(G)=(Q_1(G),...,G_K(G))$ are fixed on average, in relative proportion controlled by the free parameters $\bm{\lambda}=(\lambda_1,...,\lambda_K)$.  
These models are therefore quite general in what they can describe
 (see our discussion above and Eq.~\eqref{eq:model:ergm}).

To model higher-order interactions in the ERGM framework, one can select properties $\bm{Q}$ that measure these interactions directly.
An example of a set of statistics could be: the number of edges $Q_1$; the number of triangles $Q_2$ (a closed three-way interaction); and the number of open triangle $Q_3$  (an open three-way interaction).

The main inferential applications of the generic ERGM are the same as that of the bipartite ERGM: null modeling, and the construction of whole network description from local surveys \cite{robins2007introduction}.
There is also at least one further application of ERGMs specific to unipartite networks:  quantifying the significance of motifs.
The idea here is to fix the distribution of all small motifs, and compare the  number of \emph{larger} motifs to the expected motifs counts in the ensemble \cite{milo2002network,shen2002network}.
This method identifies motifs that are either under- or over- represented in a graph, based on what we expect from smaller connection patterns.
The method has used to argue that some small motifs are ``significant subunits'' that determine the function of the modeled system \cite{milo2002network}.\\

Another type of model of networks with motifs comes from the physics literature on spreading processes occurring on clustered networks.
These models tend to be very flexible and to reproduce quite a few structural characteristics of real systems.
They have been first and foremost used to study how structural changes affect the outcome of dynamical processes unfolding on these networks.
 
\begin{figure*}
    \centering
    \includegraphics[width=0.55\linewidth]{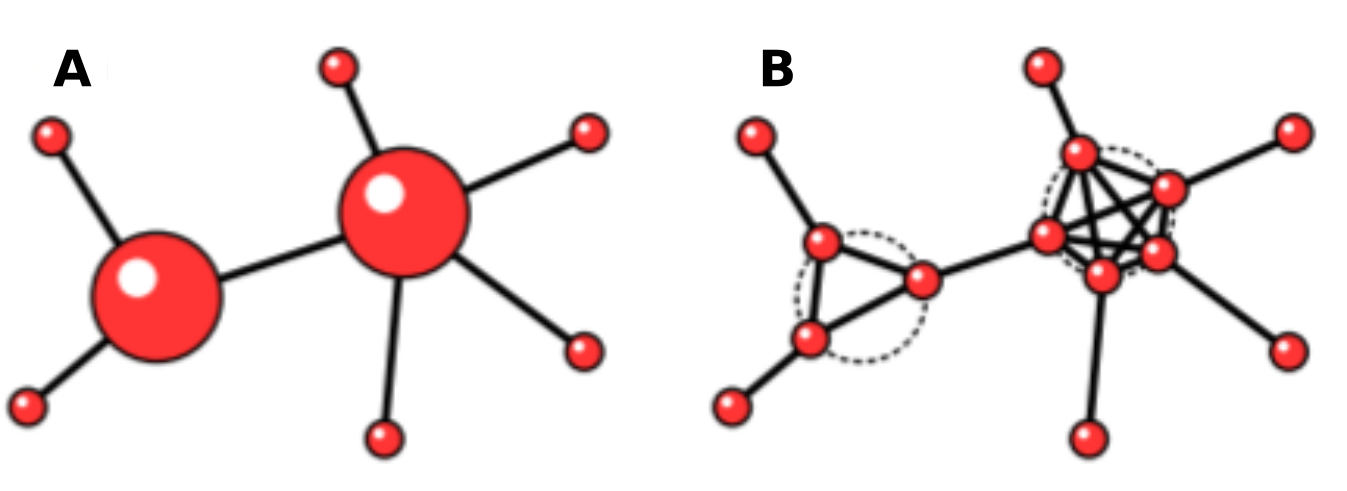}
    \caption{\textbf{Example of HOrSs generated by the model of Gleeson and Melnik \cite{gleeson2009analytical}.} In this model a fraction $g_k$ of nodes of degree $k$ of a configuration model network \cite{newman2001random} are replaced by cliques.  Figure reproduced from Ref.~\cite{gleeson2009analytical}.}
    \label{fig:models:cliquy_cm}
\end{figure*}

One of the earliest model in this category \cite{gleeson2009analytical} begins with a unipartite configuration model network, i.e., a random network that follows some arbitrary degree distribution.
After the initial network is constructed, one replaces a fraction $g_k$ of the nodes of degree $k$ with $k$-cliques for all $k\geq 3$---selecting these nodes uniformly at random.
The nodes of the new cliques are then attached to one of the edges of the node they replace, which in turn leads to a network  with the same high-level structure as the original, but with added local clustering (Fig.~\ref{fig:models:cliquy_cm}).
An earlier variant of this model also exists where one controls $g_k$ indirectly \cite{trapman2007analytical}.

Another set of model also uses the configuration model to generate HOrSs, but perhaps more directly.
In the first model in this line of work, one not only specifies the degrees $\bm{k}=(k_1,...,k_n)$ that a node should have, but also the number of triangles in which it should participate $\bm{\Delta}=(\Delta_1,...,\Delta_n)$ \cite{newman2009random,miller2009percolation}. 
A generalization of the classical stub-matching scheme \cite{fosdick2018configuring} is then used to create random network respecting these sequences.
Recall that in the classical stub-matching scheme, we first assigns $k_1,k_2,...,k_N$ ``stubs''---half-edges---to nodes $1,2,...,N$.
We then picks a random matching of the all the stubs, in which all stubs are joined in pair to form full edges. The resulting random network respects the degree sequences $\bm{k}=(k_1,...,k_N)$ by construction.
In the higher order stup-matching scheme, we should think of a node $i$ as having $k_i$ ``edge stubs'' and $\Delta_i$ `triangle stubs'  attached to it.
A network is then obtained by matching the stubs of the all nodes: 2 partial edges form an edge, and 3 partial triangles form a triangle.
One variation on this models generalizes from triangles to generic cliques of size $c>3$ \cite{gleeson2009bond}, but enforces the constraint that nodes belong to only one clique \cite{gleeson2009analytical}, i.e., that they have a single triangle or more generally a single clique of size $c\geq 3$ attached to them.
The central quantity in that model is then $\gamma(k,c)$, the probability that a node has  degree $k$ and belongs to a clique of $c$ nodes, with $k\geq c-1$.

Encompassing both the triangle and the single clique model is a generalization of the configuration model that allows for generic distribution of motifs in the neighborhood of a node \cite{karrer2010random}.
The model parameters are the number of times a motif is attached to a node, and in which way.
This can be represented, again, by "stubs"  stemming from each node, where a stub of type $(\mu,p)$ is attached to node $i$ when it participates in motif $\mu$ in position $p$.
The final graph is constructed by matching stubs of a same type to construct motifs.
Specifically, say motif $\mu$ is comprised of $n_1$ nodes in role $p_1$, $n_2$ nodes in roles $p_2$, etc.
Then one must pick $n_1+n_2+...$ stubs of the correct types at random and create a motif connecting the nodes from which the stub stem.
An even more general version of the above model assigns \emph{types} (or colors) to nodes \cite{allard2012bond,allard2015general}.
The expected number of stubs of each type for a node then depends on this type. 

Inference with these general models is challenging, because it is not clear how one should select a meaningful set of motifs to describe a given graph  \cite{karrer2010random}.
So far, the only proposed approach aiming to make this type of inference  relies on an information theoretic approach to ``subgraph covers'' \cite{wegner2014subgraph}.\\

There are a few other specifications of network models with motifs that do not fit squarely in any of the above categories.\\

One generalizes the notion of ``graphon,'' a form of latent space model in which nodes are assigned random positions $\bm{x}$ on $[0,1]$ and pairs of nodes are connected with probability parametrized by these positions.
The motif generalization \cite{bollobas2011sparse} also assigns latent positions $\{x_i\}_{i=1}^N$ to the nodes, drawn independently and identically according to some distribution on $[0,1]$.
Then, a motif $\mu$ on $r$ nodes $v_1,...,v_r$ in the collection of motifs $\bm{\mu}$ is added to the network with probability
\begin{equation}
    P(\mu| v_1,...,v_r) = \frac{\kappa_{\mu}(v_1,....,v_r)}{n^{r-1}}
\end{equation}
where $\kappa_{\mu}$ is a function of the latent positions.
Although these generalized graphons do not appear to have been used for inferential purposes, a few special cases of them (obtained by setting $r=2$ and only allowing edges) found extensive use.
For example, one can approximate $\kappa_{\mu}$ to identify the latent geometry likely to have generated a network \cite{newman2015generalized}.\\

A completely different motif mode known as the d$k$-series approach fixes local motifs and maximizes randomness otherwise \cite{mahadevan2006systematic,orsini2015quantifying}.
Formally, for a given $k$, one fixes the distribution of the number of all the motifs of $k'\leq k$ nodes centered on a node.
Hence with $k=1$, for example, one fixes the degree distribution, while with $k=2$ one fixes the joint degree distribution of pairs of nodes.
The d$k$-series model crosses into high-order specifications when $k>2$; for example when $k=3$, one fixes the distribution of wedges and triangles.
This model find use in quantifying the randomness of a network's structure \cite{orsini2015quantifying}. It fixes a null distribution where many local aspects are preserved, and helps to see whether these local constraints are enough to ``explain'' the observed large scale structure of a network.
Actually carrying out the test for any $k\geq 3$ is difficult, however \cite{orsini2015quantifying}, since constructing a single example of graphs realizing a series with $k\geq 3$ is \textsc{NP}-complete \cite{devanny2016computational}.
We note that there is related work in the social network literature, in the form of models where one specifies the local structure---the ``social neighborhood'' of nodes \cite{pattison2002}---up to a certain distance \cite{wang2013exponential}.

\subsubsection{Stochastic set models}
\label{subsec:models:equilibrium:sets}
Collections of motifs  become rapidly intractable with growing motif sizes since there are $2^{\binom{r}{2}}$ possible undirected graphs on $r$ nodes.
Many models avoid this exponential blowup by specifying the motifs \emph{stochastically}, i.e., by assigning nodes to sets of $r$ nodes and then specifying the actual motif connecting them at random.\\

\begin{figure}
    \centering
    \includegraphics[scale=0.49]{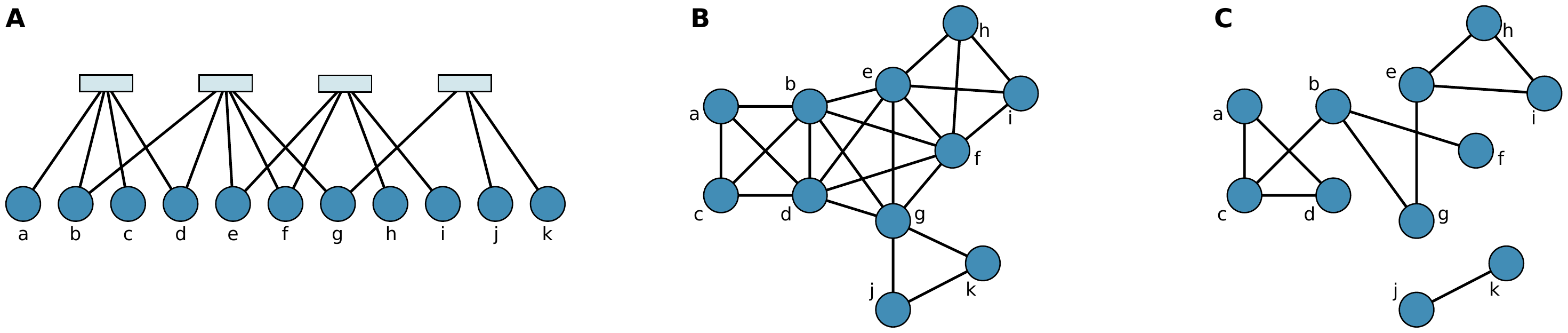}
    \caption{\textbf{Constructing random HOrSs by projecting a bipartite network of interactions \cite{newman2003mixing}.} 
    (A) Realization of the bipartite configuration model, (B) projected as a network, where (C) some edges are removed.  The construction shown in panels (A) and (B) is also known as a random intersection graph \cite{karonski1999random}. Figures adapted from Ref.~\cite{newman2003mixing}.}
    \label{fig:models:biparte_projected}
\end{figure}

A well-known family of models specified in terms of sets takes inspiration of hierarchies in social organizations, in which nodes are assigned to nested groups \cite{watts2002identity,newman2003properties}.
The groups are therefore proxy for higher-order interactions.
The simplest example of a model in this class only allows for one set of groups \cite{newman2003properties}.
In this model, a node $i$ has a membership number $k_i^{(n)}$ (number of groups it belongs to), and a group $j$ has a size $k_j^{(g)}$ (the number of nodes it has).
The rest of the structure is randomized.
Hence, in other words the node--group relationships are described by a bipartite CM \cite{newman2001random}.
Differently from the latter, however, one considers a \emph{projection} onto the nodes (see Sec.~\ref{sec:representations}), in which edges are placed at random based on the group assignments.
Specifically, one determines whether two nodes are acquainted through a group, with probability $0\leq q\leq1$, and connects them in the projection if they are acquainted through at least one group (see Fig.~\ref{fig:models:biparte_projected}).

Among the many possible generalizations of this model, some have: considered a deeper hierarchy of groups \cite{watts2002identity}; include heterogeneities by introducing a group dependent connection probability $q_r$ \cite{yang2012community}; introduce the possibility of forming a few `random' links with nodes not in the immediate group of a node \cite{hebert2010propagation}; or used a mixture of groups \cite{seshadhri2012community}.\\

An alternative point of view on these models comes from the mathematics literature, where they are known as random \emph{intersection graphs} \cite{erdos1966representation}.
A random intersection graph is formed by assigning a set on some alphabet---say $Y=1,..,m$---to each node, and connecting two nodes if their respective sets intersect \cite{karonski1999random}.
They are formally equivalent to the bipartite projection model above, since we can think of the set of a node as the higher-order interactions in which it participates, and of edges as arising because of these interactions (Fig.~\ref{fig:models:biparte_projected}).

Many models of random intersections graph exist, but are somewhat less general than the projected bipartite CM discussed above.
This is  because mathematically exact results are typically the goal of the authors studying these models, rather than coming up with models of  ``realistic'' HOrSs.
An example of one such model is dubbed $G(n,m,p)$.
It defines a distribution over HOrSs that have $m$ sets and $n$ nodes, and where every node is included independently in each set with probability $p$ \cite{karonski1999random}, yielding the likelihood
\begin{equation}
    P(\bm{B}|m,n,p) = \prod_{i=1}^n \prod_{j=1}^m (1 -p)^{{B_{ij}}}p^{{B_{ij}}},
\end{equation}
for the assignments $\bm{B}$ of nodes to sets, where $B_{ij}=1$ if node $i$ is in clique $j$ and 0 otherwise.
This model is a special case of the projected bipartite model discussed above, obtained by setting the degrees of nodes to $mp$, the size of groups to $np$, and by connecting all nodes in a shared group with probability $q=1$.
The properties of these random intersection graphs have since been studied extensively, see the detailed review of \citet{frieze2016introduction}.
Inhomogeneous generalizations  allow for some variations, such as inclusion probabilities $\bm{p}=(p_1,...,p_n)$ that are node dependent and fixed as parameters \cite{nikoletseas2008large,deijfen2009random}, or where the size of the sets are drawn from a distribution and the node in it chosen at random \cite{godehardt2003two}.

While the sets themselves are random in all of the random interaction graph models above, the \emph{interior} of sets themselves are not random---all nodes are connected and form a clique.
Other models of intersection graphs include the possibility of noise in this process \cite{davis2008clearing}, and are therefore formally equivalent to the noisy version of the node--group  models discussed above \cite{yang2012community}. 
For instance, one model assigns nodes to one or more cliques of varying sizes, and two nodes are only connected in the projection with a probability that depends on the number of shared cliques \cite{barber2012clique}.
A more exotic construction assigns nodes to the cliques with probabilities $p_i$ that are the outcome of a stochastic beta process \cite{williamson2018random}.

As we have mentioned, in the mathematical literature, random intersection graphs are studied for their structural properties \cite{karonski1999random}.
But they also find inferential application in the epidemiological literature as models of systems with higher-order interactions \cite{ball2014epidemics}.
In the statistics literature, they have found application in fitting clique-cover models to real networks \cite{williamson2018random}, in the same spirit as the cover models used to fit models of networks  with motifs \cite{wegner2014subgraph}.\\

We note in passing that some models of  overlapping communities lead to formalisms close to that of the stochastic set models mentioned here.
However, since one seldom thinks of overlapping communities as higher-order interactions---a typical community is far too big to classify as encoding an ``interaction''---we will not review them here.
The interested reader can refer to the review of Xie et al. \cite{xie2013overlapping} for an overview of models of networks with overlapping communities.

\subsubsection{Hypergraphs models}
\label{subsec:models:equilibrium:hypergraphs}

Many equilibrium models of HOrSs incorporate multi-body interactions more directly, by encoding them in hypergraphs.
Much of the work on random hypergraphs comes from the mathematical literature, where they were introduced as immediate and natural generalizations of classical models in random graph theory.

Perhaps unsurprisingly, the earliest random model of hypergraphs was an extension of the well-known Erd\H{o}s-R\'enyi (ER) model.
In the most direct generalizations of this model, every hypergraph of $m$ hyperedges of size $k$ on $n$ nodes is given the same probability \cite{de1982cardinalite}, with the ER case obtained by setting $k=2$.
The structural properties of the random hypergraphs generated in this way have been analyzed extensively, see \cite{frieze2016introduction} for a review.
A canonical variant, in which hyperedges are created independently at random with fixed probability $p$, also exists \cite{bollobas1976cliques}.

All hyperedges connect precisely $k$ nodes in the two models above.
This is an arbitrary choice and not a constraint of the hypergraph formalism.
Other uniform models do away with this constraint, and include many sizes $\mathcal{K}=\{k_1,k_2,...,k_\ell\}$ of hyperedges simultaneously (where $\mathcal{K}\subseteq\mathbb{N}$ is some choice of hyperedge sizes, with $\mathcal{K}=\{2\}$ corresponding to a graph).

One version stipulates that all hypergraphs are equiprobable, provided that they have exactly $m_{k_1}$ hyperedges of size $k_1$, $m_{k_2}$ hyperedges edges of size $k_2$, and so on for all $k\in \mathcal{K}$ (where $\bm{m}$ and $\mathcal{K}$ are chosen at deterministically \cite{schmidt1985component} or at random \cite{de2020social}).
Like in the classical ER case, hypergraphs that do not respect the constraint on the number of hyperedges are assigned a probability zero.
Yet another uniform model instead prescribes that each of the $\binom{n}{k}$ possible sets of size $k$ on $n$ nodes exists, with a probability $\lambda_k$ that depends on the size $k\subseteq \mathcal{K}$ of the set \cite{darling2005structure}.

While the specifications of these models differ from cases to cases, the underlying goal is always to study the properties of the generated hypergraphs, like their components structure, for example \cite{schmidt1985component}.
Even though these uniform hypergraph model are somewhat crude approximation of real HOrSs, they have found extensive application in technical fields like computer science, where they are used to generate the structure of idealized random decision problems (random $k$-sat) \cite{mezard2009information,dembo2008finite}.\\

Other uniform models differ from the ones above in that they ensure that the generated hypergraphs are  ``$k$-partite.''
By $k$-partite, it is meant that the $n=k \times r$ nodes of the hypergraph can be separated into $k$ disjoint subsets of $r$ nodes, such that every hyperedge comprises of precisely one node in each subset.
These $k$-partite hypergraphs are useful when one wants to encode interactions that \emph{always} involve nodes of different natures, for example, when modeling a collaborative tagging system where all hyperedges connect an element, a person, and a tag \cite{ghoshal2009random}.

One possible construction for uniform and random $k$-partite hypergraphs was introduced in the mathematical literature, with the goal of studying  ``perfect matchings'' in hypergraphs, i.e., minimal subsets of hyperedges connecting every node \cite{schmidt1983threshold,chen1996coloring}.  
The model first singles out one the subset of nodes as ``special.''
Then it stipulates that, for each node in this special subset, we should choose $d-1$ neighbors uniformly at random (one per subset) to form a hyperedge, and repeats the process $z$ times per node in the special set \cite{schmidt1983threshold}.
The resulting hypergraphs are $k$-partite by construction, and all the nodes in the special set have precisely degree $z$.
A more general but still uniform model of random $k$-partite hypergraphs, coming from the information retrieval literature, eliminates this constraint  \cite{demetrovics1998asymptotic}.
Aiming for flexibility, the model assigns a different weight to every possible hyperedges, and places the hyperedges with probability proportional to these weights.
It is shown that, within this framework, under entropy maximization constraints, one does not need all these weights: all the  probabilities are identical under the so-called ``uniform random data base model''  \cite{demetrovics1998asymptotic}.\\


Much like their graphical counterparts, the uniform hypergraph models also admit generalizations to cases where one controls the \emph{degree} of nodes, i.e., the number of hyperedges incident on each node (see Sec.~\ref{sec:measures:incidence}).
This lead to configuration models (CM) for hypergraphs.

A configuration model for $k$-partite hypergraphs  was proposed early in the network science literature, with the purpose of studying realistic \emph{folksonomies} (tagging databases) \cite{ghoshal2009random}.
A related model allowing for node features soon followed \cite{bradde2009percolation}.

As for hypergraphs where no $k$-partite structure is  enforced, there are quite a few recent generalizations, developed mostly with the goal of obtaining null models for community detection purposes.
Indeed, one of the best-known community detection methods relies on a so-called ``modularity function'' \cite{newman2004finding} to assign a quality to possible decompositions of a network in communities.
And in particular, the modularity uses a random  null model to determine whether the number edges found within a community is significant enough to warrant isolating it as a separated group.
In the case of graphs, the most popular baseline is the configuration model, and many models have since been proposed recently to fulfill the same role in the case of hypergraphs.
For example, a recently proposed model directly generalizes the configuration model of Chung and Lu \cite{chung2002connected} to the hypergraphical case \cite{kaminski2019clustering}.
In this version of the model, the number of times a node participates in any given hyperedge is drawn from a multinomial distribution.
This leads to a canonical version where the degrees of nodes are fixed on average.
Microcanonical variants are also analyzed by \citet{chodrow2019configuration}, and generalized to the case where the same node can take different roles in different edges (like broadcaster and receiver, for example), again by Chodrow et al. in \cite{chodrow2020annotated}.

Similar models have also been recently proposed in the statistics literature, where they are used for estimation purposes \cite{stasi2014beta}.
These  models are collectively dubbed $\beta$-models, and they are treated as generalizations of the $p_1$ model for graphs \cite{holland1981exponential} (an exponential random graph approach to the configuration model, for directed graphs).
They propose several flavors of the model, all making use of node \emph{propensities}, i.e., of a parameter $\beta_i\in\mathbb{R}$ to control how likely it is that node $i$ will be connected to any given hyperedge.
In the simplest proposed specification, the probability of a hypergraph $H$ is given by
\begin{equation}
    P(H|\bm{\beta}) = \prod_{i_1,...,i_j\in C_n(k)} p_{i_1, ..., i_k}^{a_{i_1,...,i_k}}(1 - p_{i_1, ..., i_k})^{1-a_{i_1,...,i_k}} \qquad p_{i_1, ..., i_k} = \frac{e^{\beta_{i_1} + ... + \beta_{i_k}}}{1+e^{\beta_{i_1} + ... + \beta_{i_k}}}
\end{equation}
where $C_n(k)$ is the set of all combinations of $n$ indexes.
They also propose a layered and general version where edges of different sizes co-exist.
The common thread shared by all the specifications of the $\beta$-model is that, the larger the parameters $\beta$, the more likely we are to see the hyperedge $a_{i_1,...,i_k}$ in the final hypergraph $H$.\\

The stochastic block model (SBM) is another network model that has been generalized to hypergraph extensively, motived by the search for random processes able to produce hypergraphs with non-trivial mesoscopic patterns.
The earliest reference to the hypergraphical SBM opts for a natural generalization from the network case \cite{ghoshdastidar2014consistency}, by parameterizing the probability of hyperedges of size $k$  with a symmetric tensor $\bm{Q}$ of dimension $k$, whose ``rows'' correspond to communities.
More precisely, in this SBM the probability that an edge exists between nodes $i_1,...,i_k$ assigned to communities $\sigma(i_1)...\sigma(i_k)$ is given by $q_{\sigma(i_1)...\sigma(i_k)}\in[0,1]$.
The likelihood of a hypergraph $H$ with adjacency tensor $\bm{A}$  is then straightforwardly:
\begin{equation}
    P(H|\bm{\sigma}) = \prod_{i_1,...,i_j\in C_n(k)} q_{\sigma(i_1), ..., \sigma(i_k)}^{B_{i_1,...,i_k}}(1 - q_{\sigma(i_1), ..., \sigma(i_k)})^{1-a_{i_1,...,i_k}} \qquad p_{i_1, ..., i_k} = \frac{e^{\beta_{i_1} + ... + \beta_{i_k}}}{1+e^{\beta_{i_1} + ... + \beta_{i_k}}},
\end{equation}
where $\sigma(i)$ is the index of the block to which node $i$ is assigned.
\citet{ke2019community} add degree-correction and modify the probability of the hyperedges $i_1,...,i_k$ is given to $q_{\sigma(i_1)...\sigma(i_k)}\prod_{j=1}^m \beta_{i_j}$ where $\beta_i>0$ is a propensity for node $i$. 
\citet{ahn2018hypergraph} consider weighted edges parametrized by the communities $\sigma$.
Finally, \citet{paul2018higher} combine the notion of communities with hyperedges of different sizes, albeit in a limited sense: the model conditions the presence of hyperedges of size $2$ (edges) and $3$ (triangles) on latent communities $\sigma$ , but does not include interactions at any higher orders.
In all cases, the models are introduced as a benchmark, to test whether partitioning methods---say spectral methods---can reliably recover the partition $\sigma$ from hypergraph generated by the model.
\\

\begin{figure}
    \centering
    \includegraphics[scale=0.7]{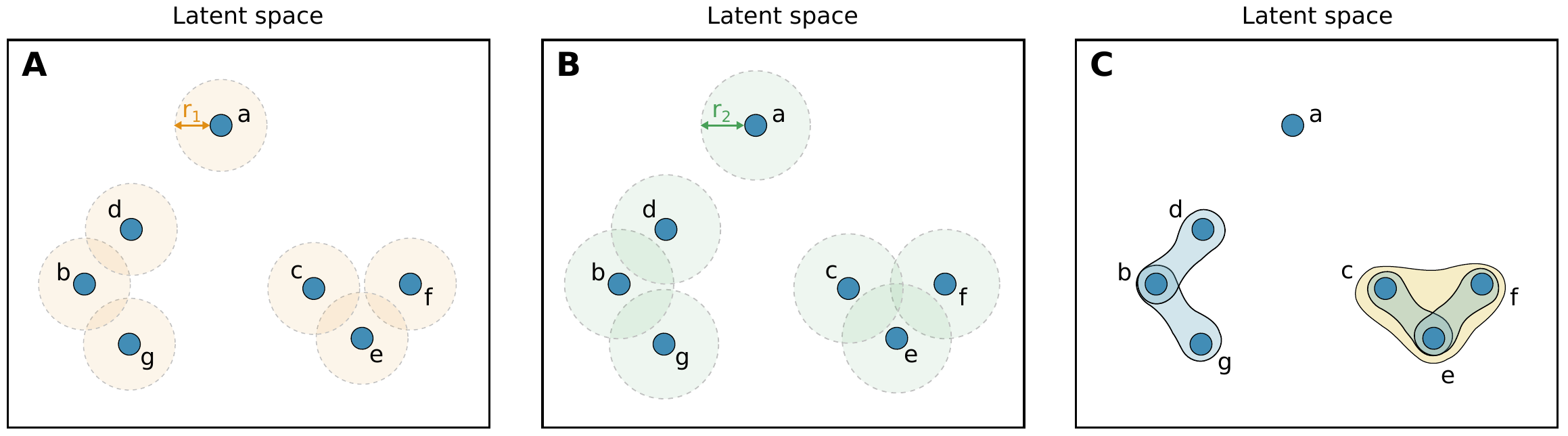}
    \caption{\textbf{Latent space hypergraphical model \cite{turnbull2019latent}.} (A,B) Nodes embedded in $\mathbb{R}^2$, with radii of length $r_1$ and $r_2> r_1$ drawn around them.  (C) Hypergraph obtained by connecting sets of nodes mutually at a distance of $d_{ij} < 2r_1$ and $d_{ij} < 2r_2$. Notice the multiple radii allows the model to create the hyperedge $\{c,e,f\}$, $\{c,e\}$ and $\{e,f\}$. A model with only one radius wouldn't be able to omit $\{c,f\}$. Figures adapted from Ref.~\cite{turnbull2019latent}. }
    \label{fig:models:latenthypergraphs}
\end{figure}

A recent approach to hypergraph modeling uses an abstract embedding space to create realistic HOrSs \cite{turnbull2019latent}, paralleling similar development in the network context  (see Sec.\ref{subsec:models:equilibrium:bipartite} above).
The idea is again that if \emph{groups} of nodes are close-by in the latent space, then they should tend to be connected.
The specific construction considered  by \citet{turnbull2019latent} embeds nodes randomly in $\mathbb{R}^{d}$, and adds hyperedges by connecting all set of nodes at a distance $d_{ij}<r_\ell$ from one another, for a few random choices of distances $r_\ell$ (see Fig.~\ref{fig:models:latenthypergraphs}).
This construction allows the model to create hyperedges included within others, such as the hyperedges $(c,e)$ and $(c,e,f)$ in Fig.~\ref{fig:models:latenthypergraphs}.
A last step is added whereby hyperedges are flipped (non-hyperedges become hyperedges and vice versa) with a small probability $\epsilon$. 
This step ensures that the model assigns a non-zero probability to all hypergraphs.\\

The last model of hypergraphs that we review generalizes the Kronecker model  \cite{leskovec2005realistic}, again first introduced in the context of networks.
In the classical model Kronecker graph model, one starts with a small matrix $\bm{P}^{(0)}$, and repeatedly takes the Kronecker product of the matrix with itself to generate increasingly large matrices $\bm{P}^{(1)}, \bm{P}^{(2)}, ..., \bm{P}^{(f)}$.
For example, supposing that $\bm{P}^{(0)}$ is a $2\times 2$ matrix we have:
\begin{equation}
    \bm{P}^{(1)} = \bm{P}^{(0)} \otimes \bm{P}^{(0)} = \begin{bmatrix} p_{11} \bm{P}^{(0)} & p_{12} \bm{P}^{(0)}\\p_{21} \bm{P}^{(0)} & p_{22} \bm{P}^{(0)}\end{bmatrix} = 
    \begin{bmatrix}
        p_{11}p_{11} & p_{11}p_{12} & p_{12}p_{11} & p_{12}p_{12}\\
        p_{11}p_{21} & p_{11}p_{22} & p_{12}p_{21} & p_{12}p_{22}\\
        p_{21}p_{11} & p_{21}p_{12} & p_{22}p_{11} & p_{22}p_{12}\\
        p_{21}p_{21} & p_{21}p_{22} & p_{22}p_{21} & p_{22}p_{22}
    \end{bmatrix}
\end{equation}
Then, once the matrix attains dimension $n\times n$, it is used to generate a graph on $n$ nodes in which edge $(i,j)$ exists with probability $p_{ij}^{(f)}$.
The hypergraph generalization, called HyperKron \cite{eikmeier2018hyperkron}, works essentially in the same way: one starts with a small $k$ dimensional tensor $\bm{P}^{(0)}$ and obtains a large final tensor of dimension $k$ and $n\times n \times .. \times n$.
One can then use the tensor to generate a random hypergraph, with hyperedges of size $k$.
The model has found application in generating large realistic graphs and hypergraph quickly.

\subsubsection{Simplicial complexes models}
\label{subsec:models:equilibrium:simplicial}

We complete our overview of equilibrium models with approaches formulated in the simplicial complex representation.
The theoretical study of models of simplicial complexes is still in its infancy \cite{kahle2011random}.
Save for some early work in the social sciences \cite{atkin1972cohomology}, the abstract simplicial complex representation has seen little applications until recent years.
As a result, the literature is so far limited, and mostly comes from mathematics and physics.\\

Before we review these models, a word of warning is in order: many authors blur the line between models of simplicial complexes and of hypergraphs.
Recall that in an abstract simplicial complex, when a \emph{facet} encodes an interaction between $k$ nodes, then implied is the existence of the $k$ \emph{faces} of $k-1$ nodes, $k(k-1)$ faces of $k-2$ nodes, and so on.
This inclusion property means that all models of simplicial complexes are specified in terms of their distribution over \emph{facets} (the top-level interactions). 
Many authors define facets as higher-order interactions, with no attention to the inclusion property.
This omission has no consequence when all interactions have the same size, but can lead to different results when they do not \cite{aksoy2019hypernetwork}.
In the interest of avoiding confusion, we have modified the nomenclature favored by the authors where necessary.\\

The  study of random simplicial complexes first started, perhaps unsurprisingly, with generalizations of the Erd\H{o}s-R\'enyi (ER) model.
The earliest model of random simplicial complexes, known as the Linial--Meshulam model, is arguably the simplest higher-order version of the ER model one can define.
In this model, one begins with a connected graph on $n$ nodes, to which some number $m$ of triangles are added to form facets of 3 nodes \cite{linial2006homological}.
The resulting object is a prototypical example of the approach favored by mathematical literature in this topic, whose focus is to find simple random objects with non-trivial \emph{homology} (see Sec.~\ref{sec:measures:homology}).
There has since been many generalization and analysis of this model, see the survey of \citet{kahle2014topology} for a summary of recent results.
Of particular interest is the natural generalization in which one begins with a complete simplicial complex of dimension $k$ on $n$ nodes and adds $k+1$ facets at random \cite{meshulam2009homological} (with the Linial--Meshulam model recovered by setting $k=1$).\\

A different form of ER-like models of simplicial complexes relies on the idea of \emph{flag complexes} (also known as clique complexes).
A flag complex is obtained by replacing all the maximal cliques of a graph by a facet. Another equivalent definition is that a flag complex is completely defined by its 1-skeleton (the underlying graph). 
In the model analyzed by Kahle \cite{kahle2009topology},  one first creates a classical ER network $G$, and then uses it to generate the associated flag complex.
The construction is related to much earlier work in graph theory \cite{bollobas1976cliques}, where the distribution of cliques in networks drawn from the ER model was analyzed.
However, in the case of \citet{kahle2009topology}, the focus is instead on the homology and homotopy of the resulting simplicial complexes.\\

To organize the rapidly growing family of models of random simplicial complexes, a model known as the ``$\Delta$--ensemble'' has been proposed in by \citet{kahle2014topology}.
In this model, one first connects the pair of nodes of a graph with probability $p_1$.
Then, all the (edge-only) triangles created in this first step are closed by a face (2-simplex), independently, with probability $p_2$.
All the empty pyramids created as a result of closing triangles are then replaced by a 3-dimensional face with probability $p_3$, and so on.
One recovers the classical ER model by setting $p_1=p$ and all other $p_d$ to 0; the model of Linial--Meshulam by setting $p_1=1$, $p_2=p$ and all other $p_d$ to 0; and the model of Kahle with $p_1=p$, $p_d=1$ for all $d>1$.
\citet{fowler2015generalized} and \citet{costa2016random} studied the homology of the simplicial complexes generated by this ensemble independently.
A closely related model, also introduced by \citet{costa2016random}, does not define the process recursively.
It instead fixes the average number of faces in all dimensions as well as the number of external faces, i.e., the way in which faces of different dimensions interact.
Restricted versions of these models have been used to generate simplicial complexes on which spreading process occurs \cite{iacopini2019simplicial} (see Sec.~\ref{subsec:social:spreading}), to predict higher-order interactions in streaming data \cite{benson2018simplicial}, and to study polymers \cite{alberici2017aggregation}.\\

The $\Delta$--ensemble is not the only general model able to encompass many models as special cases.
Indeed, a different specification, in the spirit of exponential random graph discussed above, has also been introduced recently by ~\citet{zuev2015exponential}.
In this model, one defines a series of functions $Q_{\mu}(S)$  on simplicial complexes for $\mu=1,..,K$.
These functions can be, for example, the number of 2-facets, or much more exotic functions, like the number of homological cycles of some dimension.
The \emph{exponential random simplicial  complex model} then assigns a probability
\begin{equation}
    \label{eq:model:erscm}
    P(S|\bm{Q},\bm{\lambda}) = \frac{1}{Z(\bm{\lambda})} e^{\sum_{\mu} \lambda_{\mu}Q_{\mu}(S)} \qquad Z(\bm{\lambda}) = \sum_{S} e^{\sum_{\mu} \lambda_{\mu}Q_{\mu}(S)}
\end{equation}
to simplicial complex $S$ and where $\bm{\lambda}$ is a vector of parameters controlling the relative importance of the functions $\bm{Q}$ in the ensemble.
Much like the $\Delta$--ensemble, special choices of parameters and function can be made to reproduce known models like that of Linial--Meshulam, Kahle, or even the $\Delta$--ensemble itself \cite{zuev2015exponential}.\\


\begin{figure}
    \centering
    \includegraphics[scale=0.45]{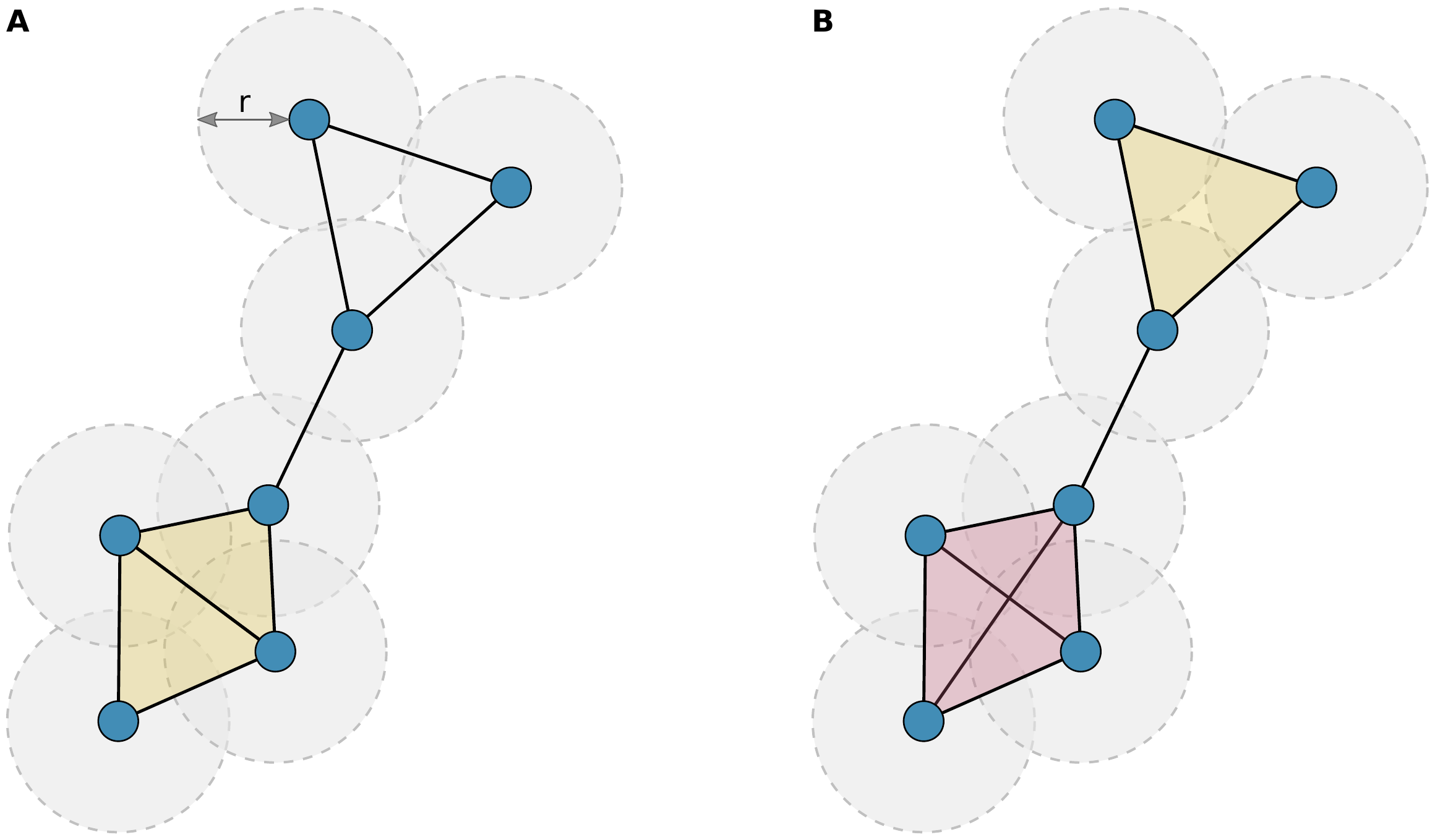}
    \caption{\textbf{Difference between (A) {\v C}ech  and (B) Vietoris-Rips complexes.} Figures adapted from Ref.~\cite{chazal2017introduction}.}
    \label{fig:models:VietorisRips_vs_cech}
\end{figure}
Another rich line of inquiry focuses on random \emph{geometric} simplicial complexes.
These models build on a long lineage of work in topological data analysis \cite{patania2017shape}, whose focus is the recovery of topological information, such as the number of holes in a surface, from noisy objects on geometric objects embedded in space as point clouds.

In these models, one typically first place nodes in some metric space, e.g. $\mathbb{R}^d$, at random by drawing from a random point process.
Then nodes are then connected based on their distance, generating a simplicial complex.
There are two canonical ways in which this last step can be done (see Fig.~\ref{fig:models:VietorisRips_vs_cech}).
In random \emph{{\v C}ech complexes} \cite{kahle2013limit}, one places a ball of radius $r$ around every node; whenever the intersection of $k$ balls is non-empty, one adds a $(k-1)$-face between these nodes.
In random Vietoris-Rips complexes \cite{kahle2011random,kahle2013limit}, instead, one connects every node at a distance at most $2r$ from one another, and then replace cliques by facets, effectively taking the flag complex of the underlying geometrical graph. 
The resulting objects have been studied for their connectivity properties and homological properties, among other things (see the thorough survey of \citet{bobrowski2018topology} for more details). 
A fundamental result in this context concerns the limiting behavior of random geometric complexes.
In particular, three regimes exist that display starkly different properties as a function of the parameter $\Lambda = n r^d$, where $n$ is the average number of points in a ball of radius $r$ in $d$-dimensional space.
The three regimes correspond to different limits for $\Lambda$: for vanishing $\Lambda$, the simplicial complexes are sparse, highly disconnected and dust-like; for constant $\Lambda$ (also called thermodynamic regime), the homology of the complexes reaches its peak growth; and for diverging $\Lambda$, higher homology displays two phase transitions, one where the homology first appears and a second where it disappears, as cycles are progressively filled in.   
Finally, \citet{fasy2014confidence} also found statistical application in the calculation of confidence interval on the results of persistent homology calculations  .\\

In the physics literature, the focus thus far has been analogs to the configuration model (CM), with the goal of introducing heterogeneities and realistic properties in the generated random simplicial complexes.
By analogy with the case of graphs, these configuration models fix the degree of nodes, defined as the number of \emph{facets} incident on them.
All other properties are randomized.
\citet{courtney2016generalized} considered a model where every facet is of size $k$ and the degree of nodes is fixed exactly or on average;  \citet{kaminski2019clustering} have later proposed a hypergraphical CM that is formally equivalent model to the previous one.
A different specification fixes the degrees \emph{exactly}, and lets the facets have different dimensions while forbidding inclusions \cite{young2017construction}.
The resulting models have been studied for their structural properties \cite{courtney2016generalized} (like the projected degree or the entropy), and have found application as a null model for the homology of real systems \cite{young2017construction}.
A recent approach to random simplicial complex \cite{bianconi2018topological,bianconi2019percolation}  follows the interpretation of the (classical) configuration model as a random branching tree \cite{newman2001random}.
In this model of random branching simplicial complexes \cite{bianconi2019percolation}, one starts with a single edge and attaches $m$ faces of dimension $k$ to each edge of the faces created in previous iterations.
Crucially, $k$ is treated as a random variable such that the dimension of every new face is random, leading to a simplicial analog to the configuration model.
It is introduced to study the percolation properties of the resulting ``simplicial tree.''\\

\subsection{Out-of-equilibrium models}
\label{subsec:outofequilibrium_models} 


All of the models we have seen thus far define distributions over static HOrSs.
In other words, these models viewed HOrSs not as dynamical, evolving objects, but instead as static systems, drawn from some  fixed distribution.
We now turn to a different approach, mainly developed in the physics and network science literature, that adopts a dynamical point of view of HOrSs.\\

Since there are several similarities between many of these out-of-equilibrium models, even across different choices of representations (see Sec.~\ref{sec:representations}), it is worth going over some general notions before we delve in.
The modeling goal motivating these models is almost always the same: finding minimal rule sets such that the typical HOrSs produced by the model reproduces the structural characteristics of empirically observed bipartite networks  \cite{guillaume2004bipartite}.
The rules are often chosen to allow for analytical calculations of properties of interest.
But the authors of these model also often try to find rules that can be motivated mechanistically i.e., that could explain \emph{why} a HOrSs evolves the way it does \cite{overgoor2019choosing}.\\

An overwhelming majority of the out-of-equilibrium models focus on \emph{growing} systems, in which nodes and edges are added as time unfolds, but never removed.
This is perhaps due to the influence of foundational work in network science, where growing models were put center stage early on \cite{barabasi1999emergence}.
Hence, with very few exceptions, these out-of-equilibrium are \emph{growth models}.
Furthermore, with few exceptions like the activity-driven models \cite{petri2018simplicial}, time is measured in discrete steps $t=1,2,...,T$, where each step is associated with an event.
Events typically involve the creation of a new node, or edge, or both.
Echoing work on the preferential attachment model for classical networks \cite{barabasi1999emergence}, many of the growth events somehow favor existing nodes.
As we will see, this is often achieved by selecting the nodes that receive new edges from a categorical distribution, with probabilities proportional to some growing function of the degree of the nodes already in the network.

\subsubsection{Bipartite models}
\label{subsec:models:outofequilibrium:bipartite}

A prototypical example of out-of-equilibrium bipartite model is the model of G. Erg{\"u}n, whose goal was to reproduce the evolution of sexual contact networks in a heteronormative society  \cite{ergun2002human}.
Recall that there are two sets of nodes (which we have called $A$ and $B$) in a bipartite network.
The model of  Erg{\"u}n \cite{ergun2002human} postulates that the evolution of the network can be explained with 3 types of events, occurring with probabilities $p,q$ and $r$ summing to 1.
At each time step, a new node arrives in set $A$ (with probability $p$), or in set $B$ (with probability $q$); or a new edge appears between sets $A$ and $B$ (with probability $r=1-p-q$). 
To avoid nodes of degree 0, all new incoming nodes are initially attached to one node in the opposite set, selected at random.
Building on the well-known preferential attachment model \cite{barabasi1999emergence}, all of the choices are made \emph{preferentially}.
That is, whenever one or two nodes must be selected to form an edge, they are selected with probability proportional to their current degree (plus some offset specific to the set, called charisma in the original model \cite{ergun2002human}).
Thus, the probability that a node $i\in A$ is chosen at time $t$ is calculated as
\begin{equation}
    p_i (t) = \frac{k_i(t) + c_A}{\sum_{j\in A} (k_j(t) + c_A)}
    \label{eq:model:attach_prob}
\end{equation}
where $k_i(t)$ is the degree of node $i$ at time $t$, and $c_A$ is the offset parameter.

Many variations on this theme have since been proposed.
For instance, a related---and this time system-agnostic---model of evolving bipartite networks \cite{guillaume2004bipartite,guillaume2006bipartite} proceeds by adding new nodes to set $A$ only.
Different from the sexual network model, the degree of the incoming node is chosen from a fixed degree distribution.
For each of its $k$ edges, the new node chooses to attach to an existing node of $B$m with probability $\lambda\in[0,1]$, or to a new one.
Hence new nodes appear in set $B$ only through their connection with incoming nodes in set $A$.
Again choices are made preferentially, using Eq.~\eqref{eq:model:attach_prob} with $c_A=0$.
\citet{ramasco2004self} make use of \emph{two} distributions instead.
After drawing the degree $k$ of the new node in $A$, a second number $\ell\leq k$ is drawn from a second distribution, determining the number of target nodes in $B$. 
These nodes are again chosen preferentially, while the $k-\ell$ remaining degrees are attributed to new nodes in $B$.
\citet{beguerisse2010competition} instead consider node sets whose evolution is independent of node creations.
In this case, time is measured in terms of edge creation events, connecting the two sets.
Different from the models above, it mixes the preferential attachment probabilities appearing in Eq.~\eqref{eq:model:attach_prob} with \emph{uniform attachment}, in which nodes are chosen uniformly from the node set, i.e.,
\begin{equation}
    p_i(t) = \frac{1}{N(t)}
    \label{eq:model:attach_prob_unif}
\end{equation}
where $N(t)$ is the size of the target set \cite{beguerisse2010competition}.

In all of the models above, the modeling goal is to reproduce the structure of  empirically bipartite systems, be it sexual \cite{ergun2002human}, collaborative \cite{ramasco2004self} or competitive \cite{beguerisse2010competition}.
Connections tend to be concentrated in these human systems, and preferential probabilities like the one appearing in Eq.~\eqref{eq:model:attach_prob} are used to induce such a skewed distributions of degrees \cite{barabasi1999emergence}.
Uniform probabilities like the one appearing in Eq.~\eqref{eq:model:attach_prob_unif}, on the other hand, favor more equitable distributions, which can also be found in some bipartite systems \cite{guillaume2004bipartite,beguerisse2010competition}.\\

\begin{figure}
    \centering
    \includegraphics[scale=0.4]{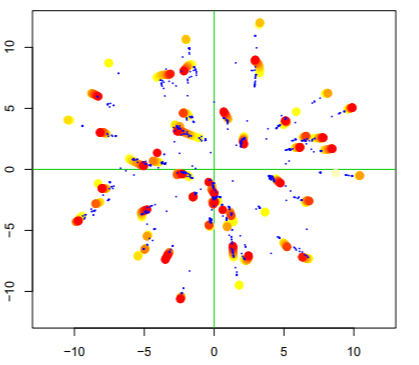}
    \caption{\textbf{Boards and directors moving about a latent space.} Board are represented as large colored dots, going from 2003 in yellow to 2013 in red. Directors are shown as small blue dots. Edges are not shown. Figure reproduced from Ref.~\cite{friel2016interlocking}.}
    \label{fig:models:latent_dynamics_bipartite}
\end{figure}

Not all out-of-equilibrium models of HOrSs can be modeled with straightforward growth models like the one above.
For instance, in one model \cite{sneppen2004simple} that is closer to the literature of self-organized criticality \cite{bak1987self}, nodes re-arrange their edges following connection events, and replicate the behavior of  nodes to which they are connected. 
As another example, a different approach by \citet{friel2016interlocking}  relies on  a dynamical formulation of latent space models, in which nodes in the two sets $A$ and $B$ move in a latent space, and in which edges and edges / non-edges tend to perpetuate themselves  (Fig.~\ref{fig:models:latent_dynamics_bipartite}).\\

Other out-of-equilibrium models focus on \emph{rewiring}, a process by which the edges of a  bipartite network are reorganized, all the while preserving the node set.
For example, \citet{evans2007exact} start from some configuration where the nodes in one of the sets  $A$ have exactly one neighbor in the other.
At each time step, a node in set $A$ is chosen using an arbitrary selection process, and the edge connecting it to its sole neighbor in $B$ is disconnected.
A new target in $B$ is then chosen, again arbitrarily, and a new edge is formed.
It turns out that the sequence of generated bipartite networks can be described exactly at all times \cite{evans2007exact}, and that the model allows for a number of useful generalizations, including a version of the models where the choices are driven by a superimposed unipartite network or node types \cite{evans2007exact_b}.
We note that for some choices of rewiring mechanism, averages over the rewiring process can be thought of as averages over a static ensemble of bipartite networks, such that these rewiring processes straddle the boundary between out-of-equilibrium and equilibrium models \cite{fosdick2018configuring}.

\subsubsection{Stochastic set models}
\label{subsec:models:outofequilibrium:cliques}

A different category of out-of-equilibrium models focuses on how the membership of nodes to sets evolve through time.
There has been relatively little work in this representation, since models that reproduce the evolution of set memberships are ironically most often couched in the language of bipartite networks \cite{friel2016interlocking}, hypergraphs \cite{holland1976local}, or simplicial complexes \cite{wu2015emergent}.
That said, a few models have made explicit use of the set representation, because it turns out to be a natural choice for higher-order interaction of limited scope, and for two-mode data \cite{atkin1974mathematical}.\\

These set-based models again build on the observation that the distribution of the number of sets per node and of the number of nodes per set are often heavy-tailed \cite{guillaume2004bipartite}.
Hence, the evolution of these sets can be plausibly reproduced with a process in which rich-get-richer \cite{barabasi2011network}, an observation that has been supported by the empirical analysis of evolving sets \cite{pollner2005preferential}.
\citet{zhou2008weighted} harness these observations to create a model of evolving networks with communities overlain, in which nodes join communities based on their size and create connections with nodes chosen preferentially in large communities.
A simpler model, known as the structural preferential attachment (SPA) model \cite{hebert2011structural,hebert2012structural,young2016growing}, incorporates the rich-get-richer mechanism without any explicit need for a network; it instead models the evolution of the sets directly.
In this particular model, every time step consists of a node joining a set, with both the node and set being either new or old (such that are $2\times 2=4$ possible outcomes).
The particular type of event is determined from a categorical distribution, and every choice of nodes and sets is made preferentially with respect to the set size / membership numbers of the nodes, see Eq.~\eqref{eq:model:attach_prob}.
A hierarchical extension of the model exists \cite{hebert2015complex}, where sets themselves belong to larger overlapping sets, and preferential attachment is applied at all levels.\\

The SPA model is closely related to the Chinese restaurant process  \cite{aldous1985exchangeability} and the related Indian Buffet Process\cite{griffiths2011indian}, both developed to create distributions over set membership relationships, in the statistics and machine learning literature.
In the Chinese restaurant process, for instance, a new element (node) is created at each time step, and either join an existing set with probability $1 - 1/(t+1)$  or create a new set by itself with probability $1/t$.
The specific set to which the incoming element is attributed is chosen preferentially.
The Indian Buffet Process is defined in similar terms but allows for multiple set memberships.
These process have found wide statistical applications because they satisfy an exchangeability property.
Here, exchangeability means that the probability of an observed collection of sets is independent of the order in which the ``growth'' events actually occurred.  
This property greatly simplifies, for instance, the calculation of expectations over the process.

\subsubsection{Hypergraphs models}
\label{subsec:models:outofequilibrium:hypergraphs}

There has been some work on out-of-equilibrium  models of HOrSs in the hypergraph representation as well, mostly confined to the physics literature.
In this literature, the evolving hypergraphs are often called \emph{hypernetworks}, but the underlying concepts are nonetheless the same.\\

One of the earliest out-of-equilibrium model of hypergraphs considers the very same type of hypergraphs that were analyzed in the first \emph{equilibrium} models of hypergraphs, in the physics literature: $k$-partite hypergraphs \cite{ghoshal2009random}, used as models of folksonomies in which users tag items.
In the model in question \cite{zhang2010hypergraph}, users have intrinsic \emph{activities}, corresponding to the likelihood that they will be the next user to tag an item. 
At each time step, one picks a random user proportionally to this activity, and then decide on both the type of tag to apply and the item to tag.
The specific ways in which the choices are implemented allow for a rich-get-richer phenomenon, and the creation of new items.\\

This particular model is somewhat unique in that most models of evolving hypergraphs focus on more general hypergraphs that need not be $k$-partite.
Work on these general out-of-equilibrium-models of hypergraph was initiated by \citet{wang2010evolving}, where a prototypical model of growing hypergraphs was proposed.
The model takes motivation in the study of how co-authorship systems evolve (with nodes being authors and hyperedges being papers).
In this model, $k$  new nodes are added at every time step, and they form a hyperedge with precisely $1$ node present in the extant hypergraph, chosen proportional to its degree (the number of hyperedge incident on nodes, see Sec.~\ref{sec:measures:incidence}).
By construction, the model generates hypergraphs in which every hyperedge has size $k$.

There have since been countless variations on these rules, all leading to slightly different models.
For instance, \citet{liu2012social} allow the sizes of the hyperedges to vary by specifying these sizes as input (the sequence can be generated at random or deterministically).
Hu et al. explore alternatives where both the size and the composition of new hyperedges to vary at random \cite{hu2019hypernetwork}.
\citet{guang2013local} introduce the notion of a ``local world,'' by forming the new hyperedges with nodes selected in a small subset of nodes that is itself selected at random.
In \citet{wu2014synchronization}, choices are not made preferentially but instead proportionally to the ``joint degree'' of nodes, i.e., the number of hyperedges they share with the hyperedges that are already in the set.
Finally, yet another model uses a complicated choice function to decide which nodes should be involved in new hyperedges \cite{guo2016non}, namely
\begin{equation}
    p_i(t) = \frac{f(k_i(t))}{\sum_j f(k_i(t))}
\end{equation}
where $k_i(t)$ is the degree of node $i$ at time $t$, and $f(k_i(t))= (k_i(t) + c)^\gamma$ include both an offset $c$ and a non-linear exponent $\gamma$ in the spirit of the classical model of \citet{krapivsky2000connectivity}.
See also \cite{guo2015brand} for a version of the function that allows for node dependent offsets and includes a node-dependent multiplicative term.\\

Paralleling the bipartite case (see Sec.~\ref{subsec:models:outofequilibrium:bipartite}), these models have all been introduced to reproduce some set of characteristics of empirical systems, e.g. collaboration hypergraphs.
We note, however, that the analysis of these models has so far been limited to reproducing the degree distribution of the nodes, with matching numerical simulations.

\subsubsection{Simplicial complexes models}
\label{subsec:models:outofequilibrium:simplicial}
The last set of out-of-equilibrium models that we review are specified in the simplicial complex representations.
Similar to the equilibrium models of simplicial complexes, some of these models can be interchangeably thought of as hypergraphs models, when they do not make use of the inclusion property explicitly.
Hence, we have again altered the nomenclature favored by the authors when most appropriate.

\begin{figure}
    \centering
    \includegraphics[scale=0.4]{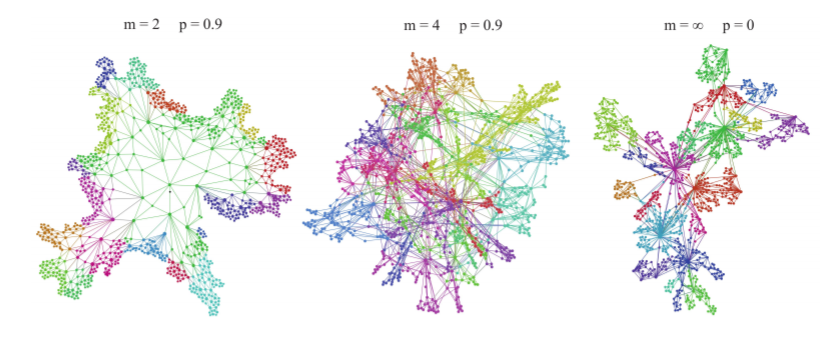}
    \caption{\textbf{Different growing geometrical networks produced by the simple model of Wu et al.}
             Parameter $m$ controls the maximal number of triangle per edges, and $p$ controls closure. Figure reproduced from Ref.~\cite{wu2015emergent}.}
    \label{fig:models:wu2015}
\end{figure}

There are a few approaches to dynamical models of simplicial complexes.
The first work that uses simplicial complexes to model hyperbolic network geometry comes from the physics literature, where it is studied under the name of ``Complex Quantum Network Manifolds'' \cite{wu2015emergent}.
These models are motivated by geometric considerations \cite{bianconi2015interdisciplinary}, the modeling goal being to specify models of how a wide variety of discrete spaces, represented as simplicial complexes, may arise.
The first model in this line of work tracks the evolution of a growing simplicial complex, made exclusively of triangles \cite{wu2015emergent}.
At each step, a new node comes in and forms a triangle with two existing nodes, chosen uniformly from the set of all connected nodes that have less than $m$ triangles together, where $m$ is called the saturation bound.
At each time step, with probability $p\in[0,1]$, one also closes a triangle, a process that is implemented by choosing an edge uniformly from the set of unsaturated edges, and choosing an unsaturated edge at random in the neighborhood of $e_1$.
With this simple model, one can create many quantitatively different outcomes, for example: planar graphs for low $m$, or complex geometries in the limit $m\to\infty$, see Fig.~\ref{fig:models:wu2015}.

Many variations on this model followed \cite{bianconi2015complexa,bianconi2015complexb}, culminating into a general model of ``Network Geometry with Flavor (NGF)'' \cite{bianconi2016network}, in which facets have all the same dimension, but are not necessarily triangles anymore.
The existing facets have a latent (quenched) energy $\varepsilon$ that is a function of the nodes they connect to, and incoming $k$ facets are connected to an existing $k-1$ facet $\alpha$, chosen with probability
\begin{equation}
    p_{\alpha}(t) = \frac{e^{-\beta \varepsilon_\alpha}(1+s n_{\alpha})}{\sum_{\alpha'} e^{-\beta \varepsilon_{\alpha'}}(1+s n_{\alpha'})}
\end{equation} 
where $s\in\{-1,0,1\}$ is called the flavor of the model, $\beta$ is a temperature, and $n_{\alpha}$ is a the number of facets incident on $\alpha$, minus one.
The models that precede the NGF model are all special cases of it.
For example, the model of \citet{bianconi2015complexa} focused on the case of triangles $k=2$, with  saturation parameter $m=2$, flavor $s=0$, and facets that have latent energies.
A later model of \citet{bianconi2015complexb}  is similar, but focuses on the flavor $s=1$.

Some models that have followed borrow much of the mechanisms from the NGF model. For instance, the model of \citet{courtney2017weighted} functions more or less in the same way, but adds a mechanism to track the evolution of weights on the simplices, while that of \citet{fountoulakis2019dynamical} also works in fixed dimension but is otherwise very general, introducing the possibility to remove smaller facets when adding new ones.\\

Other out-of-equilibrium models of simplicial complexes favor approaches that are not related to the NGF model. 
For instance, one set of methods \cite{sizemore2018knowledge,blevins2020reorderability} by Sizemore et al. favors the flag complex approach that was also used in the context of equilibrium models by \citet{kahle2009topology} (see Sec.~\ref{subsec:models:equilibrium:simplicial}).
In these approaches, sequences of growing graphs are first generated with a series of standard network models, like for example preferential attachment \cite{barabasi1999emergence}.
Then, one takes the flag complexes of these graphs by replacing every clique with a facet, yielding a sequence of growing simplicial complexes.
These models have been used to study the evolution of the topological invariants of the resulting growing simplicial complexes \cite{sizemore2018knowledge} and their sensitivity to changes in the sequence of events \cite{blevins2020reorderability}.\\

Another recent---and extremely general---model moves away from the ``simplicial-complexes as generalized networks''  metaphor, and instead focuses on the properties of the manifold they describe, like their Hausdorff and spectral dimensions \cite{da2018complex}.
Many basic mechanisms by which these complex manifold may evolve are explored and classified.\\

A different model, introduced by \citet{courtney2018dense}, focuses on \emph{directed} triangular simplicial complexes.
In this model, the simplices are either created or reinforced by first selecting a source node, proportionally to its out-strength (the total weight of triangles for which it is the source), with some probability that the node is new.
To determine whether the event leads to the creation of a new simplex or to reinforcement, one then selects an edge at random in the simplicial complex, and reinforces the weight of the triangle this edge forms with the source, if its exists, or creates the triangle when it doesn't. 
In this case, the modeling goal is to obtain dense simplicial complexes with scale-free degree distributions.\\

Finally, there are some models that consider \emph{dynamical} simplicial complexes---not simplicial complexes that merely grow.
These models are event-based, focusing on what happens in a time slice of a continuous process \cite{kim2018hyperedge}.
The work of \citet{petri2018simplicial} is prototypical in this regard.
In this model, each node $i$ is endowed with an activity rate $a_i$ drawn from some distribution treated as a parameter of the model.
Then, when a node $i$ fires (at rate $a_i$) in continuous time, a $(k - 1)-simplex$ is created with with $k-1$ other nodes chosen uniformly at random.
The simplex then disappears after $\Delta t $, a parameter.
In this case, the model is introduced to study the structural property of the generated simplicial complexes like the degrees aggregated over time, as well as and dynamics taking place \emph{on} the simplicial complexes generated by this model (see Sec~\ref{sec:social}).


\section{Diffusion}
\label{sec:diffusion}

In the previous sections we have focused on the structure of higher
order interactions. We have introduced higher order systems (HOrS) and
showed their versatility in describing the structural properties of
complex systems with more than pairwise couplings among their
components.  We will now discuss how the dynamical processes
traditionally defined on networks can be extended to higher order
systems.  We will start in this section with diffusion, a linear
process that, despite its simplicity, is of high relevance in many
different contexts, and also provides an useful first approximation in
the case of nonlinear dynamical systems.

Under the name of diffusion we usually indicate two distinct processes
that is important to distinguish. The first one is the pure (or standard)
process of
{\em diffusion}, also known as the ``fluid model'', in which the
quantity of interest moves from one region to another following the
gradient of concentration. The second one is the so-called {\em
  continuous-time random walk}~\cite{masuda2017random}.
Here, we will first discuss these two
processes in the context of networks, focusing on their similarities
and differences. We will then show how to implement them on HOrS in
Secs.~\ref{sect:diff1} and ~\ref{sec:rw}, respectively.

\bigskip
In standard diffusion on a network, a (material or immaterial) substance
is allocated to the nodes of a graph and flows over each of its edges
from the node with higher concentration to the node with lower
concentration~\cite{aldous2002reversible,samukhin2008laplacian,hoffman2012generalized}. The process produces a redistribution of the substance, which finally
leads to a state where all the nodes have the same concentration. 
This state of the system, which also takes the name of
consensus~\cite{degroot74reaching}, represents a stable equilibrium of
the process, subject to detailed balance condition~\cite{boltzmann1964lectures,tolman1979principles}. 
If we indicate as $x_i(t)$, with $i=1,2,\ldots,N$, the concentration 
at the generic vertex $v_i$ at time $t$, the time evolution of the
network state is governed by the following system of $N$
coupled linear differential equations: 
\begin{equation}
\dot x_i(t) = \sum_j a_{ij} (x_j(t) - x_i(t)) = -\sum_j (k_i \delta_{ij} - a_{ij}) x_j(t) = -\sum_j (L_0^{\rm D})_{ij} x_j(t)
\label{eq_diff}
\end{equation}
where $A= \{a_{ij} \}$ is the adjacency
matrix of the network, which we assume here for simplicity to be a
binary and symmetric matrix (although it is straightforward to extend
the formalism to directed and weighted networks), and $k_i = \sum_j
a_{ij}$ is the degree of node $i$. 
In the last equality, we have defined the diffusion Laplacian matrix:
\begin{equation}
(L_0^{\rm D})_{ij}=\begin{cases}
k_i \text{ if } i=j\\
-1 \text{ if } v_i \sim v_j\\
0 \text{ otherwise}
\end{cases}
\end{equation}
where $ v_i \sim v_j$ indicates that vertices $v_i$ and $v_j$ are 
adjacent. In matricial form, we can write 
$ L_0^{\rm D} = D - A$, with $D$ the diagonal degree matrix.
Equation~\eqref{eq_diff} can then be written as 
${\dot {\bd x}}(t) =   - L_0^{\rm D} {\bd x}(t)$, where we have defined the concentration 
vector ${\bd x} = (x_1,x_2,\ldots x_N)$.  
The fact that the homogeneous solution $\bd x^{(\infty)} = \bd 1 \sum_i x_i(0) / N$
represents a stationary equilibrium for the 
process is easily proven by observing that the Laplacian, by
definition, is characterized by having all the rows summing to zero,
$\sum_j (L_0^{\rm D})_{ij}=0$. For undirected networks, characterized by symmetric Laplacian, this implies from Eq.~\eqref{eq_diff} that the total concentration $\sum_i x_i$ is conserved, and in particular 
$\dot {\bd x}^{(\infty)} = 0$. 
The stability of such equilibrium is governed by the
spectral properties of the Laplacian matrix. 
In fact, the solution of Eq.~\eqref{eq_diff} can be written by
projecting on the Laplacian eigenvectors, which forms a basis in the
case of a connected network:
\begin{equation}
x_i(t) = \sum_{\alpha=1}^N c_{\alpha}(0) e^{-\lambda_{\alpha}t} \phi^{(\alpha)}_i = c_{1}(0) \phi_i^{(1)} + \sum_{\alpha=2}^N c_{\alpha}(0) e^{-\lambda_{\alpha}t} \phi^{(\alpha)}_i
\label{t_sol}
\end{equation}
where $\lambda_\alpha$ and $\bd{\phi}^{(\alpha)}$, with
$\alpha=1,2,\ldots N$ are the
$\alpha$-th eigenvalue and eigenvector of $ L_0^{\rm D}$, while the coefficients 
$c_{\alpha}(0)$ depend on the initial conditions. We have used the
fact that one of the eigenvalues of the Laplacian is zero, because of
the zero-row-sum property, which also implies that the corresponding
eigenvector $\bd \phi^{(1)}$ is homogeneous.
If the network is connected the zero eigenvalue is unique,
and all the other eigenvalues are positive by
definition~\cite{newman2010networks}. Hence, 
we can see from Eq.~\eqref{t_sol} that $\bd x(t)$ will
always converge to the homogeneous solution for $t \rightarrow \infty$,
the convergence time being given by the inverse of 
the smallest eigenvalue different from zero, $\lambda_2$. 

\bigskip
A qualitatively different class of processes arises when one considers continuous-time random walk.
In this  stochastic process a single walker jumps from one node
to one of its neigbours on the network, and  cannot divide or distribute itself over more than one node, as happens in standard diffusion. 
In this case, we consider various realizations of the process and we   
describe the state of the sytem by a vector $\bd q(t)$, representing the  
probability for each node to be occupied by the walker
at a given time $t$. 
The probability that the walker moves from the
generic node $v_j$ to $v_i$ in one step is given by the $(i,j)$ entry
of the transition matrix $\Pi = \{ \pi_{ij} \}$. In an unbiased
random walk this is given by $\pi_{ij}= a_{ij} /k_j$, representing the
fact that the walker on node $v_j$ can choose equally among $k_j$ neighbors.
The time evolution of the occupation probability is
consequently governed by a set of differential equations: 
\begin{equation}
\dot q_i(t) = \sum_j \pi_{ij}q_j(t) - \sum_j \pi_{ji}q_i(t) = -\sum_j (\delta_{ij}-\pi_{ij}) q_j(t) = -\sum_j (L_0^{\rm RW})_{ij}q_j(t).
\label{rw0}
\end{equation}
which is similar to that in Eq.~\eqref{eq_diff}. 
The last equality defines the random walk Laplacian $L_0^{\rm RW}$, which
is related to the diffusion Laplacian by $L_0^{\rm RW} = L_0^{\rm D} D^{-1}$, 
and for this reason is also called {\em normalized Laplacian}.
In matricial form, Eq.~\eqref{rw0} can be written as
${\dot {\bd q}}(t) =   - L_0^{\rm RW} {\bd q}(t)$. 
\\
As well as diffusion, a random walk process is characterized by a
stationary distribution where the flows of probability in each direction
equal each other, and a detailed balance is reached. 
In this case the stationary state ${\bm q}^{(\infty)}$ corresponds to a
probability distribution that is proportional to the degree of nodes: 
$q_i^{(\infty)}=k_i/2K$, implying that, at the equilibrium, it is more likely
to find the walker on the network hubs. As for standard
diffusion, also the dynamics of a random walk
process is intimately related to the spectral properties of its own 
Laplacian operator. Again, the stationary state can be found as the
Laplacian eigenvector $\bd \phi^{(1)}$ associated to the unique (if the
network is connected) eigenvalue $0$,  
which is indeed proportional to the vector of node degrees.  
The time dependent solution of Eq.~\eqref{rw0} can
be written analogously to Eq.~\eqref{t_sol}, and it is thus clear that also
in a random walk the velocity at which the stationary state is reached
depends on the second eigenvalue (the smallest eigenvalue different from
zero) of $L_0^{\rm RW}$.
Summing up in both cases of standard diffusion and random walk, 
the Laplacian matrix encodes the structure of the network
and its spectral properties are related to the dynamical features
which ultimately govern the time evolution of the state
vector~\cite{chung1997spectral}.
\\

\bigskip 
When it comes to generalizing diffusive processes to higher order 
structures, many authors have attempted to extend the microscopic mechanism
known to underlying diffusions in pairwise networks to larger groups
of nodes. Importantly, traditional diffusion is a linear process and consequently the simplest generalization to higher-order structures can always be reduced to equations involving only pairwise couplings. For instance, as shown by
Neuh\"{a}user et al in \cite{neuhauser2019multi}, an extension of 
Eq.~\eqref{eq_diff} to 3-body interactions can be formalized as:
\begin{equation}
\dot x_i(t) = \sum_{jk} a^{\triangle}_{ijk} [(x_j(t)-x_i(t))+(x_k(t)-x_i(t))]
\label{3body_diff}
\end{equation}
where $\bd{A}^{\triangle} = \{ a^{\triangle}_{ijk} \}$ represents a
tensor whose entry $(i,j,k)$ is equal to 1 only if there is a 
triangle involving the three nodes $i$, $j$ and $k$.
It is easy to see that Eq.~\eqref{3body_diff} can be rewritten
in terms of a new Laplacian matrix $L^{\triangle}= \{ \ell^{\triangle}_{ij} \}$ 
as  
$\dot x_i = - 2 \sum_{j} \ell^{\triangle}_{ij} x_j$,
where
$\ell^{\triangle}_{ij} = \delta_{ij} \sum_{kj} a^{\triangle}_{ijk} -
\sum_k a^{\triangle}_{ijk}$.
The system in Eq.~\eqref{3body_diff} can thus be reduced to a system
with pairwise interactions, where the pairwise interactions are weighted
according to the organization of the HOrS under study.

To see the effects of multi-body coupling
we need to insert a non-linearity in the equations, see
Sec.~\ref{sec:social}. 
There are many ways in which this can be done.
As we have seen in Sec.~\ref{subsec:ho-laplacian},
it is possible to introduce generalized Laplacians both for
diffusion and for random walk. In the next two subsections 
we will see how such mathematical concepts can turn useful to
talk of diffusion on HOrS, when the role played by the nodes
for the network Laplacian is, at higher orders, played by the
edges, the triangles, the tetrahedra, etc.

\subsection{Higher-order diffusion}
\label{sect:diff1}

\begin{figure*}
\centering
\includegraphics[width=0.8\textwidth, keepaspectratio = true]{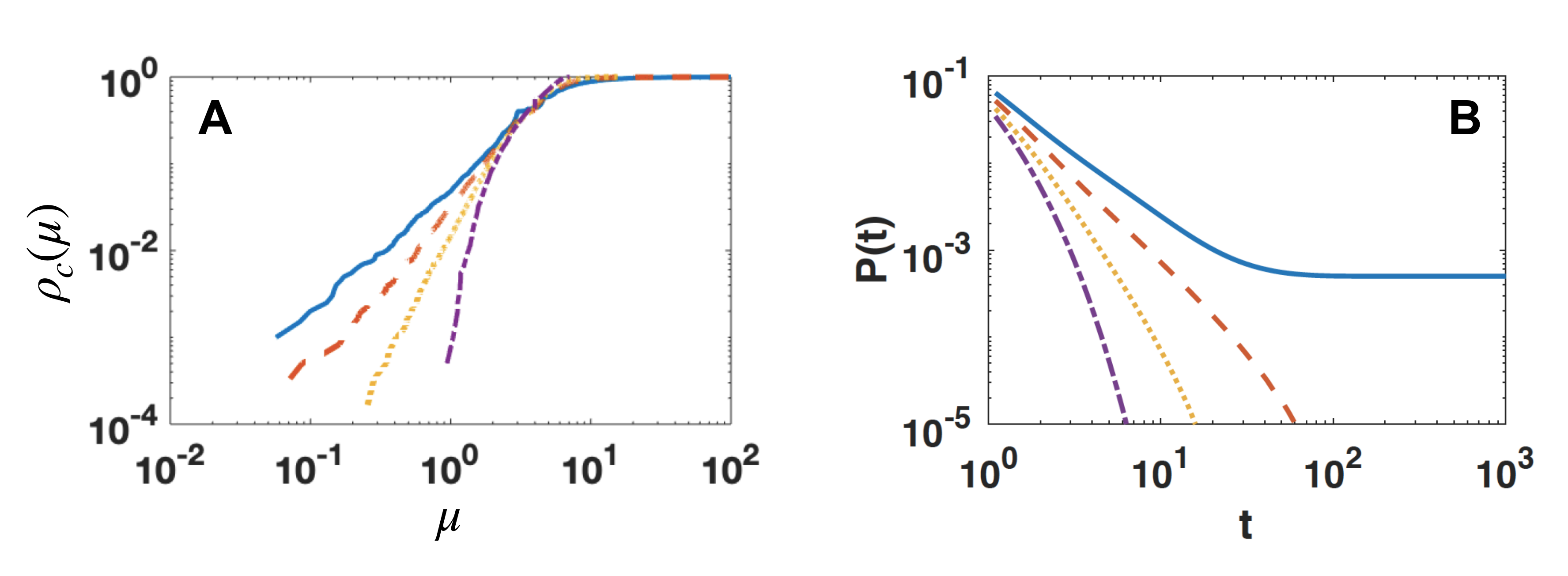}
\caption{{\bf Laplacian spectrum and return-probability in higher-order diffusion according to \citet{torres2020simplicial}.}
(A) Cumulative density of the eigenvalues of the combinatorial
  Laplacian of order $k$, with $k=0$, standard Laplacian (blue solid line),
  $k=1$ (red dashed line), $k=2$ (yellow dotted line) and $k=3$ (purple
  dot-dashed line) for a symplicial complex with 2000 nodes
  generated by means of the NGF model with $d=3$ and flavour
  $s=-1$ (see Sec.~\ref{subsec:models:outofequilibrium:simplicial}).
  (B) Return-probability for the same system. Figures adapted from 
  Ref.~\cite{torres2020simplicial}.}
\label{figGinestra}
\end{figure*}

A simplicial complex can be studied at different orders $k \ge 0$,
since for each
order we can define a $k$-Laplacian with its spectrum
and consequently its own diffusive dynamics. As a consequence, the
same simplicial complex can sustain different types of diffusion,
depending on the order $k$, or in other words, depending on the
dimension of the simplices over which the diffusion is defined.
The idea is to indicate as $x_{\sigma}(t)$ the concentration,  
at time $t$, at the generic simplex $\sigma$ of order $k$, and to 
consider the following set of coupled dynamical equations:  
\begin{equation}
\dot x_{\sigma}(t) = -\sum_{\sigma' \in X_k} (L_k^{\rm D})_{\sigma \sigma'} x_{\sigma'}(t)
\label{diff_Ginestra}
\end{equation}
where  $L_k^{\rm D}$ is the combinatorial
Laplacian based on the upper and lower matrices of dimension
$k$ that we have introduced in Sec.~\ref{subsec:ho-laplacian}
\cite{muhammad2006control}.  
Notice that we have $N_k$ of these equations where $N_k = | X_k |$, 
and $X_k$ represents the set of all
simplices of dimension $k$ in the simplicial complex under study. 
The system in Eq.~\eqref{diff_Ginestra} generalizes the one in 
Eq.~\eqref{eq_diff} and reduces to the latter when $k=0$. 
In this more general setting the substance that is diffusing
is not bounded to live at the nodes of a network, but depending 
on the value of $k \ge 1$, 
is located at the edges (when $k=1$), or the triangles ($k=2$),
or higher-order simplices of a HOrS, respectively. 
In analogy with Eq.~(\ref{t_sol}), the general solution to  Eq.~\eqref{diff_Ginestra} can be written as \cite{torres2020simplicial}: 
\begin{equation}
  x_{\sigma}(t) = \sum_{\alpha=1}^{N_k} e^{-\mu{\alpha} t}\phi_{\sigma}^{(\alpha)}\sum_{\sigma' \in X_k}\phi_{\sigma'}^{(\alpha)} x_{\sigma'}(0)
\end{equation}
where $\mu{\alpha}$ and $\boldsymbol{\phi}^{(\alpha)}$ are respectively
the eigenvalues and eigenvectors of the Laplacian $L_k^{\rm D}$.
The authors of Ref.~\cite{muhammad2006control} have  
estimated how fast the system relaxes to equilibrium. If 
$\bd x(t)$ indicates the vector whose $N_k$ entries
represent the concentrations at time $t$ 
at the $N_k$ simplices of order $k$, the bounds they have found, 
when the systems is started at $\bd{x}(0)$, 
read:  
\begin{equation}
\frac{1}{\mu_2}|| L_k^{\rm D} \bd x(t)||\leq ||\bd{x}^{(\infty)} - \bd{x}(t) || \leq N_k \exp(-\mu_2 t) ||\bd x(0)||
\end{equation}
where $\bd{x}^{(\infty)} = \lim_{t \to \infty} \bd{x}(t)$, $\mu_2$ is the
smallest non-zero eigenvalue of  $L_k^{\rm D}$, and $||\cdot||$ is
the usual Euclidean norm.
\\
\citet{torres2020simplicial} have studied Eq.~\eqref{diff_Ginestra} on
simplicial complexes generated by the NGF model (see Sec.~\ref{subsec:models:outofequilibrium:simplicial}), and have focused on
the spectral (and thus dynamical) differences between different orders
$k$. In particular, they have generalized to HOrSs the concept of
spectral dimension. In a network, the spectral dimension is the
dimension of the network ``as seen by a diffusion process'' and is
defined from the exponent of the power-law scaling of the density
$\rho(\mu)$ of the eigenvalues of the standard Laplacian (the
network spectral density) when $\mu \ll
1$~\cite{burioni1996universal, millan2019synchronization}.
Figure~\ref{figGinestra}A shows that a similar definition can be
adopted for HOrSs, and that the spectral dimension
of simplicial complexes generated by the NGF model 
increases with the order $k$. 
Moreover, the authors of Ref.~\cite{torres2020simplicial} have found
an analytical relation between the spectral density and the
return-time probability, reported in Fig.~\ref{figGinestra}B. 
The latter is defined in HOrSs as the
probability that, starting from an initial state which is
localized on a given simplex $\sigma$, the diffusion process comes
back to the
simplex $\sigma$ at time $t$, averaged over all simplices $\sigma$. 
Such a relation strengthens the link between spectral dimensions 
and the dynamical behaviour of high-order diffusion processes
on HOrSs. 


\subsubsection{Edge-flows}
\label{sect:diff2}

The special case $k=1$, the so-called edge-based Laplacian, is a
particularly important case of the $k$-th order Laplacian $L_k^D$ we
have discussed above. Edge-flows turn in fact very useful in many
contexts, ranging from graph-based machine learning to signal
analysis.

In graph signal processing, although the
basic approach is to consider the signals at the nodes of a graph, an
edge-based approach becomes important when we need to analyze a
flow (of mass, energy, information, traffic, etc.). In such cases, a
vertex-based analysis cannot take into account notions like the
orientation of flows, which is instead considered in the edge
Laplacian by the sign of the entries. 
The natural generalization has been studied by Schaub and Segarra, who have used the
spectrum of the combinatorial Laplacian $L_1^{\rm D}$ to decompose the
space of edge-flows into harmonic and gradient flows~\cite{schaub2018flow}.
Within this framework the authors have been able to address the problem of
denoising and smoothing of flow signals by means of a series of filters that
enforce approximate flow-conservation in the processed signals. They 
also show an application of their methods to denoise vehicular 
traffic flows in street networks.

\begin{figure*}
\centering
\includegraphics[width=0.6\textwidth, keepaspectratio = true]{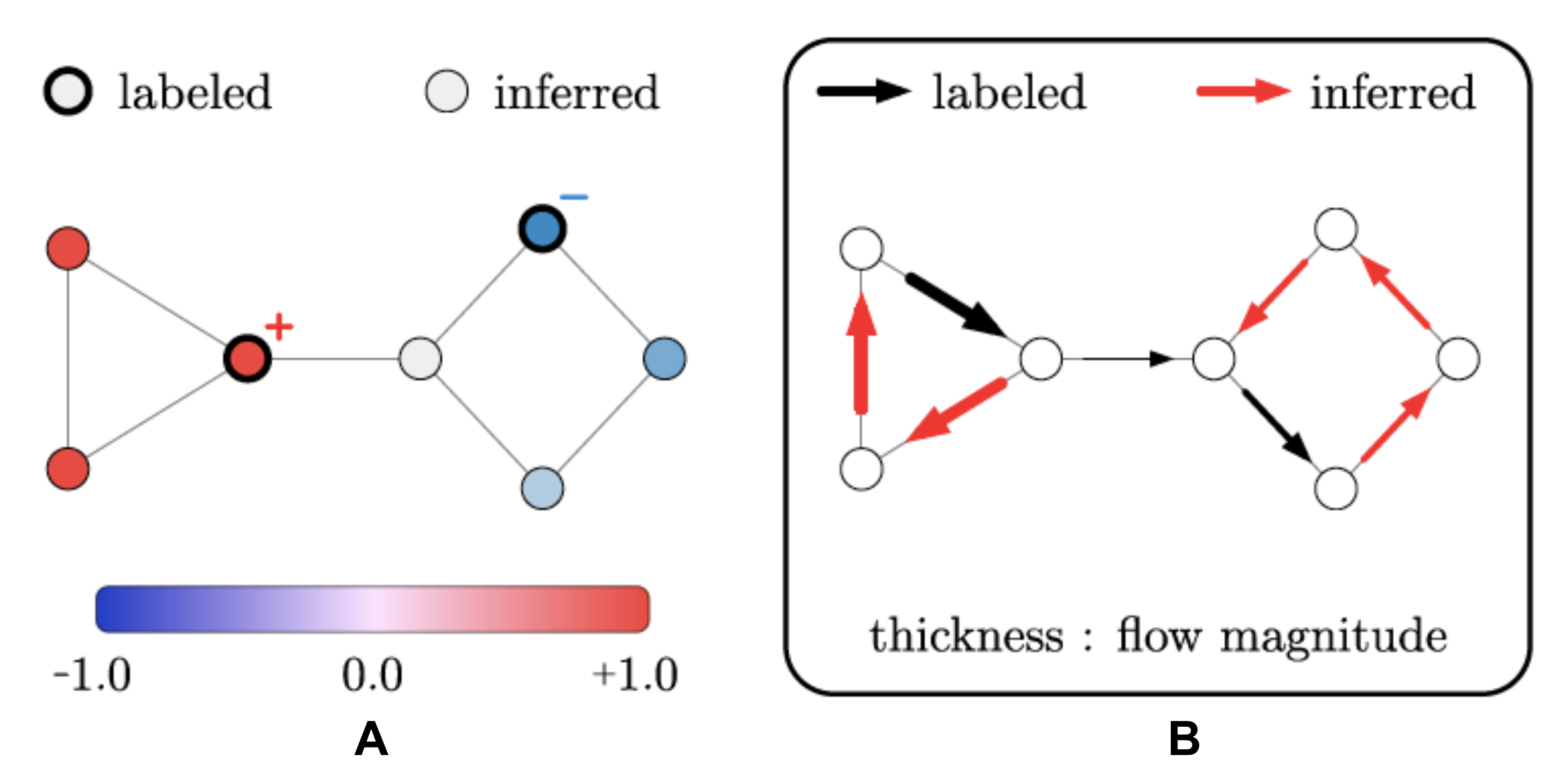}
\caption[]{{\bf Semi-supervised learning: vertex and edge perspective.} (A) In the standard graph-based semi-supervised learning, 
the structure of data points is encoded in a similarity graph, 
where each node is a data sample and the edges represent the similarity
between pairs of nodes. (B) In the semi-supervised learning for
edge-flows proposed in Ref.~\cite{jia2019graph} data points are
instead assigned and inferred on the edges of a graph. 
Figures reproduced from Ref.~\cite{jia2019graph}.
}
\label{figJia}
\end{figure*}

The authors of Ref.~\cite{jia2019graph} have instead considered the
problem of semi-supervised learning (SSL) for edge-flows in networks.
Given a graph $G(V,E)$
edge-flows are defined by a set of real-valued functions $f: V \times V \rightarrow
\mathbb{R}$, such that:
\begin{equation*}
f(i,j) = \begin{cases}
- f(j,i) \hspace{2mm} \forall (i,j) \in E\\
0 \text{ otherwise.}
\end{cases}
\end{equation*}
The authors make use of the edge-Laplacian and focus on
divergence-free flows (i.e. cases in which the flow is approximately
conserved) in order to define a process where, given a set of labeled
edge-flows it is possible to infer the unlabeled edge-flows. In
particular they study how to select the fraction of most informative
edges in order to accurately infer the remaining ones.
Figure~\ref{figJia} illustrates the relation between a vertex-based and
the edge-based SSL. 
An application of the method to real-world street networks proves its
superiority with respect to traditional alternatives.

\subsection{Higher-order random walks}
\label{sec:rw}

\subsubsection{Random walks on simplicial complexes}

An example of higher-order random walk in which the walkers populate
the edges instead of the nodes has been proposed by Schaub et
al.~\cite{schaub2020random}.  The Laplacian $ L^{\rm D}_1$ defined in 
Sec.~\ref{subsec:ho-laplacian} is characterized by positive and
negative entries that depend on the edge orientations. Therefore, differently from the standard random walk Laplacian, the non-diagonal entries do not reflect the transition probabilities among nodes. 
It is however possible to define a normalized
variant $ L^{\rm RW}_1$ of the Laplacian $ L^{\rm D}_1$ which can be
related to an edge-based random walk process. The idea is to consider
a random walk in a higher dimensional lifted state space.  In other
words, instead of considering $ L^{\rm D}_1$ applied to an edge-flow
function $f$ of the co-chain space $C^1$, Schaub et al. propose a
sequence of three operations: the lifting of $f$ into a higher
dimensional space, the action of a linear operator (playing the role
fo $ L^{\rm RW}_1$) and the projection back down to the original state
space. The linear transformation in the lifted space can be normalized
so as to represent a random walk. This procedure allows to interpret
the components of the flow in the following way: the magnitude
represents the volume of the flow, while the sign indicates the
orientation, which can be aligned or anti-aligned with the chosen
reference orientation.
The higher space where the co-chain vector is lifted up has double the
size of ${C}^1$ since both possible orientations for each edge are
present.
The effect of the Laplacian can be split in two contributions,
taking into account connections at the upper and lower order
respectively. At each time step, the walker takes a
step via either the upper or the lower adjacency connections, with
equal probability.  If the lower connection space is chosen, the
walker has probability 1/2 of moving along the reference edge
orientation and 1/2 of moving against. The transition probability
towards a target edge is, in each case, proportional to the upper
degree (or weight) of the edge.  If instead the walker makes a step
via the upper adjacency connections, there are two cases: if the edge
has no upper adjacent faces, the walker stays at the same edge and can
change orientation with probability 1/2; if instead the edge has upper
adjacent faces, then the walker will transition to an upper adjacent
edge with a different orientation with respect to their shared
face. In other words, the walker performs a random walk with adjacency
matrix $ A^u$, unless there is no upper adjacent connection, in which
case the walker can stay put or move to the edge with reverse
orientation.  It is important to observe that the eigenvectors of $
L^{\rm RW}_1$ relative to the eigenvalue $\mu=0$ are associated 
to harmonic functions. Schaub et al. propose an application
of their random walk trajectory embedding and
simplicial PageRank \cite{schaub2020random}.

\bigskip
A second example of a higher-order random walk in which the walkers
live instead on the triangles of a HOrS has been proposed by Mukherjee
et al.~\cite{mukherjee2016random}. The authors have considered a
special case of a simplicial complex where every edge is contained in
at most two triangles. At each time step $t$, a walker at a
triangle $\sigma$ remain still with probability 1/2,
or otherwise move to a triangle on the
other side of one of the three edges of $\sigma$. 
In other words, at each time step, the walker can move to a
triangle that is lower adjacent to the current triangle. 
The transition matrix $\Pi = \{ \pi_{\sigma \sigma'} \}$ of the walker reads: 
\begin{equation}
\pi_{\sigma \sigma'} =\begin{cases}
\frac{1}{2} \text{ if } \sigma = \sigma' \\
\frac{1}{6} \text{ if } \sigma \text{ and } \sigma' \text{ share an edge}\\
0 \text{ otherwise}.
\end{cases}
\end{equation}
and is possible to show that it can be expressed in terms of the
higher order Laplacian $L_2^{\rm D}$ as $\Pi = I - L_2^{\rm D}/6$.


\subsubsection{Random walks on hypergraphs}

The hypergraph formalism is simple enough to allow to consider random
walks on hypergraphs of arbitrarily large order. 

\begin{figure*}
\centering
\includegraphics[width=1\textwidth, keepaspectratio = true]{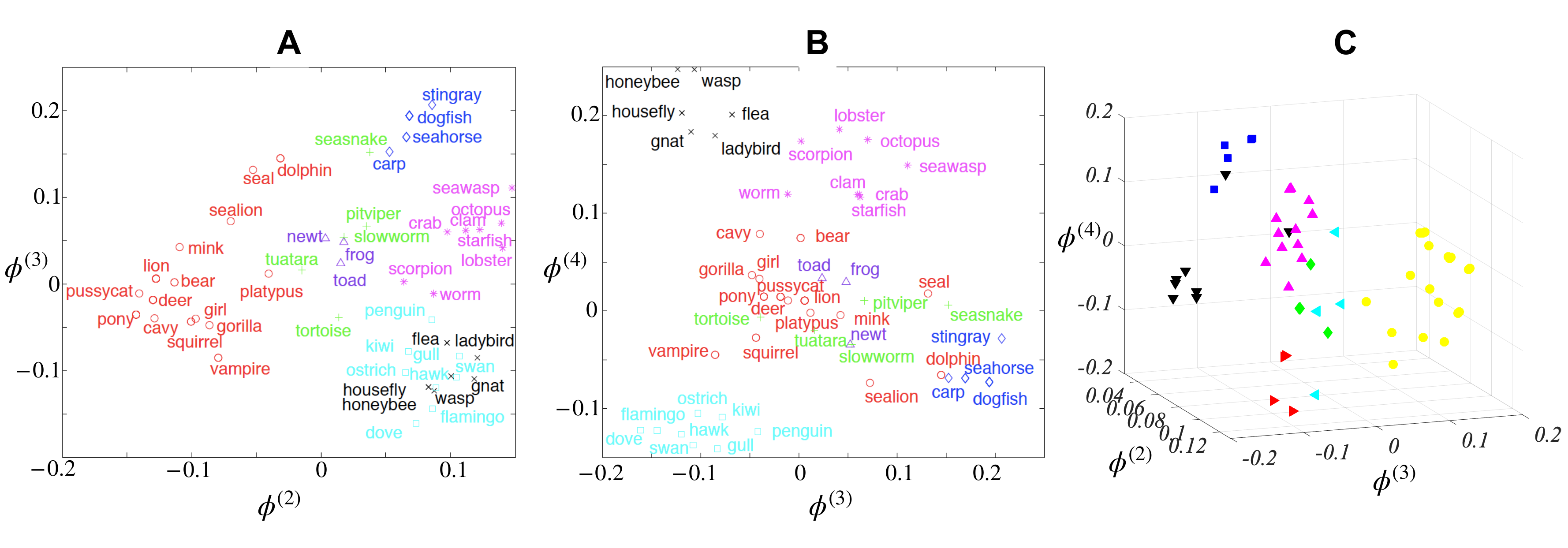}
\caption[]{{\bf Classification methods based on random walks on hypergraph.}
  The performance of both  \citet{zhou2007learning} and
  \citet{carletti2020random} models of random walk on hypergraphs is tested 
  on a classification task performed on a zoo dataset. Reported are the
  embeddings of the nodes of the hypergraph in a Euclidean space
  built from the Laplacian eigenvectors. Different symbol colors and shapes
  represent different animal classes. (A,B) Results obtained with
  eigenvectors corresponding to the 2nd and 3rd smallest
  eigenvalues, and the 3rd and 4th smallest eigenvalues respectively,
  in the model by Zhou at al. (C) Embedding based the eigenvectors
  corresponding to the three smallest eigenvalues  in the model by
  Carletti et al. Figures adapted from Ref.~\cite{zhou2007learning} and
  Ref.~\cite{carletti2020random}.
  }
\label{fig_zoo}
\end{figure*}

A first basic model of random walk on hypergraphs has been proposed by Zhou
et al. \cite{zhou2007learning}. In their implementation, the walker
selects one of the hyperedges of the current node, and then chooses to
move to one of the nodes of the selected hyperedge with a uniform
probability. In principle, hyperedges can be weighted according to
different criteria. However,
in their numerical
experiments the authors adopt the case in which each hyperedge  is assigned a weight equal to 1.
Zhou et al. use their random walk model to perform a classification
task. They make use of a dataset \cite{dua2019UCI}, in which the animals
of a zoo are associated to a set of features (tail, hair, legs and so on),
and build a hypergraph where each animal is a node and two or more nodes are
in the same hyperedge if the corresponding animals have a 
common feature. The eigenvectors associated to the smallest non-zero
eigenvalues  $\mu_{\alpha}$  of the random walk Laplacian 
on this hypergraph are then used in order to embed each animal
in a two-dimensional Euclidean space.
Figures~\ref{fig_zoo}A,B show the
classification performance of the method when
the 2nd and 3rd smallest
eigenvalues, $\mu_{2}$ and $\mu_3$, 
or the 3rd and 4th smallest eigenvalues, are used
respectively. 

\bigskip

\begin{figure*}
\centering
\includegraphics[width=1\textwidth, keepaspectratio = true]{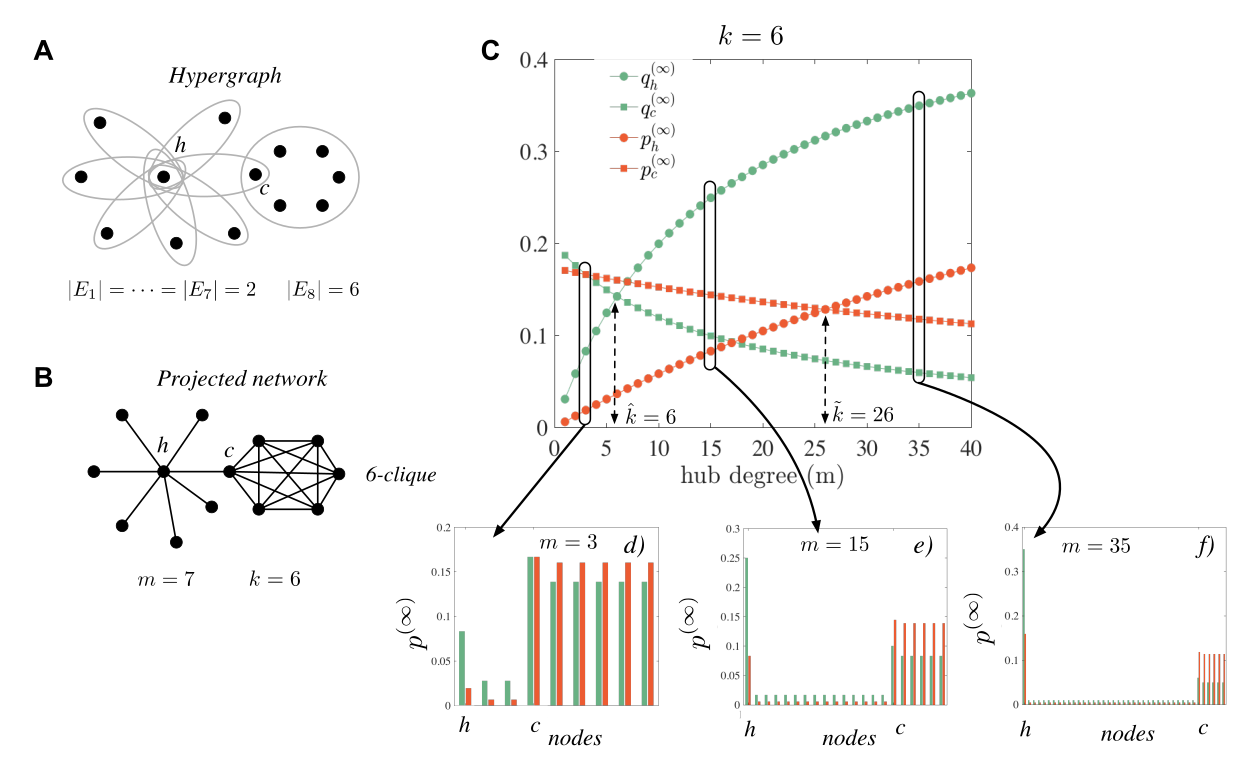}
\caption[]{{\bf Example of random walk on hypergraphs.} (A) An
  hypergraph with $m=7$ hyperedges of size $k=2$ and one hyperedge of
  size $k=6$, and (B) its corresponding projected network. (C) 
  Probability of finding the walker on node $h$ (circles) and $c$
  (squares) for a random walk on the
  hypergraph (red) and on the projected network (green), and for different
  size $m$ of the hub. Figure reproduced from
  Ref.~\cite{carletti2020random}.  }
\label{figCarletti}
\end{figure*}

Carletti et al. have instead proposed a random walk model in which the
walkers spend more time inside larger communities
\cite{carletti2020random}. In practice, this is obtained by
considering the basic model of Ref~\cite{zhou2007learning} and
assigning to each hyperedge a weight that is proportional to its
size. In this way, the transition probability reflects the interplay
between the walker's willing to explore the network and the
attractiveness of large hyperedges. The authors of
Ref.~\cite{carletti2020random} have provided an analytical description
of the process and of its stationary state. They make use of a
different Laplacian operator suitably defined to generalize the
standard random walk, and which reduces to the traditional pairwise
Laplacian when all hyperedges are of size 2 and the hypergraph reduces
to a graph.
The results obtained for this random walk on a given hypergraph are compared
with those obtained by a traditional random walk on a projected network,
where the hyperedges of the original hypergraph are transformed into
cliques. This is illustrated with the example reported in
Fig.~\ref{figCarletti}. The hypergraph in Fig.~\ref{figCarletti}A is composed
of one hyperedge of size $k$, where one of the nodes,
denoted as $c$, is also connected to a hub node, $h$, which is at the
center of $m$ hyperedges of size 2.
The corresponding projected network is shown in Fig.~\ref{figCarletti}B.
The difference between the HOrS and its projection clearly emerges from 
the node rankings based on the stationary probability distribution
of the walkers in the two cases. In the case of the projected
network, the equilibrium distribution $\bd{q}^{(\infty)}$ 
is proportional to the degree of the nodes. This means that
$q^{(\infty)}_h \propto m$ and $q^{(\infty)}_c \propto k$,
which implies that the hub node $h$ is the top node in the
ranking when $m>k$. 
The stationary distribution of the random walk on the hypergraph, denoted as
$\bd{p}^{(\infty)}$,  gives instead $p^{(\infty)}_h \sim
m$ and $p^{(\infty)}_c \sim 1+(k-1)^2$\cite{carletti2020random}.
Consequently, the hub node $h$ results the node with the highest
occupation probability only when $m > 1+(k-1)^2$. 
In general, at fixed hyperedge order $k$, the
two processes provide the same ranking 
for $m<k+1$ or $m>1+(k-1)^2$. 
Conversely, at intermediate
values of $m$, the hub $h$ is the top node in the 
projected network, while  node $c$ is  
ranking first in the hypergraph. The inversion is graphically
shown in Figs.~\ref{figCarletti}C-F.
The authors of Ref.~\cite{carletti2020random} have proposed concrete
applications of their random walk model to rank nodes in large systems. For
instance they have produced a ranked list of  
scientists based on the hypernetwork representing 
co-authorship relations in published articles (see Sec.~\ref{subsec:application:social}
for scientific collaborations as HOrSs) . 
The same model has also been used for classification task.
Figure~\ref{fig_zoo}C shows the three-dimensional embedding (defined by the
eigenvectors associated to the first three eigenvalues) obtained
for the zoo data set. 


\bigskip

Chitra and Raphael~\cite{chitra2019random} go beyond the previous approaches,  
and propose a model of random walks on hypergraphs with edge-dependent
vertex weights. Namely, they consider a transition probability assigning
different weights even to nodes belonging to the same hyperedge. In their
random walk, a walker at node $v_j$ at time $t$ move to a node $v_i$ chosen
in the following way. 
First, a hyperedge $\sigma$ is selected among all the hyperedges of $v_j$ 
with probability $\omega(\sigma)/ d(v_j)$,
where $\omega(\sigma)$ is the weight of the hyperedge and 
and $d(v_j)$ is the degree of $v_j$.
Then a node $v_i$ is selected from hyperedge $\sigma$
with a probability $\gamma_{\sigma} (v_i)/ \delta(\sigma)$, where
$\gamma_{\sigma} (v_i)$ is the weight of vertex $v_i$ and
$\delta(\sigma)= \sum_{v_i \in \sigma} \gamma_{\sigma} (v_i)$ 
is the degree of the hyperedge $\sigma$.  
The novelty of this random walk with respect to previous models on 
hypergraphs with edge-independent vertex weights consists in the second
step, where the probability is not uniform over all the vertices of $\sigma$. 
Chitra and Raphael claim that for a random walk on hypergraphs with
edge-independent weights is always possible to find a choice of edge
weights such that the process on the hypergraph is equivalent to a
traditional random walk on the corresponding projected network with
such chosen weights. The proof of equivalence is based on the
time-reversibility of the process, which is a typical property of the
associated Markov chain. Such property cannot be extended to
hypergraphs with edge-dependent weights, which happen to be not
time-reversible. Therefore, random walks on hypergraphs with edge-dependent
vertex weights cannot be reduced to traditional
random walks on weighted networks.
This generalizes a result already found in~\cite{agarwal2006higher} for
$k$-uniform hypergraphs.
\\
Notice that, without using edge-dependent weights, another way to
obtain a higher order random walk which is not reducible to the
traditional model is, similarly to what happens for diffusion, to
insert a non-linearity in the equations, as it has been done by
\citet{chan2018spectral}, and by Li and Milenkovic in 
\cite{li2017inhomogeneous}, and
\cite{li2018submodular}.
\\
In the model by Chitra and Raphael the stationary state $\bd{p}^{(\infty)}$, 
can be analytically computed. If for the traditional processes this can be written as
by $p^{(\infty)}_i = \rho \sum_{\sigma \in E(v_i)} \omega(\sigma)$, with $E(v_i)$ the set
of hyperedges containing $v_i$, the process proposed by Chitra and
Raphael brings to $p^{(\infty)}_i = \sum_{\sigma \in E(v_i)} \rho_{\sigma} \omega(\sigma)
\gamma_{\sigma}(v_i)$, i.e. the proportionality constant $\rho$ depends on the
hyperedge and each term in the sum is multiplied by the vertex weight.
\\
By making use of this definition the authors also provide an
application example, again based on scientific collaborations.
They obtain the ranking of different
scientists according to a hypergraph where hyperedges represent
articles and the authors have weights which reflect their appearance
order (first/last or middle authors), and they compare such ranking
to that obtained by a random walk on the corresponding edge-independent
hypergraph. 
Other applications of edge-dependent hypergraphs include e-commerce
\cite{li2018tail}, text ranking \cite{bellaachia2013random}, image
visualisation and processing
\cite{ding2010interactive,huang2010image,ducournau2014random,zeng2016learn,zhang2018dynamic}.\\


Random walks on hypergraphs have also been used to study graph
expansion. Louis in~\cite{louis2015hypergraph} introduces a non-linear
Laplacian operator on hypergraphs as a generalization of the random
walk Laplacian operator on graphs and studies its spectrum. He in
particular proves that the second smallest eigenvalue is related to
the expansion of the hypergraph, generalizing the Cheeger's
inequality.  A similar process is analysed by Chan et
al~\cite{chan2015spectral} considering a diffusive flow from the node
with maximum density to the one with minimum density within a
hyperedge. The two works are merged together
in~\cite{chan2018spectral}. A generalization of this process is
provided in~\cite{chan2019generalizing}, while in
\cite{chan2019diffusion} the framework is extended to directed
hypergraphs. Li and Milenkovic~\cite{li2018submodular} answer the same
questions on a different kind of higher order network, named
submodular hypergraph, for which they define a different Laplacian and
analyze its spectrum. 

Higher order random walks have also been used for applications such
as node ranking~\cite{bellaachia2013random}, community structure
detection~\cite{billings2019simplex2vec}, topological data
analysis~\cite{salnikov2018simplicial}, machine
learning~\cite{tran2015combinatorial, satchidanand2015extended}, and
even quantum walks have been extended beyond pairwise
connections~\cite{liu2018quantum}.
The traditional definition of
cover time in random walk, i.e. the maximum expected time to visit all
the vertices of a graph, has been extended to hypergraphs in
\cite{cooper2011cover}. While  \citet{avin2010radio} define a higher-order random walk suitably designed to describe communication over a wireless shared channel, which is not well captured by a traditional network. 

Finally, another interesting extension would be to consider
reaction-diffusion systems based simplicial complexes or
hypergraphs, which historically represent
the basic processes for pattern
formation~\cite{turing1952chemical,nakao2010turing,asllani2014theory,cencetti2018pattern}. For instance, higher-order
interactions have been considered in~\cite{harush2017dynamic}, where
the cumulative response of nodes to a specific signal is investigated
by means of the flow patterns stemming from the interplay between
topology and dynamics in a coupled reaction system.


\section{Synchronization}
\label{sec:synchronization}

Synchronization is the emergence of
order in populations of two or more coupled dynamical systems.
It shows up in many physical, biological and social systems, and at
different scales, with typical examples including the synchronized motion
of weakly coupled pendulum clocks~\cite{huygens1986pendulum}, the clapping of an audience~\cite{neda2000sound} or
the flashing of fireflies~\cite{buck1988synchronous}. Synchronization has been an active research topic in the last decades with successful applications ranging from
neuroscience to climatology and engineering~\cite{boccaletti2002synchro,pikovsky2003synchronization,strogatz2004sync,boccaletti2018synchronization}.
Interactions play a key role in the emergence of synchronization, which is why
network science provides natural and powerful tools to inquire about
the nature and the underlying mechanisms of synchronization.
In synchronization on networks, each node of a graph is a dynamical system,
and its dynamics is influenced by its neighbours through pairwise interactions. 
Synchronization occurs when the interactions are such
that all, or a macroscopic fraction of, the  
oscillators reach a coherent state. Historically, this was made 
clear by Kuramoto and his mathematically tractable model of all-to-all
coupled phase oscillators~\cite{kuramoto1984chemical}.
The Kuramoto model paved the way for others to study
the effects of complex
topologies~\cite{acebron2005kuramoto,arenas2008synchronization,rodrigues2016kuramoto}. Key results have 
shed light on the relationship between network
synchronizability and topology, e.g. the improved synchronizability of
small-world networks~\cite{barahona2002synchronization}, and have 
revealed different routes to synchronization, such as abrupt 
synchronization in scale-free networks~\cite{gomez2011explosive,boccaletti2016explosive}. In
addition, different types of synchronized behaviors have been identified 
and linked to specific properties of the network structure,
including remote synchronization~\cite{nicosia2013remote}, cluster states~\cite{pecora2014cluster}, chimera~\cite{abrams2004chimera} and
Bellerophon~\cite{bi2016coexistence} states.
The existence and properties of these phenomena depend
on the type of interactions, but also on the topology of
the network. See Refs. \cite{arenas2008synchronization,rodrigues2016kuramoto,boccaletti2018synchronization} for comprehensive reviews.

In this Section, we discuss how coupled dynamical systems can be
extended to higher-order systems (HOrSs), and how the presence of
high-order interactions can affect synchronization in both phase
oscillators and nonlinear dynamical systems.  
As we will show below, higher-order interactions can change the nature of
the transition to synchronization, can favor the emergence of certain
collective phenomena, e.g. cluster states, and can even 
give rise to new dynamical regimes.

\subsection{Phase oscillators}
\label{subsec:phase_oscllators}

In this section, we report on studies that investigate synchronization
in populations of phase oscillators: the most basic oscillatory unit which is fully described by a phase.

\subsubsection{Higher-order Kuramoto model}

The Kuramoto model captures the essence of the
emergence of synchronization in a mathematically tractable
setting~\cite{kuramoto1984chemical}. In the original 
model, the state of oscillator $i$ ($i=1,\ldots,N$) at time $t$
is described by its phase $\theta_i(t) \in [0,2 \pi[$. The dynamics 
of each oscillator is governed by its interactions with all other
oscillators (all-to-all interactions) according to:
\begin{equation}
\dot \theta_i = \omega_i + \frac{K_1}{N} \sum_{j=1}^{N} \sin (\theta_j - \theta_i) ,
\label{eq:kuramoto_original}
\end{equation}
where $\omega_i$ denotes the natural frequency associated to $i$, drawn
from a given distribution $g(\omega)$, and $K_1>0$ is the (pairwise)
coupling constant. In this model, two
opposing driving forces are at play: heterogeneity in the natural
frequencies pushes the oscillators away from synchronization, while
interactions favor synchronization. As a result of this, we
observe a phase transition.
Above a critical value, $K_1^*$, of the coupling strength, the 
oscillators synchronize their frequency, so that their phase difference
does not change in time (this is known as phase locking).
This phase transition from incoherence to synchronization can be
captured by the usual complex order parameter $Z_1(t) = R_1(t) e^{i \Phi_1 (t) } =
\frac{1}{N} \sum_{j=1}^{N} e^{i \theta_j(t)}$, a macroscopic quantity that 
characterizes the collective dynamics of the entire system. 
The modulus $0 \le R_1(t) \le 1$ measures the
phase coherence, while $0 \le \Phi (t) < 2 \pi$ is the average phase
of the whole population of oscillators.
When $K_1<K_1^*$, the oscillators behave incoherently and $R_1 \approx
0$. When $K_1$ is above the critical value, instead, the oscillators
synchronize and $R_1 \neq 0$. As $K_1$ tends to large values, $R_1$ tends
to 1, corresponding to all oscillators having exactly identical phases,
a case that can be obtained for identical oscillators, i.e. when $\omega_i =
\omega~ \forall i$~\cite{watanabe1993integrability, watanabe1994constants}.
To make the driving of the mean field explicit, the 
system in Eq.~\eqref{eq:kuramoto_original} can be rewritten as: 
\begin{equation}
\dot \theta_i = \omega_i + K_1 R_1 \sin (\Phi_1 - \theta_i) .
\label{eq:kuramoto_meanfield}
\end{equation}
About ten years ago, Ott and Antonsen proved that the dynamics of the 
system in Eq.~\eqref{eq:kuramoto_meanfield} reduces to a stable low-dimensional
synchronization manifold, by applying self-consistency arguments to
the evolution of the distribution of the oscillator 
phases~\cite{ott2008low}.

The Kuramoto model can be extended from the original all-to-all
interactions to the case in which the oscillators are the nodes of a
network and their couplings are governed by the adjacency matrix
$A = \{a_{ij}\}$ of the network. Formally, this is equivalent to consider
the equations:  
\begin{equation}
\dot \theta_i = \omega_i + \frac{K_1}{N} \sum_{j=1}^{N} a_{ij} \sin (\theta_j - \theta_i) .
\label{eq:kuramoto_complex}
\end{equation}
The Kuramoto model has been studied on different types of complex
networks, and particular attention has been devoted to investigate how
the structural properties of a network affect synchronization.  A
review of the main results obtained can be found in
Refs.~\cite{acebron2005kuramoto,arenas2008synchronization}.  In what
follows, we will review more recent works that extend the Kuramoto
model to higher-order networks, starting with a short detour on
synchronization on motifs.

\begin{figure}[t]
	\centering 
	\begingroup
	\setlength{\tabcolsep}{12pt}
	\renewcommand{\arraystretch}{1.5}	
	\begin{tabular}{l| ccc}
		\toprule 
		\textbf{Motif} & 	\raisebox{-.2\height}{\includegraphics[width=1cm]{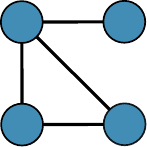}} & 
		\raisebox{-.25\height}{\includegraphics[width=1cm]{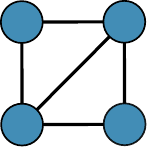}} &
		\raisebox{-.25\height}{\includegraphics[width=1cm]{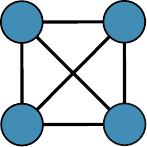}}  \\
		\textbf{\# links} & 4 & 5 & 6 \\
		\textbf{$\tilde K_1$} & 0.22 & 0.18 & 0.14 \\
		\bottomrule
	\end{tabular}
	\endgroup
	\caption{\textbf{Critical coupling and motifs.} In motifs, i.e. small graphs recurrent in various
		biological, social and technological networks, 
		the critical coupling strength $\tilde K_1$ for synchronization
		decreases as the number of links increases. Figure adapted from Ref.~\cite{moreno2004fitness}.}
	\label{tab:moreno2004fitness_1}
\end{figure}

\bigskip
Motifs are small graphs that appear in a statistically significant
way in real networks~\cite{alon2007network}. This is why studying 
synchronization in motifs can help understanding the dynamics of
large networks and is also the first step towards an explicit
treatment of synchronization in a HOrS. 
\\
A first study by Moreno et al.~\cite{moreno2004fitness} investigated
if some motifs of Kuramoto oscillators are more synchronizable
than others. For each undirected small graph of $N=3$ and 4 nodes,
the authors evaluated its probability to synchronize by
considering many random realizations of the set of natural frequencies 
$\{ \omega_i \}_{i=1}^N$ drawn from a given distribution $g(w)$, and computing
the fraction of realizations for which the graph synchronizes. 
As expected, the probability of synchronization increases with 
the coupling strength $K_1$. Additionally, they have
evaluated a threshold coupling strength $\tilde K_1$, 
as the value of
$K_1$ above which the probability of synchronization is greater than
0.5. The main result of the study is that, at fixed
number of nodes, the larger the number of links is,
the lower the value of $\tilde K_1$, i.e. 
the easier it is for the graph to synchronize. This is shown in 
Fig.~\ref{tab:moreno2004fitness_1} for graphs with four nodes. 
The authors suggest this could help understand why some motifs, which 
appear in biological networks and are more conserved across evolution,
have a higher link density than others.

In a following study, D'Huys et al. have investigated the influence of
time delays in the synchronization of motifs~\cite{dhuys2008synchronization}. 
The authors studied the existence and stability of several types of
synchronized behaviors, in unidirectionally and bidirectionally
coupled small rings, pairs of oscillators, and open chains. The work 
showed that delays tend to induce multistability in
unidirectional rings, in contrast to what happens in 
bidirectional rings.

\begin{figure}[t]
	\centering
	\includegraphics[width=0.6\linewidth]{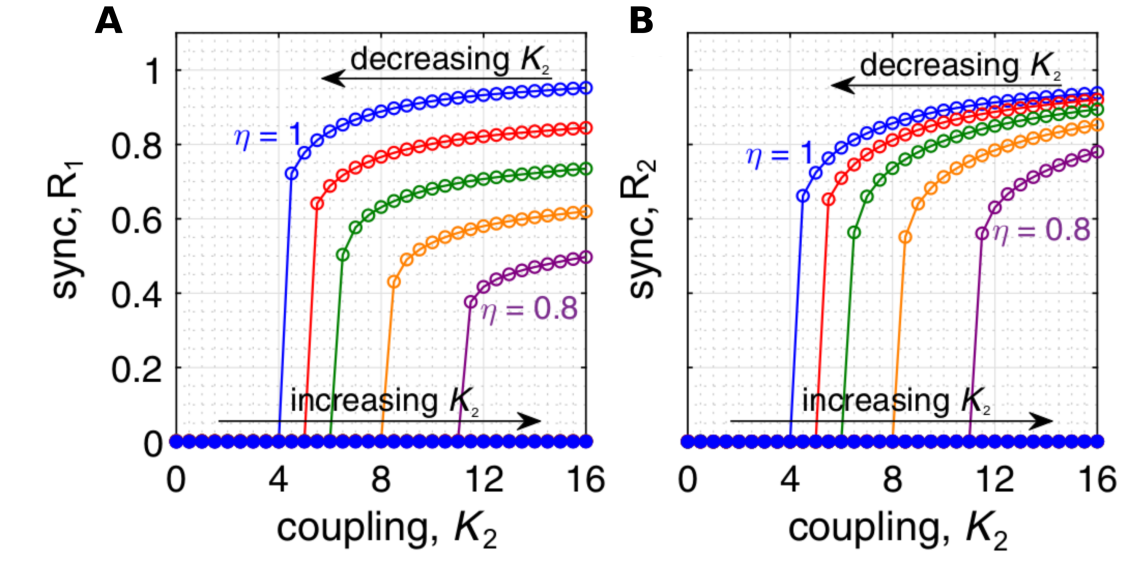}
	\caption{\textbf{Abrupt desynchronization induced by higher-order
			interactions in the model in Eq.~\eqref{eq:skardal_system}.}
		The two order parameters $R_1$ and $R_2$ are shown as a function
		of the three-body coupling strength $K_2$, in (A) and (B) respectively.   
		The system exhibits
		multistability, and each stable branch represents a
		two-cluster state with a proportion of $\eta$ oscillators in
		the first cluster. Figures adapted from Ref.~\cite{skardal2019abrupt}. }
	\label{fig:skardal2019_1}
\end{figure}

\bigskip
More recent works have studied variations of the Kuramoto model that 
explicitly include higher-order interactions.
Skardal and Arenas~\cite{skardal2019abrupt} have investigated what
happens when all-to-all pairwise interactions are replaced
by pure three-body interactions, considering the system: 
\begin{equation}
\dot \theta_i = \omega_i + \frac{K_2}{N^2} \sum_{j=1}^{N} \sum_{k=1}^{N} \sin (\theta_j + \theta_k - 2 \theta_i) ,
\label{eq:skardal_system}
\end{equation}
which is a direct generalization of the original Kuramoto
model in Eq.~\eqref{eq:kuramoto_original}. 
Now, the dynamics is ruled by interactions of all possible 
triplets of oscillators $(i,j,k)$ in the system (corresponding to
a hypergraph with all 3-node hyperedges). The 
value of the coupling strength $K_2$ tunes the strength of such
interactions. Notice that the number of three-body interactions 
node $i$ is involved in, scales as $N^2$, which explains the presence
of the normalization factor $1/N^2$ to ensure a smooth
thermodynamic limit. The system in 
Eq.~\eqref{eq:skardal_system} can be cast into the following form:  
\begin{equation}
\dot \theta_i = \omega_i + K_2 R_1^2 \sin [2(\Phi_1 - \theta_i)] ,
\label{eq:skardal_meanfield}
\end{equation}
where $R_1(t)$ and $\Phi_1(t)$ are, as in Eq.~\eqref{eq:kuramoto_meanfield},
the modulus and the phase of the complex order parameter $Z_1(t)$.
By arguments of continuity and self-consistency of the oscillator
density, the authors
then derived analytical formulae for the dynamics of 
$Z(t)$ and of a second order parameter $Z_2(t) = R_2(t) e^{i \Phi_2
	(t) } = \frac{1}{N} \sum_{j=1}^{N} e^{i 2 \theta_j(t)}$,
which is a typical indicator of 2-cluster states. Indeed, $R_2 \approx 0$
for incoherent states, while $R_2 \approx 1$ for 2-cluster states when the 
two clusters have a phase difference equal to $\pi$.
Two main novel dynamical phenomena were identified. Firstly, the
three-body interactions give rise to an abrupt desynchronization
transition, which is not present in the classical Kuramoto model of
Eq.~\eqref{eq:kuramoto_original}. In other words, as shown in
Fig.~\ref{fig:skardal2019_1}, as the coupling
strength $K_2$ is decreased from large values, the order parameter $R_1$
drops from positive values to zero. On the other hand, increasing
$K_2$ back does not yield a transition to coherence since the
incoherent state is stable for any coupling strength. Second, the
three-body interactions yield multistability: for large enough values
of the coupling strength $K_2$, there exist infinitely many 
stable coherent branches that correspond to 2-cluster solutions. Each of
these stable 2-cluster solutions can be distinguished by the
relative size $\eta$ of the two clusters, i.e. the relative number of
oscillators in the two clusters.
Interestingly, each of the stable
branches undergoes an abrupt desynchronization transition at a
different value of $K_2$, so that there is actually a continuum of
transitions. Very recently, Xu and coworkers have studied the same
system using bifurcation theory and the symmetries of the $SO_2$ group
to reveal scaling properties
of the transitions~\cite{xu2020bifurcation}. The authors show that their analysis 
generalizes to
a coupling scheme where natural frequencies are correlated to the coupling strength.

\begin{figure}[t]
	\centering
	\includegraphics[width=0.5\linewidth]{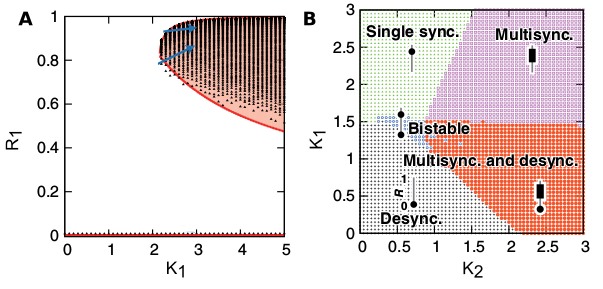}
	\caption{\textbf{Phase diagrams of the model with higher-order interactions in Eq.~\eqref{eq:tanaka_system}. }
		(A) Case $K_1=0$ when only three-body interactions are present. (B)
		General case with both two- and three-body interactions. Vertical black lines represent the interval
		$[0,1]$ of values the order parameter $R_1$ can take. On these lines,
		black circles and boxes represent the value of $R_1$ for all coexisting
		stable states. Figures adapted from Ref.~\cite{tanaka2011multistable}.}
	\label{fig:tanaka2011_3}
\end{figure}

That higher-order interactions favor multistability is a
general trend, and this is also in agreement with the results of some earlier
studies~\cite{tanaka2011multistable,ashwin2016hopf, komarov2015finitesizeinduced}. Already in 2015, Komarov and Pikovsky~\cite{komarov2015finitesizeinduced} studied Eq.~\eqref{eq:skardal_system} in great detail for identical and distributed frequencies, with or without common noise. The main focus of their study is different to Ref.~\cite{skardal2019abrupt}: here, the authors show that while the incoherent state $R_1=0$ is always stable in the thermodynamic limit, finite-size fluctuations can induce a transition to synchrony $0< R_1 \le 1$. The authors explained this transition by showing that, remarkably, the order parameter scales as $R_1 \sim \sqrt{N}$ with the size of the system, which vanishes in the thermodynamic limit. Even earlier, in 2011, motivated by the
importance of many-body interactions in signal transmission between
neurons, Tanaka and Ayogi~\cite{tanaka2011multistable}
studied a system similar to the one in Eq.~\eqref{eq:skardal_system}. 
They have considered a coupling function of
the form $\sin(\theta_j - \theta_i) \cos (\theta_k - \theta_i)$, which 
indeed corresponds to the first of the two terms to which the original 
coupling function in Eq.~\eqref{eq:skardal_system} reduces when using the 
trigonometric identity for the sine of a sum~\cite{tanaka2011multistable}.
As shown in Fig.~\ref{fig:tanaka2011_3}A, the model exhibits
multistability similar to that in Fig.~\ref{fig:skardal2019_1}.
Additionally, Tanaka and Ayogi have studied
the effect of having, at the same time, both all 2- and all 3-body interactions,
with respective coupling strength $K_1$ and $K_2$:  
\begin{equation}
\dot \theta_i = \omega_i +  \frac{K_1}{N} \sum_{j=1}^{N} \sin(\theta_j - \theta_i) + \frac{2 K_2}{N^2} \sum_{j=1}^{N} \sum_{k=1}^{N} \sin(\theta_j - \theta_i) \cos (\theta_k - \theta_i) .
\label{eq:tanaka_system}
\end{equation}
In our language this means considering all possible 2-simplices.   
The phase diagram of this more general model is reported in
Fig.~\ref{fig:tanaka2011_3}B and shows the existence of different
regimes, many of them exhibiting multiple coexisting stable
states. For example, in the region indicated as ``Bistable'', both the
incoherent state with $R\approx0$ and the synchronized state with
$R\approx1$ are stable.

\begin{figure}[t]
	\centering
	\includegraphics[width=0.8\linewidth]{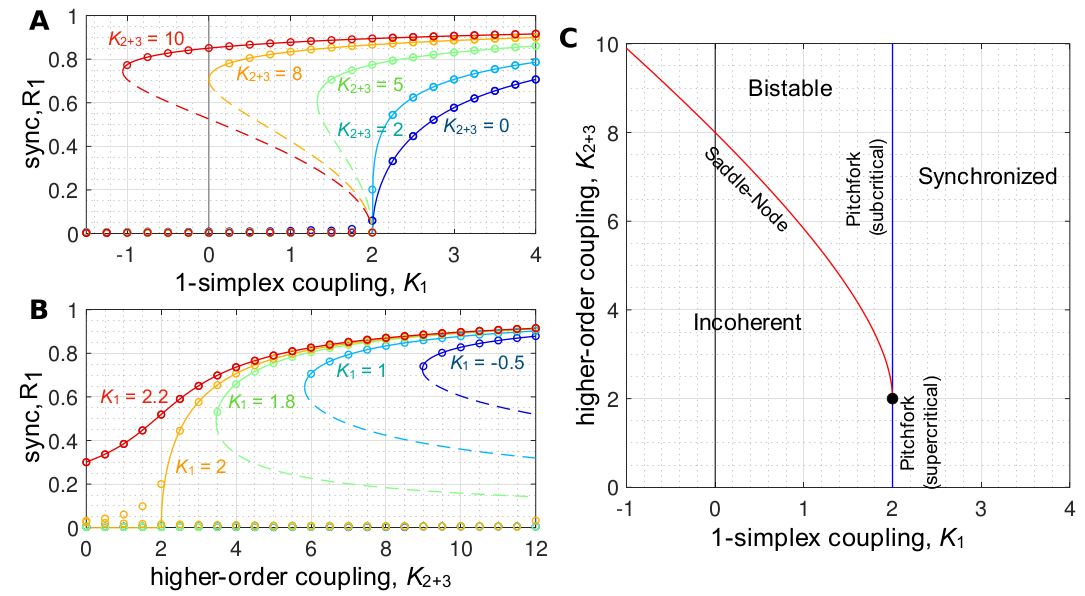}
	\caption{
		\textbf{Phase diagrams of the simplicial complex oscillator model of
			Ref.~\cite{skardal2019higher}.}
		(A) Higher-order interactions induce
		abrupt transitions between incoherence and 
		synchronization. (B) Moreover, higher-order interactions can
		stabilise the synchronized state, even when pairwise (1-simplex)
		interactions are repulsive ($K_1 <0)$. (C) Two-parameter bifurcation diagram
		identifying the region of bistability. Analytical predictions
		(solid curves) are in agreement with the numerical simulations (open
		circles). Figures adapted from Ref.~\cite{skardal2019higher}.
	}
	\label{fig:skardal2019b2ac}
\end{figure}

In a second study, Skardal and Arenas added two layers of complexity
to the previous models: (\textit{i}) a combination of interactions of different orders up to
four-body interactions,(\textit{ii}) a microscopic description of such interactions
in the form of a simplicial complex \cite{skardal2019higher}.    
They showed numerically that higher-order
interactions were sufficient to induce an explosive transition from
incoherence to synchronization in a real-world higher-order system.  
In practice, they have used a Macaque brain data set with 248 nodes 
and pairwise connections in which any 3-node clique was promoted
into a 2-simplex, and any 4-node clique into a 3-simplex.
We refer to simplicial complexes constructed
this way as ``maximal'', because from $a_{ij}$ all possible
$q$-simplices are constructed. A maximal simplicial complex 
cannot thus have, e.g. any empty triangle (three 1-simplices), but 
only filled triangles (three 1-simplices and one
2-simplex). The resulting network has 1-, 2-, and 3-simplex
interactions, with the 2-simplex interactions term for oscillator 
$\theta_i$ written as
\begin{equation}
\frac{K_2}{2! \langle k^{(2)} \rangle} \sum_{j=1}^N  \sum_{k=1}^N a_{ijk} \sin (2 \theta_j - \theta_k - \theta_i) ,
\label{eq:skardal2019_higher_system}
\end{equation}
and where the 1-
and 3-simplex interaction terms have similar structure. Note the choice of an asymmetrical
coupling function in Eq.~\eqref{eq:skardal2019_higher_system} which is different from 
the symmetrical choice in Eq.~\eqref{eq:skardal_system}. This will be discussed in more details in 
the next paragraph. Here,
$\bd{A}=\{a_{ijk}\}$ denotes the 2-simplex interaction tensor with
$b_{ijk}$ equal to 1 if there is a three-body interaction among
oscillators $i$,$j$ and $k$, and 0 otherwise.
The $\langle k^{(2)} \rangle$ represents the average degree of
2-simplex degree, i.e. the average number of distinct 2-simplices
nodes are part of. In general, the average $q$-simplex degree is
written $\langle k^{(q)} \rangle$, and the coupling coefficient $K_q$ is
rescaled by $q! \langle k^{(q)} \rangle$.
The complexity of the model makes it hardly tractable analytically. However,
by means of self-consistency arguments similar to those used in Ref.~\cite{skardal2019abrupt},
the authors were able to obtain a closed equation for the order
parameter $R_1$ in a simplified all-to-all version of the model.
They obtained the bifurcation diagram reported
in Fig.~\ref{fig:skardal2019b2ac}, which shows the following 
two main findings. Firstly, 
higher-order interaction are able to induce an abrupt transition from
incoherence to synchronization. Indeed, above a critical value of
$K_{2+3} = K_2 + K_3 = 2$, the system becomes bistable and exhibits a
hysteresis cycle, yielding abrupt transitions. Since the
incoherent and synchronized states have been linked to resting and
active states, respectively, such abrupt transitions provide a
potential mechanism for fast switching between brain states. Second,
strong enough higher-order interactions (large enough values of $K_{2+3}$) can
stabilise the synchronized state even when the 1-simplex interactions
are repulsive, i.e. when $K_1<0$. Notice that 
explosive synchronization was first observed in global coupling with
evenly spaced frenquencies~\cite{pazo2005thermodynamic} and then in
scale-free networks with frequency-degree
correlations~\cite{gomez2011explosive}. Several other structural or
dynamical ingredients have then been identified that can also lead to 
explosive synchronization, such as multilayer couplings \cite{nicosia2017intertwined}
or time delays \cite{dsouza2019explosive}. Here, Skardal and Arenas have been
able to show that higher-order interactions are sufficient to induce
explosive synchronization. Explosive synchronization and 
chimera states have also been studied in simplicial complexes by 
Berec in Refs.~\cite{berec2016chimera, berec2016explosive}.

The choice of coupling function is known to affect the dynamics of coupled
dynamical systems~\cite{stankovski2015coupling}. In coupled oscillators, 
a typical choice at order 1 is the sinusoidal function $\sin(\theta_j - \theta_i)$
that vanishes when oscillators are synchronized and that is $2\pi$-periodic in its phases, 
as in the Kuramoto 
model of Eq.~\eqref{eq:kuramoto_original}. For interactions of more than two oscillators, 
however, there are more choices of functions that satisfy those requirements. Indeed, 
even if we restrict ourselves to the simplest case,
i.e. $2\pi$-periodic sinusoids with no harmonics and no phase-shift,
we are faced with two possibilities for the functional form accounting
for three-body interactions in the equation for oscillator $i$,
namely $\sin (\theta_j + \theta_k - 2 \theta_i)$, or $\sin (2 \theta_j
- \theta_k - \theta_i)$. The first choice was used for example in Eq.~\eqref{eq:skardal_system}. 
This function is the natural generalization of the pairwise function above, in that it is 
symmetric in $i$, meaning that it is invariant under any permutation of the other indices.
This choice was also made in Ref.~\cite{lucas2020multiorder} where it was generalized to all 
possible orders. The second choice, which was used for example in Eq.~\eqref{eq:skardal2019_higher_system}, 
is asymmetric, in the sense described above. It can, however, arise naturally from the phase reduction 
of nonlinear oscillators, see for example Eq.~\eqref{eq:ashwin_hopf} and Refs.~\cite{leon2019phase}. For coupling functions
of larger numbers of oscillators, there is always one symmetric choice, but
the number of asymmetric choices increases.  However, to the
best of our knowledge, the implications of the precise choice of
higher-order coupling functions on the dynamics have not been
investigated systematically as deserved so far. We hope this review will encourage
works in this direction.

\begin{figure}[t]
	\centering
	\includegraphics[width=0.5\linewidth]{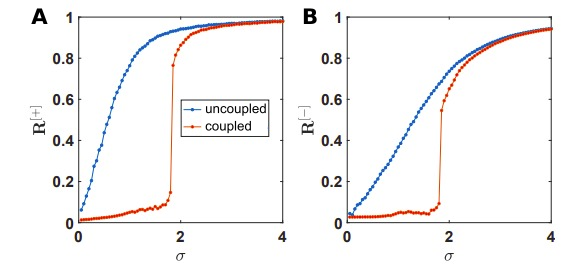}
	\caption{\textbf{Explosive synchronization in the higher-order Kuramoto model.}
		In Ref.~\cite{millan2019explosive}, the oscillators are
		associated not to the nodes, but to the $q$-simplices
		(with $q\ge1$) of a simplicial complex. 
		$R^{[+]}$ and $R^{[-]}$ denote the order parameter of
		the dynamics projected on $(q-1)$- and $(q+1)$-simplices,
		respectively shown in (A) and (B). The model has been implemented on 
		simplicial complexes constructed by a configuration model
		(see Sec.~\ref{sec:models}). When the
		projections are uncoupled the transition is
		continuous, while when the projections are adaptively coupled, 
		the transition is explosive.
		Figures adapted from Ref.~\cite{millan2019explosive}.
	}
	\label{fig:ginestra20192}
\end{figure}

A radically different approach from those discussed above has been
proposed by Mill\'{a}n et al., who have formulated a higher-order
extension of the Kuramoto model in which the oscillators are placed
not only on the nodes but also on the higher-order simplices, such as the links
or the triangles, of a simplicial complex~\cite{millan2019explosive}.
The authors showed that the dynamics defined on $q$-simplices can be 
projected, through Hodge decomposition (also see Sec.~\ref{subsubsec:combinatorial-laplacians}), on the dynamics defined on $(q-1)$ and $(q+1)$
dimensional simplices. This means, for instance, that the dynamics on
edges can be projected on nodes and triangles. Interestingly, when 
the two projected dynamics on $(q-1)$- and $(q+1)$-simplices are
adaptively coupled the transition to synchronization is explosive,  
while it is continuous in the uncoupled case. This implies for
instance that a dynamics defined on links can induce a simultaneous
explosive synchronization on the dynamics projected on nodes and
triangles. The phenomenon is illustrated in Fig.~\ref{fig:ginestra20192},
which reports the results of numerical simulations on a simplicial complex
generated by a configuration model (see Sec.~\ref{sec:models}).

\bigskip 
The three authors of Ref.~\cite{millan2019explosive} have
also studied synchronization in so-called Complex Network
Manifolds (CNMs), which are growing simplicial complexes of dimension
$d$, i.e. built with $d$-simplices
~\cite{millan2018complex,millan2019synchronization}
(see Section \ref{subsec:models:outofequilibrium:simplicial}).
Note that simplicial complexes were only used here to  
construct the adjacency matrix $\{ a_{ij} \}$ of networks with an
easily tunable spectral dimension,
in addition to being small-world and having a highly modular
structure.
Indeed the goal of these studies was to study the effect of 
the spectral dimension of the underlying graph on network
synchronization. The spectral dimension of a graph describes
the power-law scaling of the eigenvalue density of the standard Laplacian $L$.  
Most complex networks have a spectral gap, i.e. the second eigenvalue
$\lambda_2$ of the standard Laplacian $L$ does not tend to zero as the
size of the network is increased. In CNMs, however, $\lambda_2$ does
tend to zero, so that a spectral dimension $d_S$ of the network can be
defined. Remarkably, in CNMs, the eigenvalue density scales
roughly with the dimension $d$ of the simplicial complex,
$d_S \simeq d$. Hence, one can investigate the effect of $d_S$
just by changing the dimension of the simplicial complex used to
construct the network. 
In~\cite{millan2018complex}, the authors have reported the observation
of frustrated synchronization, a dynamical regime in which the order
parameter is non-stationary and exhibits large fluctuations even
at large times. In~\cite{millan2019synchronization}, they have further
shown via a linear approximation that the synchronized state is
thermodynamically stable only in networks with $d_S > 4$.
In other words, for smaller spectral 
dimensions, the average fluctuations of the phases diverge as the
network size $N \to \infty$. Hence, frustrated
synchronization is only possible for $d_S < 4$.
Additionally, $d_S > 4$ is a necessary condition for a stable
synchronized state in the thermodynamic limit, as also
confirmed numerically.

\bigskip
Higher-order interactions can yield interesting dynamics even when the
oscillators have identical frequencies, as shown by two very recent
studies~\cite{gong2019low, lucas2020multiorder}. 
\\
In the first of these works~\cite{gong2019low}, Gong and Pikovsky have considered 
the following sytem of all-to-all coupled oscillators
with identical frequencies and higher-order harmonics:   
\begin{equation}
\theta_i = \omega(t) + \Im [H(t) e^{-iq \theta_j} ] ,
\end{equation}
with pure harmonics of order $q \ge 2$. 
The quantity $H(t)$ in the equations represents the coupling and
depends on the generalized complex order parameters $Z_q(t) = R_q(t)
e^{i \Phi_q (t) } = \frac{1}{N} \sum_{j=1}^{N} e^{i q \theta_j(t)}$.
The model reduces to the original Kuramoto model in
Eq.~\eqref{eq:kuramoto_original} when $H = Z_1$ and only the first
harmonic $q=1$ is considered. With $H = Z_2$ and only the second order
harmonic $q=2$, one obtains a coupling function of the type $\sin(2
(\theta_j - \theta_i))$.  This does not imply higher-order
interactions, which can instead be obtained by considering nonlinear
meanfield couplings, i.e. powers of $Z_q$. For example, by taking $H =
Z^2_1$ and $q=2$, one recovers the system in
Eq.~\eqref{eq:skardal_system} with pure three-body interactions.
Higher-order harmonics and higher-order interactions are linked in
that they both allow $q$-cluster states at order $q$, for
example. Equation \eqref{eq:skardal_meanfield} can also be seen as
driven by the meanfield with a second order harmonic. \\
In their work, the authors extend the Watanabe-Strogatz theory to
account for any pure higher-order \textit{harmonics}, $q\ge2$, for a
general form of $H$ that can include higher-order interactions. The
Watanabe-Strogatz theory provides a lower-dimensional description of
all-to-all coupled oscillators with identical
frequencies~\cite{watanabe1993integrability,watanabe1994constants},
similarly to the Ott-Antonsen theory for distributed
frequencies~\cite{ott2008low}. As an example, Gong and Pikovsky have applied 
their theory to the cases of pure 3-body and pure 5-body interactions,
and have been able to study the basin of attractions of the
3- and 5-cluster states, respectively. It is worth mentioning here 
three other earlier but related studies by Pikovsky and coworkers~\cite{rosenblum2007self,
	pikovsky2009self, burylko2011desynchronization}. In 
Refs.~\cite{rosenblum2007self,pikovsky2009self}, had already considered the effect of 
nonlinear mean field coupling, i.e. where $H$ is a function of powers of $Z_1$, 
as opposed to only the first power as in the Kuramoto model~\eqref{eq:kuramoto_original}.
The authors found that this additional nonlinearity yields richer dynamics, including self-organized
quasiperiodic behavior. Similarly, in Ref.~\cite{burylko2011desynchronization}, 
the authors consider coupling functions
that depend nonlinearly on $R_1$.

In the second of these works~\cite{lucas2020multiorder}, Lucas et al. have focused directly on
higher-order interactions rather than harmonics, and proposed a natural
generalization of the usual Laplacian formalism to account for complex
topologies (instead of all-to-all) and with a mix of orders. 
They introduced a multi-order Laplacian and applied it to simplicial
complexes (though the formalism is valid for hypergraphs in general) 
of identical phase oscillators. This Laplacian allows to assess the
stability of the fully synchronized state $\theta_i(t)= \theta^S(t)= \omega t$ for all $i$. 
Formally, the authors considered the following system: 
\begin{equation}
\begin{aligned}
\dot \theta_i = \omega &+ \frac{K_1}{\langle k^{(1)} \rangle} \sum_{j=1}^N a_{ij} \sin(\theta_j - \theta_i) 
+ \frac{K_2}{2! \langle k^{(2)} \rangle} \sum_{j,k=1}^N a_{ijk}  \sin(\theta_j + \theta_k - 2 \theta_i) \\ 
& + \frac{K_3}{3! \langle k^{(3)} \rangle} \sum_{j,k,l=1}^N a_{ijkl}  \,  \sin (\theta_j + \theta_k + \theta_l - 3 \theta_i) 
+ ... \\
& + \frac{K_{q_{\text{max}}}}{{q_{\text{max}}}! \langle k^{({q_{\text{max}}})} \rangle} \sum_{j_2, \ldots, j_{{q_{\text{max}}}+1} =1}^N \, \, \, a_{ij_2 \dots j_{q_{\text{max}}+1}} \sin \left( \sum_{m = 2}^{{q_{\text{max}}}+1} \theta_{j_m} -  {q_{\text{max}}} \, \theta_i \right) ,
\end{aligned}
\label{eq:lucas_system}
\end{equation}
which is a natural extension of the Kuramoto model with all oscillators
having identical frequencies $\omega$, and $K_q$ denoting the  
coupling strength of $q$-simplex interactions.
Here, the complex topology, which is as in
Ref.~\cite{skardal2019higher} remains mathematically tractable
because oscillators have identical frequencies. All cliques in the graph 
defined by the adjacency tensors ${\bf A}^{(q)} = \{ a_{ij_2 \dots j_{q+1}} \}$ with $q$ indices: for example, at order $q=2$, $a_{ijk}=1$ if there is
a triplet interaction $(i,j,k)$ and 0 otherwise.
Coupling functions at
each order are also a natural generalization of the standard pairwise
sine function, as they are chosen to be symmetric with respect to
oscillator $i$, meaning that any permutation of the other indices
leaves the coupling function invariant. 
The authors have shown that each term in Eq.~(\ref{eq:lucas_system})  
can be rewritten in terms of a newly defined Laplacian matrix, and all
such matrices can combined into a multi-order Laplacian 
controlling the dynamics of the whole system.
The authors denote as $k_i^{(q)}$ the connectivity of order $q$,
i.e. the number of distinct $q$-simplices that node $i$ is part of, and
as ${\hat A}^{(q)} = \{\hat a_{ij}^{(q)} \}$ the adjacency matrix of order $q$, i.e. the number of
distinct $q$-simplices that the pair $(i,j)$ is part
of. These definitions recover the usual definitions for $q=1$. Note that the adjacency matrices ${ \hat A}^{(q)}$ are different objects than the adjacency tensors ${\bf A}^{(q)}$: the former have dimension 2 and can take any integer value, whereas the latter have dimension $q+1$ and only takes binary values. Infinitesimal heterogeneous perturbations $\delta \theta_i$ around the synchronized state $\theta^S(t)= \omega t$, defined as 
$\delta \theta_i = \theta_i - \omega t$, evolve according to the linearized dynamics. In a system with $q$-simplex interactions, this dynamics is determined by a  Laplacian $L^{(q)} = \{ l^{(q)}_{ij} \}$ of order $q$ defined as: 
\begin{equation}
l^{(q)}_{ij} =  q \, k^{(q)}_i \delta_{ij} - \hat a^{(q)}_{ij} ,
\label{eq:laplacian_q}
\end{equation}
which is a natural generalization of the usual pairwise Laplacian. 
For instance, in the case of pure 3-simplex
interactions, $K_q = 0$ for all $q\neq3$, the linearized system reads:
\begin{equation}
\dot{\delta \theta_i} = - \frac{K_3}{\langle k^{(3)} \rangle} \sum_{j=1}^N l^{\text{(3)}}_{ij} \delta \theta_j .
\end{equation}
The stability of the synchronized state is then measured by the second
Lyapunov exponent $\lambda_2^{(3)}$, which is proportional to the
second eigenvalue $\Lambda_2^{(3)}$ of the Laplacian $L^{\text{(3)}}$.
When interactions of all different orders are present, it is practical
to define a
\textit{multi-order Laplacian} $L^{\text{(mul)}} = \{ l^{\text{(mul)}}_{ij} \}$
as 
$l^{\text{(mul)}}_{ij} =
\frac{K_1}{\langle k_1 \rangle} l^{(1)}_{ij} + \frac{K_2}{\langle k_2
	\rangle} l^{(2)}_{ij} + \dots + \frac{K_{q_{\text{max}}}}{\langle
	k_{q_{\text{max}}} \rangle} l^{   (q_{\text{max}})   }_{ij}$.
so that the linearized
equations read:  
\begin{equation}
\dot{\delta \theta_i} = - \sum_{j=1}^N l^{\text{(mul)}}_{ij} \delta \theta_j ,
\end{equation}
and the stability of the synchronized solution can be assessed
by simply computing the second Lyapunov exponent 
$\lambda_2^{\text{(mul)}}$, which is proportional to the second
eigenvalue $\Lambda_2^{\text{(mul)}}$ of the multi-order
Laplacian $L^{\text{(mul)}}$.
\begin{figure}[t]
	\centering
	\includegraphics[width=0.9\linewidth]{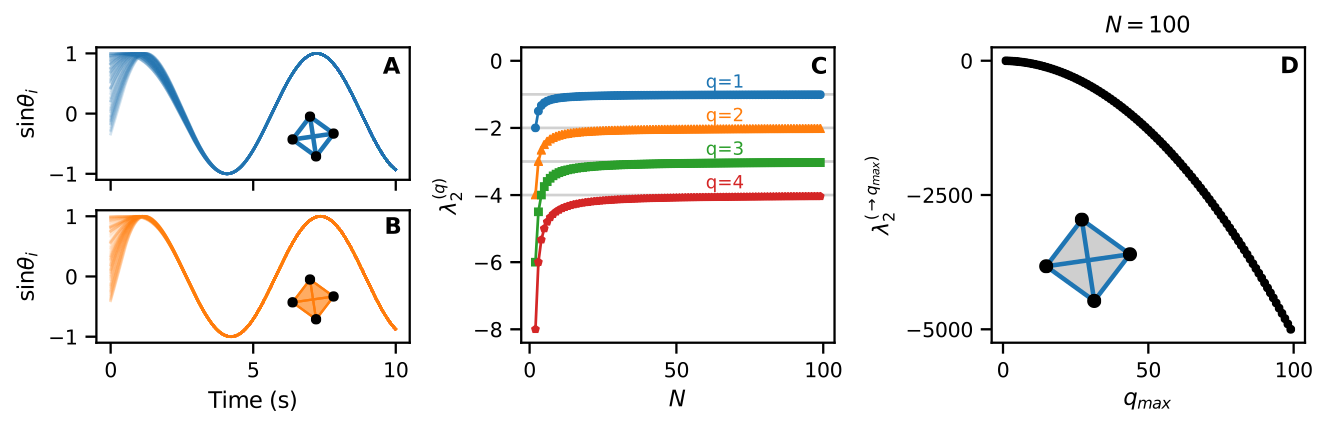}
	\caption{\textbf{Higher-order oscillator model
			of Ref.~\cite{lucas2020multiorder} in the case of all-to-all higher-order interactions.}
		The higher the order of
		interactions taken into account, the more stable the
		synchronized state. Convergence of $N=100$ oscillators with
		(A) only 1-simplex interactions and (B) only 2-simplex
		interactions. Convergence is faster in the second case. (C)
		This is confirmed by the analytical first non-zero Lyapunov
		exponent at each order $q$, which is proportional to
		$q$. Here, it is plotted against $N$. (D) Multi-order
		Lyapunov exponent more negative as $q_{\text{max}}$ is
		increased. Figures reproduced from Ref.~\cite{lucas2020multiorder}.}
	\label{fig:lucas2020fig2}
\end{figure}
The authors have then applied the multi-order Laplacian
framework to simplicial complexes of increasing complexity.
In the
case of all-to-all higher-order interactions,
where all possible interactions take
place at each order, a full analytical spectrum can be
obtained.
When only attractive couplings are
present, the spectrum indicates (\textit{i}) that the higher the order 
of pure interactions, the more these interactions stabilize 
synchronization, as shown in Figs.~\ref{fig:lucas2020fig2}A-C, and (\textit{ii}) that the more orders are
taken into account, the more stable synchronization is, 
as shown in Fig.~\ref{fig:lucas2020fig2}D.
When instead the coupling is attractive at some orders, and
repulsive at others, the interplay of the different terms
can either lead to stability or instability, which confirms and
extends a result found in Ref.~\cite{skardal2019higher}.
Decaying couplings strength have also been considered in analogy with 
higher-order phase reduction techniques discussed in the next
section. The multi-order Laplacian is also applied to other 
models of simplicial complexes, such as the simplicial star-clique
model, and real-world brain system. The multi-order Laplacian is a general tool that can be
used to investigate the effects of higher-order interactions in other
oscillatory systems.

\subsubsection{Higher-order interactions from phase reduction}
\label{sec:phase_reduction}

Real-life oscillators are often nonlinear, and their dynamics is more
complex than that of a simple phase oscillator. However, one can
obtain phase models that approximate the original dynamics by using
phase reduction
techniques~\cite{nakao2016phase,pietras2019network}. This makes phase
models very powerful since they can capture the dynamics of networks
of general nonlinear oscillators, as long as they are weakly
coupled. For example, the Kuramoto-Sakaguchi model can obtained via a
phase reduction of the mean-field complex Ginzburg-Landau
equations~\cite{leon2019phase}. The Kuramoto-Sakaguchi model is an
extension of the original Kuramoto model obtained
by the addition of a phase shift $\alpha$ in the
coupling function $\sin (\theta_j - \theta_i + \alpha)$
in Eq.~\eqref{eq:kuramoto_original}. 
However, the Kuramoto-Sakaguchi model is still a first-order 
phase approximation in the weak coupling parameter limit, and it
only displays full incoherence or synchronization
(see Refs.~\cite{watanabe1993integrability,watanabe1994constants} for the
case of identical oscillators). Other nontrivial dynamics such as
chaos, cluster states, or weak chimeras, have only been observed by
introducing different symmetries or more
harmonics in the coupling function~\cite{ashwin2016identical}. Ashwin et al. have noted
ahead of time that, even though adding harmonics does unfold
degeneracies, there might be some that will only unfold by considering
non-pairwise interactions~\cite{ashwin2016identical}.
More recent results indicate, however, that many-body interactions emerge
naturally when considering phase reductions including higher-orders
terms in the weak coupling
parameter~\cite{ashwin2016hopf,leon2019phase,matheny2019exotic}. Moreover,
these studies show that the inclusion of higher-order terms unlock
nontrivial dynamical regimes as well as transitions between them. In
this section, we will only review studies approaching phase
reduction of higher-order networks of oscillators from a theoretical
point of view. Section~\ref{sec:inference} will then be complementary to this 
section, as phase reduction will there be approached from the inverse-problem
point of view of network inference.

Ashwin, Bick, Rodrigues, and coworkers have produced a series of important
contributions for phase reduction of populations of identical
oscillators~\cite{ashwin2016hopf,bick2016chaos,bick2018heteroclinic,bick2019heteroclinic,
	bick2019heteroclinic2}. The authors of Ref.~\cite{ashwin2016identical} 
have suggested that interactions beyond pairwise might be the only
way to unfold some degeneracies and unlock nontrivial
dynamics~\cite{ashwin2016identical}. In a paper published in
the same year~\cite{ashwin2016hopf}, Aswhin 
and Rodrigues have shown that the application of phase reduction
to a systems of generic nonlinear identical systems with global
symmetric coupling yields the Kuramoto-Sakaguchi at the lowest order,
but at the next order, terms including 2-, 3-, and 4-body interactions
naturally emerge: 
\begin{equation}
\dot \theta_i = \tilde \Omega (\theta, \epsilon) 
+ \frac{\epsilon}{N} \sum_{j=1}^{N} g_2 (\theta_j - \theta_i)
+ \frac{\epsilon}{N^2} \sum_{j,k=1}^{N} g_3 (\theta_j+ \theta_k - 2 \theta_i)
+ \frac{\epsilon}{N^2} \sum_{j,k=1}^{N} g_4 (2 \theta_j+ \theta_k - \theta_i)
+ \frac{\epsilon}{N^3} \sum_{j,k,l=1}^{N} g_5 (\theta_j+ \theta_k - \theta_l - \theta_i)
\label{eq:ashwin_hopf}
\end{equation}
The inclusion of such terms allows for a wider range of
dynamical behaviors such as cluster states, and predictions that are
valid for longer timescales. Moreover, the authors have been able to
demonstrate that non-pairwise interactions yield
multistability via the coexistence of
many two-cluster states, with varying cluster size, just as
in Refs.~\cite{tanaka2011multistable,skardal2019abrupt}.
Here, the many-body interactions terms are traced back to cubic
nonlinearities in the original nonlinear system.

Bick et al. have further investigated the dynamics of the phase
reduced system of Eq.~\eqref{eq:ashwin_hopf}, looking in particular
for the possibility of chaos in small networks~\cite{bick2016chaos}.
Previously, it was 
thought that for populations of identical oscillators, higher
harmonics in the coupling functions, or nontrivial amplitude dynamics,
were necessary to observe chaos in small networks ($N=4$ is smallest
theoretical size to exhibit chaos). In fact, for pure pairwise
interactions, for $N=4$, the only known coupling function to yield
chaos has a minimum of four nontrivial harmonics (for larger networks, less
harmonics are needed)~\cite{bick2016chaotic}. Remarkably, the authors of 
Ref.~\cite{bick2016chaos} were able to show that two nontrivial
harmonics are sufficient
to see chaos, even in the case $N=4$, when
many-body interactions are considered.
\begin{figure}[t]
	\centering
	\includegraphics[width=0.4\linewidth]{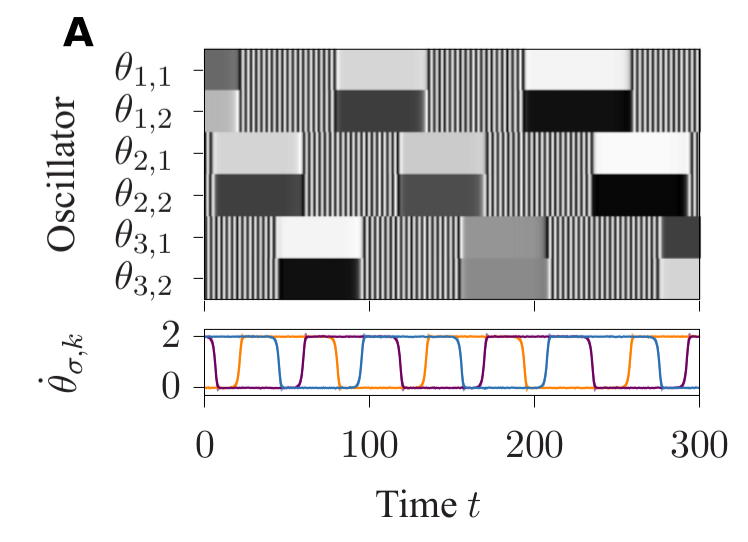}
	\caption{\textbf{Switching dynamics in the Bick model of Ref.~\cite{bick2018heteroclinic} with three populations of two oscillators each.} The oscillators in the three populations intermittently
		synchronize and desynchronize. Figure reproduced from Ref.~\cite{bick2018heteroclinic}.}
	\label{fig:bick2018_1c}
\end{figure}
In the last two years, Bick and coworkers have produced three more
pieces of work in the same direction, focusing in particular on
heteroclinic
cycles~\cite{bick2018heteroclinic,bick2019heteroclinic,bick2019heteroclinic2}.
In the first paper 
Bick has considered $M$ populations of $N$ oscillators each,
with pairwise interaction between oscillators 
in the same population, and non-pairwise interactions among 
oscillators of different populations \cite{bick2018heteroclinic}.
This setup builds on a previous work by Komarov and Pikovsky
who considered populations of oscillators with
distinct frequency and focused on a specific three-population
resonance~\cite{komarov2013dynamics}.
In Bick's setup, all oscillators
have identical frequencies, and it is the presence of
nonpairwise interactions that yields the heteroclinic
connections joining the weak chimeras.
Indeed, as shown in Fig.~\ref{fig:bick2018_1c} switching between states of
so-called localized frequency synchrony were
observed~\cite{bick2018heteroclinic}.  
Bick has further built on that paper to 
formally prove the existence of those heteroclinic cycles
in Ref.~\cite{bick2019heteroclinic}, and finally to assess their
stability together with Lohse in Ref.~\cite{bick2019heteroclinic2}.

More recently, Le\'on and Paz\'o have produced a very thorough
study of the higher-order phase reduction of the mean-field complex
Ginzburg-Landau equation (MF-CGLE)~\cite{leon2019phase}. To date, the
phase reduction of MF-CGLE was only known up to the first order, and
produces the standard Kuramoto model of
Eq.~\eqref{eq:kuramoto_original}. The Kuramoto model exhibits 
two dynamical regimes only, synchronization and
incoherence, but is unable to reproduce more exotic regimes of
the MF-CGLE, such as quasiperiodic partial synchronization and cluster
states. To remedy that, the two authors have proposed a isocron-based phase
reduction method up to the second and third order. Crucially, the
authors demonstrate the accuracy of their technique by showing that
the obtained phase model does exhibit the exotic dynamics of the
weakly coupled MF-CGLE.

\begin{figure}[t]
	\centering
	\includegraphics[width=0.6\linewidth]{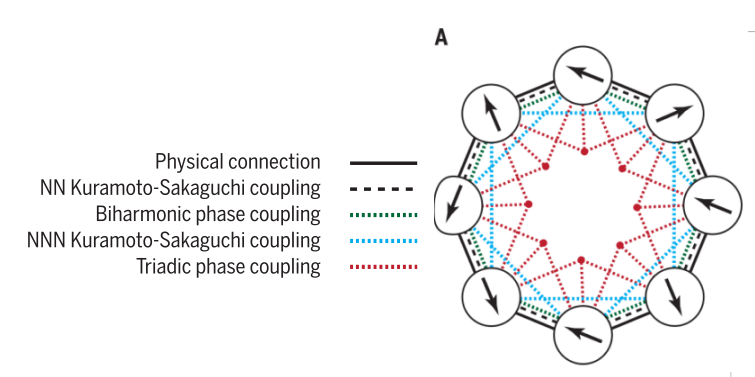}
	\caption{\textbf{Ring of eight nano-electromechanical nonlinear oscillators from Ref.~\cite{matheny2019exotic}.}
		The solid black lines represent physical connections between the oscillators. Dashed and dotted lines represent effective higher-order coupling that appear in the phase reduced model. This systems exotic complex dynamics. Figure reproduced from Ref.~\cite{matheny2019exotic}.}
	\label{fig:matheny20191a}
\end{figure}

Finally, Matheny et al. have published a very complete study combining
experiments, numerics, and analytics~\cite{matheny2019exotic}.  They
have performed experiments with a ring of eight nano-electromechanical
nonlinear oscillators, and have observed exotic and complex dynamics
including weak chimeras, decoupled states, traveling waves, and
inhomogeneous synchronized states. Then, they have obtained a phase
approximation of their initial nearest-neighbor (i.e. pairwise
interactions) network of so-called saturated oscillators, up to the
second order in the coupling strength. As in the other studies, terms
with three-body interactions naturally emerged in the phase
description, with the addition of next-nearest neighbor and biharmonic
terms as illustrated in Fig.~\ref{fig:matheny20191a}. The strength of their phase model is that it is able to qualitatively reproduce all the complex dynamics regimes observed both in the experiments and in the initial nonlinear model. 

In conclusion, phase reduction naturally yields higher-order
interactions between the phases. Progress in analytical phase
reduction up to higher-order can potentially helps us
understand the exotic and complex dynamics of network of general
nonlinear oscillators, which cannot be captured by first-order phase
models. Moreover, it will certainly be helpful in the future to the
network inference area presented in Sec.~\ref{sec:inference}. Finally,
it is also worth mentioning that some efforts has been made
to compute phase reductions numerically, when no analytical
derivation is available~\cite{rosenblum2019numerical}.

\subsection{Nonlinear oscillators}
\label{subsec:nonlinear_oscillators}

\subsubsection{Chaotic oscillators}

When chaotic oscillators are coupled, they can synchronize in various
ways. The study of chaotic synchronization started with two seminal
papers published in the
90s~\cite{pecora1990synchronization,rosenblum1996phase}. In the first
of these works, Pecora and Carroll discovered the phenomenon of
\textit{complete synchronization}, a regime in which coupled chaotic
oscillators converge to an identical state and evolve with the same
trajectory \cite{pecora1990synchronization}. 
Then, Rosenblum, Pikovsky and Kurths, showed that chaotic oscillators
can \textit{phase synchronize}, while their amplitudes vary
chaotically and are uncorrelated~\cite{rosenblum1996phase}.
The richness of the dynamics of chaotic oscillators allows for more
complex behaviors in networked systems than those observed with
phase oscillators, and has been
the object of interest in the scientific 
community~\cite{pikovsky2003synchronization}. 
However, all the studies on chaotic synchronization, with very few
exceptions~\cite{wu1998synchronization,krawiecki2014chaotic,gambuzza2020master},
have considered pairwise interactions. 
We will therefore start our discussion of high-order effects with a series
of studies related to network motifs. 

In Ref.~\cite{lodato2007synchronization} Lodato et al. have
investigated whether there can be an underlying dynamical reason
to explain the existence of network motifs. In particular they
have focused on synchronization and have evaluated analytically
the stability of the synchronous state in all directed and
undirected 3- and 4-node graphs of chaotic oscillators.
They have then compared the results with the known abundance of 
these small graphs as subgraphs (motifs) of real
biological systems. 
Interestingly, the authors were able to show that 3- and 4-node graphs
exhibiting more stable synchronous states in general coincide with
network motifs preserved across evolution, while the bifan motif, one
of the three most relevant biological motifs~\cite{alon2007network},
was not compatible with synchronization for any type of chaotic
dynamics. In another study~\cite{soriano2012synchronization}, Soriano et
al. have investigated the link between generalized synchronization and
correlation between the oscillators present in a motif, remarkably
showing that it is possible to construct small graphs of oscillators that
synchronize but at the same time do not exhibit correlations. This
result, important for network inference, stresses how indirect
connections might be systematically underestimated. Finally, it is worth
mentioning here one more study on
motifs~\cite{krishnagopal2017synchronization}. Although the
oscillators considered were not chaotic, a fractal topology  
was used. The authors were able to analytically compute the stability
of various dynamical regimes. As an example, they showed that
oscillation death was possible in that setting, even with a symmetric
coupling function.

We continue with four studies of synchronization in bipartite setups
representing two types of populations of oscillators, with
inter-population links. The first 
study considered a complete bipartite
network of nonlinear maps (chaotic or periodic)~\cite{amritkar2005synchronized}.
The existence and stability of synchronization and cluster synchronization
was assessed via Lyapunov exponents through the Master Stability Function (MSF)
formalism~\cite{pecora1998master,barahona2002synchronization,boccaletti2006complex}.
Similarly, the second study investigated chaotic synchronization in
coupled Bernoulli maps by using the same formalism~\cite{englert2011synchronization}.  
The bipartite setting was one of
the settings investigated, and the study concluded that
synchronization was not possible in that setting.
In the third study~\cite{sorrentino2007network}, Sorrentino and Ott used a bipartite setting
as a special case of multiple interacting populations of oscillators.
The authors found that adding intra-population links could enhance the stability of
so-called multisynchronous states.
Finally, bipartite networks (together with random and
tree-like networks) were one of three settings in which Pecora et
al. showed the possibility for cluster synchronization and so-called
isolated desynchronization~\cite{pecora2014cluster}.
The latter is a dynamical regime in which
one or more clusters desynchronize while the other clusters
remain synchronized.

It was as early as 1998 that the synchronization of coupled chaotic
circuits was studied by Wu on hypergraphs~\cite{wu1998synchronization}. The system under
consideration consisted of indentical chaotic circuits (the nodes of
the hyphergraph) coupled via multi-terminal resistance-devices,
effectively yielding multi-circuit interactions (the hyperlinks).
For the sake of mathematically tractability the author restricted his 
study to a case with only triplet interactions. By looking  
at the algebraic connectivity of the hypergraph, defined as the 
smallest nonzero eigenvalue of the hypergraph Laplacian,  
he was then able to derive sufficient conditions for the
complete synchronization of the circuits. 
Indeed, a large algebraic connectivity and linear but passive
coupling were required to yield synchronization.
Note that all computations rely on the hypergraph structure, even if
the system was represented as a bipartite network in the figures of
the original paper. 
A more recent study by Krawiecki has gone much further in
complexity~\cite{krawiecki2014chaotic}. In particular, the author has
considered a system of identical chaotic (Lorenz) oscillators placed
on the nodes of scale-free $q$-hypergraphs (with $q\ge 2$), i.e.
hypergraphs that exhibits only ($q-1$)-simplex 
interactions, i.e. in which each hyperedge connect exactly $q$ nodes,
and the number of hyperedges attached to a node follows a
power-law distribution. He found out that a state of complete
synchronization can be achieved and coexists with a state of oscillation
death~\cite{koseska2013oscillation}. Remarkably, the traditional
Master Stability Function formalism was generalized to
hypergraphs so as to investigate the stability of the complete
synchronization state. Furthermore, the study has reported the existence
of other dynamical regimes such as partial anti-synchronization.

In another study~\cite{wu2014synchronization}, Wu et al. have derived
analytical criteria for the synchronization
of Chua oscillators in $q$-hypergraphs. The hypergraphs have a power
law distribution of hyperdegrees such that when $q=2$, the structure
reduces to that of a Barab\'{a}si-Albert network.
Recently, Mulas et al.~\cite{mulas2020coupled} used the Master Stability formalism to derive 
stability criteria in coupled nonlinear oscillators on hypergraphs
with the particularity that they are directed. 

\begin{figure}[t]
	\centering
	\includegraphics[width=0.7\linewidth]{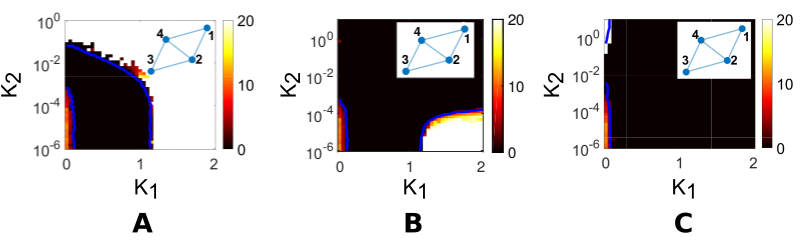}
	
	\caption{\textbf{Coupling functions affect synchronization in simplicial
			complexes of coupled chaotic oscillators as in the framework of Ref.~\cite{gambuzza2020master}.} Synchronization phase diagram of a system of
		four R\"{o}ssler systems are coupled in pairs and triplets according to
		the simplicial complex sketched. A baseline case (A) is compared to (B)
		where the pairwise coupling function is changed, and (C) where the
		triplet coupling function is changed. The predictions of the 
		Master Stability Function formalism (blue lines) are in good
		agreement with the regions of synchronization
		obtained by numerical simulations (black).
		Figures adapted from Ref.~\cite{gambuzza2020master}.}
	\label{fig:gambuzza2020master}
\end{figure}

Finally, very recently, Gambuzza et al.~\cite{gambuzza2020master} have
generalized the Master Stability Function formalism to the most general
case of simplicial complexes. They have considered a system of $N$ 
dynamical units, which are placed on the nodes of a simplicial complex of any
dimension $q_{\text{max}}$ and can be involved in $q$-simplex interactions, 
with $q=1,2,\ldots, q_{\text{max}}$, as described by the structure of the
simplicial complex.
The equations of motion of the system read: 
\begin{equation}\label{eq:general}
\begin{array}{lll}
\dot{\mathbf{x}}_i & = & {\mathbf f}(\mathbf{x}_i ) + \sigma_{1}\sum_{j_1=1}^{N} a_{ij_1}^{(1)} \: \mathbf{g}^{(1)}(\mathbf{x}_i, \mathbf{x}_{j_1}) + \sigma_{2} \sum_{j_1=1}^N \sum_{j_2=1}^N a_{ij_{1}j_{2}}^{(2)} \: \mathbf{g}^{(2)}(\mathbf{x}_i, \mathbf{x}_{j_1}, \mathbf{x}_{j_2})+\ldots\\
& & +
\sigma_{q_{\text{max}}} \sum_{j_1=1}^N ... \sum_{j_{q_{\text{max}}}=1}^N a_{ij_1....j_{q_{\text{max}}}}^{(q_{\text{max}})} \: \mathbf{g}^{(q_{\text{max}})}(\mathbf{x}_i, \mathbf{x}_{j_1}, ...,  \mathbf{x}_{j_{q_{\text{max}}}}),
\end{array}
\end{equation}
\noindent where $\mathbf{x}_i \equiv \mathbf{x}_i(t) $ is the $m$-dimensional
vector describing the state of node (dynamical unit)
$i$ at time $t$, the real valued
parameters $\sigma_{1},..., \sigma_{q_{\text{max}}}$ tune the strength of the interactions 
at the different orders $q=1,...,q_{\text{max}}$, and $a_{ij_1...j_q}^{(q)}$ are the entries
of the adjacency tensor ${\bf A}^{(q)}$ representing the structure of the
simplicial complex. Furthermore, $\mathbf{f}: \mathbb{R}^m
\longrightarrow \mathbb{R}^m$ is the most general function describing
the local dynamics, which is
assumed to be identical for all units, while $\mathbf{g}^{(q)}:
\mathbb{R}^{(q+1)\times m} \longrightarrow \mathbb{R}^m$, with
$q=1,....,q_{\text{max}}$, are the functions 
governing the interaction forms at different orders.
The authors have been able to study analytically the stability of the complete
synchronized state  $\mathbf{x}_i(t) = \mathbf{x}^S(t) ~ \forall i$,
under the only assumption that the coupling functions are
synchronization non-invasive, 
i.e. that $\mathbf{g}^{(q)}(\mathbf{x}^S, \mathbf{x}^S, ..., \mathbf{x}^S) \equiv 0 \ \forall q$. 
Based on a set of Laplacian matrices similar to those in
Eq.~\eqref{eq:laplacian_q}, they have derived a Master
Stability Function with the negativity of the maximum Lyapunov
exponent as the stability criterion.
The method has been illustrated on simplicial complexes
of coupled chaotic oscillators, such as R\"{o}ssler and Lorenz dynamical
systems, with both pairwise and triplet interactions.
When nonlinear oscillators are
coupled, the Lyapunov exponents depend nonlinearly on the eigenvalues
of the Laplacian of the system. Hence, it is possible to have
bounded regions of the parameter space where synchronization is
stable, as defined by a negative maximum Lyapunov exponent. The authors
have investigate how the region of stability depends on the structure of
the simplicial complex and on the coupling functions.
Figure~\ref{fig:gambuzza2020master} show an example of 
coupling functions for which the region of
synchronization is respectivey bounded (panel (A))
and unbounded (panels (B) and (C)), for a given node dynamics
and structure of the simplicial complex.
This framework will hopefully be
used in further studies along those lines, for various oscillators,
topologies, and coupling functions.

\subsubsection{Neuron models}

The brain provides a very rich and important terrain to study synchronization of neurons with higher-order interactions. However, so far, little is known about synchronization of neuron models in higher-order networks from the theoretical side. In the theoretical study of neuronal networks, various oscillators models are used to represent neurons, depending on the context and the goals of the modeler~\cite{dayan2001theoretical, gerstner2002spiking}. Historically, the most famous model is that of Hodgkin and Huxley, dating back from 1952 and for which they won a Nobel Prize in Physiology or Medicine. Their model describes how the neurons spike, and consists of a set of nonlinear differential equations for the membrane potential. Others models such as neural mass models have been developed since and have been used successfully to understand synchronization phenomena in the brain. Here, we report the different higher-order settings in which synchronization of neuron models has been carried out.

We start with a bipartite setting that was investigated in~\cite{bian2011adaptive}. Here the authors use that setting as a mean to study the interactions between two populations of neurons, and in fact study two coupled bipartite networks. The authors concluded, by analytical and numerical calculations, that the two networks can be synchronized with the help of adaptive feedback. 

\begin{figure}[t]
	\centering
	\includegraphics[width=0.5\linewidth]{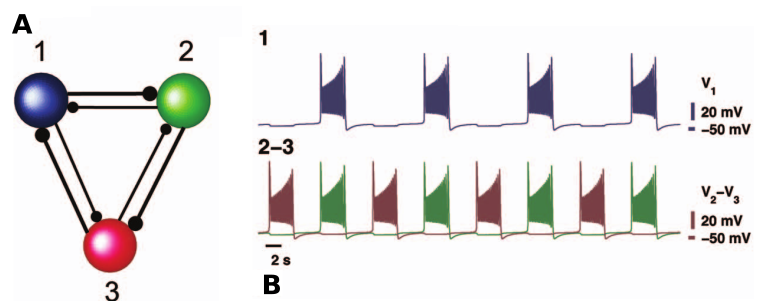}
	\caption{\textbf{Synchronization pattern in inhibitory motif from Ref.~\cite{shilnikov2008polyrhythmic}.} The pacemaker neuron (blue) is the one with the longer interburst time. Figure adapted from Ref.~\cite{shilnikov2008polyrhythmic}.}
	\label{fig:shilkinov20085}
\end{figure}

To the best of our knowledge, three studies have considered synchronization in motifs of neurons~\cite{shilnikov2008polyrhythmic,matias2011anticipated, gollo2014mechanisms}. Shilnikov and collaborators~\cite{shilnikov2008polyrhythmic} provide a detailed analysis of synchronization in motifs of both inhibitory and excitatory, and inhibitory-only neurons, see for example Fig.~\ref{fig:shilkinov20085}. The authors showed that the neurons  can self-organize to designate the pacemaker among them by shortening the burst duration of the (secondary) driven neuron. This effect hold in inhibitory-only motifs, but the synchronous patterns can coexist with asynchronous patterns. The authors show that the addition of excitatory links ensure synchronous patterns of bursting. Finally, the inhibitory-only motifs exhibited multistability with as much as eight coexisting attractors. The authors suggest these attractors could be associated with patterns known to control certain animal and human locomotion activities. 

In the second study~\cite{matias2011anticipated}, stable so-called anticipated synchronization was shown to exist in biologically plausible 3-neuron motifs including inhibitory and excitatory synapses with time delays. All parameters have a clear biological interpretation, and the authors identified a transition from delayed synchronization to anticipated synchronization when synaptic conductances are increased withing physiological ranges. 

In the third study~\cite{gollo2014mechanisms}, Gollo et al. identified a mechanism by which zero-lag synchronization is facilitated. Zero-lag synchronization is of widely accepted importance in brain studies. Indeed, experimental evidence shows zero-lag synchronization between distant regions of the brains. A mechanism called ``dynamical relaying'', related to a specific motif, was proposed to account for such phenomenon in the presence of conductance delays. However, Gollo and coworkers showed that the motif was not always a reliable condition for zero-lag synchronization. Instead, they narrowed down a ``resonance pair'' of reciprocally connected neurons, which ensures zero-lag synchronization. They did so by systematically analyzing synchronization in small network motifs of Hodgkin-Huxley, neural mass, and Izhikevich neurons.

Finally the few studies that considered synchronization in larger networks of neurons relied either on numerics or phase reduction techniques, such as 
\cite{tanaka2011multistable}, and were described in the corresponding sections above. For more applied studies on the higher-order structure of brain networks in neuroscience, see Sec.~\ref{sec:neuroscience}. 

\subsection{Inference of nonpairwise interactions in coupled oscillators}
\label{sec:inference}

In networks of coupled dynamical units, the standard approach is to 
fix the structure of the network and to study how the system evolves. However, 
when modeling real oscillatory networks, we very often have
to face the so-called inverse-problem too: can we reconstruct
the underlying network structure by just looking at
the dynamical evolution of the system? This is a problem
with relevant applications. 
In neuroscience, for example, one wants to infer the brain
neuronal network from EEG data, i.e. times series of the electrical
activity of neurons or cortical areas.
This task, known as network inference, is highly non trivial, and
three main approaches have been proposed in the literature to reconstruct directional
pairwise interactions based on information theory~\cite{smirnov2005detection},  state-space approaches~\cite{frenzel2007partial}, and phase dynamics~\cite{rosenblum2001detecting}. 
Here, we report on the few studies that have addressed 
the inference of higher-order interactions, i.e. of interactions  
between three or more oscillators, based on the third approach using phase dynamics.

Following~\cite{kralemann2011reconstructing}, let us start by considering the general system of $N$ coupled dynamical systems:
\begin{equation}
\dot{\vect{x}}_i = \vect{G}_i(\vect{x}_i) + \epsilon \vect{H}_i (\vect{x}_1, \ldots, \vect{x}_N) ,
\label{eq:full_network}
\end{equation}
with $i=1,\ldots, N$, where function $\vect{G}_i$ determines the local (uncoupled)
dynamics of oscillator $i$, $\epsilon$ is the coupling strength, and
$\vect{H}_i$ is the function describing the structural coupling. It is assumed that the $\vect{G}_i$ are such that each uncoupled oscillator has a stable limit cycle, which can be parametrized by a phase $\theta_k$. The \textit{structural
	couplings} are directed physical couplings, e.g. synapses in the case of neurons. If
$\vect{H}_i$ does not depend on $\vect{x}_k$, we say there is no
structural coupling from $k$ to $i$. In general, $\vect{H}_i$ can
contain terms of pairwise coupling, $\vect{H}_{ij}(\vect{x}_i,
\vect{x}_j)$, triplet couplings, $\vect{H}_{ijk}(\vect{x}_i,
\vect{x}_j, \vect{x}_k)$, 
or coupling between any larger numbers of
oscillators. For example, one can write $\vect{H}_i= \sum_j
\vect{H}_{ij}(\vect{x}_i, \vect{x}_j)$, if only pairwise interactions
are present in the network.
In the case of weak coupling, i.e. in the small $\epsilon$ limit, each individual limit-cycle is perturbed weakly enough so that system~\eqref{eq:full_network} has an attracting $N$-torus solution which can be written in terms of $N$ phases
$\theta_1, \ldots, \theta_N$~\cite{kralemann2011reconstructing} as:
\begin{equation}
\theta_i = \omega_i + h_i(\theta_1, \ldots, \theta_N) .
\label{eq:phase_description}
\end{equation}
The new coupling functions can then be written as an expansion in the small
coupling parameter $\epsilon$
\begin{equation}
h_i(\theta_1, \ldots, \theta_N) = \epsilon  h_i^{(1)}(\theta_1, \ldots, \theta_N) + \epsilon^2  h_i^{(2)}(\theta_1, \ldots, \theta_N) + \ldots ,
\end{equation}
by performing a perturbative reduction of Eq.~\eqref{eq:full_network} (see Sec. \ref{sec:phase_reduction}). We
refer to the couplings functions $h_i$ in system~\eqref{eq:phase_description} as
{\em effective phase couplings}. It is important to note that, as in
Sec. \ref{sec:phase_reduction}, the effective phase couplings $h_i$ differ
from the original structural couplings $\vect{H}_i$. More specifically, there can
be an effective phase coupling from $j$ to $i$ even if there is no
structural coupling, but the opposite is not true.  So, there are more
effective links than structural links. However, if $\epsilon$ is
sufficiently small, structural and effective couplings are practically
identical. Indeed, the additional effective links appear only in the
second-order terms $h_i^{(2)}(\theta_1, \ldots, \theta_N)$ which is rescaled by $\epsilon^2$, and in the higher-order terms which scale in higher powers of $\epsilon$.

\begin{figure}[t]
	\centering
	\includegraphics[width=0.95\linewidth]{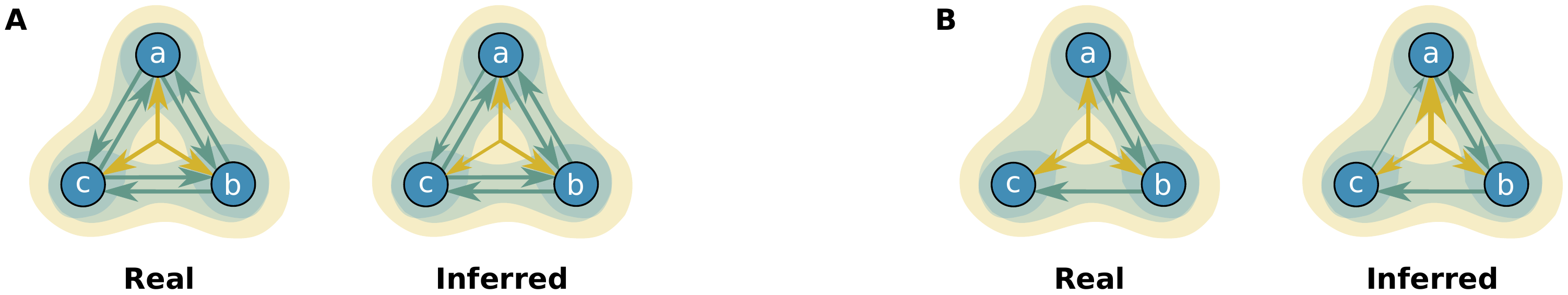}
	\caption{\textbf{Inference of directed pair and triplet couplings among three Van der Pol oscillators from the method in Ref.~\cite{kralemann2011reconstructing}.}
		An arrow from the center to a node $i$
		indicates a directed triplet interaction from $(j,k)$ to
		$i$. Panels (A) and (B) show two examples of the original
		structural network and the reconstructed one. 
		While the inference in the first case is good, the method yields 
		a spurious links from node $c$ to $a$. 
		Figures adapted from Ref.~\cite{kralemann2011reconstructing}.}
	\label{fig:kralemann2011_5}
\end{figure}

The goal of the methods proposed in 
Refs.~\cite{kralemann2011reconstructing,kralemann2014reconstructing}
is to reconstruct the effective coupling
of the system in
Eq.~\eqref{eq:phase_description}, including couplings beyond pairwise,
starting from at least one scalar time series for each node of the original system in Eq~\eqref{eq:full_network}. A situation where only a scalar time series is available for each node is common for example in neuroscience with EEG recordings, and is such cases, the information can be sufficient to reconstruct the phase system~\eqref{eq:phase_description} but not the original system~\eqref{eq:full_network}.  Such a goal is achieved in two main steps:
first, by reconstructing phases from the original time series,
and second, by reconstructing the effective phase couplings $h_i$
from the phases. Finally, weighted directed links are extracted from
the coupling functions by measuring the so-called partial norms~\cite{kralemann2011reconstructing} associated to each link. Indeed, since the $h_i$ are $2\pi$-periodic functions of $N$ phases, they can be decomposed into a Fourier expansion for the phases. Then, the pairwise interaction from $i$ to $j$ is determined by the coefficients in the expansion that depend only on $i$ and $j$, and it can be measured as the sum of the square of those coefficients.
More specifically, in Ref.~\cite{kralemann2011reconstructing}
Kralemann et al. have generalized this method
of spectral decomposition of the
effective coupling functions from Ref.~\cite{kralemann2008phase} 
to the case of interactions among more than two oscillators.
Hence, each term corresponds to a coupling among two or
more oscillators, and its associated partial norm 
indicates its strength. The numerical 
method has been successfully tested on networks of size 3, 5, and 9 of van
der Pol oscillators. Figure~\ref{fig:kralemann2011_5} shows an example
of the results obtained in the case of networks with three nodes.  
In general, the effective links detected by the method reliably reveal the true
structural links (pairwise or higher-order). However, 
the method also detects some additional links, as
expected: some are true higher-order links of the phase description, while
some others are spurious links, due to a systematic error of the
method. Unfortunately, no analytical derivation of the phase reduction
exists yet for these systems (see Sec.~\ref{sec:phase_reduction} for
other systems), so that there is no clear way to distinguish, of those
additional links, which are true and which are spurious, to date. The
method also successfully avoids detecting functional links. Two
oscillators are said to be {\em functionally coupled} if their dynamics is
correlated. Functional coupling is a concept typical of brain activity studies, and it may only be loosely related to structural and effective coupling~\cite{tass1998detection}.

It is worth mentioning here that Rosenblum and Pikovsky have developed 
in parallel a numerical phase reduction beyond the first
order to remedy the missing analytical
derivations and complement the existing ones~\cite{rosenblum2019numerical}.
For such a method to work, the available time series data must satisfy 
two criteria: they must not come from a fully synchronized trajectory,
and it must be of sufficient length. Indeed, to infer the coupling
functions, the phases must cover the $N$-torus. Hence, first, when the
system is synchronized, however, the dynamics only happens on a
limit-cycle, and no information can be extracted. Second, the time
series must be long enough to cover the $N$-torus. Because of this, 
the method, is practically useless for large networks, as 
the data needed to cover a $N$-dimensional space rapidly grows
with $N$. To be able to deal with networks larger than $N>3$, only partial phase
dynamics reconstruction is possible. Typically, one assumes pairwise
interactions, and only considers the phases $\theta_i$ and $\theta_j$
to infer the coupling functions $h_{ij}$ from $j$ to $i$. The idea is
to do so for all pairs of oscillators, and instead of reconstructing
the full system in Eq.~\eqref{eq:phase_description}, one only allows for
pairwise links to be inferred. However, this method yields spurious
effective phase links that do not exist in the full
system in Eq.~\eqref{eq:phase_description} or the original structural
connectivity of system in Eq.~\eqref{eq:full_network}.

To overcome this, Kralemann et al. have extended their previous method
discussed above, and have proposed a partial triplet
analysis~\cite{kralemann2014reconstructing}. They have considered all
the triplets of phases of the type $\theta_i, \theta_j, \theta_k$ to
reconstruct the coupling functions $h_{ijk}, h_{jki},
h_{kij}$.  From these functions, pairwise connections have been obtained from the partial norm of the spectral decomposition, as above. The authors have showed
that triplet partial analysis performed better than the pairwise
partial analysis in networks of 3 and 4 Van der Pol oscillators.
Indeed, the true links were detected equivalently well, but the
triplet partial analysis successfully avoided detection of
spurious links produced instead by the pairwise
analysis.

As mentioned above, all these methods need input data that do not come from 
synchronized trajectories. This can be checked by evaluating the usual pairwise
$n:m$ synchronization index $\gamma_{j,k} = | \langle e^{i (n\theta_j
	- m \theta_k) } \rangle |$, where $\langle \cdot \rangle$ indicates
temporal averaging, and where $n$ and $m$ are integers~\cite{pikovsky2003synchronization}.
This index is
close to 1 if oscillators $j$ and
$k$ are phase locked. Complete network synchronization can 
be checked by computing the index for all pairs of oscillators. However, triplet
synchronization is not revealed by this index. Indeed, it is well possible
that three phases $\theta_i, \theta_j, \theta_k$ satisfy the relation
$n \theta_i + m\theta_j + l \theta_k = \text{constant}$ for three integers
$n,m,l$, even though the pairwise index of each pair is not close to
1. To reveal triplet synchronization from data, the following triplet synchronization
index has been introduced in Ref.~\cite{kralemann2013detecting}: 
\begin{equation}
\gamma_{i,j,k} = | \langle e^{i (n\theta_i + m \theta_j +  l \theta_k) } \rangle | .
\end{equation}
Notice that a value of $\gamma_{i,j,k}$ close to 1 is not a sufficient
condition for synchronization. Indeed, it only indicates its
possibility, since a large value of the index can also be the
consequence of other types of interdependence between the
phases. Finally, as mentioned by the same authors, the index can be
readily extended to higher-order synchronization indices, valid for
quandruplets and higher-resonances. Jia et
al. have built on that, showing 
experimentally the existence of states where triplets are
synchronized but pairs are not~\cite{jia2015experimental}.  

In Ref.~\cite{stankovski2015coupling}, Stankovski et al. have proposed 
another method to reconstruct the effective phase connectivity of a
network. The specificity of their method is that it works for network 
of oscillators with time-varying coupling and frequencies, and that
are subject to noise. The method is based on dynamical Bayesian
inference~\cite{duggento2012dynamical}, and detects coupled pairs,
triplets, and quadruplets of oscillators. The method computes the
values of a set of parameters that fully determine the couplings. The
values of the parameters are inferred by making use of Bayes' theorem,
which takes a prior distribution and evolves it into a posterior
distribution, by using time series of the system and building a
likelihood function. The authors have demonstrated the accuracy of the
method on a simulated 5-oscillator network, as well as on real
multi-channel EEG data. The method has been shown to outperform inference
based on pairwise interactions only. Unfortunately, similarly to the other
methods presented above, this method works for relatively small networks.

For applications of inference methods of higher-order interactions
on real brain networks data, see Sec.~\ref{sec:neuroscience}.


\section{Spreading and social dynamics}
\label{sec:social}

Dynamical processes that emulate human behaviors have been the focus of many studies, where social relationships and interactions are typically considered as an underlying structure. Social interactions are a natural testing ground for higher-order approaches. Since individuals can interact in pairs or groups, the dynamics should in turn account for the higher-order effects that the non-pairwise interactions might lead to. In this section, we review a broad variety of models, initially introduced and studied on graphs, that have been extended as dynamical processes on HOrSs. We start reviewing \textit{spreading processes}, historically embedded within the literature of epidemics on networks~\citep{pastor2001epidemic, pastor2015epidemic}, but recently revisited to fit the dynamics of social contagions~\citep{centola2018behavior}. We then continue with a wider class of models of \textit{social dynamics} mostly devoted to the formation of opinions and consensus \citep{sen2014sociophysics, castellano2009statistical, baronchelli2018emergence}.

\subsection{Spreading in higher-order networks}
\label{subsec:social:spreading}

The study of spreading processes on networks is one of the branches of network science that attracted more attention among the community. Building on top of classical epidemiological compartmental models \citep{kermack1927contribution, anderson1992infectious, hethcote2000mathematics}, the recent success of these models is partially due to the increasing availability of large scale data that opened up new research avenues in which researchers make use of the newly available data sources to inform the models, which on turn allow us to forecast and possibly control the disease spreading \citep{zhang2017spread, y2018charting, viboud2019future, kucharski2020early, Kraemereabb4218}. In light of these new advancements, network scientists have been slowly, but extensively, introducing more and more details into the modeling framework in order to increase its accuracy and ultimately its predictive power. 

In this scenario, two of the most studied compartmental models are the Susceptible-Infected-Recovered (SIR) and the Susceptible-Infected-Susceptible (SIS). In both models, susceptible individuals (S) can get infected by mean of an interaction with infectious ones (I). This SI process always leads, by construction, to the absorbing state in which all individuals are infected. The introduction of an additional transition leads to richer phenomena. More specifically, in the case of the SIS, individuals can switch multiple times between the S and I states, eventually reaching a steady state in which the epidemic is sustained by a non-zero number of individuals. Contrarily, in the SIR, individuals gain immunity to reinfections after a certain amount of time, or with a given probability per unit time. These immune individuals are then called recovered (R) and do not participate anymore to the spreading dynamics. This type of models is therefore used when it comes to modeling infectious diseases such as Ebola, or seasonal influenza, in which individuals can acquire immunity against reinfections. As a consequence, the SIR presents also the disease-free state as an absorbing state.

\medskip

Many theoretical approaches have been developed to analytically describe, with increasing level of complexity, the dynamics of epidemic spreading on complex networks. An accurate analytical description should include the interplay between the structure of the contact patterns and the dynamics of the spreading process on top. Here,  instead of going through the assumptions, advantages and drawbacks of all the possible descriptions, from the mean-field (MF) and the heterogeneous mean-field (HMF), to the most accurate microscopic Markov-chain approaches, we refer the interested reader to \citep{gleeson2011high, gleeson2013binary, pastor2015epidemic, wang2017unification, kiss2017mathematics} and references therein. 

\medskip

While the aforementioned models have been widely used to study the spread of diseases, there's a variety of other domains where they have been successfully applied. 
Indeed, another long tradition of modelers that have been using similar frameworks to characterize the spreading of social phenomena, such as the diffusion of rumors and fads or the adoption of novelties and technological innovations \citep{daley1964epidemics, bass1969new, bikhchandani1992theory, rogers2010diffusion}. 
However, in all these situations the social nature of the contacts that mediate these processes calls for ad-hoc modeling adjustments that are not present in simple disease epidemics models. These approaches, developed under the name of \textit{complex contagion}, are meant to include additional ingredients, such as mechanisms of social influence and peer pressure, already widely studied within the social sciences~\citep{wasserman1994social, centola2007complex, centola2010spread}. The requirement of these new features, not needed when dealing with the spreading of a pathogen, gave rise to a plethora of models that have been already extensively reviewed in Ref.~\citep{guilbeault2018complex}.

Here, keeping the focus on the dynamics of social contagion, we shift the attention towards the structural aspect of the social contacts on top of which the dynamics evolves. Moving from pairwise to higher-order structures, we investigate the dynamical effects brought by the novel representations. There is a matter of discussion whether social relationships could be better modeled by using simplicial complexes rather than hypergraphs. In the end, depending on the situation, it might be reasonable or not to assume that in a group interaction all the sub-interactions among the group members should be considered as well~\citep{kee2013social}.
In what follows, we distinguish between the two approaches and discuss recent developments towards the inclusion of HOrSs in the modeling approach. Our limited goal is to introduce some of the recent efforts in this direction without imposing selective constraints on how a ``pure" higher-order dynamics should be defined on these new structures. This leads to a mixture of models based on higher-order and not-so-higher-order dynamics on HOrSs. We explicitly distinguish between the HOrSs, starting from spreading processes that take place on simplicial complexes and then moving to hypergraphs. 

\subsubsection{Spreading on simplicial complexes}
\label{subsubsec:social:spreading:simplicial}

\begin{figure*} 
	\centering
	\includegraphics[width=\textwidth, keepaspectratio = true]{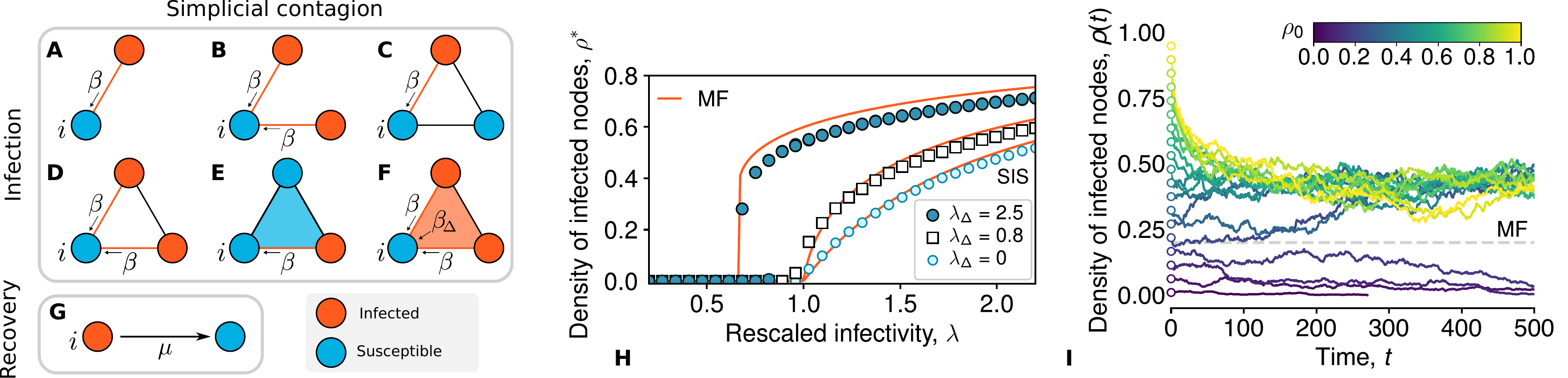}
	\caption[]{\textbf{Simplicial contagion model ($D=2$)~\citep{iacopini2019simplicial}.} (A-F) Different channels of
		infection for a susceptible node $i$ are shown. Notice (F), where node $i$ can get the infection from each of the two 1-simplices with probability $\beta$, and also from the 2-simplex with probability $\beta_{\Delta}$.
		Behavior on synthetic random simplicial complexes: In (H) the average fraction of infected obtained by means of numerical simulations is plotted against the rescaled infectivity $\lambda = \beta \langle k \rangle / \mu$ for different values of $\lambda_\Delta$ ($\lambda_\Delta=0$ gives results for the standard SIS model without higher-order effects). The red lines correspond to the analytical MF solution described by Eq.~\eqref{eq:simplagion:MF}. When $\lambda_\Delta$ ($\lambda_\Delta=2.5$ we observe a discontinuous transition with the formation of a bi-stable region where healthy and endemic states co-exist. (I) Temporal evolution of the densities of infectious nodes in the bi-stable region ($\lambda=0.75$, $\lambda_\Delta=2.5$). Different curves—and different colors—correspond to different values of $\rho_0$, the initial density of infectious nodes. The dashed horizontal line corresponds to the unstable branch of the MF solution, separating the two basins of attraction. Figures adapted from Ref.~\citep{iacopini2019simplicial}.}
	\label{fig:spreading_1}
\end{figure*}

In the simplicial contagion model proposed by \citet{iacopini2019simplicial}, a simplicial complex is used to represent the social structure on top of which the contagion dynamics takes place. By definition, all the sub-interactions contained in each group interaction are considered. Therefore, the dynamics of the model specifically relies on the different channels of infections (1-simplex, 2-simplex, etc.) through which, with different transmission rates, a contagion can happen. The SIS-like model of order $D$ is controlled via a set of control parameters $\beta_1, \beta_2, \dots,\beta_{D}$, whose elements represent the probability per unit time for a susceptible node $i$ that participates to a simplex $\sigma$ of
dimension $D$ to get the infection from each one of the infectious sub-faces composing $\sigma$ (sub-faces in which all nodes but one are infected). 
At order $D=2$, one has $\beta_1$ and $\beta_2=\beta_{\Delta}$ corresponding respectively to the probability that a susceptible node $i$ receives the infection from an infected node $j$ through the link $(i,j)$ and to the probability of receiving it from an infectious 2-simplex $(i,j,k)$ incident on $i$. The recovery dynamics is controlled by the standard recovery probability $\mu$, which, being node-dependent, does not ``feel" the higher-order structure (Figs.~\ref{fig:spreading_1}A-G).

Even the inclusion of only the lowest higher-order interactions (2-simplices) dramatically changes the nature of the spreading process, going from a continuous to a discontinuous phase transition in the prevalence as a function of the 1-simplex infectivity $\lambda$ (Fig.~\ref{fig:spreading_1}H). Notice that the nature of the transition depends on the 2-simplex infectivity $\lambda_\Delta$. This behavior is confirmed by numerical simulations on empirical data obtained from the Sociopatterns collaboration~\citep{sociopatterns} and on synthetic random simplicial complexes (see Sec.~\ref{subsec:models:equilibrium:simplicial}), where a bi-stable region in which healthy and an endemic states co-exist appears. This is illustrated in Fig.~\ref{fig:spreading_1}I, in which different curves represent different realizations of the model starting from seeds of infectious nodes of different sizes (colors). Further analytical insights on this bi-stability can be found in Ref.~\cite{cisneros2020multi}.

The authors further explained the observed phenomenology through an analytical investigation based on an extension of the standard mean-field (MF) approach for networks, specifically adapted to the case of HOrSs. In this case, the general equation for the evolution of the stationary density of infected $\rho(t)$ reads

\begin{equation}\label{eq:simplagion:MF}
	d_{t}\rho(t) = -\mu\rho(t) + \sum_{d=1}^{D} \beta_{d}\langle k_{d}\rangle \rho^{d}(t)\bigl[1-\rho(t)\bigr]  
\end{equation}

with $\langle k_{d}\rangle$ denoting the average generalized degree, i. e., the number of $d$-dimensional simplices incident on average on each 1-dimensional simplex $\alpha$: $\langle k_{d}\rangle=\langle k_{d,1}(\alpha) \rangle_\alpha$ (see Sec.~\ref{sec:measures:degree_centrality}).
This approach confirmed the results obtained on synthetic random simplicial complexes, showing that the steady-state dynamics, the position, and the nature of the transition can be predicted analytically on social structures characterized by homogeneous degree distributions. This is shown in Fig.~\ref{fig:spreading_1}H, where the MF curves (red) are compared to the simulated results (points), and in Fig.~\ref{fig:spreading_1}I, where the dashed gray line---corresponding to the unstable solution of the MF approach---correctly detects the two basins of attraction that split the simulated curves.

\medskip

Further developments of the simplicial contagion model based on probabilistic descriptions showed that more complex analytical formulations, namely the microscopic Markov-chain approach~\citep{gomez2010discrete} and the link equation~\citep{matamalas2018effective}, can improve the accuracy of predictions \citep{matamalas2020abrupt}. Differently from the MF, these approaches can indeed be used to analytically describe the contagion dynamics on higher-order heterogeneous structures.

\subsubsection{Spreading on hypergraphs}
\label{subsubsec:social:spreading:hypergraph}

 In contrast with simplicial complexes, hypergraphs can be used to describe interactions that only take place in groups, lifting the constraint of having to include all the sub-interactions within the groups themselves. 
 Therefore, hyperedges can efficiently be used to represent clusters or communities, when such sub-interactions are unlikely to be relevant ingredients in the description of social HOrSs~\citep{girvan2002community, newman2003social}. Previous results on spreading dynamics on networks have already showed the impact that the presence of clusters, communities and sub-graphs might have on the epidemic threshold and on the final epidemic size ~\cite{miller2009percolation, miller2009spread, hebert2010propagation, karrer2010random, ritchie2014higher, o2015mathematical, hebert2015complexPNAS, stonge2020master, hebert2020spread}. Hypergraphs have been also used to model knowledge diffusion in collaboration networks \cite{yang2015knowledge, wang2015improved, peng2019hypernetwork}.
 
 The idea of modeling communities as hyperedges was first proposed by \citet{bodo2016sis}, who used the nodes of a hypergraph to represent individuals and hyperedges to represent the different communities a node belongs to, such as a household or a workplace~\cite{ghoshal2004sis, house2008deterministic, ball2015seven}.
The authors studied the behavior of an SIS model on hypergraphs under a continuous time Markov chain formalism in which both infection and recovery are governed by Poisson processes. However, while the recovery is a spontaneous process controlled by a fixed recovery rate $\gamma$, the rate of infection $r$ takes into account the higher-order connectivity patterns. In particular, they defined the probability for a susceptible individual to become infected as $1-\exp(-r\Delta t)$, with $r$ being $r=\tau\sum_e f(i_e)$. The summation runs over all the hyperedges -the communities- containing the susceptible individuals, while  $f(i_e)$ denotes a generic function of $i_e$, the number of infected nodes in the hyperedge $e$. Bod\'o et al. chose $f$ to be a piece-wise linear function, with the idea of not increasing the infection pressure for a susceptible node when the number of infected neighbors is higher than a given threshold. This is conceptually different from the conventional threshold mechanism---largely exploited by the complex contagion literature---in which thresholds are used in the opposite way, e. g., to set the critical amount of exposures from the peers that an individual needs in order to adopt a new technology~\citep{granovetter1978threshold, karsai2014complex}. 
Simulations on hypergraphs having hyperedges of different sizes showed that heterogeneous structures might significantly fasten the initial phase of the spreading when compared to regular hypergraphs, while leading to slightly smaller values of prevalence in the stationary state.

\medskip

Later, \citet{suo2018information} investigated a similar SIS model on hypergraphs particularly designed to study the differences between two different spreading strategies. In the global one, at each time step an infected node $i$ can infect with a probability $\beta$ all the susceptible neighboring nodes that are connected to $i$ via a hyperedge (global). In the local one, an infected node $i$ randomly chooses $e$, one of its hyperedges, and then tries to infect with $\beta$ all the susceptible nodes composing $e$ (local). This is inspired by the different ways in which an individual might decide to share a content on a social media platform, either to all the contacts or exclusively targeting a particular group.
Notice that, differently from the higher-order models previously introduced, here the HOrS is used as a structure, but the global contagion dynamics does not specifically ``feel" it. Hence, the global spreading strategy would in principle be equivalent to the one defined on the 1-skeleton of the hypergraph, in which each hyperedge is a clique instead.

The two different strategies lead to different long term behaviors, with a vanishing epidemic threshold in the global strategy. Contrarily, the particular positioning of the initial seed of infectious nodes---either on high or low hyperdegree nodes---seems to affect only the early evolution of the process: as expected, choosing nodes with a high hyperdegree as seeders can significantly speed up the contagion in the early times. No differences in the stationary states were found. 

\medskip

The two models just presented made use of the HOrS to define the neighborhood of a node that might be responsible for a contagion event, but no explicit mechanism of peer pressure was included in the modeling framework. Here, we briefly discuss two following works in which, differently from before, the higher-order representation explicitly enters into the contagion dynamics to account for reinforcement effects that might occur at the group level.

\medskip

\begin{figure*} 
	\centering
	\includegraphics[width=0.9\textwidth, keepaspectratio = true]{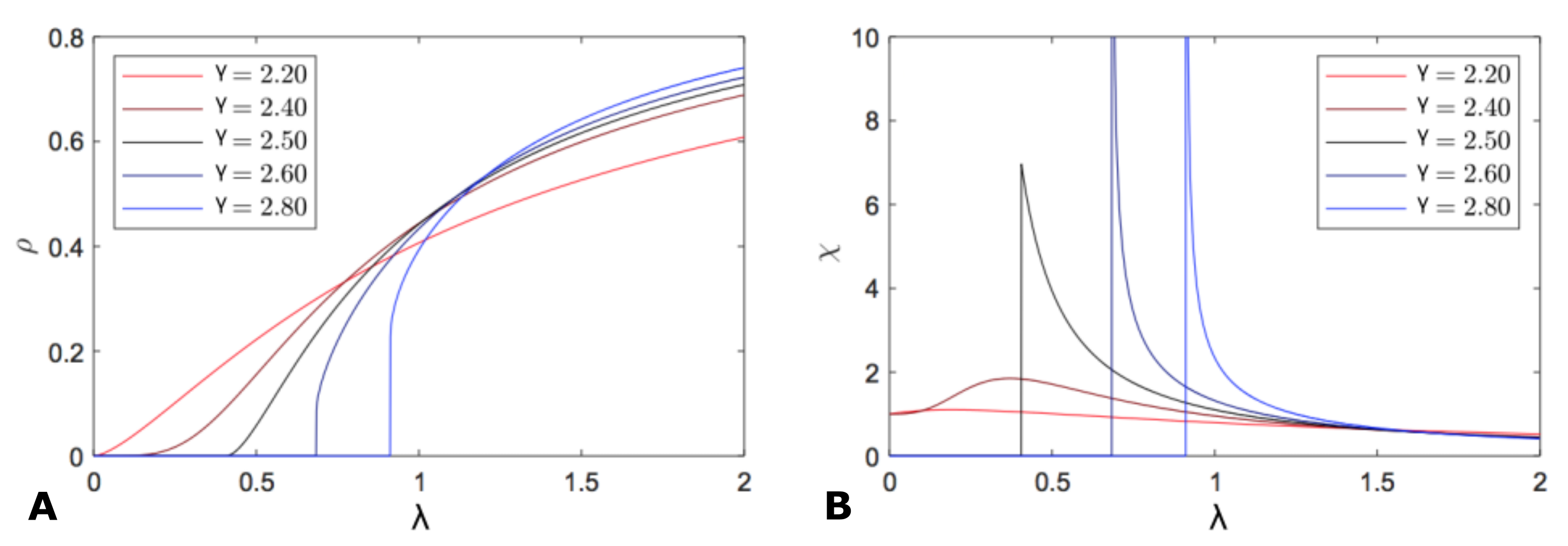}
	\caption[]{\textbf{Behavior of the higher-order contagion model on scale-free uniform hypergraphs ($D=3$)~\citep{jhun2019simplicial}.} (A) Stationary density of infected nodes against control parameter $\lambda\equiv\beta/\mu$ for different values of the SF exponent $\gamma$. For $\gamma=2.2$ and $2.4$, the epidemic threshold vanishes ($\lambda_c=0$), while $\lambda_c$ is finite for higher values of $\gamma$. For $\gamma=2.2,\ 2.4$ and $2.6$, the transition is second-order, and for $\gamma=2.6$ and $2.8$ the transition is hybrid. (B) Susceptibility $\chi$ versus $\lambda$. For $\gamma\leq \gamma_c$, $\chi$ converges to a finite value $1 + d(d-2)$. In contrast, for $\gamma> \gamma_c$, the susceptibility diverges as $\lambda \rightarrow \lambda^{+}$. Figures adapted from Ref.~\citep{jhun2019simplicial}.}
	\label{fig:spreading_2}
\end{figure*}

\citet{jhun2019simplicial} extended the simplicial contagion model discussed in Sec.~\ref{subsubsec:social:spreading:simplicial} to the more general case of hypergraphs. The SIS-like model works exactly as the one proposed by \citet{iacopini2019simplicial}, but this time the spreading process takes place on top of $d$-uniform hypergraphs in which all the hyperedges have the same size $d$. As for the simplicial model, a susceptible node that is part of a hyperedge $\alpha$ of size $d$ can get an infection from $\alpha$, with rate $\beta_d$,  only if the remaining $d-1$ nodes composing $\alpha$ are infectious. As for the recovery, the standard recovery probability $\mu$ is used. The authors considered the case of scale-free (SF) uniform hypergraphs. Notice that even if all the hyperedges have the same size, different nodes can belong to a different number of these hyperedges. In this sense, the heterogeneity is given by the number of hyperedges a node belongs to, which is distributed as $\sim P(k)^\gamma$. The heterogeneous mean-field formalism (HMF)---in which nodes of the same hyperdegree class as considered equivalent~\citep{pastor2001epidemic}---leads to the following equation for the evolution of the stationary density of infected nodes of hyperdegree $k$:

\begin{equation}\label{eq:simplagion:HMF}
	d_t\rho_k = -\mu\rho_k + \beta_k(1-\rho_k)k\Theta^{d-1}
\end{equation} 

The contagion term on the r.h.s.~considers the probability that a susceptible node of hyperdegree $k$ gets the infection from one of the hyperedges. This is, as usual, proportional to the infection rate $\beta_k$, the number of hyperedges $k$, and the probability $\Theta^{d-1}$ to be connected to a hyperedge having all the other nodes infected.
A comparison of Eq.~\eqref{eq:simplagion:HMF} with Eq.~\eqref{eq:simplagion:MF} highlights the difference in the representation used. Indeed, differently from the simplicial case, here the contagion term does not dependent on the lower order sub-faces.

The resulting phase diagram of the model for a $3-$uniform hypergraph is reported in Fig.~\ref{fig:spreading_2}A, where the stationary density of infected nodes $\rho$ is plotted against the re-scaled control parameter $\lambda\equiv\beta/\mu$ for different values of the SF exponent $\gamma$. The system presents a characteristic exponent $\gamma_c=2+1/(d-2)$ of the degree distribution that determines the nature of the transition. In particular, for $\gamma<\gamma_c$ the epidemic threshold vanishes ($\lambda_c=0$), as it is confirmed by the finite value of the susceptibility $\chi$ reported in Fig.~\ref{fig:spreading_2}B. By contrast, if $\gamma=\gamma_c$ a second order transition appears, that becomes hybrid when higher values of $\gamma$ are considered (see curves for $\gamma=2.6$ and $2.8$). The associated values of the susceptibility diverge at the transition point, as expected~\citep{lubeck2004universal, ferreira2012epidemic}. These results are consistent with simulations on SF uniform hypergraphs, confirming the validity of the HMF approach on such topologies. 

\medskip

Another version of the higher-order social contagion model on hypergraph was recently proposed by \citet{de2020social}. Based on a similar SIS framework, the fundamental difference with respect to the other models relies on the explicit inclusion of a critical-mass dynamics into the contagion process that generalizes the one in Ref.~\citep{iacopini2019simplicial}. In the simplicial contagion model previously discussed, a susceptible node $i$ belonging to hyperedge $\alpha$ (or a simplex) of size $d$ could get the infection from $\alpha$ only if all the remaining $d-1$ nodes composing it are infected. Here, the authors relax the constraints by (i) moving from simplicial complexes to hypergraphs and (ii) allowing a hyperedge $\alpha$ to be potentially infectious for $i\in\alpha$ if the number of infected nodes composing $\alpha$ is greater or equal to a given threshold $\Theta_\alpha$. The standard SIS model is then recovered by restricting this threshold mechanism to hyperedges of size greater than two, so that a contagion through active links can always happen (no threshold). This model reveals a similar phenomenology to the one on simplices, characterized by the appearance of first and second-order transitions and hysteresis. Further insights on the conditions for the continuity of the phase transition and the stability of general dynamical process on hypergraphs have been subsequently given in Ref.~\cite{de2020phase}. In addition, the authors provide further analytical results on two particular regular hypergraphs, namely a hyperblob (a random regular network with one hyperedge containing all the nodes) and a hyperstar (a star network with one hyperedge containing all the nodes). The critical values analysis is then extended with the introduction of the concept of a ``social latent" heat, interpreted as the fraction of individuals to be added or removed to move the dynamics from one solution to the other. 

 These findings provide a possible phenomenological explanation for some apparently contradictory results previously obtained. In fact, experimental work has shown different values of critical mass levels needed to initiate a social change, i.e., to revert an existing equilibrium to a new one by mean of a committed minority~\citep{dahlerup1988small, grey2006numbers, centola2018experimental}. These threshold values, spanning from $10\%$ to $40\%$, could be consistently seen as the effect of the interplay between a global critical mass and the local thresholds as given by the $\Theta_\alpha$, which also depend on the size of the interacting group.

\medskip

\begin{figure*} 
	\centering
	\includegraphics[width=\textwidth, keepaspectratio = true]{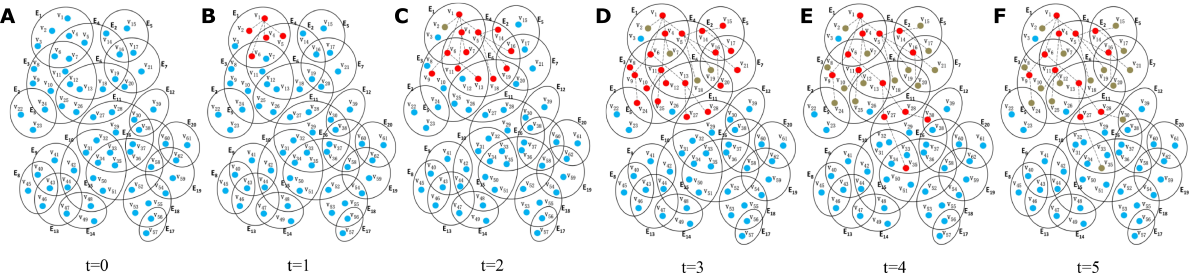}
	\caption[]{\textbf{SIR model on hypergraphs~\citep{ma2018study}.} Ignorant (S), spreader (I), and stifler (R) nodes are respectively depicted in blue, red, and green. At each time step a spreader can transmit the information to ignorant nodes within the same hyperedge with a given probability. Figures adapted from Ref.~\citep{ma2018study}.}
	\label{fig:spreading_3}
\end{figure*}

Finally, \citet{ma2018study} introduced different extensions of the SIR model to hypergraphs. In their model, placed within the framework of rumor spreading \citep{dietz1967epidemics, moreno2004dynamics}, individuals are divided into the three standard classes corresponding to ignorant (S), spreader (I), and stifler nodes (R). In this particular context, the spreading process wants to model the transmission of information between different members of an enterprise, whose internal structure is characterized by informal organizations (spontaneous groups). In particular, Ma and Guo consider different variations of the mechanism of information transmission. Figure \ref{fig:spreading_3} reports an illustrative example of the probabilistic transmission, as defined by the authors. At $t=0$ (Fig.~\ref{fig:spreading_3}A), all the nodes are ignorant (blue). At $t=1$ (Fig.~\ref{fig:spreading_3}B),  the information starts to spread, with a given probability, from a randomly selected spreader node (red) to the other nodes within the same hyperedge. Subsequently, the rumor reaches more and more nodes while some nodes become stifler (green) and can no longer spread the information. Other variations where the information passes to the entire hyperedge at once or to a constant number nodes within the same hyperedge are also considered.

\subsection{Opinion and cultural dynamics beyond pairwise interactions}
\label{subsec:social:opinion} 

In this subsection, we will review some of the best known models of opinion and cultural dynamics. At the core of these agent-based models, often referred to as spin models in the physics literature or as interacting particle systems in the mathematics literature, there is the idea of describing a social dynamics by relying on simple---yet sufficient---rules. Many different variations and extensions have been proposed and extensively studied, therefore we will limit our focus to those for which, in the spirit of this review, an higher-order extension exists.

\subsubsection{Voter model}
\label{subsubsec:social:opinion:voter} 

\begin{figure*} 
	\centering
	\includegraphics[width=\textwidth, keepaspectratio = true]{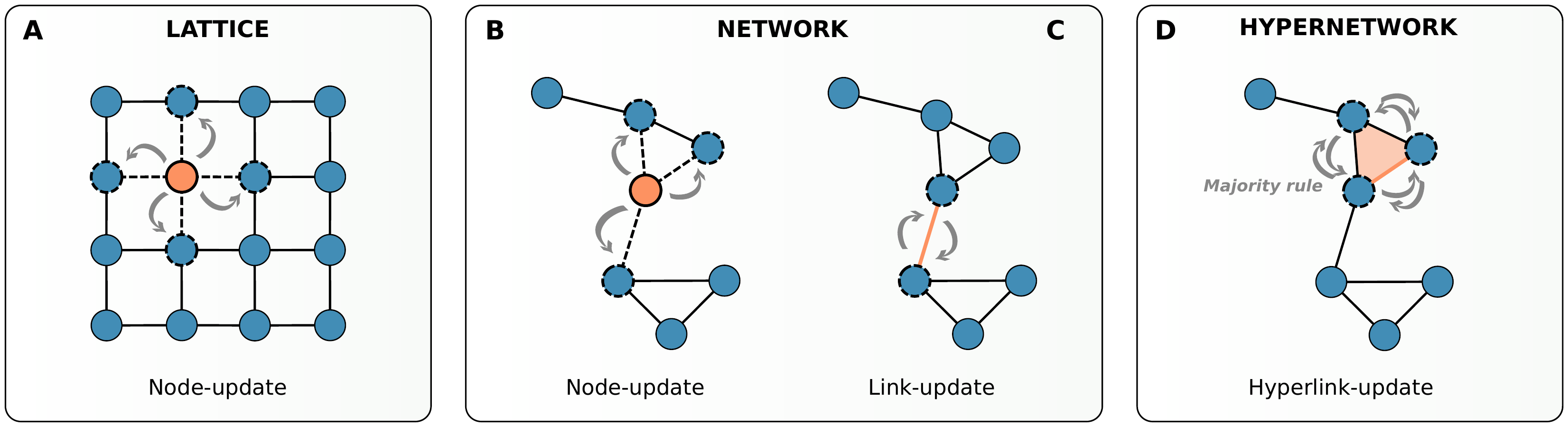}
	\caption[]{\textbf{Voter dynamics on different structured populations.} (A) Node-update rule on a lattice and (B) on a network: a randomly selected node copies the opinion of a randomly selected neighbor. (C) Link-update rule on a network: one of the two nodes of a randomly selected edge adapts its opinion to the one of the other. (D) Hyperlink-update rule on a hypergraph/simplicial complex: the nodes of a randomly selected simplex incident on a randomly selected edge adapt their opinion to the one of the majority.}
	\label{fig:spreading_4}
\end{figure*}

With its origins deeply rooted in the statistical physics literature, the voter model is one of the simplest models of opinion dynamics \citep{liggett2012interacting}. In the most basic version, it consists in a population of $N$ interacting individuals located on the sites of a lattice, each endowed with a binary variable (spin) $\sigma_i=\{-1,1\}$, $i=1,\dots N$, representing an opinion, or a vote. The fundamental mechanism of the model relies in the node-update rule, according to which, at each time step, a randomly selected node copies the opinion of a randomly selected neighbor (Fig.~\ref{fig:spreading_4}A). This dynamics is iterated until one of the absorbing states of full consensus is reached. Despite its simplicity, this model presents a non-equilibrium dynamics leading to non-trivial behaviors \citep{shao2009dynamic}. Extending the voter dynamics from a lattice to a network requires a change of perspective. Indeed, in order to maintain the average magnetization of the system, one has to move from the aforementioned node-update rule to a link-update rule \citep{suchecki2004conservation} (Figs.~\ref{fig:spreading_4}B,C). 
This is because the degree heterogeneity potentially present in a network biases the random selection of the neighboring nodes in favor of the most connected ones, ultimately making the ``order of play" matter. 
As for many other dynamical processes on structured populations, the interplay between structure and dynamics has been the focus of many studies \citep{suchecki2005voter}. 
Introducing the many different---and sometimes multi-layered~\citep{diakonova2016irreducibility}---variations of the voter model and their applications \citep{fernandez2014voter} goes beyond the scope of this review, therefore we now focus on the particular version that has been generalized in order to account for higher-order interactions. 

\medskip

If we let the node variables take more than two values---going beyond binary opinions---we can call colors these different opinions, and as a consequence the voter dynamics becomes a coloring \textit{coordination game} (see Sec.~\ref{sec:games}). Motivated by coloring game experiments \citep{kearns2006experimental, judd2010behavioral} in which agents make informed decisions based on some local information available, i.e., a subset of nodes, Chung and Tsiatas used hyperedges as a natural way to encode these group interactions \citep{chung2014hypergraph}. In their model, voters have non-pairwise relationships which are represented as hyperedges of a hypergraph, and the dynamics obeys to the following hyperedge-update rule (Fig.~\ref{fig:spreading_4}D). At each time step a hyperedge is selected, and the nodes composing it simultaneously change their vote according to a given probability. In the simplest case this probability does not depend on the status of the voters at the time of the interactions. Further insight on the process can be gained by using the duality of the process with a random walk on an associated weighted graph whose weights encode the transition probabilities among different voting configurations. By performing this mapping, Chung and Tsiatas were able to study the dynamics of the memoryless game in terms of the spectral properties of the random walk. 

\medskip

Another prominent research topic in network science deals with the investigation of dynamical processes that directly affects the structure of the network. This co-evolution of network structure with the dynamics that takes place on it is particularly relevant when it comes to modeling social systems, in which the states of the nodes or a behavioral change might force the network to react by changing its connectivity patterns. Voter models with binary opinions have been widely extended in this direction, and many versions of adaptive voter models have been proposed~\citep{redner2019reality}. The minimal adaptive voter model on networks works as follows~\citep{vazquez2008generic}. At each time step an edge $e$ is randomly selected. If $e$ is inactive, i.e., it connects nodes with the same opinions, nothing happens. Contrarily, if $e$ is active, there are two possible mechanisms, both leading to the death of this active link in favor of a inactive one. More precisely, with probability $p\in[0,1]$ a rewiring happens. In this case a randomly selected node between the two composing $e$  rewires to a random node in the network that shares its own opinion. Otherwise, with probability $1-p$ one of the two nodes of $e$, selected at random, adapts its opinion to the one of the other, ultimately making $e$ inactive. Recently, Horstmeyer and Kuehn have extended this model in order to account for higher-order structures, such as simplicial complexes~\citep{horstmeyer2020adaptive}. In their model, studied for simplicity up to the level of the 2-simplices, an edge $e$ (1-simplex) is randomly selected. If $e$ is not part of a 2-simplex the standard rules just defined on networks apply. Alternatively, if $e$ is active and part of at least one 2-simplex then a new mechanism of peer pressure might appear. This is controlled via a second parameter $q\in[0,1]$ representing the probability of the higher-order structure affecting the dynamics through a majority rule. When this happens, with probability $q$, one of the 2-simplices containing $e$ is selected at random, and all its nodes adopt the opinion of the majority with probability $p$. Similarly to the standard case, with probability $1-p$ a rewiring happens, but this time on the 2-simplices. In particular, all the 2-simplices containing $e$ are ``downgraded'' to three standard 1-simplices and an equal number of new 2-simplices is created by randomly ``promoting'' triangles of the network formed solely by 1-simplices. However, this mechanism combined with the rewiring leads to a natural depletion of $2$-simplices that determines the stopping criterion for the simulations (when there are no triangles left to ``promote''). This models reduces to the standard one when $q=0$, while for $q>0$ the simplicial structure plays a relevant role on the dynamics. In this last regime, the characteristic behavior of the adaptive voter model does not change and one can still observe a transition at a critical value $p_c$ between an active phase with a nonzero stationary density of active links and a frozen phase in which the system breaks up into two disconnected components having opposite opinions. However, the higher-order structure has several effects on the model dynamics. When $q>0$ the speed at which the dynamics reaches the two phases is increased, the critical $p_c$ separating the phases is shifted towards lower values of rewiring, and the stationary density of active edges in the first phase is lowered due to the peer pressure mechanism introduced.

\subsubsection{Majority models}
\label{subsubsec:social:opinion:majority}

A similar class of models is the one of the {\it majority vote models}, originally proposed by \citet{galam2002minority}. In the most basic formulation, individuals are endowed with binary state variables denoting opinions, and interact without specific topological constraints. These models behave similarly to the voter-like models discussed above, but with one fundamental difference in the updating rule. In fact, as the name suggests, here the copying mechanism is replaced by a deterministic majority rule according to which, at each time step, a subset of $n$ individuals is chosen and their opinion is set to the one of the majority within the subset. An additional bias---justified as social inertia---that favors a particular opinion is usually introduced to resolve ties when $n$ is even. Typical quantities of interests are the probability of reaching a particular type of consensus as a function of the initial configuration and the time required to reach it. Many studies have been conducted in this regard by using models on lattices and graphs, highlighting the key role played by the dimensionality of the system, together with changes in the dynamics when finite or infinite systems are considered.  In all these cases, i.e., when a structured population is used, the node-update consists in adopting the opinion of the majority of the neighboring nodes. In this sense, the majority rule model can be seen as a special case of a threshold model in which the threshold parameter of each node is set to half of the number of neighbors \citep{watts2002simple}. 

The spatial version of the majority rule model proposed by Lanchier and Neufer \citep{lanchier2013stochastic} is based on the idea that social groups are better defined in terms of hyperedges rather than dyadic interactions. Thus, they extended the majority rule model on HOrSs defined as hypergraphs, so that nodes within each hyperedge simultaneously change their opinion to the majority opinion of the hyperedge they are part of. By focusing on a particular regular social structure, in which a hyperedge consists of a $n\times \cdot\cdot\cdot\times n$ block on a lattice, they were able to show through analytical results and simulations that, for each dimension, the model dynamics behaves similarly to the voter model when a fixed odd number $n$ of interacting nodes is selected. This means that for hyperedges of even size $n$ the system always reaches a consensus where all the nodes have the opinion favored by the bias, while if $n$ is odd, the system presents growing clusters that eventually reach consensus. This is radically different from the voter model in high dimensions ($d\ge 3$), in which the system reaches a stationary state in which the two opinions co-exists (see Sec.~\ref{subsec:ecology}).

\medskip

A popular variation of the majority rule is the {\it majority vote model}~\citep{liggett2012interacting}, in which a parameter $q\in[0,1]$ is introduced so that a node changes its state to the one of the majority of its interacting nodes with probability $(1-q)$. Since $q$ controls the randomness, sometimes the model goes under the name of majority-vote model with noise, and it obeys the following update-rule. At each time step, an individual $i$ is selected, and its opinion $\sigma_i$ is flipped with probability 

\begin{equation}\label{eq:majority_vote}
w(\sigma_i)=\frac{1}{2}\Biggl(1-(1-2q)\sigma_i \ \text{sgn}\Bigl(\sum_{j}\sigma_j\Bigr) \Biggr)
\end{equation}

where the sum $\sum_{j}\sigma_j$ runs over all the nodes $j$ that are interacting with $i$ so that $\text{sgn}(\cdot)$ takes either the sign of the argument or is equal to 0 in case of a lack of majority (zero sum). 

Results on regular lattices show that the model undergoes an order-disorder phase transition at a critical value $q_c$ \citep{de1992isotropic}, with critical exponents falling within the universality class of the corresponding equilibrium Ising model, with $q$ acting as a temperature. A similar behavior was found on Erd\H{o}s-R\'enyi random graphs \citep{pereira2005majority}. Contrarily, in small world networks the position of the transition point was found to be a function of the rewiring probability~ \citep{campos2003small, luz2007majority}, with critical exponents not belonging to the same universality class of the corresponding Ising model. The same holds for directed and undirected networks with heterogeneous degree distributions $P(k)\propto k^{-\alpha}$~\citep{lima2006majority,lima2007majority}.\medskip

Gradowski and Krawiecki extended the majority vote model to the case of hypergraphs~\citep{gradowski2015majority}. They introduced two different higher-order versions, a first one based on a node-update rule and a second one based on a hyperedge-update rule. According to the hyperedge-update dynamics, a hyperedge, representing a group, is randomly selected. The majority is then checked within the group. With the node-update dynamics, a random node is selected at random and then the majority rule acts on one of its hyperedges, selected at random as well. In both cases, when a hyperedge $\alpha$ is selected the opinions of all the nodes in $\alpha$ are updated according to the standard rule, as in Eq.~\eqref{eq:majority_vote}, but with the sum  $\sum_{j}\sigma_j$ running over all the nodes $j\in\alpha$ instead (including the node of the node-update rule). The main difference between the two rules is in the interaction dynamics. While in the first case each node at each time step interacts with all the other nodes in a shared group, in the second case just the nodes of a single group are considered. Obviously, this difference becomes significant when heterogeneous structures are considered.
Thus, Gradowski and Krawiecki studied the model on SF hypergraphs having a hyperdegree distribution that follows a power law $P(k)\propto k^{-\gamma}$ with $\gamma=1+\frac{N}{N-m}$, constructed by using a growth model with preferential attachment~\citep{wang2010evolving} (see Sec.~\ref{subsec:models:outofequilibrium:hypergraphs}). Both dynamics present a similar behavior, with a second-order phase transition appearing at a finite critical value $q_c$. Notice that here the transition appears even for a SF exponent $2<\gamma<3$, while the Ising model on SF networks presents a phase transition at finite temperature only for $\gamma>3$~\citep{hong2007finite}. As expected, the heterogeneous structure affects the two dynamics in different ways. In the case of the hyperedge-update, the topology does not have a strong influence on the critical exponents, and the hyperedge dynamics locally behaves as a mean-field for the Ising model. In contrast, when the node-update rule is considered the hypergraph topology matters, and the values of the critical exponents strongly differ from the corresponding equilibrium Ising model on SF networks having the same $\gamma$.

\subsubsection{Continuous models of opinion dynamics}
\label{subsubsec:social:opinion:non-linear}

The different models discussed so far describe the dynamics of interacting agents having discrete opinions. This approaches are suitable in those cases in which an individual can only have a clear and well-defined opinion on a subject, such as in politics, where one could be for or against the introduction of a given policy or the adoption of a given strategy. However, to model the more general dynamics of political orientation, discrete opinion variables might be too restrictive, leaving only ``black or white'' polarized options. In these cases, the opinion of an individual might be better represented by a continuous variable $x_i\in [0,1]$ spanning between two extremes \citep{deffuant2000mixing, lorenz2007continuous}. 
\citet{neuhauser2019multi} have investigated the effects of non-linear interactions in a model of continuous opinions dynamics on HOrSs. Starting from the formalism of dynamical systems on networks (see Sec.~\ref{sec:diffusion}), they have proposed a generalization with (3-body) higher-order interactions that captures the two important social mechanisms of peer pressure and homophily \citep{asch1951effects, mcpherson2001birds}. In their model, nodes interacts through the hyperedges of size 3 of a hypergraph. The evolution of the $N$ dynamical variables $x_i$ is given by:
\begin{figure*}[t]
	\centering
	\includegraphics[width=0.9\textwidth, keepaspectratio = true]{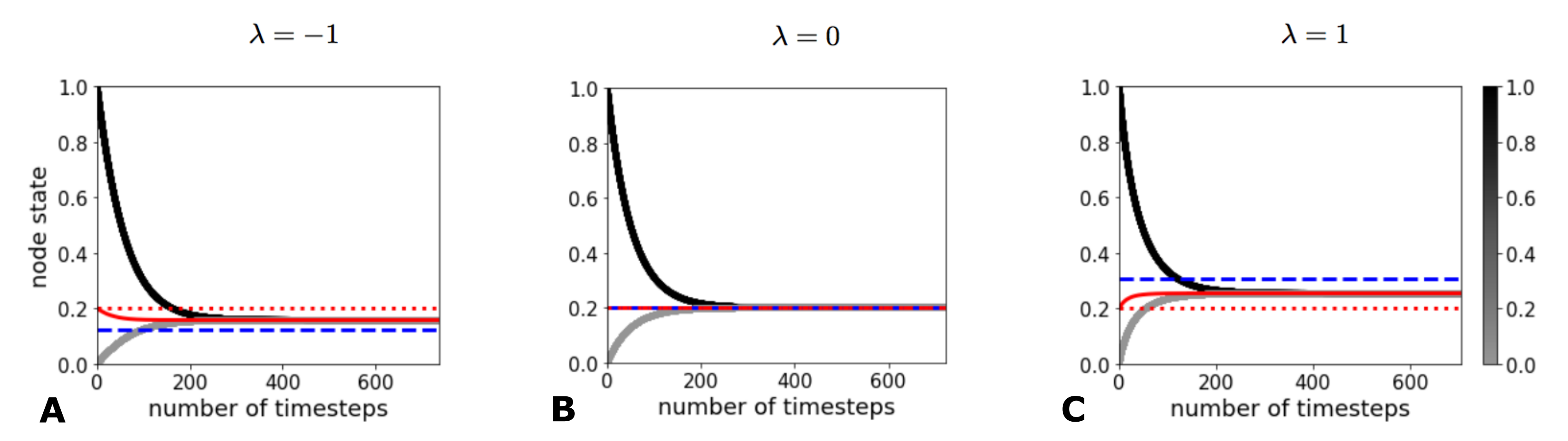}
	\caption[]{\textbf{Average node state for the 3-body non-linear consensus dynamics with continuous-valued opinions on a fully-connected hypergraph.} At $t=0$ an asymmetric opinion initialization is considered, such that $\bar{x}(0)=0.2$. The interaction function is$ s(x) = \exp(\lambda x)$. Dotted red lines indicate the initial value of the average node state. Black and grey solid lines represent the evolution of the state of nodes in the two initial configurations, one and zero respectively. Dashed blue lines denote the approximated final state. (A) If $\lambda < 0$ (similar node states reinforcing each other) the asymptotic average opinion state drifts towards the majority opinion. (C) The opposite effect is observed for $\lambda > 0$, where the dynamics shows a drift towards balance. (B) When $\lambda = 0$ the linear dynamics with a conserved average state is recovered. Figures adapted from Ref.~\citep{neuhauser2019multi}. }
	\label{fig:spreading_5}
\end{figure*}
\begin{equation}
	\dot{x}_i=\sum_{j,k=1}^N A_{ijk}\underbrace{s(|x_j-x_k|)}_{\text{Influence function}}\underbrace{[(x_j-x_i)+(x_k-x_i)]}_{\text{Linear term}}
\end{equation}

where the adjacency tensor $A_{ijk}$ restricts the interactions between nodes that share a hyperedge (see Sec.~\ref{subsec:adjacency}). Notice that the standard linear term, denoting the influence of the two nodes $j$ and $k$ on $i$, is modulated by an additional influence function $s(x)$ which depends on the difference between the states of the other nodes composing the hyperedge. According to the choice of $s(x)$, this model can reproduce the reinforcing or inhibitory effects that two nodes of a 2-simplex can have on the third one. For example, with a form $s(|x_j-x_k|)=\exp(\lambda|x_j-x_k|)$ and $\lambda<0$, similar states of $j$ and $k$ can accelerate the dynamics of $i$ (or decelerate if $\lambda>0$).
As expected, if the influence function is constant, i. e. $\lambda=0$, the standard linear case is recovered and the dynamics conserves the average state at time $t$, typically defined as $\bar{x}=\frac{1}{N}\sum_i x_i(t)$. Interestingly, this is not true when non-linear interactions are considered. Neuh\"auser et al. have showed that in the mean-field approach the higher-order interactions may produce a shift on the average state of the system depending on the initial state of the nodes. In particular, for an unbalanced binary initialization ($\bar{x}(0)\neq0.5$) the asymptotic average state is shifted towards the majority if $\lambda<0$ (Fig.~\ref{fig:spreading_5}A) or towards balance if $\lambda>0$ (Fig.~\ref{fig:spreading_5}C). These results are confirmed by numerical simulations on fully connected hypergraphs. Further analyses on modular hypergraphs have highlighted the additional role played but local sub-graphs in driving the system towards an asymmetric dynamics when non-trivial topologies are considered.

\subsubsection{Cultural dynamics}
\label{subsubsec:social:opinion:axelrod}

In the previous Section, we reviewed models that made use of scalar variables to represent evolving opinions. There is a general agreement in calling models with such characteristics {\it opinion models}. A separate class of {\it cultural models}, originally proposed by Robert Axelrod \citep{axelrod1997dissemination}, define the cultural profile of an individual as a vector rather than a scalar. This approach, useful to model the emergence of multi-culturality, incorporates two basic mechanisms of homophily and social influence into what is called now, unsurprisingly, the Axelrod model. In this model, individuals interact through the links of a social network by imitating each other, that is by copying an element of the feature vector of a neighbor. The imitation probability is proportional to the so-called {\it cultural overlap} among the two nodes, which in the original model corresponds to the fraction of common cultural features. The model has been extended to multi-layer networks, where interactions among individuals on different topics, such as religion, sport or politics, happen on different layers~\citep{battiston2017layered}. This approach allows for interaction patterns that are topic specific, therefore limiting the social influence among two individuals to the subset of features on which there is an actual social interaction (a link in the topic layer).\\

Even if a ``proper" extension of these model to HOrSs is still missing, HOrSs have still found their way into these modeling approaches. For example, \citet{maletic2014consensus} proposed to move away from the vectorial representation of cultural features and adopt a higher-order representation instead. In this case, an opinion can be represented as a set of interconnected judgments, so that different judgments forming an opinion represent the vertices of a simplex. In this formulation, overlapping opinions sharing arguments or judgments correspond to simplices sharing faces and ultimately forming a simpicial complex of opinions. This framework opens up new research directions in which overlapping opinions can then be used to shape social interactions~\citep{maletic2018hidden}.


\section{Evolutionary games}
\label{sec:games}

Imitation is an important mechanism to model social dynamics, at the heart of many processes described in Section~\ref{sec:social}. Yet, in several cases individuals do not make decisions simply based on peer pressure and social influence~\cite{roca2009evolutionary}. In many contexts, they can set and update their behavior based on strategic choices. In biology, for instance, the selection of a specific physical trait among the many alternatives typically occurs because of the beneficial effects which it brings to the survival of the species. Similarly, human decisions, or \textit{dilemmas}, are frequently based on computing and evaluating a trade-off between the positive and negative consequences of different scenarios.

Games are often studied in a simple dyadic setting, where pairs of individuals are given the chance to pursue either a selfish strategy and defect (D), or a cooperative choice (C), with the selfish strategy being the more rewarding unless both of them undertake it. In the most general set-up, pairwise games can be defined according to a payoff matrix:
\begin{equation}
\bordermatrix{
  & C & D \cr
C & R & S \cr
D & T & P \cr}
\label{tab:payoffMatrixgames}
\end{equation}
where \emph{R} is the \textit{reward} obtained by a cooperator playing against another cooperator, \emph{S} is the \textit{sucker} payoff received by a cooperator when its opponent is a defector, \emph{T} is the \textit{temptation} a defector has to resist when it plays against a cooperator, and \emph{P} is the \textit{punishment} that a defector receives when it plays against another defector. Games where more than two strategies are possible, such as the rock-paper-scissors~\cite{szabo2004rock, szolnoki2014cyclic}, are not discussed in this section. Games can be categorized into different classes, based on the relative order of the four payoffs previously introduced.

The most famous game is the prisoner's dilemma~\cite{axelrod1984evolution, rapoport1966taxonomy}, where two members of a criminal gang are arrested and isolated from each other. As each prisoner could be convicted of a small charge, but there are no sufficient evidence to convict them for the main greater charge, both prisoners are given the opportunity to bargain, i.e. defecting by stating that the other committed the crime, and being in exchange set free. Cooperation is harmed by a high \textit{temptation} $T$, as in this setting the payoff associated to defect against a cooperator yields a higher payoff than the \textit{reward} $R$ associated to both players cooperating and staying silent. Besides, cooperating against a defector gives the lowest earning, typically known as \textit{sucker} $S$ payoff, and the payoff ordering of the game is $T > R > P > S$. As no defecting player can benefit by changing strategy if the others keep theirs unchanged, this makes the decision of both players to defect, i.e. \textit{punishment} $P$, the Nash equilibrium of the system, despite being a less rewarding situation than full cooperation. 

Many alternative real-life situation have been described as social dilemmas. The Stag-Hunt game describes the dilemma of two hunters, which must cooperate to kill a stag and avoid going hungry~\cite{rousseau1997discourses, luce1957games}. It is described by the ordering $R > T > P > S$, and suitable to formalize conflicts between cooperation and safety. Differently from the prisoner's dilemma, this game has two distinct pure Nash equilibria, full cooperation, leading to the highest payoff, and full defection, which is risk dominant as it prevents from the risk to be the only hunter involved in the attempt to kill the animal. The Stug-Hunt is a coordination game, as it requires the two individuals to coordinate in order to converge towards the payoff dominant equilibrium. 

Other games are described by the payoff ordering $T > R > S > P$. This ordering is associated to the chicken game, where two individuals drive towards each other looking for a free way at the risk that both may die in the crash, but hoping that the other swerves away (acting cowardly like a poultry). Outside the political sciences, the same ordering of payoff is typically referred to as the snowdrift game. In this game, a snowdrift is blocking the way, and at least one of two individuals has to shovel away snow to free the road. This setting describes well situations where defectors benefit from cooperators without paying a cost for accomplishing a given task, but at least a cooperator is needed for the task to be performed. Here the player's optimal choice depends on what their opponent is doing, as one should yield only if the opponent fails to. For this reason, the chicken game is an anti-coordination game~\cite{gui2005economics}. It is worth to notice that the same ordering is also associated to the hawk-dove game, where two players compete for a resource to be shared, an outcome which is possible without damage only when two doves meet~\cite{smith1982evolution, smith1976logic, cressman1995evolutionary}. All these games have two pure Nash equilibria, in which each player plays one of the pair of strategies, and the other player chooses the opposite one. 

These dilemmas were introduced as static games. Almost fifty years ago the pioneering work of John Maynard Smith~\cite{smith1972game} first considered the dynamics of a population with repeated strategic interactions, a discipline now known as evolutionary game theory. In the well-mixed scenario, where all agents have equal probability to interact with each other, the evolution of the fraction of cooperators $x_c$ can be tracked by the set of differential equations describing the so-called replicator dynamics ~\cite{diederich1989replicators,hofbauer1998evolutionary,opper1999replicator,chawanya2002large}
\begin{equation}
\dot x_c = x_c ( 1 - x_c ) [ \bar \pi_C - \bar \pi_D ]
\end{equation}
where $\bar \pi_C$ is the average pay-off of a cooperator, and $\bar \pi_D$ the average pay-off of a defector. The replicator equations are also widely used to model species interactions, as we will discuss in Section \ref{subsec:ecology}.

Numerically, the dynamics of games are often studied as agent-based models~\cite{perc2010coevolutionary}. The importance of numerical simulations, in particular in the case of agents placed on a network, is also linked to the limitations achieved by analytical methods. A first approach is the so-called best response, where individuals choose the strategy which produces the best outcome for them taking the strategies of the other players as given. However, more complex update processes are often considered, where players can update their strategy by imitating the behavior of the most successful individuals, where the higher copying probability the larger the difference in earnings. This allows to consider more realistic scenarios, where agents can also make mistakes.
 
For the prisoner's dilemma, in well-mixed population (where at each round individuals have the same probability to play with any other agent in the population) the evolutionary dynamics brings the system into a state of full defection~\cite{hofbauer1998evolutionary}. This is in spite of the low payoff associated to the outcome, a situation sometimes referred to as the tragedy of the commons~\cite{hardin1968tragedy}. However, when populations are structured, meaning that interactions between agents---often limited by spatial constrains---can be described by a network of relationships, cooperators are able to emerge even in adverse settings. First discovered by Nowak and May by placing agents at the nodes of a simple square lattice, repeated games between the same pairs of individuals allow for \textit{network reciprocity}, i.e. the creation of robust mutual interactions based on trust, even if the temptation to defect would prove to be more rewarding in a single individual round~\cite{nowak1992evolutionary}. Graphs which are heavy-tailed~\cite{santos2005scale, gomez2007dynamical} or clustered~\cite{assenza2008enhancement} provide the best conditions for the emergence of cooperation, exploiting the beneficial effect of prosocial hubs and the presence of tightly connected communities to sustain the formation of trust among players.

In the snowdrift game the relative values of the sucker payoff and punishment are inverted, as a lone cooperative shoveler still has a better cost benefit ratio than an individual in a pair where both agents defect. This appearently small difference generates the emergence of a stable state with coexisting cooperators and defectors in structuredless population, differently from the prisoner's dilemma~\cite{hofbauer1998evolutionary}. Surprisingly, spatial structures were shown to be detrimental for cooperation in the snowdrift game~\cite{hauert2004spatial}.

In the last 15 years the fields of evolutionary game theory and network science have become significantly closer, and we are now witnessing an explosion of contributions at the boundary of these topics~\cite{nowak2006five, szabo2007evolutionary, wang2015evolutionary}. In the following of this Section, going beyond the traditional pairwise scheme, we provide an overview of the main results on evolutionary games in networks with group and higher-order interactions.

\subsection{Multiplayer games on networks}

\subsubsection{Public goods game}

\begin{figure*} 
	\centering
	\includegraphics[width=\textwidth, keepaspectratio = true]{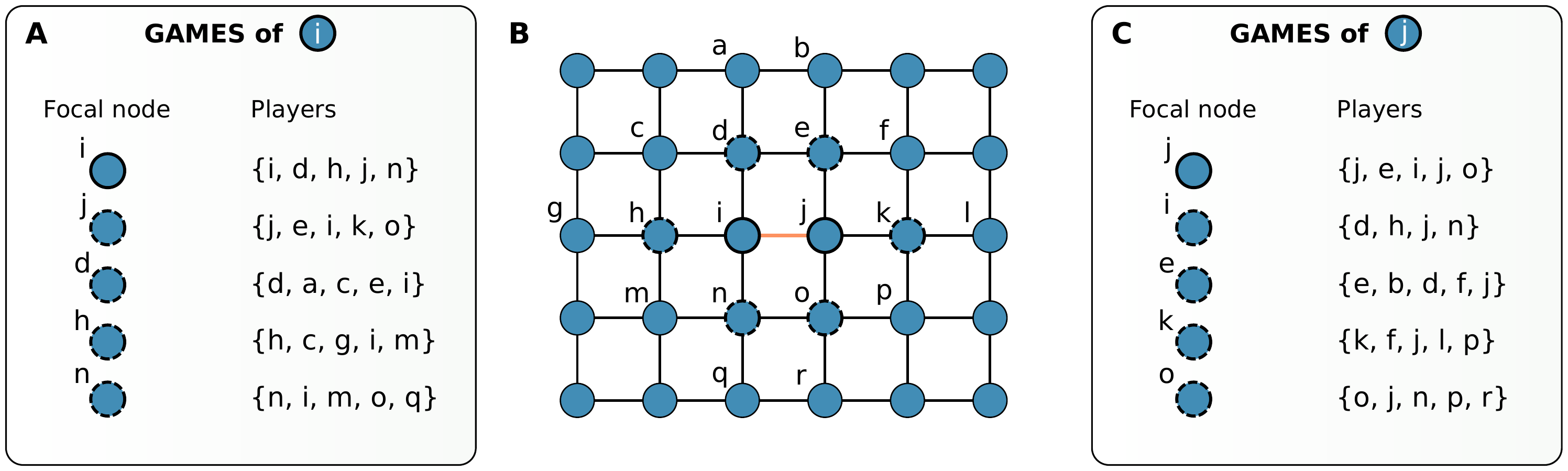}
	\caption[]{\textbf{Traditional graph implementation of a multiplayer game. }At each elementary step a player $i$ and one of its neighbors $j$ are chosen at random. Each individual  accumulates earnings by playing all games in which it is involved, namely the game in which it is the focal individual of the group, and the $k$ games where it participates in a group centered on one of its neighbors. All groups in which $i$ and $j$ participate are listed in panels (A) and (C), for a simple two-dimensional lattice (B). Finally, $i$ compares its payoff to that of $j$, updating its strategy by imitating the strategy of the neighbor with a probability which depends on the relative difference of the payoff. The presence of links among neighbors of $i$ and $j$ (clustering) does not affect the definition of the groups.}
	\label{fig:games_1}
\end{figure*}

Many dilemmas do not involve pairs of individuals, but occur at the level of groups. This is the case of taxes for welfare state, which are beneficial from an individual perspective only if most individuals are willing to contribute. The public goods game is the paradigmatic game to describe social dilemmas in the case of group interactions~\cite{sigmund2010calculus, archetti2012review, perc2013evolutionary, perc2017statistical}, and it is considered the generalization of the prisoner's dilemma to $N>2$ players. In the most simple implementation, players, belonging to a group of size G, are asked to contribute to a common pool. Cooperators contribute with a token $t$ whereas defectors do not contribute at all. The tokens are then multiplied by a synergy factors $R$, with $R<G$, and shared evenly across the population no matter the strategy of the agents. For this reason, if we indicate with $N_c$ the number of cooperators in the group, cooperators earn a payoff $\pi_c = t (N_c R / G -1)$, whereas defectors obtain $\pi_d = t (N_c R / G)$. To simplify the payoffs and without loss of generality, $t$ can be set equal to $1$. The game is fully controlled by the effective parameter $r=R/G$, which is known as the reduced synergy factor. The traditional implementation of a multiplayer game on a graph is illustrated in Fig.~\ref{fig:games_1}. As explained later on, different implementations are necessary if one wants to explicitly take into account the real pattern of higher-order interactions among individuals.

Similarly to the prisoner's dilemma, the network structure affects the emergence of cooperation also in the public goods game~\cite{santos2005scale,pena2016evolutionary}. Simulations on lattices first showed that structured interactions sustain cooperation for values of the synergy factor well below the critical condition $R=G$~\cite{szabo2002phase, brandt2003punishment}. Yet, interest in the public goods game sparked when more realistic network structures with heterogeneous degree distributions were considered~\cite{santos2008social}. In heterogeneous networks, at every round each agent is involved in $k+1$ games (where $k$ is the degree of each node), meaning that hubs play significantly more games than agents placed on poorly connected vertices. In the implementation known as \textit{fixed cost per game}, in particular, $(k+1)$ is also the total contribution of each agent after a full round. Because of this inequality, another set-up has also been widely investigated, where each agent has a \textit{fixed cost per individual} $t$, and hence each game contributes to its total payoff as $1/(k+1)$. Fixed costs per individual are considered a suitable setting to model cumulative costs, such as in the case of taxes, while fixed costs per individual are best suited to model a scenario in which the resource associated to the cost is finite, and it is equally distributed among players. In scale-free networks cooperation was found to be significantly enhanced through network reciprocity if cooperators pay a fixed cost per individual, whereas the same effect is reduced when prosocial individuals pay a fixed cost per game~\cite{santos2005scale}. The enhanced cooperation of the first scenario is due to the disproportionately higher payoff obtained by those agents participating in a very large amount of groups.

The beneficial effect from interacting according to a scale-free networks can be lost if the underlying structure is characterized by positive degree-degree correlation, reducing the evolutionary advantages of individuals with high-degree who decide to cooperate~\cite{rong2009effect}. Similarly to pairwise games, tight community structure and clustering can sustain positive feedback and the survival of cooperators for low values of the synergy factor also in the public goods game~\cite{rong2010feedback}. Prosocial behavior may be further promoted by allowing heterogeneous contributions, for instance by making them proportional to the cooperation of each group~\cite{gao2010diversity, vukov2011escaping}. However, interestingly, increasing group size does not always lead to the dynamics of well-mixed populations in such a multiplayer game~\cite{szolnoki2011group}. 

Multiplayer games are considered to be inherently different from the corresponding pairwise games, as the emergent collective behavior in the case of group interactions might be different~\cite{perc2013collective}. This is rooted in the formation of indirect links between players who belong to the same group but are actually disconnected. As an important consequence of this, the details of the local topology of the network of interactions for multiplayer games are often irrelevant for the final outcome, as discussed in Ref.\cite{szolnoki2009topology}. Besides, multiplayer games also show qualitatively different evolutionary dynamics, giving rise to new forms of self-criticality which have not been observed in pairwise games~\cite{szolnoki2013correlation}. Among those, we report the emergence of new temporal and spatial patterns of dominance, such as the so-called indirect territorial competition, first reported in a modified public goods game in Ref.~\cite{helbing2010evolutionary}.

More recently, the role of multiplexity has also been taken into account. In this setting, different layers are associated to different games (different values of the synergy factor) and individuals may have different neighbors depending on the layer. While not making possible direct strategy exchanges between the networks, interconnectedness can affect the utility function of players, who do not have access to the earnings of their neighbors on each of the layers but only to a (possibly non-linear) combination of them~\cite{wang2012evolution}. Interestingly, this simple payoff coupling was shown to further enhance cooperation in the whole system, through a mechanism dubbed \textit{interdependent network reciprocity}~\cite{wang2013interdependent}. Despite the presence of the same synergy factors at different layers, the system naturally self-organizes into a configuration where one layer is more cooperative than the others through spontaneous symmetry breaking~\cite{wang2013interdependent}.
However, interdependent network reciprocity is not a universal properties of all interconnected systems, but strongly depends on the structure of the layers. In particular, interdependent network reciprocity is proportional to the fraction of edges shared by the different layers (i.e. edge overlap~\cite{battiston2014structural}), and the beneficial effects of multiplexity completely disappear when the overlap between the network structures goes to zero, no matter the number of layers~\cite{battiston2017determinants}. In this detrimental configuration, cooperators can appear in the system only if the synergy factors of all layers are at least as high as the critical conditions associated to each network in isolation~\cite{battiston2017determinants}. Further coverage on the effects of multiplexity and network interdependence on evolutionary games is provided in Ref.~\cite{wang2015evolutionary}. 

Much attention has also been devoted to the case of coevolutionary games, where the structure of the networks of social interactions may change over time as a result of the outcome of the strategic interactions among players~\cite{perc2010coevolutionary}. When individuals can alter the connections in their social network in response to unsatisfactory interactions, coevolution between cooperation and spatial organization in the public goods game naturally leads to increased social cohesion~\cite{roca2011emergence} and prosocial behavior~\cite{pichler2017public}.
A different setting is considered in Ref.~\cite{ren2018coevolution}, where a survival cost parameter is introduced, and agents with low payoff are replaced in the games by new random players. With this mechanism, cooperation emerges if the synergy factor is higher than the average degree, and the population self-organizes into a scale-free network of interactions, naturally more beneficial to sustain prosocial behavior. Finally, coevolutionary rules have been implemented in the case of interdependent network of players, leading to complex social structures where strong inter-layer links were promoted around agents performing the best~\cite{shen2018coevolutionary}.

As most of these results rely on numerical simulations, it is important to highlight the importance of proper simulations practices. An important contribution to the topic is found in Ref.~\cite{perc2018stability}, which investigates the stability of observations from agent-based model simulations going beyond traditional finite-size scaling, or on the role of noise~\cite{javarone2016role}. In particular, only a complete stability analysis of all subsystems solutions (solutions that are formed by a subset of all possible individual strategies) can be explicitly linked to the existence of a phase transition in the thermodynamic limit in multiplayer games where competing strategies are more than two. For a complete review on the public goods game on networks, we refer the interested reader to the work by Perc et al. ``Evolutionary dynamics of group interactions on structured populations: A review''~\cite{perc2013evolutionary}.

\subsubsection{Other multiplayer games}
\label{subothergames}

The public goods game is not the only multiplayer game whose evolutionary dynamics has been studied on graphs. An interesting alternative is the generalized multiplayer snowdrift game, where individuals receive a benefit $b$ if the task is performed by one or more agents belonging to their same group, no matter their strategy~\cite{zheng2007cooperative}. In this game cooperators share the workload and have a payoff $\pi_c = b - c / N_c$, while defectors have a payoff $\pi_d = b$ as long as there is at least a cooperator in the group, otherwise $\pi_d = 0$. In well-mixed populations cooperation decreases with high cost-to-benefit ratio, as well as a function of the number of agents in a group~\cite{zheng2007cooperative}. When network structures are considered, the introduction of an underlying homogeneous graph steadily promotes prosocial behaviors, similarly to the corresponding game played on well-mixed populations. By contrast, heterogeneous networks typically generate multiple new internal equilibria~\cite{santos2012dynamics}. The introduction in the game of dynamical grouping, where agents are placed in groups of different sizes at different times, and players of different strategies are dynamically mixed was found to greatly enhance prosocial behavior~\cite{ji2011effect}. A common modification of the traditional multiplayer snowdrift games links the benefit of accomplishing a task to the existence of a minimum threshold of cooperative individuals~\cite{pacheco2009evolutionary, souza2009evolution, santos2011risk}.
Despite its peculiarity, the game is sometimes considered as a particular type of public goods game.

More recently, a generalization of the hawk-dove game to multiple interacting players was proposed~\cite{chen2017evolutionary}. In particular, whereas in the corresponding pairwise settings the game is considered equivalent to the snowdrift game, this is not true when interactions occur between groups of players. In particular, while in the snowdrift game the accomplishment of the task benefits all individuals, in the hawk-dove problem only strategists of a particular kind (hawks) benefit from the shared resource, excluding the opponent type (dove) from the distribution, unless a sufficiently high number of cooperative doves is present in the group. The emerging evolutionary dynamics is very rich: different scenarios associated to dominating hawks, coexistence, bi-stability, multiple interior equilibria and dominating doves can be obtained as a function of the dynamical parameters describing the resource to be shared, cost and minimum number of doves in a group for prosocial individuals to benefit.

The games discussed so far are multiplayer generalizations of pairwise dilemmas described by the payoff matrix in Eq.~\eqref{tab:payoffMatrixgames}. It is worth to mention that also different families of pairwise games have been investigated in the setting of wider group interactions. An example is that of the ultimatum game, where two players, one acting as a proposer and one as a receiver of an offer, bargain to split a sum of money~\cite{guth1982experimental, sinatra2009ultimatum}. In the multiplayer case, offers can have an arbitrary number of receivers, who can reject or accept the proposal individually. Similarly to a threshold model~\cite{granovetter1978threshold}, the offer is accepted and shared equally among responders only if the number of individual acceptances is above a given threshold. The game is significantly affected by the value of this parameter, with higher values associated to more generous and fair outcomes~\cite{santos2015evolutionary}.

\subsection{Games with higher-order interactions}

\subsubsection{Public goods game on bipartite networks}

\begin{figure*} 
	\centering
	\includegraphics[width=\textwidth, keepaspectratio = true]{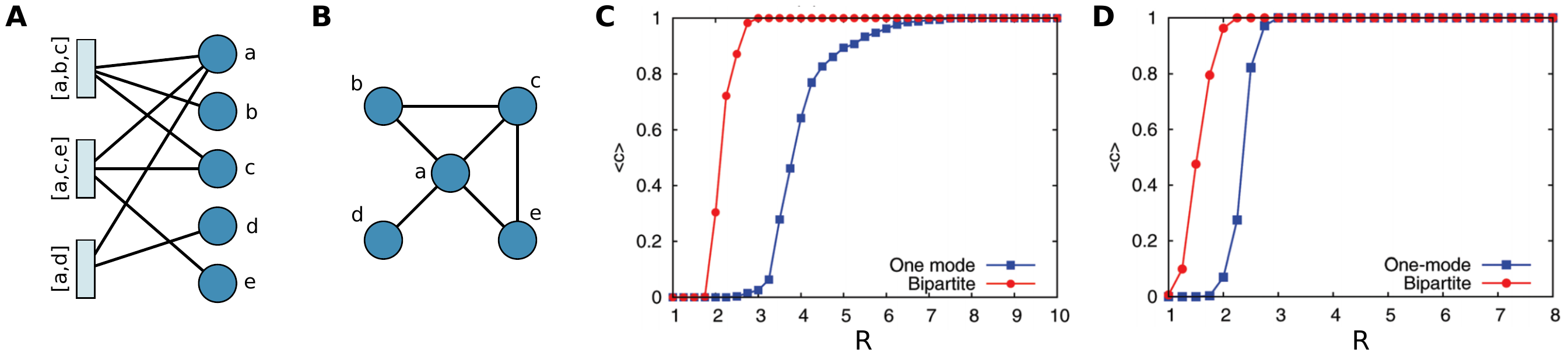}
	\caption[]{\textbf{Public goods game on bipartite graphs.} Information on the exact group structure encoded in a bipartite network (A) is lost when its one-mode projection is considered, where two individuals are directly linked if they both participate in at least one group (B) . Cooperation is enhanced when the game is played by considering the real group structure instead of the projected graph, both for the fixed cost per game (C) and fixed cost per individual (D) implementation. Prosocial behavior is greater when fixed costs per individual are considered. Figures (C) and (D) reproduced from Ref.~\cite{gomez2011evolutionary}.}
	\label{fig:games_2}
\end{figure*}

Networked implementations of multiplayer games discussed so far lack control on the real higher-order structure of interactions. 
For instance, a scale-free degree distribution  $P(k) \sim k^{- \gamma}$ generates a similarly heavy-tailed distribution of group sizes $P(G)$. This is not a realistic feature of many real social networks, such as collaboration networks, where
individuals tend to collaborate together in fairly small groups of homogeneous size~\cite{ramasco2004self}.

Motivated by this finding, in Ref.~\cite{gomez2011evolutionary} G{\'o}mez-Garde{\~n}es et al. studied the public goods game on an empirical bipartite network of scientific collaborations, where the two sets of nodes describe respectively scientists and papers. The authors compare the results with what obtained by implementing the game on the corresponding one-mode projection, acting as a null-model, where scientists who co-authored at least a paper are linked together, as shown in Figs.~\ref{fig:games_2}A,B. The main finding is that---no matter the details of the updating rule for the evolutionary dynamics---cooperation is systematically enhanced by considering the real higher-order structure described by the bipartite network. This is due to the interplay between the heterogeneous distributions of the number of games in which each player is involved, and the homogeneous distribution of the groups. Taken together, these two features allow for the existence of a fairly high number of agents with high payoff involved in small groups, making this scenario responsible for further promoting prosocial behavior.

Enhanced cooperation with respect to the one-mode projection is observed consistently in both settings of fixed cost per game (Fig.~\ref{fig:games_2}C) and fixed cost per individual (Fig.~\ref{fig:games_2}D): the degree of cooperation is always greater when fixed costs per individual are considered rather than fixed costs per game, consistently with what found for classical monopartite networks, though the differences are now smaller. The boost in cooperation for fixed cost per individual is linked to the long tail in the distribution of degrees and number of games. However, this effect is mitigated when the real structure is considered. This can be understood by considering the agents who participate in a single collaboration, whose contribution is the same in the bipartite network both for fixed cost per individual and fixed cost per game scenarios, while this is not the same in the projected one-mode network where a node in general participates in $k+1$ groups. Interestingly, the authors remark that the final cost per individual is a better setting to model collaborations, where researchers have a limited amount of time to be shared among parallel projects, and where those involved in only a few collaborations tend to take most of the workload. Finally, increasing the number of members of a group leads to a decrease of the level of cooperation in a population.

In a following work, G{\'o}mez-Garde{\~n}es et al. deepen their investigations of higher-order structure by studying the effect of bipartite networks which have groups of the same sizes, but where the number of groups in which which a player can participate is a tunable parameter~\cite{gomez2011disentangling}.
Surprisingly, the average level of cooperation achieved with homogeneous connectivity  is remarkably larger than that for scale-free substrates.
This finding indicates that the ability of scale-free networks to outperform the promotion of cooperators in homogeneous structures, first discussed in Ref.~\cite{santos2005scale}, is not directly linked to the fat-tail in the distributions of number of games per player. At difference, instead, it rather depends on the entanglement of social and group heterogeneities which is unavoidable in the one-mode projection. Further analyses on the impact of different distributions for group sizes and the number of individual contacts reveal the importance of overlap between groups for cooperation, similarly to the role of clustering on one-mode projected networks~\cite{pena2012bipartite}.

An interesting modification of the game considers the possibility that information on the earnings is shared between groups. This can be done by introducing an effective payoff, where the normal payoff associated to a member playing in a group is combined with the earnings of the same agents from the other groups in which it participates. The strength of this second term can be tuned  with a parameter $\alpha$, which describes the degree of cross-information among groups. Interestingly, information exchange is positively correlated with an enhancement of the cooperation in the system~\cite{gracia2014intergroup}. The positive effect induced by cross-information is analogous to that of interdependent network reciprocity for multiplex networks, where individual payoffs are aggregated across the layers of the system.

\subsubsection{Public goods game on hypergraphs}

\begin{figure*} 
	\centering
	\includegraphics[width=0.9\textwidth, keepaspectratio = true]{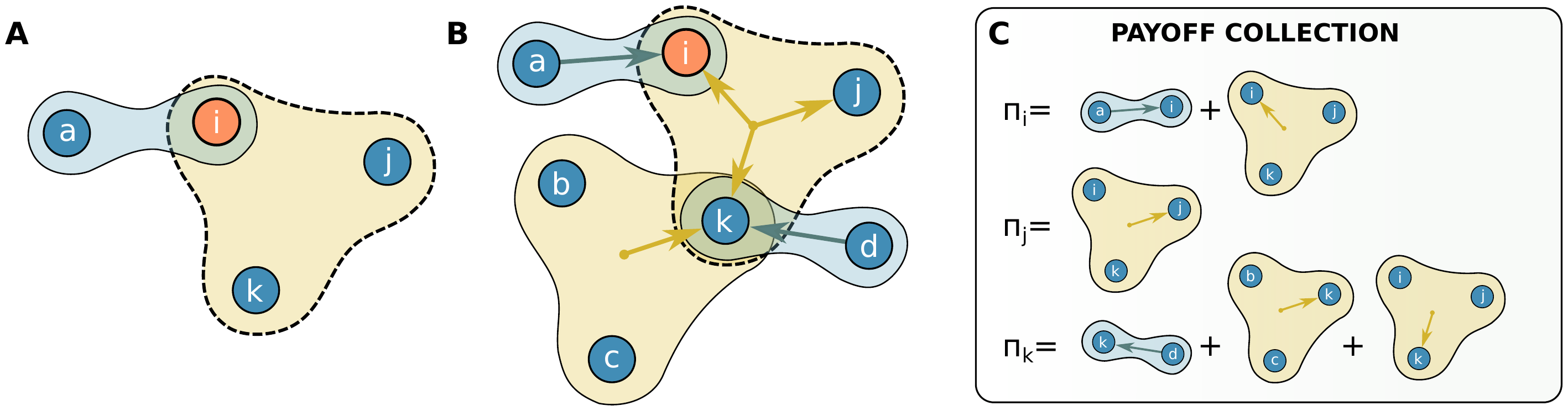}
	\caption[]{\textbf{Hypergraph implementation of strategic group interactions.} (A) At each time step, a node $i$ is chosen randomly,  and one of the hyperlinks to which it participates is selected. (B) All the members of the hyperlink play a game for each of the hyperlinks they are part of, and (C) accumulate payoffs accordingly.}
	\label{fig:games_3}
\end{figure*}

Recently, Alvarez-Rodriguez et al. introduced a new formalism to describe the evolutionary dynamics of higher-order interactions~\cite{alvarez2020evolutionary}. In this set up, groups of individuals are described by the hyperlinks of a hypergraph, making explicit the lift of interaction networks to the case of non-dyadic interactions.
The Monte Carlo implementation of the dynamic is illustrated in Figs.~\ref{fig:games_3}A-C. In the manuscript, the authors finally normalize the payoffs by the number of played games, in an implementation reminiscent of the fixed cost per game scenario. The update process is also modified, and $i$ imitates the strategy of its best performing neighbor $k$ with a probability which depends on the difference $\pi_i - \pi_k$.

\begin{figure*} 
	\centering
	\includegraphics[width=\textwidth, keepaspectratio = true]{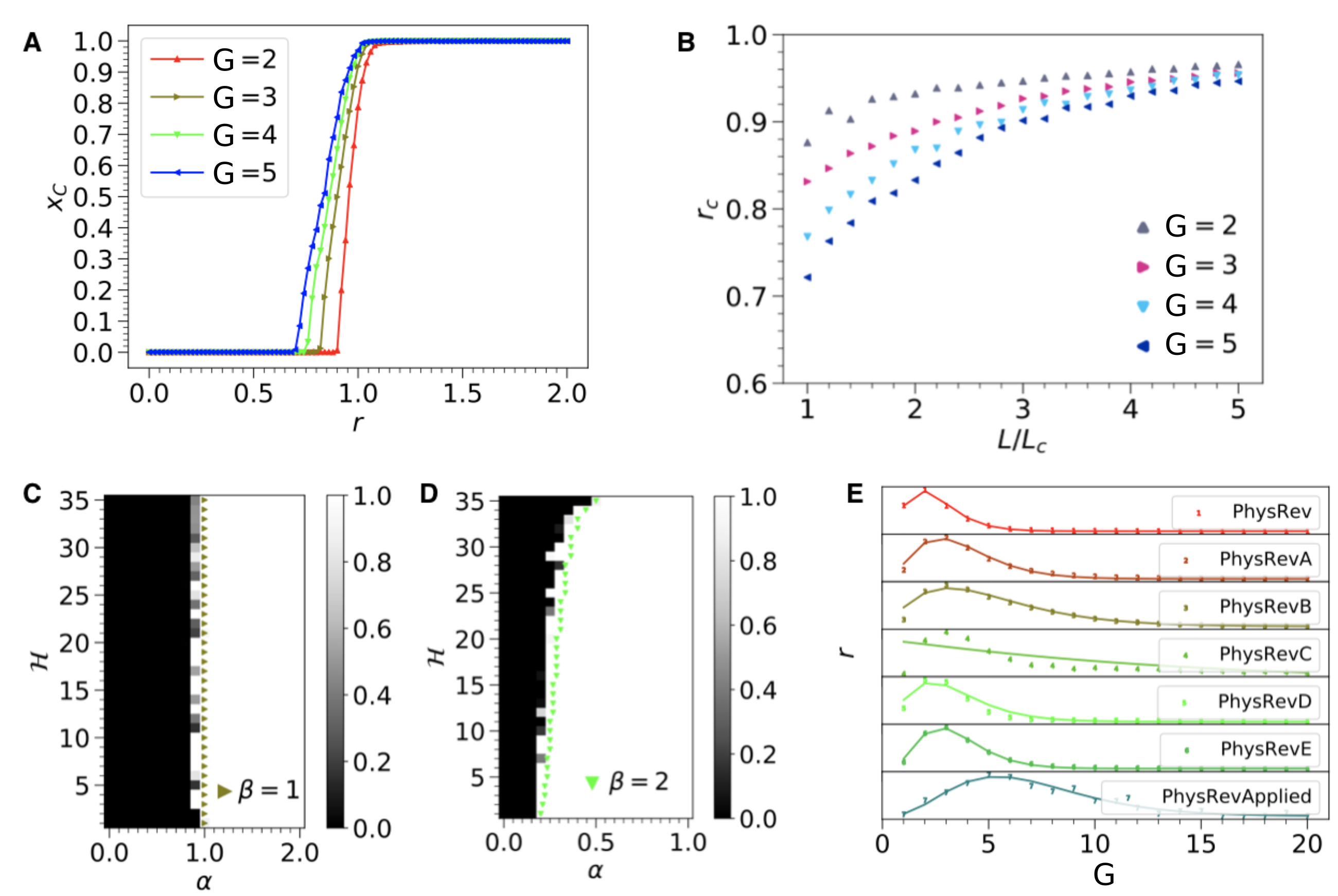}
	\caption[]{\textbf{Cooperation in the public goods game on hypergraphs.} (A) Average fraction of cooperators as a function of the reduced synergy factor $r$ for homogeneous hypergraphs with interactions of different order $g$. (B) Critical value $r_c$ for the emergence of cooperation as a function of the density of the hypergraphs $L/L_c$, where $L_c$ is the critical number of hyperlinks for a connected hypergraphs, for different values of $g$. Classes of $35$ different heterogeneous hypergraphs $\mathcal H$ with hyperlinks of orders $g = \{2, 3, 4, 5\}$ are considered. The synergy factor $R(\alpha, \beta)$ scales according to Eq.(\ref{eq:games_heter}). Results for two values of the exponent, $\beta=1$ (C) and $\beta=2$ (D), are shown. (E) Dependency of synergy factors from hypergraphs describing co-authored publications in journals of the American Physical Society assuming the collaboration process is optimal. Figures reproduced from Ref.~\cite{alvarez2020evolutionary}.}
	\label{fig:games_4}
\end{figure*}

The stable state achieved from the evolutionary dynamics is first studied on uniform random hypergraphs, where all players are involved in the same number of hyperlinks of equal size, $G=2, \ldots, 5$. Figure~\ref{fig:games_4}A shows the fraction of cooperators $x_c$ as a function of the reduced synergy factor $r$, where the hyperdegree of each node is the minimum one to guarantee a connected hypergraph. As shown, the presence of larger groups promotes the onset of cooperation in more adverse conditions described by a low value of $r$. When density increases---no matter the group size---the hypergraph implementation of the public goods game converges to the limit of well-mixed population, as shown in Fig.~\ref{fig:games_4}B. Besides, the relaxation time $T$ of the dynamics can be computed with good analytical accuracy with a mean-field approach.

The introduced formalism also allows for an analytical treatment of heterogeneous random hypergraphs, where nodes participate in groups of different order with a weight described by a probability distribution. For simplicity, the authors focus on heterogeneous hypergraphs with a different frequency of groups $f_G$ of order from 2 to 5 and $\sum_G f_G = 1$, giving rise to 35 different hypergraphs classes. In this set-up, it is interesting to investigate the collective outcome of the game in scenarios where larger or smaller collaborations can be more or less effective. These different conditions are easily described by setting the synergy factor $R$ to be a function of the group size $G$, i.e. 
\begin{equation}
R(G) =\alpha G^{\beta}, 
\label{eq:games_heter}
\end{equation}
where $\alpha>0$ and $\beta \ge 0$ and different from 1. The emergence of cooperation is in general affected by both parameters $\alpha$ and $\beta$. In the particular case $\beta=1$, all classes of hypergraphs show the same behavior as a function of $\alpha$ (Fig.~\ref{fig:games_4}C). This is not true for different values of the exponent, such as $\beta=2$ (Fig.~\ref{fig:games_4}D). Besides, the average relaxation time $T$ of the dynamics as a function of the critical point $\alpha_c$ for the emergence of cooperation scales linearly with $\alpha$ if $\beta \neq 1$. Interestingly, while a degeneracy is observed for $\beta \le 1$, this is broken for the superlinear case. This means that it is possible to exploit this additional degree of freedom by choosing an adequate hypergraph to set independently a chosen critical point and relaxation time.

Finally, the authors considered several datasets describing synergies and group tasks in the real-world. By imagining that these collaborations---ideally described by a public goods game---evolved over time to produce an optimal hypergraph structure and that a coordination cost is added to sustain too large collaborations, they inferred the ideal synergy factor associated to prosocial behavior. As an example, results for collaborations among physicists publishing in different journals of the American Physical Society are shown in Fig.~\ref{fig:games_4}E). Journals in experimental and applied physics typically have an optimal synergy factor for larger values of the group size $G$.

To conclude, even more than for the dynamical processes considered in Sections (5-7), the landscape of HOrSs in the field of social dilemmas is still widely unexplored. In the future, we foresee that an explicit treatment of higher-order interactions could be applied to the many games discussed in Section~(\ref{subothergames}), as well as others such as the naming game~\cite{baronchelli2006sharp, baronchelli2006topology}, the sender-receiver game~\cite{gneezy2005deception, capraro2019evolution,capraro2020lying}, or problems of collective risk~\cite{milinski2008collective}.


\section{Applications}
\label{sec:applications}

Non-pairwise interactions are common in various types of systems in the
real world. Important examples include group interactions in both offline
and online social networks
\cite{wasserman1994social,freeman1980qanalysis,andjelkovic2015hierarchical},
multi-authors scientific collaborations \cite{patania2017shape},
network motifs in transcription networks \cite{mangan2003feedforward}
and trigenic interactions in gene regulatory networks
\cite{kuzmin2018threegenic}, 
beyond pairwise mechanisms of species coexistence in ecological
communities \cite{levine2017beyond}, and higher-order correlations  
in neuronal \citep{schneidman2003network} and whole-brain functional patterns
\citep{petri2014homological,ibanez2019topology}. 
In this section, we present a selection of some of the
possible applications of HorSs in fields spanning 
from social sciences to neuroscience and biology.

\subsection{Social systems}
\label{subsec:application:social}

\begin{figure*}
\centering
\includegraphics[width=\textwidth, keepaspectratio = true]{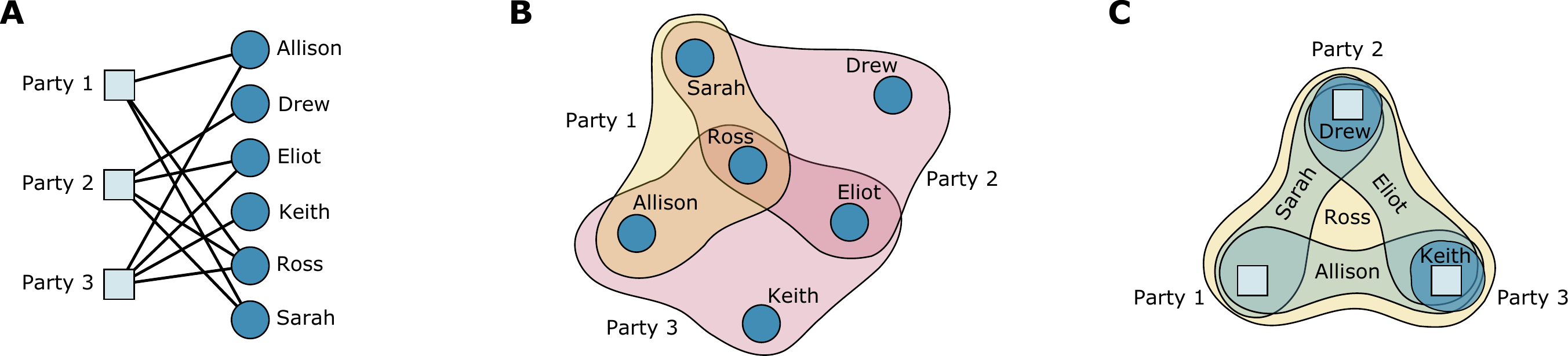}
\caption[]{\textbf{
  Affiliation network of six children and three parties.}
  The interactions are shown respectively as a bipartite network actor-events, as the hypergraph with
  the six children as the nodes, and as the dual hypergraph with the three
  parties as the nodes. Figures adapted from Ref.~\cite{wasserman1994social}}.
\label{fig_social1}
\end{figure*}

Social scientists have realized since long time the importance of
hypergraphs and simplicial complexes to describe and study
affiliation data \cite{wasserman1994social}. {\em Affiliation networks},
also known as {\em membership networks}, are a special kind of two-mode
social networks representing the affiliation of a set of $n$ actors to a
set of $m$ events, or social occasions. As an example, Fig.~\ref{fig_social1}A
shows the bipartite network of the attendance of $n=6$ children
to $m=3$ birthday parties.    
This system can be naturally described as a hypergraph whose nodes
are the children and the hyperedges are the set of events (panel b). 
Notice that the data can be represented equally well by the dual
hypergraph, obtained by reversing the roles of nodes and edges (panel
c). In this latter case the three nodes are the three parties, while
each child is a hyperdegree. 

\bigskip
%
The literature on applications of hypergraphs and
simplicial complexes in the social sciences is vast. We will then
limit our survey to the pioneering works and to some recent results.
Ref.~\cite{mcpherson1982_hyper} presents one of the
earliest use of hypergraphs to investigate affiliation to
voluntary organizations. The work focuses on the issue of sampling and
proposes estimators for the number and size distribution of
organizations in cities, the density of relations among individuals
generated by organizations, and the amount of membership overlap among
organizations. Such estimators are then applied to data from a sample
of individuals in different towns of the state of Nebraska. Results
show that, while the mean affiliation rate does not systematically
vary from city to city, the number of inter-organizational links per
organization increases with the size of the city.
Hypergraphs have also been used to capture the characteristic
fluidity of urban social structures arising from collections
of overlapping subsets such as voluntary associations,
ethnic groups, action sets, and quasi-groups~\cite{foster1982urban},
and to study participation of Thai households
to ritual celebrations~\cite{foster1984thai}.

Ref.~\cite{faust1997centralityaff} discusses the conceptualization,
measurement, and interpretation of centrality in affiliation
networks. The main underlying assumption is that, in an affiliation
network, also the events can acquire and transmit centrality.
Ref.~\cite{bonacich2004hyper} shows how centrality can be adapted to HOrSs
and can turn very useful to capture important properties of real-world
systems. The measures proposed in this work are extensions of the Bonacich
eigenvector centrality
\cite{bonacich1987power,bonacich1972factoring,bonacich1991simultaneous}
to the case of hypergraphs (see also Section \ref{sec:measures:centrality}).
The basic idea is simple. Let $I$ be the incidence
matrix of a hypergraph.
If we indicate as vectors $x$ and $y$ the centrality scores for the rows and
columns, representing the hyperedges and the nodes of the hypergraph
respectively, we can write that:
$ I^T x =  \lambda y$  and  $I y  =  \lambda x$, which means assuming
that individuals acquire their centrality by attending important events,
and important events are attended by central individuals.
The vectors $x$ and $y$ are then eigenvectors of two different matrices,
although both are associated with the same eigenvalue $\lambda^2$:
\begin{equation}
I I^T x = \lambda^2 x  \qquad I^T I y =  \lambda^2 y 
\end{equation}
The final outcome is a measure of centrality for events, as
well as an improved measure of centrality for actors. The authors
of the paper show an application to study data describing
56 attacks on European settlements occurred between the years 1509
and 1700 and involving Caraibe from 22 different islands.
Such data require a description considering more-than-dyadic
interaction, as attacks could involve more than two islands
at the same time. 

\bigskip
%
Simplices and simplicial complexes provide an alternative way to describe
and study membership networks using methods from algebraic topology. This
approach draws heavily on the pioneering ideas of Ron Atkin
\cite{atkin1974mathematical,wylie_1976} and on his {\em q-analysis}, which
makes use of a geometric interpretation of the relationships between
actors and events. Atkin's framework to study social systems is based
on a fundamental distinction between what he calls the ``backcloth''
of social action, namely the structure of ties among the events, and
the ``traffic'' of social activities that can take place over the backcloth,
such as the formation of pairwise acquaintanceship between actors.
The backcloth is a simplicial complex and the q-analysis is designed
to describe the patterns of relations among its constituents. 

\begin{figure*}
\centering
\includegraphics[width=0.4\textwidth]{fig39a.pdf}
~
\includegraphics[width=0.5\textwidth]{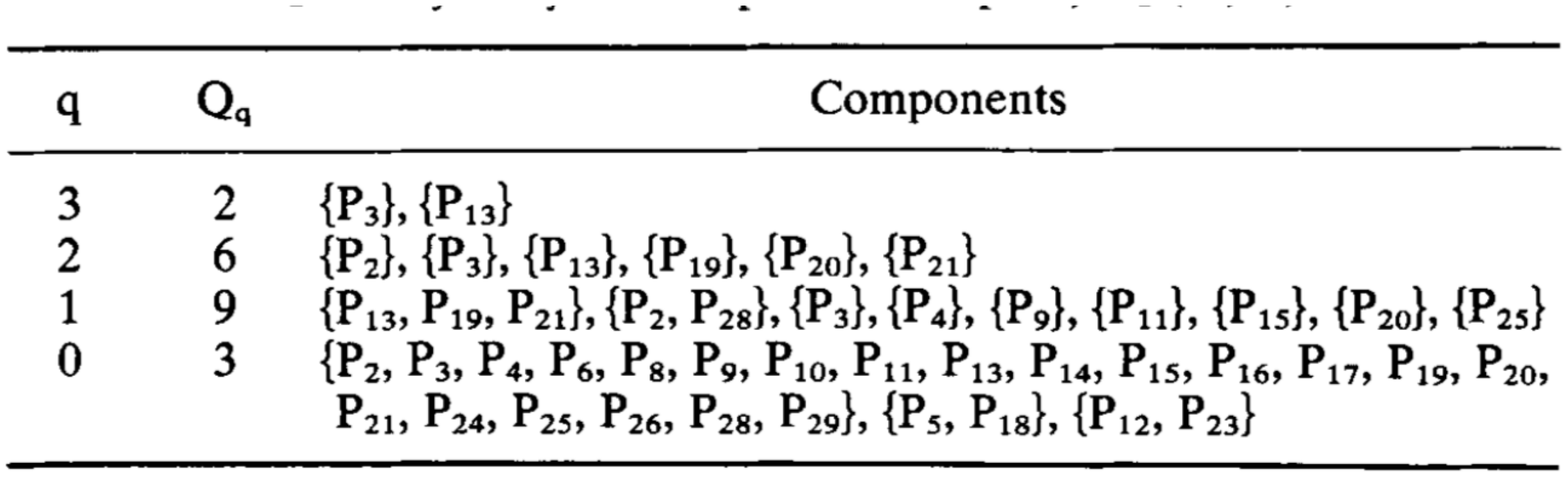}
\caption[]{\textbf{Early simplicial representation of interactions among social science researchers.}
(A) Simplicial complex showing the pattern of links
  between social science researchers, $P_1$-$P_{29}$, through
  shared linking events, nodes 1-19, indicating participation to 
  events or affiliation to university departments. (B) Table showing
  the q-analysis of the simplicial complex. The first column is 
  the dimension of components made up of chains of simplices.
  The second column is the number of chains at each level, while
  the third column reports the names of the simplices
  making up each chain.
  Figure and table reproduced from Ref.~\cite{freeman1980qanalysis}}. 
\label{fig_social2}
\end{figure*}

To give a concrete example, we will discuss here an application of
Atkin's framework to study the formation of friendship in a scientific
community against the backcloth of shared contacts
\cite{freeman1980qanalysis}.  In his analysis of friendship among a
set of 29 social science researchers, Linton Freeman looked at 19
linking events corresponding to scientists been located in the same
university department at the same time, or having attended conferences
together.  Each person is then represented as a simplex of the linking
events in which has been involved. 25 of the 29 persons participated at
least to one of the 19 events, with 13 of them involved in only one
event. These persons are 0-simplices, but there are also 
six 1-simplices (pairs of events), and four 2-simplices (persons involved
in three linking events). 
The highest order simplices found
are two 3-dimensional tetrahedrons.
These simplices form the building blocks for the
construction of a backcloth for social action.  The basic idea is that
conferences and universities provide the perfect setting in which
friendships between academics can be developed. Therefore to
understand the formation of social ties it is important to analyze  
how the simplices are intertwined into a larger
structure, i.e. into a simplicial complex. 
The simplicial complex corresponding to the system under study 
is shown in Fig.\ref{fig_social2}A.
The linking events are the nodes of the simplicial complex and are
labeled with a number from 1 to 19.  
The persons  are instead indicated with the symbols $P_1 - P_{29}$ and can either be represented as nodes,
links, or higher order objects.  
Notice the two tetrahedrons respectively corresponding to persons 
$P_{13}=\{11,13,15,16\}$ and $P_3 =\{2, 8, 9, 19 \}$. These are the
highest-order simplices present in the simplicial complex, which can
then be well represented in three dimensions. Two persons can share
one or more common linking events. For instance, person $P_{20}$ is
linked at dimension 0 to person $P_4$ since they have only a node
(event 3) in common. Person $P_{13}$ and $P_{21}$ are instead linked by two
events (nodes 8 and 9). Hence they are 1-connected, as they are
glued together by a 1-dimensional line (the edge 8-9). What
matter are not only direct connections between simplices, but also
chains of connections. For instance, the three simplices $P_4$,
$P_9$ and $P_{20}$ are linked in a chain of connection of order 0, as 
$P_4$ is 0-connected to $P_{20}$, and $P_{20}$ is 0-connected to $P_9$.
Instead $P_{13}$, $P_9$ and $P_{20}$ are in the same 1-connected
component. Atkin's q-analysis describes the  
sets of chains of connection and their dimensions.
The table in Fig.\ref{fig_social2}B tells us that there
are three components at dimension 0, with the largest one
containing 21 of the 25 persons. Moreover, there are nine
components at dimension 1: the largest one made by
$P_{13}$, $P_9$ and $P_{20}$, the second one by $P_{2}$ and $P_{28}$,
and seven isolated 1-dimensional simplices (edges).  
No connections (of any type) are instead observed among
simplices of dimension larger than 1.
Another interesting feature of the simplicial complex is the existence
of a 0-dimensional $q$-hole (akin to a $H_1$ cycle, see Section \ref{sec:measures:homology})  
involving the four persons $P_9$, $P_{11}$, $P_{13}$ and $P_{20}$ and
shown as a shaded area in Fig.~\ref{fig_social2}A.  This corresponds to a cycle of four
nodes only pairwise connected and is an indication of an obstruction
to the free flow of social traffic on the backcloth.
The main purpose of Freeman was to correlate the formation
of personal friendships to the structure of the backcloth. He
had data of 12 close mutual friendships reported by
the 29 researchers. And he was indeed able to show that
the structure of the backcloth constrains the choice of personal friends. 
In fact, none of the 12 existing social links  
was among the pairs prohibited by the obstructions
in the backcloth. Moreover, 11 of these 12 pairs  
were among the 31 adjacent pairs which are 0-adjacent in the backcloth.

A similar application of q-analysis to study the evolution of social
groups has been published by Patrick Doreian in the same years
\cite{doreian1979evolution}. Doreian had data on the participation of
18 women in 14 events through time, and used q-connectivity to trace
the group structure over time.  His primary objective was to
investigate conflict within the group and eventually predict the
observed split of the group of women into two subgroups. Differently
from Ref.\cite{freeman1980qanalysis}, the approach here is dynamic:
the q-connectivity analysis is applied to an enlarging set of of
successive events in time. Although, on the one hand, the work
confirmed that algebraic topological approaches are flexible enough to
provide a description of structural changes, on the other hand the
results showed that the data used were not rich enough to explain the
observed changes. This also pointed to the importance of collecting
high-resolution temporal data \cite{holme2012temporal}.

Certainly one of the most original applications of simplicial
complexes to social networks is the structural analysis of a team
sport presented in Ref.\cite{gould1977liverpool}. In their work, Gould
and Gatrell used Atkin's q-analysis to define and characterize
intuitive notions of structure in a soccer match. They focused on the
England FA Cup Final between Liverpool and Manchester United played on
21 May 1977 at Wembley Stadium, London.  Although commentators
generally rated the play of Liverpool as superior, Manchester United
won the match by 2 goals to 1. The authors considered the 22 players
and defined a relation in this set using a variable threshold on the
number of times the ball passed from one player to another. Then they
examined the internal structure of the two teams separately, and also
the relations between the two sets of players, defined by the loss of
the ball by one team to the other. The analysis is able to show the
relevance and role of different players and group of players. The
results also indicate that the injection of q-holes by the defense of
the Manchester United, created an obstruction to the free flow,
contributing to the fragmentation and loss of the Liverpool.  It could
be very instructive to apply and validate this type of analysis on a
larger scale, now that soccer analytics is attracting increasing
interest and detailed data on all the spatio-temporal events (passes,
shots, fouls, etc.) occurring during a match are available for 
entire seasons and different soccer competitions 
\cite{pappalardo2019soccer}.

More recently, simplicial complexes have also been used to investigate
online social networks \cite{andjelkovic2015hierarchical,gao2019privacy} 
and social resource sharing systems \cite{cattuto2007folksonomies}.
For instance, the authors of
Ref.~\cite{andjelkovic2015hierarchical} have assessed the role of an
individual in MySpace computing what they name the node structure
vector, which allows to characterize the topological space around a
node. This is defined, for a node $i$, as the vector ${\bf Q}^i= \{
Q^i_0,Q^i_1,\ldots,Q^i_{d_{\rm max}} \}$ whose $d_{\rm max} + 1$
components denote respectively the generalized node degree
$k_{d,0}(i)$ of order $d$ introduced in Section
\ref{sec:measures:centrality},
i.e. the number of $d$-dimensional simplices, with $d=0,1,\ldots,
q_{\rm max}$, to which node $i$ participates, and $d_{\rm max}$ is the
dimension of the largest simplex in the complex. 
The study of the simplicial node degree $k(i) = \sum_{d=0}^{d_{\rm max}} Q^i_d =
\sum_{d=0}^{d_{\rm max}} k_{d,0}(i)$, also known as the node topological
dimension, provides a good measure of the social capital of the corresponding
individual. 
In fact, it has been found that the so called Simmelian brokerage \cite{latora2013simmelian}, which quantifies a node's
ability to act as a broker in a community, scales as a power of the
node topological dimension.
Moreover, the analysis of the components of the node vector over the
different social layers and communities of MySpace reveals that
influential individuals connects higher-order simplices and 
build their social capital by combining their
connections in different layers. 

\bigskip
%
Authorship of scientific articles is a 
particularly interesting type of affiliation networks, as it provides
important insights on patterns of collaboration within the academic
community. In this case, the two sets of nodes represent scientists
and their publications, respectively.  The basic units of scientific
collaborations and of the social network of acquaintances among
scientists are co-authored publications, which often involve groups of
authors rather than just two \cite{milojevic2014teams, xiao2016node}.  Hence, in
their study on the ``shape of scientific collaborations'', the authors
of Ref.~\cite{patania2017shape} have proposed to complement the
results that have been obtained by methods of network analysis
\cite{newman2001ascientific,newman2001bscientific,newman2001structure} with an approach
based on a simplicial description of scientific publications and on
the use of tools from algebraic topology. They have considered all the
papers posted on the arXiv, a repository of electronic preprints
spanning from physics to quantitative biology and mathematical
finance, in the period 2007-2016, and have constructed 18 different
simplicial complexes, one for each of the different categories of
arXiv. Each paper with $k$ authors corresponds to a $(k-1)$-simplex,
and only papers with author sets not fully contained in the author
sets of other papers have been retained in the construction of the
simplicial complexes in order to preserve their basic structural
properties.
Both the size distribution of facets (maximal simplices, see Section \ref{sec:representations}) 
$P(s)$ and the simplicial node degree distributions $P(d)$ of the
complexes display broad tails, indicating the
presence of large collaborations and of authors with a large number of
different collaborations, respectively. The 18 different categories
can then be grouped in only two large classes based on their $P(d)$, showing
that the number of collaborations to which an author is able to
participate is quite well conserved across fields. 
Also, all the 1-dimensional homological cycles, i.e. the
two-dimensional holes bounded by edges, of the various co-authorship
simplicial complexes have been studied. In particular, focusing on the
shortest possible cycles, triangles, and counting how many of the set
of three edges arranged in a triangle are covered by a full triangle
(2-simplex), allows to investigate the concept of simplicial closure
(the extension of triadic closure to simplicial complexes \citep{bianconi2014triadic}) in the data. 
Results indicate the presence of very strong simplicial closure for all categories of arXiv, meaning that in the great
majority of cases whenever three authors have collaborated in pairs,
they also have collaborated on a paper together. An application to  collaboration networks of a similar extension
to hypergraphs of the concept of clustering coefficient, and of that  of subgraph centrality can be found in Ref.~ \cite{estrada2006subgraph}.

\subsection{Neuroscience and brain networks}
\label{sec:neuroscience}
Lively debated over the last decade, the question of whether
high-order correlations --in addition to the basic pairwise
interactions-- were needed to properly account for brain function was
met with strong evidence of a positive answer. Using higher-order
connected correlation functions \citep{schneidman2003network},
\citet{schneidman2006weak} revealed that high-order correlations exist
in neural populations.  Similarly, \citet{ganmor2011sparse} and
\citet{yu2011higher} provided evidence that introducing higher-order
interactions between neurons allowed to improve the predictions at
mesoscopic scales, e.g. for cortical dynamics such as neuronal
avalanches in the awake monkeys or visual responses in the
anesthetized cats.  More recently, further research in neural spike
trains provided methods to measure the strengths of multi-spike
interactions \cite{shimazaki2012state}, and showed their importance in
shaping the dynamics of cortical columns \cite{koster2014modeling} and
in population coding
\cite{shimazaki2015simultaneous,cayco2015triplet}. \\

The models used to estimate higher-order interactions in the cases mentioned above are usually tailored after the generalized Ising model \cite{schneidman2006weak}.
In these models, the probability of observing a pattern of firing neurons $(\sigma_0, \sigma_1, \ldots, \sigma_n)$ is given by 
\begin{equation}
P(\sigma_0, \sigma_1, \ldots, \sigma_n) = \frac{1}{Z} \exp(\sum_i \alpha_i \sigma_i + \sum_{i<j} \beta_{ij} \sigma_i \sigma_j  + \sum_{i<j<k} \gamma_{ijk} \sigma_i \sigma_j \sigma_k + \ldots)
\end{equation}
where $\alpha, \beta, \gamma, \ldots$ control 
the self-, pairwise- and third-order interactions
among firing units.  These models, while very powerful, have several 
limitations. First, they are designed for systems with
discrete states. Neurons are usually considered to be
firing or quiescent so are well described by 
Ising spins, which   
take values $\pm 1$. Conversely, continuous data (e.g. local field
potentials, EEG or BOLD signals) need to be binarized to be amenable
to this type of analysis, and this can represent an important problem
when we want to deal with macroscopic brain networks.
Second, and more importantly, these models neglect
the information encoded in the spatial and temporal structure of
the interactions. Third, their scalability to large networks is made more
complicated by the requirement of large amounts of data (e.g. long
timeseries) to estimate model parameters.
\\

\begin{figure}[t]
\centering
\includegraphics[width=.80\textwidth]{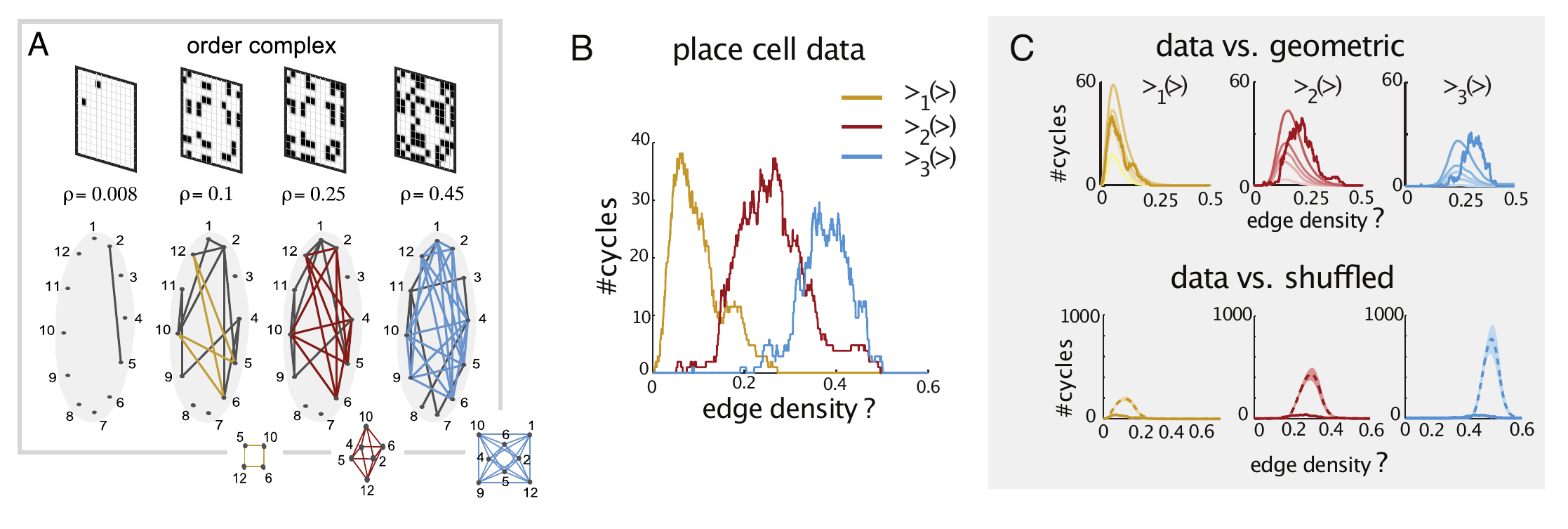}
\caption{\textbf{Topology of hyppocampal cells' activations encodes geometrical information about the environment.} (A) an example of construction of order complex from a full correlation matrix. At each step the order complex (top row) encodes the topology of the density-filtered correlation graph (bottom row). 
(B) Betti curves of the pairwise correlation matrix for the activity of N = 88 place cells in hippocampus during open-field spatial exploration. 
(C) The same Betti curves from B (bold lines) shown overlaid on the mean
  Betti curves from random geometric complexes (top) and from complexes built from shuffled correlation matrices (bottom). 
  Note the differences in when Betti numbers emerge in the case of random geometric complexes and in the magnitude itself for shuffled weight complexes. Figures adapted from Ref~\cite{giusti2015clique}.}
\label{fig:giusti}
\end{figure}

Against this context, \citet{giusti2015clique} studied how the
correlations of spike trains can be used to detect intrinsic
structures in neural activity, without recurring to external stimuli
or receptive fields, and how they relate to the topology and geometry
of the animal's space.  In particular, they computed pairwise
correlations from the cross-correlograms of pyramidal neurons in
freely roaming mice.  Each correlation matrix was then transformed
into an order complex.  This is a filtration of simplicial complexes,
obtained in their stead from a sequence of progressively denser
graphs.  At each density, only the strongest edges until the fixed
density were retained.
In such a way, for each density a binary graph was built and the
corresponding clique complex computed, that is, in each graph all the
cliques were considered as simplices (Fig.~\ref{fig:giusti}A).
\citet{giusti2015clique} then compared the Betti curves of the real
order complex with those obtained from randomized versions, which were
built by reshuffling the original correlation matrices.  They found
that the Betti curves, which encode the topological complexity of the
cell activation patterns, displayed consistently lower values than
those from the randomized models.  These observations implied that the
correlation structure of hippocampal neurons intrinsically represented
the low dimensional input space (a two-dimensional roaming space in
this case). \\

These results confirmed previous evidence on the role of hippocampal
place cells in encoding primarily a space's topological qualities
rather than its geometry
\citep{dabaghian2012topological,dabaghian2014reconceiving}, but also
showed that some coarse geometrical information can be encoded in the
fabric of correlations.  Even more interestingly,
\citet{babichev2018robust} extended previous hypotheses on the
topological nature of the hippocampal map
\citep{dabaghian2012topological,dabaghian2014reconceiving} to account
for the temporal nature of interactions among place cells.  Indeed,
the mammalian hippocampus is thought to be able to learn an
internalized cognitive map representing the ambient space.  It is
however unclear how such a map is conserved in time and updated due to
the transient nature of synaptic connections and of the downstream
neuronal networks.  To investigate this mechanism,
\citet{babichev2018robust} represented the instantaneous state of the
internalized map as a coactivity complex, in which simplices represent
groups of coactive place cells.  At the beginning only few groups of
place cells, hence simplices, are present.  In time however by
accumulation of activity, the coactivity complex should approximate
the topology of the underlying space in which the animal moves.  An
open question however is how it is possible to preserve a consistent
representation of a space while the animal is moving, place cells are
constantly remapping and in general the population coding fluctuates.
Studying the robustness of the topological features of coactivity
complexes, \citet{babichev2018robust} strongly suggested that the
temporal stability of the hippocampal map is a generic phenomenon
stemming from a compensatory mechanism in which neuronal activity
compensates deterioration in the network structure to preserve
hippocampal function.

\begin{figure}
\centering
\includegraphics[width=.80\textwidth]{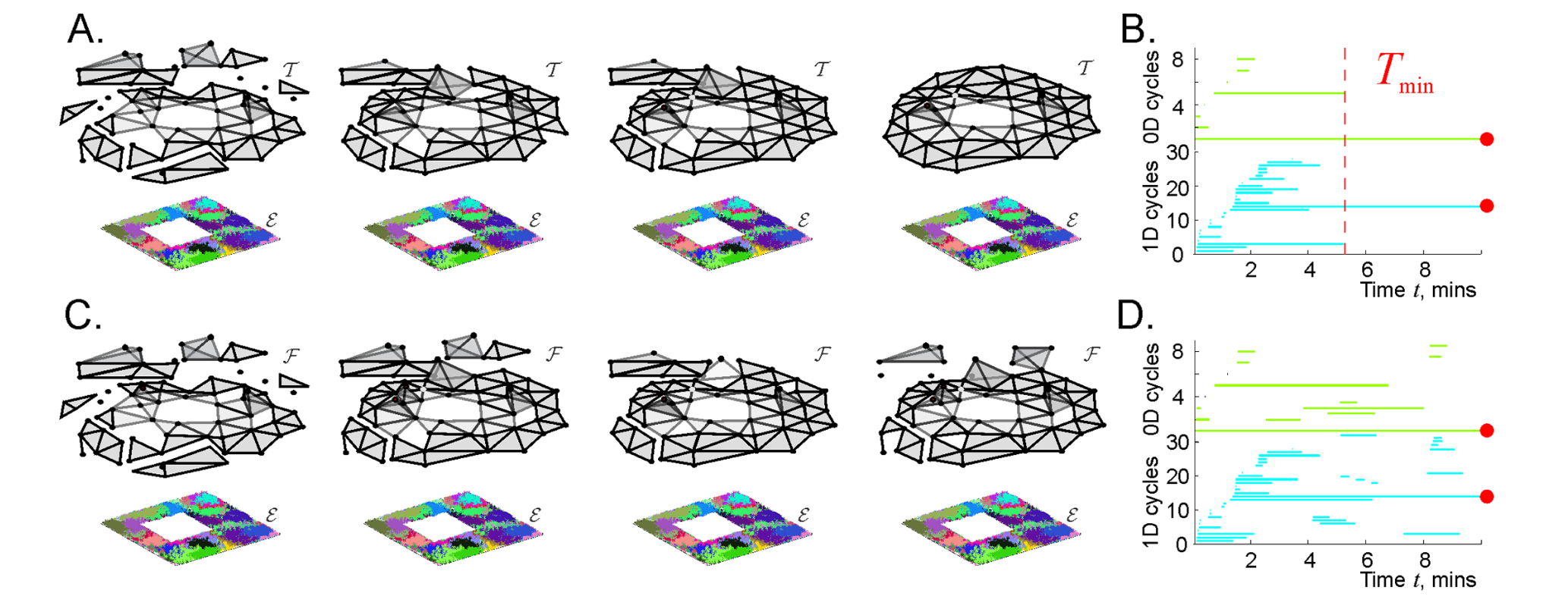}
\caption{ \textbf{The structure of coactivation complexes.} 
(A) Simulated place field map $M(\epsilon)$ of a small planar environment $\epsilon$ with a hole in the center. 
The series of snapshots illustrates the temporal dynamics of the coactivity complex: the complex goes from being from small and fragmented, in the early part of the
exploration, to becoming a stable representation of the shape of the underlying
environment.  (B) The timelines, encoded as barcodes, of topological
persistent $H_0$ and $H_1$ cycles in the coactivity simplicial
complex: 0-dimensional persistent generators are shown in light-blue
lines, 1-dimensional ones in light-green.  Most loops correspond to
accidental, short-lasting structures, effectively representing noise
in the complex.  The persistent topological loops (marked by red dots)
represent physical features of the environment.  The time to eliminate
the spurious cycles can be used as a theoretical estimate of the
minimal time needed to learn the path connectivity of $\epsilon$.  (C)
Simplices can also disappear, and hence the coactivity complex may be
flickering, instead of stable.  (D) The timelines of the topological
cycles in such complex may remain interrupted by opening and closing
topological gaps produced by decays and reinstatements of its
simplices. Figures reproduced from Ref.~\cite{babichev2018robust}.}
\label{fig:dabaghian}
\end{figure}

\citet{reimann2017cliques} investigated the topology of excitation networks built from simulated activity on reconstructed cortical micro-circuitry. They found that different stimuli elicited a surprisingly large number of high-dimensional directed cliques and created a wide variety of high-dimensional homological holes. 
In particular, simulations on a variety of synthetic and null models did not display such an array of topological responses, suggesting these topological metrics do not emerge from traditional constraints on graph structure (e.g. degree sequences, clustering, etc..), but rather from particular species-specific coordination among links.
Moreover, they observed also that, in response to sensory stimuli, pairwise correlations grew with the number and order of the simplices to which the neurons belonged, suggesting that the hierarchy in physical structure results in hierarchically correlated activities.
\\

At the macroscopic brain network level, the question of the importance
(or lack thereof) of higher-order interactions appears less settled.
On the one hand, for example, \citet{huang2017weak} suggested that
weak higher order interactions might indeed be present in macroscopic
functional networks, but also that, due their weakness, pairwise
interactions are dominant in shaping brain activities, hence
justifying functional connectivity descriptions based on pairwise
interactions alone.  On the other hand instead, higher order features
were shown to be reliable under test-retest analysis
\cite{zhang2017test}, and important as indicators of aberrant
connectivity in mental disorders \citep{plis2014high} and mild
cognitive impairment \citep{zhang2016topographical}.  Also, higher
order interactions were useful in the inference of the parameters of
coupled oscillator models of EEG signals (for example
\cite{kralemann2014reconstructing}), which we discussed in more detail
in \ref{sec:inference}.

\begin{figure}[h!]
\centering
\includegraphics[width=\textwidth]{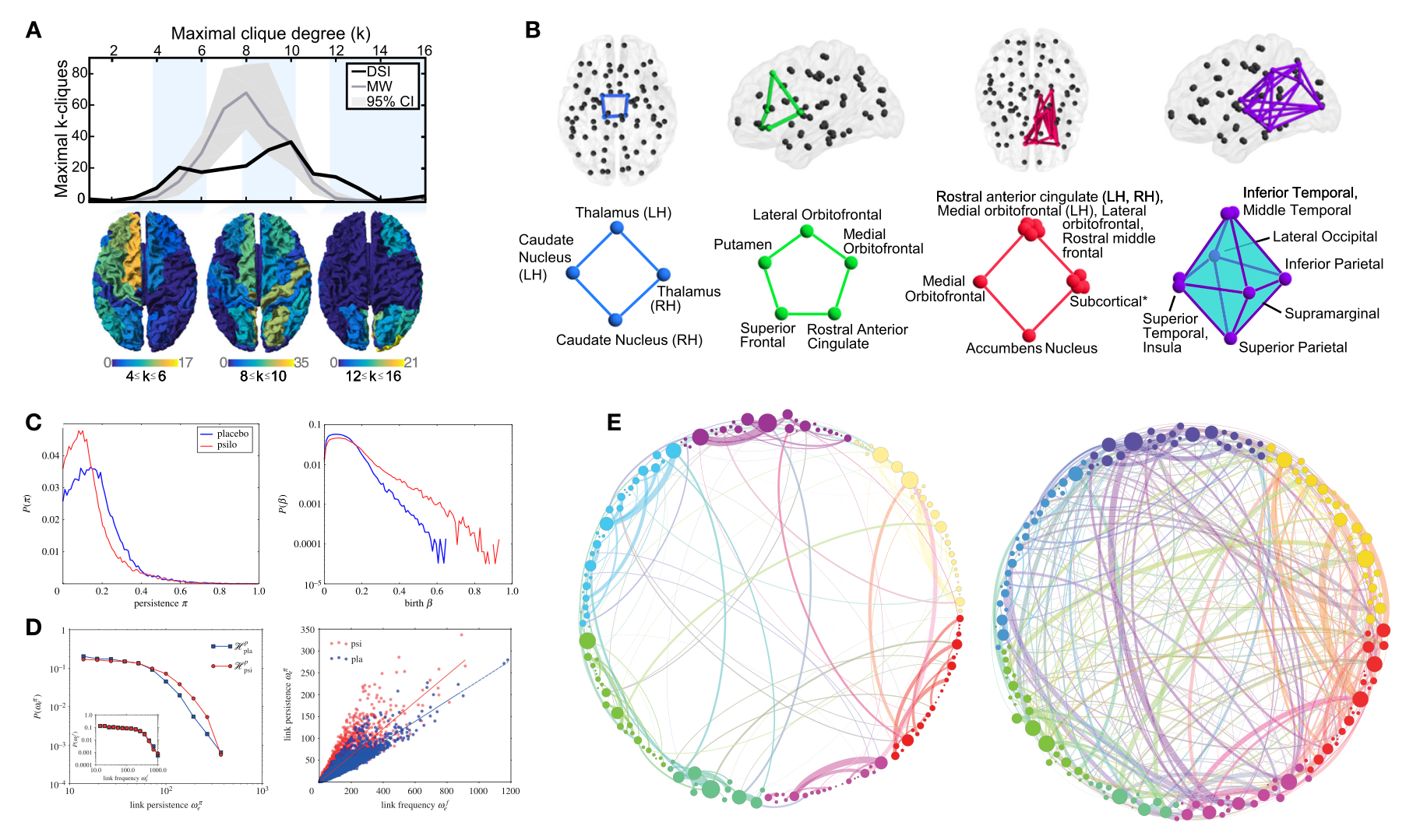}
\caption{
\textbf{Persistent homology of structural and functional brain connectivity.}
(A) Distribution of maximal cliques in the average DSI (black) and
  individual minimally wired (gray) networks, thresholded at an edge
  density of $\rho$ = 0.25. Heat maps of node participation shown on the brain surfaces for a range of clique degrees equal to 4-6 (left), 8-10
  (middle), and 12-16 (right).
(B) Minimal cycles representing each persistent cavity at the density
  at birth represented in the brain (top) and as a schematic (bottom) (adapted from
  \citep{sizemore2018cliques})..
(C) Comparison of persistence p and birth b distributions.  Left, H1
  generators’ persistence distributions for the placebo group and
  psilocybin group.  Right, distributions of homology cycles' births.
(D) Statistical features of group homological scaffolds.  Left,
  probability distributions for the edge weights in the persistence
  homological scaffolds (main plot) and the frequency homological
  scaffolds (inset).  Right, scatter plot of the scaffold edge
  frequency versus total persistence for both placebo and psilocybin
  scaffolds.
(E) Simplified visualization of the persistence homological scaffolds
  for subjects injected with placebo (left) and with psilocybin
  (right).  Colours represent communities obtained by modularity optimization
  on the placebo scaffold and display the departure of the psilocybin
  connectivity structure from the placebo baseline. Figures adapted from
  Ref.~\cite{petri2014homological}.
}
\label{fig:cliques_scaffold}
\end{figure}

Recent seminal research has shown the potential and impact of
topological approaches, in particular those inspired by topological
data analysis.  Structurally, persistent homology techniques were
adopted to describe and discriminate healthy and pathological states
in developmental \citep{lee2017integrated} and neurodegenerative
diseases \citep{lee2014hole}.  \citet{sizemore2018cliques} described
the white matter network of fibers between brain regions as a weighted
network and then studied both its dense portions, in terms of cliques,
and its cavities, in terms of homology.  They found that large cliques
are much more frequent than expected in an appropriate randomized
model built using biologically-inspired principles of parsimonious
wiring (Fig.~\ref{fig:cliques_scaffold}A).  These cliques were
interpreted as local dense units able to perform rapid processing, and
were found to be positioned around topological cavities, which in turn
acted as obstructions and guides for the information flows.  These
cavities were also reproducible across subjects and appeared to
connect regions belonging to different phases of brain evolutionary
history (Fig. \ref{fig:cliques_scaffold}B).  Similarly,
\citet{bendich2016persistent} described the morphology of brain
arteries using topological observables.  In particular, using
persistent homology of trees, they characterized arterial morphology
using the structure of branching and looping of vessels at multiple
scales, and found distinctly different patterns at different ages.

Topological differences have been also found at both population and individual levels in functional connectivity in healthy and pathological subjects \citep{lee2012persistent,lee2011discriminative}. 
Higher dimensional topological features have been employed to detect differences in brain functional configurations in neuropsychiatric disorders and altered states of consciousness relative to controls \citep{petri2014homological,chung2017exact}, and to characterize intrinsic geometric structures in neural correlations\citep{giusti2015clique,rybakken2019decoding}. 

As an example, \citet{petri2014homological} compared the topology of
the functional connectivity of subjects that had been subministrated
with psilocybin, a psychedelic drug, with their own under placebo.
They found that the topological structure of the two conditions was
very different and such difference could be quantified already at the
level of persistence diagrams (Fig.~\ref{fig:cliques_scaffold}C).
While the difference between topological summaries. obtained from persistent homology, was already discernible, it provided little information
on how topological information mapped back to the underlying brain
regions.  The authors solved this problem by defining a topological
backbone, called \emph{scaffold}, built on approximated minimal
homological generators (Fig.~\ref{fig:cliques_scaffold}D), which
allowed them to show that altered states of consciousness induced by
psilocybin (and likely, other psychedelics) stem from very different
patterns of information integration and importance of the brain
regions~\citep{lord2016insights} with respect to the normal state
(Fig.~\ref{fig:cliques_scaffold}E).
Other examples can be found in the following series of works. 
\citet{lee2011discriminative} have proposed methods 
to discriminate between cohorts of children with
attention deficit hyperactivity disorder, autism spectrum disorder and
pediatric control subjects on the basis of their functional topology.
\citet{lee2019harmonic} instead represented the topological substructure of
brain networks through the eigenvectors of the corresponding Hodge
Laplacians and used it to discriminate between mild and progressive
cognitive impairments, and Alzheimer’s disease. \citet{chung2019exact}
described the heritability of differences in whole-brain functional
topology in a cohort of twins. \citet{ibanez2019topology} related the
topological functional structure of EEG data during imagery to
functional equivalence in a population of skilled versus unskilled
imagers \citep{ibanez2019spectral}.

\bigskip
Going beyond functional connectivity, 
\citet{saggar2018towards} constructed a simplified topological
backbone representation of the full fMRI activation space and showed
that the properties of these backbones associated with behavioral
performance in a series of cognitive tasks.  In the context of
event-related fMRI, \citet{ellis2019feasibility} investigated the
feasibility of topological techniques for recovering signal
representations from BOLD signals. In particular, they embedded
specific signal configurations by a convolution of the signal with the
hemodynamic response, showing that the persistent homology was able to
recover the signal topology with high accuracy.

Finally, moving from neuroimaging to applications in cognitive neuroscience, 
\citet{sizemore2018knowledge} mapped the evolution of early semantic
networks in toddlers by identifying words with nodes and considering
higher order interactions among them. They found that sparse
regions of the resulting HOrSs displayed remarkable similarities at
the topological level across subjects, and the timing of their
disappearance was more closely to the patterns of connections among
words than to their actual semantic content, thus suggesting that
knowledge acquisition might generally happen via filling knowledge
gaps.  For an extended review of the current research on the effects
of non-pairwise interactions in neuroscience, we refer the reader to
the following references
\citep{giusti2016two,sizemore2019importance,expert2019topological}.

\subsection{Ecology}
\label{subsec:ecology}

\begin{figure*}[htb]
\centering
\includegraphics[width=0.5\textwidth, keepaspectratio = true]{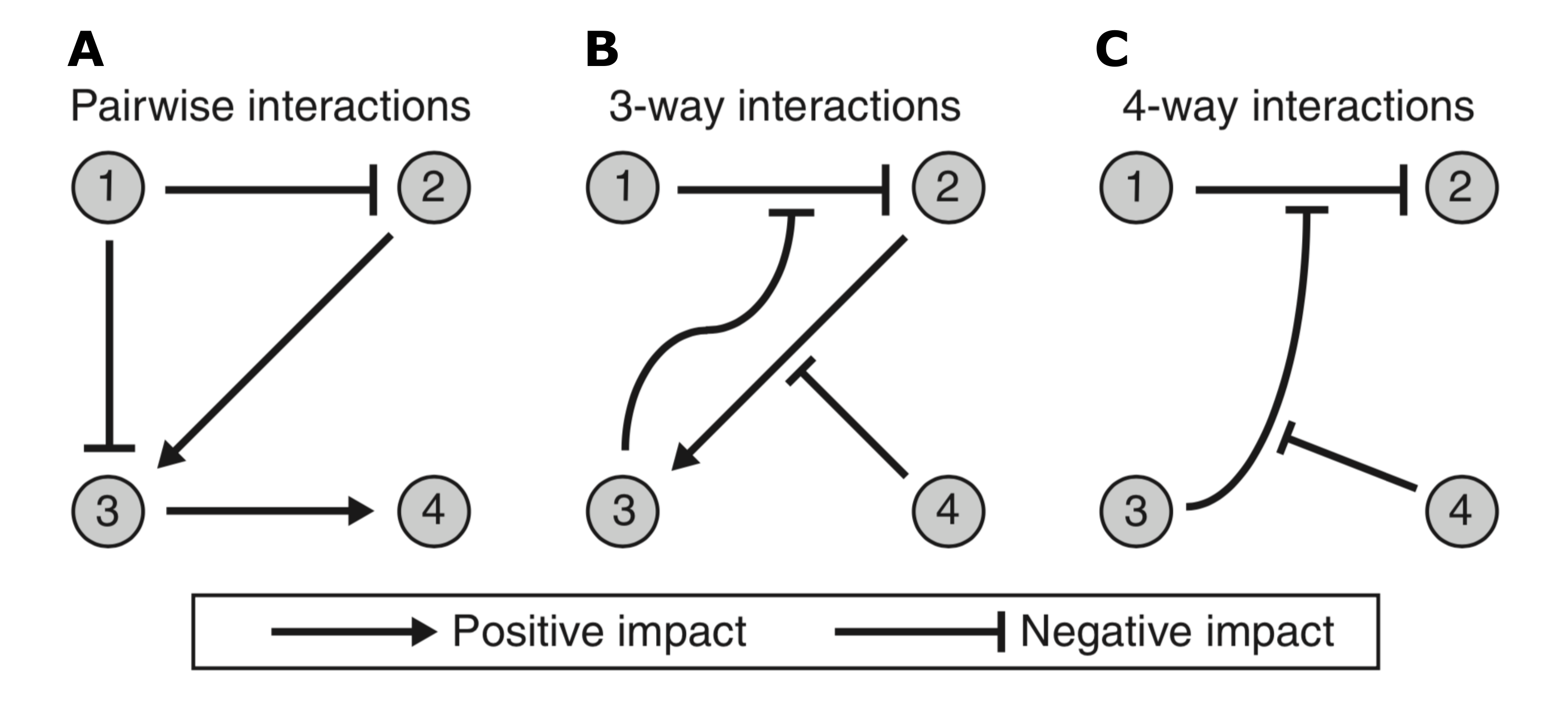}
\caption[]{\textbf{Pairwise and high-order interactions in ecological systems.}
 (A) Direct
  pairwise positive and negative interactions among species.
  (B) Three-ways interactions: for example species 3 attenuates the direct inhibitory
  effect of species 1 on species 2. (C) Four-way interactions:
  species 4 inhibits the inhibition produced by species 3 on the
  interaction between 1 and 2. Figures adapted from Ref.~\cite{bairey2016high}.
}
\label{fig_ecology-bairey_0}
\end{figure*}

Higher-order interactions have been studied for decades in the
context of ecological models
\cite{case1981testing,abrams1983arguments,kareiva1994foil}. However,
only very recently it has been highlighted their crucial role for
the stability of large ecological communities and for the remarkable
biodiversity observed in nature.
While pairwise interactions in ecology consider the direct effect,
either positive or negative, of a species on another, as shown 
in Fig.~\ref{fig_ecology-bairey_0}A, high-order interactions include
all the cases where the relation between two species can be modified
by the presence of other species, which may also be not directly
affecting the former \cite{billick1994higher, wootton1993indirect}.
For instance, this happens when there is a microbial species that
produces an antibiotic to interfere with a competing species, and a
third species produces an enzyme which degrades the antibiotic thus
reducing the strength of the interaction among the other two
~\cite{bairey2016high}. This is illustrated in
Fig.~\ref{fig_ecology-bairey_0}B, where species 1 inhibits species 2,
and species 3 produces the enzyme that reduces the direct inhibitory
effect of species 1 on species 2. Such a mechanism gives rise to a
so-called ``trait-mediated indirect interactions'' (TMIIs) between
species 3 and 2, which is different from a possible direct pairwise
interaction between 3 and 2 (also reported in the figure), being
intrinsically a three-species interactions. 
As finally shown in Fig.~\ref{fig_ecology-bairey_0}C, 
the enzyme produced by species 3 can in turn can be inhibited by a compound 
introduced by a species 4, creating a
four-species inseparable/entwined interaction, 
and so on~\cite{kelsic2015counteraction,perlin2009protection,abrudan2015socially}.
Similar effects can also arise from
an adaptive behavior, for instance a predator which changes its target
prey because another prey becomes available~\cite{koen2007process}. In
this case there is no direct interaction between the two preys but 
they are part of a hyper-interaction, and considering only pairwise
interactions would not allow to correctly take into account this
effect.

Analyses of ecological networks almost often omit non-pairwise
interactions, many classes of which are instead fundamental to 
the structure, the function and the resilience of ecosystems. 
It has been shown that the class of
three-species interactions in which one of the three species has the effect of
mitigating the negative interaction between the other two can have a
stabilizing effect \cite{kelsic2015counteraction},  
while it has been found that increasing the order
of the interactions reduces the fraction of extinct species \cite{deoliveira2000replicators} and increases the variance of species abundances at
equilibrium~\cite{yoshino2008rank}.
The literature is growing and many scientists have shown that higher-order
interactions can have a stabilizing effect under many particular condition
settings.

\begin{figure*}[t]
\centering
\includegraphics[width=\textwidth, keepaspectratio = true]{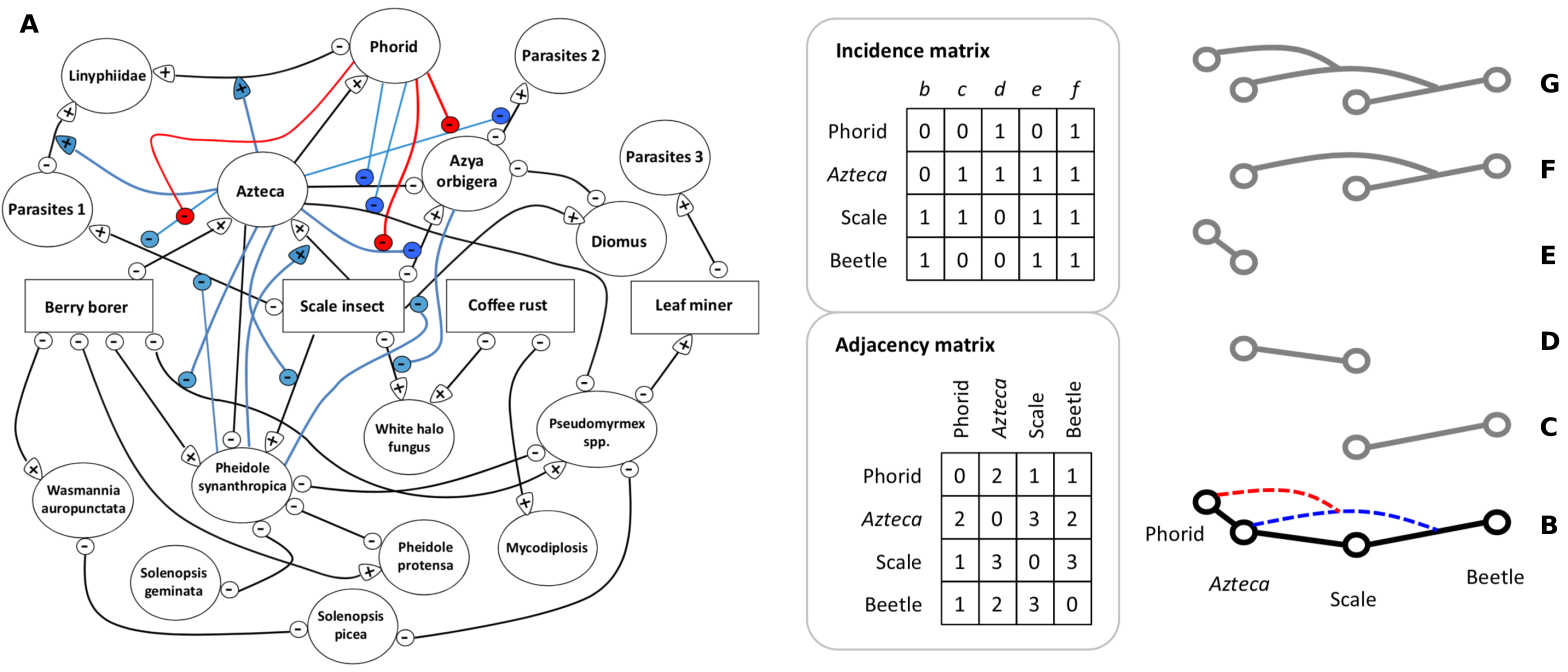}
\caption[]{\textbf{Hypergraph description of the coffee agroecosystem in
  southern Mexico~\cite{golubski2016ecological}.} Nodes in (A) represent the different agricultural pests, while lines
  indicate indicate direct effects (black), modifications of direct
  effects (blue), and modifications of those modifications (red). 
  Key interactions (B) among four nodes of the system, namely Phorid, Azteca, Scale
  and Beetle (Azya orbigera), are represented in the
  form of a hypergraph whose hyperedges (C-G) and incidence and adjacency matrices
  are reported in right panel. Figures adapted from Ref.~\cite{golubski2016ecological}.
}
\label{fig_ecology-golubski}
\end{figure*}

\bigskip
It has been shown that hypergraphs can be a very useful mathematical
framework to represent and take into account non-pairwise ecological
interactions, such as TMIIs 
\cite{sonntag2004competition,estrada2006subgraph,golubski2016ecological}.
As an illustrative example of the value of hypergraphs in describing
ecological communities, the authors of
Ref.~\cite{golubski2016ecological} have studied a real-world coffee
agroecosystem in southern Mexico, in which resistance to agricultural
pests depends upon a large number of TMIIs.  Based on field studies,
they have assembled the intricate web of interactions among
agricultural pests that is reported in Fig.~\ref{fig_ecology-golubski}A. Black arrows indicate direct
effects, while blue and red arrows represent modifications of those
direct effects, and modifications of those modifications,
respectively. In particular, some protective effects attributable to
TMIIs imposed by ants of the genus Azteca (blue lines from node Azteca),
some of which are further
modified by ant-parasitizing phorid flies (three red lines from node
Phorid), have been shown to be crucial
for controlling agricultural pests. 
Interactions within this system can be well represented by an
hypergraph. The hypergraph corresponding to key interactions involving
the four nodes Phorid, Azteca, Scale and Beetle (Azya orbigera) is
reported in Figs.~\ref{fig_ecology-golubski}B-G,
together with its adjacency and incidence matrices. Notice that
positive and negative direct effects, or strengthening and weakening
TMIIs are respectively indicated by triangular and circular
arrow-heads in the interaction web in the left panel. However, this
information has been omitted to produce the simpler case of an
undirected system shown in (B). Direct effects
represented as black lines in (B), and corresponding
to the three edge of the hypergraph, include beetles preying on scale
insects (C), Azteca ants consuming energy from Scales (D), and Phorid flies
parasitizing Azteca ants (E). Indirect effects, consisting of Azteca ants
reducing the magnitude of the interaction between Scale and Beetle 
[dashed blue line in (B)], and Phorid flies reducing the magnitude of the
effect of Azteca ants on the Scale-Beetle interaction [dashed red line in ()B)], 
are instead represented as the two hyperedges reported in (F) and (G),
respectively. 
This example immediately illustrates the straightforward way in which
hyperedges can represent TMIIs. The authors of
Ref.~\cite{golubski2016ecological} further elaborate on how the
analysis of hypergraph topology, and concepts such as shortest
hyperpaths and hypergraph centrality measures, can turn very useful
for studying important aspects of ecological systems (such as how a
species is affected by the removal of other species from the system)
that a network description only based on pairwise interactions alone
can fail to faithfully represent.

\bigskip
Bairey et al. have investigated the stabilizing role of higher-order
interactions in replicator equations ~\cite{bairey2016high}. They 
have proposed a mathematical model based on random
replicator dynamics to study ecosystems when both
pairwise and higher-order interactions are present. 
An ecosystem is described at each time by its 
state ${\bd x} = (x_1,x_2,...,x_N)$, where $x_i\equiv x_i(t)$ denotes the abundance of
species $i$ at time $t$, with the physical constraint $\sum_i x_i(t) =1 \forall t$. 
The temporal evolution of the abundances are governed by the
following set of differential equations:  
\begin{equation}
\dot x_i = x_i [f_i(\bd x) - \sum_{j=1}^N x_j f_j(\bd x)] \qquad i=1,\ldots,N
\label{bairey_eq1}
\end{equation}
in the usual form of replicator
dynamics~\cite{diederich1989replicators,hofbauer1998evolutionary,opper1999replicator,chawanya2002large}, already introduced in Section~\ref{sec:games} to model the evolutionary dynamics of strategic interactions. 
The first term in bracket, $f_i$ represents the fitness
of species $i$, which depends on the effect of the other species through
the system state $\bd x$, while the second term is the average population
fitness. The key point is that the fitness function here adopted: 
\begin{equation}
  f_i(\bd x) = - x_i + \sum_{j=1}^N a_{ij} x_j
  + \sum_{j=1}^N \sum_{k=1}^N b_{ijk}  x_j x_k
  + \sum_{j=1}^N \sum_{k=1}^N \sum_{l=1}^N c_{ijkl} x_j x_k  + \ldots
\label{bairey_eq2}
\end{equation}
includes pairwise but also higher-order interactions. 
Entry $a_{ij}$ of matrix $A$ determines the effect of species $j$ on
species $i$. Three-dimensional (or third-order) tensor $B$ rules 
three-species interactions, with the value of the entry $b_{ijk}$
determining the joint effect of species $j$ and $k$
on species $i$, and so on with the four-dimensional tensor $C$, etc. 
Notice that the negative sign of the first term, $-x_i$, implies the
stability of the system when interactions are turned off
(species are self-limiting in high concentrations). 
Limiting the analysis to hyper-interactions not involving 
more than four species, random perturbations of different order to
a stable ecosystem are then modeled by setting $A= \sqrt{\alpha} \tilde{A}$,  
$B= \sqrt{\beta}  \tilde{B}$
and
$C= \sqrt{\gamma}  \tilde{C}$. Here 
$\tilde{A}$, $\tilde{B}$ and $\tilde{C}$ are
a random matrix (a two-dimensional tensor), and  
a random three-dimensional and four-dimensional 
tensors respectively. The elements of $\tilde{A}$, $\tilde{B}$ and $\tilde{C}$ 
can either be positive or negative 
and are drawn from a Gaussian distribution
with mean $0$ and variance $1$. The
important parameters of the model are then the
values of $\alpha$, $\beta$ and $\gamma$, which 
represent the strength of the pairwise, three-species and
four-species interactions, respectively.
\begin{figure*}
\centering
\includegraphics[width=0.85\textwidth, keepaspectratio = true]{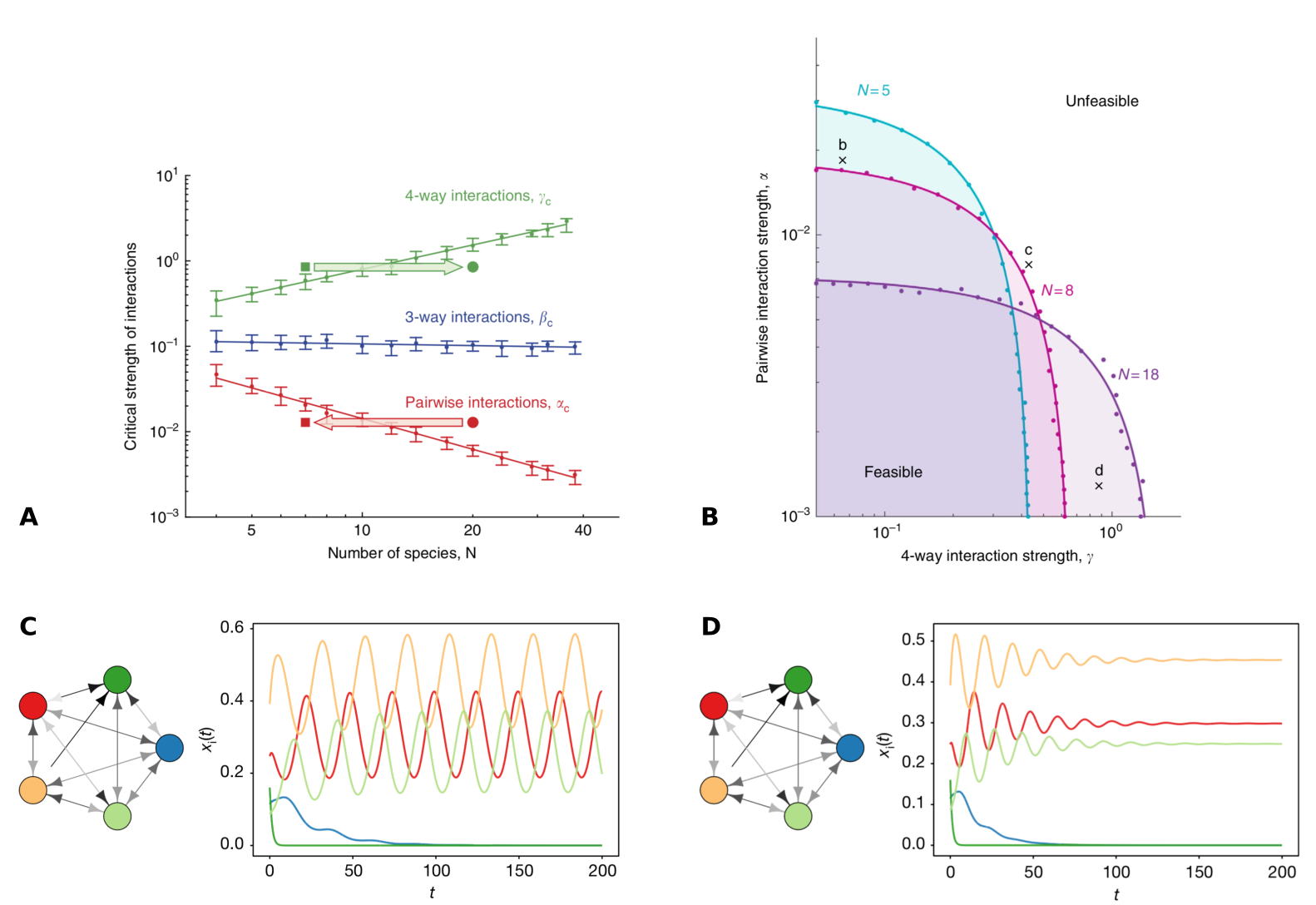}~~~~
\caption[]{\textbf{Dynamical effects of higher-order interactions in ecological systems.}
(A) Critical strength of interactions in the Bairey
  et al.~\cite{bairey2016high} model in
  Eq.~\eqref{bairey_eq1} and \eqref{bairey_eq2} beyond which the coexistence of
  species is lost as a function of the number of species $N$.  The
  three curves represent the case of only pairwise, three-species and
  four-species interactions, respectively.  (B) Regions of
  stability for ecosystems with $N=5, 8$ and $18$ species in the
  ($\gamma$, $\alpha$) space (assuming $\beta=0$). (C) Temporal evolution of
  the abundances of five different species as given by
  the Grilli et al.\cite{grilli2017higher} model for the competition matrix $H$
  reported in the five node graph, and when only pairwise interactions
  are considered. (D) Same as in the previous panel, but with sampling
  three seedlings at a time instead of two.
  Figures adapted from Ref.~\cite{grilli2017higher} and Ref.~\cite{bairey2016high}.}
\label{fig_ecology2}
\end{figure*}
As expected, in a system with only pairwise interactions ($\beta=0$,
$\gamma=0$), species exhibit extinctions when the strength $\alpha$ is larger
than a critical value, and such value decreases with the system diversity,
i.e. with the number of species $N$. Reported in Fig.~\ref{fig_ecology2}A
is the threshold $\alpha_c$ at which coexistence is lost in $5\%$ of
the simulations.
When the number of species $N$ is increased, the critical threshold $\alpha_c$  
decreases as $1/N$, in agreement with the result by May \cite{may1972stable}.   
The situation changes when higher order interactions are considered.
In particular, if only three-species interactions are present
($\alpha=0$, $\gamma=0$), then the value of the critical strength
$\beta_c$ is not affected by the number of species.
Conversely, if only four-species interactions are considered, the
threshold $\gamma_c$ increases with the system diversity. This is a
striking result: for a fixed strength of four-species interactions, an
ecosystem is stabilized, rather than destabilized, as the number
species increases. Thus, while pairwise interactions introduce a
higher bound on diversity, a lower bound is instead created by
high-order interactions. Or in other words while ecological
communities with a large number of species become sensitive to
pairwise interactions, communities with a small number of species are
sensitive to high-order interactions.

Bairey et al. have also considered the more interesting ``mixed case'' in
which a combination of the three types of interactions can be
present at the same time. The resulting stability region for
three different sizes $N$ of the system is graphically represented in
Fig.~\ref{fig_ecology2}B.
in the space ($\gamma$, $\alpha$) of pairwise and four-species interaction
strengths, in the case of no three-species interactions,  
i.e. assuming $\beta=0$. Notice that there is a small area  
where an ecosystem with $N=8$ species is feasible, while systems of both sizes
$N=5$ and $N=15$ are unstable. Since it is plausible that, in the most general
case, an ecosystem is characterized by interactions of 
different orders and by a set of values $(\alpha, \beta, \gamma, \ldots)$
for the corresponding strengths, this will imply the existence of both a 
lower and upper bound for the number of species, for
which Bairey et al. provide an analytical estimation. They also found that,
if the total strength $\alpha+\beta+\gamma$ is increased, then upper and lower
bounds get closer, restricting the range of allowed diversity of the
ecosystem.

\bigskip
Grilli et al.~\cite{grilli2017higher} have instead
studied the role of high-order interactions in a model of interacting
competitors
\cite{vandermeer1969competitive,neil1974competition,dormann2005coexistence,weigelt2007competition}. Although the proposed framework is quite general, the model describes the
dynamics of a forest with a large but fixed number of trees, in which
$N$ different species of trees compete for space.  As in the case of
the model by Bairey et al., the state of the system is described by
the vector  ${\bf x}(t)$, i.e. by the proportion $x_i(t)$ of trees
of each species $i$ at time $t$, with
$\sum_i x_i(t)=1$ $\forall t$. The dynamics stems from the fact that
at each time step a tree, selected at random (with all the species
having the same death rate), dies leaving an empty space in the
canopy, which can be filled by a new tree.  That is when the
competition among seedlings begins.  The simplest way to model this
mechanism is through a pairwise competition: two species are randomly
selected and the winner of the competition will fill the gap.
Pairwise competitions are characterized by matrix
$H$ whose entry $h_{ij}$ represents the winning probability of
species $i$ on species $j$. In particular, Grilli et al.
considers the most general case of a matrix of randomly generated
positive numbers between 0 and 1, with $h_{ij}+h_{ji}=1$, which
represents an extension of previous
works~\cite{allesina2011competitive,kerr2002local}. The dynamics of the $N$
species is ruled by the following set of differential equation: 
\begin{equation}
  \dot x_i =  -x_i + 2 x_i  \sum_{j=1}^N h_{ij} x_j   =  x_i  \sum_{j=1}^N p_{ij} x_j
\label{grilli_eq1}
\end{equation}
where the negative terms $-x_i$ describes the death process, while the
positive term $2 x_i  \sum_j h_{ij} x_j$ gives the probability of selecting
two seedlings of species $i$ and $j$, with $i$ winning the
competition. Notice that the competition process can be seen as a
game~\cite{hofbauer2003evolutionary,nowak2004evolutionary}, and the right hand side of the equation can be rewritten in the
form of a replicator equation 
for a zero-sum, symmetric matrix game with two players,
where the payoffs $p_{ij}$ are the entries of the skew-symmetric payoff matrix 
$P = H - H^T$~\cite{taylor1978evolutionary,hofbauer2010note}, similar to the payoff matrix introduced in Eq.~\label{tab:payoffMatrix} in Sec.~\ref{sec:games}.
Independently from the initial conditions ${\bf x}^*(0)$, after
an initial transient, Eq.~\eqref{grilli_eq1} drives the system to a
state where some of the $N$ species go extinct, while the remaining
ones cycle around a unique equilibrium point ${\bf x}^*$. This is
shown for a case with $N=5$ and a particular random choice of matrix
$H$ in Fig.~\ref{fig_ecology2}C. By changing $H$ the
model can lead to arbitrarily many species coexisting, and can generate any
possible species-abundance distribution empirically observed.
However, the neutral cycling around the equilibrium is problematic, 
as such cycles are not observed in nature. In addition to this, the main
issue with the model is that the equilibrium is highly unrobust:
any deviation from perfectly identical death rates 
destabilizes the dynamics and leads to just one species surviving.
Grilli et al. have shown that the problem can be solved by going beyond the pairwise
interactions and considering the simultaneous competition among more than
two species when a new empty space appears in the canopy. They propose
an extension of the model in Eq.~\eqref{grilli_eq1}  
where three seedlings are picked at random,
the first competes with the second and the winner with the third.
The equations now read:
\begin{equation}
  \dot x_i =  -x_i +  x_i  \sum_{j=1}^N  \sum_{k=1}^N
  \left(2 h_{ij} h_{ik} + h_{ij} h_{jk} + h_{ik} h_{kj} \right)  x_j x_k 
  =  x_i  \sum_{j=1}^N  \sum_{k=1}^N p_{ijk} x_j x_k
\label{grilli_eq2}
\end{equation}
where $h_{ij} h_{ik}$ is the probability that $i$ beats both $j$ and $k$,
$h_{ij} h_{jk}$ is the probability that first $j$ beats $k$, and then $i$ beats $j$,
and $h_{ik} h_{kj}$ is the probability that $k$ beats $j$, and then $i$ beats $k$.
Also in this case the equations can be rewritten in the form of a replicator dynamics
for a three-player game with a three-dimensional payoff tensor $P$ whose entry   
$p_{ijk} = 2h_{ij} h_{ik} - h_{ji} h_{jk} - h_{ki} h_{kj}$ gives the payoff of
the first player 1 playing strategy $i$ when player $2$ plays $j$ and player $3$ plays
$k$.
Surprisingly, the evolution of this new model leads to globally stable fixed
points instead of cycles. As shown in Fig.~\ref{fig_ecology2}D
for the same matrix $H$ as in Fig.~\ref{fig_ecology2}C, 
the system converges to a fixed point characterized by the same vector $x^*$
that was the center of the oscillation in the model with only pairwise
interactions. In addition to this, the fixed point is now globally stable. 
Hence, sampling three seedlings at a time instead of two 
produces stability in a system of competitors. 
The same authors have also proven that the inclusion of fourth- or
higher-order terms does not change the equilibrium but
simply accelerates the convergence to it.  
Moreover, when transforming this deterministic. 
model into a stochastic one, the presence of higher-order interactions
delays the extinction time, allowing a prolonged coexistence of
species. Summing up, the model in Eq.~\eqref{grilli_eq2} clearly indicates
that the inclusion of higher-order interactions in competitive networks
stabilizes dynamics, making species coexistence robust
to perturbations. 

\bigskip
Mayfield et al.~\cite{mayfield2017higher} have pointed out the role of
high-order interactions to another important aspect, that of
estimating the values of fitness in ecological models. They have
shown that it is quite difficult to explain the empirically observed
fitness outcomes by considering only pairwise interactions. The
inclusion of higher-order interactions, defined as changes to the
interactions between two species mediated through a third species, can
instead improve the ability to perform such an estimation.

\bigskip
Very recently, Valverde et al~\cite{valverde2020coexistence} have applied a
HOrS framework to the analysis of environmentally mediated
host-pathogen infections. The hyperlinks of a hypergraph are used to
depict three-way associations between plants (hosts), viruses
(pathogens), and different habitats. Projecting this information, it
is possible to study the interactions between viruses and different
host ecotypes, explicitly including the spatial context in which
host-pathogen interactions take place. By building a neutral model for
the evolution of host–pathogen networks across multiple habitats, the
study showed that real ecosystems live in a continuum between nested and
modular networks, going beyond the traditional dichotomy between
modularity and nestedness in ecological networks~\cite{mariani2019nestedness}. The model has been 
empirically validated by the analysis of different ecosystems in an
agricultural landscape in Spain.

\bigskip
For a more complete review of the current research on the effects of non pairwise
interactions on the mechanisms to maintain biodiversity in ecological systems we refer the
reader to the review ``Beyond pairwise mechanisms of species coexistence in complex communities'' by Levine et al.~\cite{levine2017beyond}.

\subsection{Other biological systems}

Over the last decades network science has become an established
framework to describe and understand interactions between biological
agents, including proteins, metabolites and
genes~\cite{oltvai2002life,aittokallio2006graph,vermeulen2020exposome}. Yet,
the complexity of biological processes can only rarely be decomposed
as a sum of pairwise interactions. For
instance, metabolic reactions often involve multiple partners, and
proteins typically interact with each other in small groups known as
complexes.  As a consequence, traditional pairwise approaches, which neglect
the presence of higher-order structures, are
at risk of oversimplify the
complexity of biological systems.

\begin{figure*}
\centering
\includegraphics[width=1\textwidth, keepaspectratio = true]{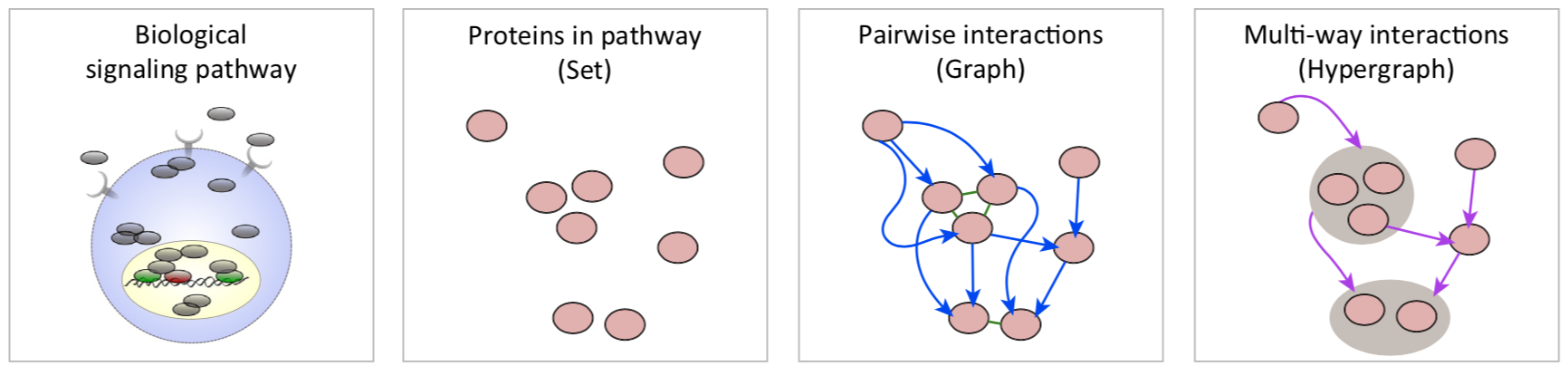}~~~~
\caption[]{\textbf{Different representations of biological signaling
  pathways.} In the simplest representation, a signaling pathway is
  simply a set of proteins, with no additional information. Networks
  can only capture pairwise interactions between proteins. Hypergraphs
  naturally encode multilateral interactions and reactions. Figure
  reproduced from Ref.~\cite{ritz2014signaling}.}
\label{fig_bio1}
\end{figure*}

\begin{figure*}
\centering
\includegraphics[width=1\textwidth, keepaspectratio = true]{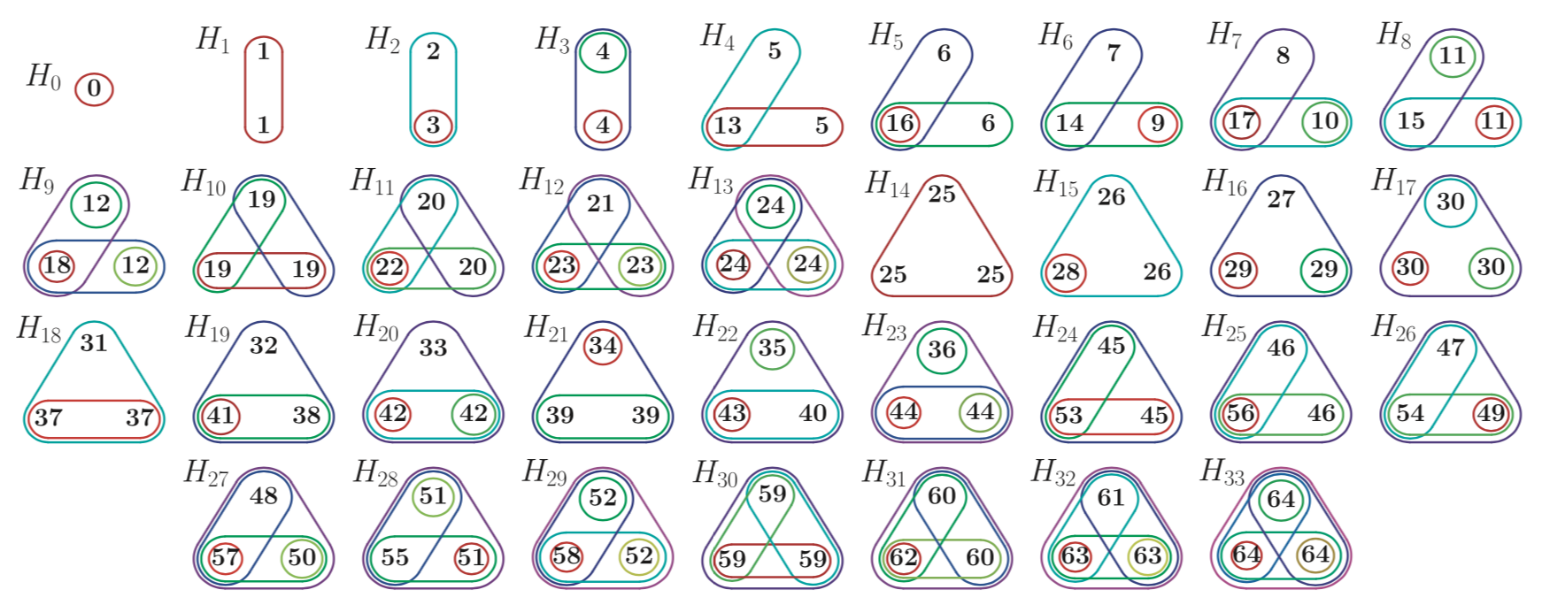}~~~~
\caption[]{
\textbf{Hypergraphlet representation of local connectivity patterns in hypergraphs.}
Complete illustration of the 65 orbits associated to hypergraphlets of order 1, 2 and 3. More than 6000 orbits are associated to hypergraphlets of order 4, and more than a hundred thousands to hypergraphlets of order 5. Figure reproduced from Ref.~\cite{gaudelet2018higher}.}
\label{fig_bio2}
\end{figure*}

Among the first higher-order analyses in biology were studies showing
that the hypergraphs corresponding to mammalian protein
complexes~\cite{ruepp2010corum} had a scale-free distributions of both node
degrees and hyperedge sizes~\cite{wong2008evolutionary}.
The full potential of using HOrS frameworks, in particular
hypergraphs, to characterize complex biological processes taking place
in biomolecular systems, was already clear more than ten years
ago~\cite{klamt2009hypergraphs}.
An important range of applications is  
that of signaling pathways in cell biology, where group of molecules
have to work together to efficiently control cell functions, such as
death or division~\cite{ritz2014signaling}. Different representations of signaling pathways are summarized in Fig.~\ref{fig_bio1}.
\citet{gaudelet2018higher} have investigated protein interaction
hypergraphs by extending the concept of graphlets~\cite{prvzulj2004modeling}, to 
the case of higher-order networks. Hypergraphlets were defined
as small induced sub-hypergraphs of a given large hypergraph, and an orbit identifies each different set of automorphic nodes.
An example of all the hypergraphlets of 1, 2 and 3 nodes is 
illustrated in
Fig.~\ref{fig_bio2}. Similarly to motifs~\cite{alon2007network}, hypergraphlets allow to
characterize wiring patterns of higher-order networks at the
local scale. By focusing on the case of yeast and human pathways, the
authors showed that modeling protein interactions as hypergraphs
allows for better functional predictions than a description in terms
of graphs with pairwise interaction only.

\citet{franzese2019hypergraph} have challenged the current approaches to
molecular connectivity, which they found either too permissive or too
restrictive. As an alternative, they have proposed
an intermediate optimal solution that   
interpolates between graph and hypergraph approaches and 
allows to better capture the importance of small molecules
involved in many distinct reactions.
More recently, \citet{klimm2020hypergraphs} have used
hypergraphs to investigate multiprotein complex data, showing how a
pairwise (network) projection produces a  
hierarchical structure, that is instead not observed when polyadic
interactions are considered.
After comparing the protein complexes
with appropriate null models, the authors found that larger complexes tend to
be more essential, with a hyperdegree that better correlates with
gene-essentiality information than the standard graph degree. All
these results suggest the importance of considering the inherent
higher-order structure of protein complexes to reveal complementary
information.

Redundancy is an important property of biological systems that
guarantees their functionality in the case of misfunctioning of some
local components. ~\citet{pearcy2016complexity} have used hypergraph
percolation to assess the robustness of empirical bacterial metabolic
higher-order system to random failures. In particular, they 
have used site percolation, in which a hyperedge (describing a
reaction) is activated only when all metabolites involved
in the hyperedge are active.
Results showed that interacting systems that have evolved in
environments with a higher degree of variability are 
more robust, and that, similarly to their simple network counterparts,
also metabolic hypergraphs are characterized by the presence of a
core-periphery structure. In another study, metabolic networks have
been characterized by the spectrum of a symmetric tensor associated to
their hypergraph connectivity, successfully capturing the chemical
information of enzymes and structural changes of compounds to define
novel classes of functional reactions ~\cite{shen2018genome}. Other
higher-order topological operators such as combinatorial Laplacians (see
Section \ref{subsubsec:combinatorial-laplacians}) were shown to provide a
more complete characterization of chemical reaction
networks~\cite{jost2019hypergraph}.

The complexity of biological systems entails that the data we obtain
from experiments are often incomplete or characterized by a limited
accuracy. For this reason, various works have concentrated on the
problem of reconstructing the hypergraphs associated to different 
kinds of cellular processes where higher-order interactions are at
play~\cite{tian2009hypergraph, battle2010automated,
  sumazin2011extensive, rahman2013reverse}. Recently,
higher-order inference frameworks have also been extended to
deal with dynamic correlations of abundance levels of genes,
transcripts and metabolites changing over time. The results
provide a better picture of the global dynamic correlation patterns of the
investigated biological systems~\cite{kong2019hypergraph}.

Higher-order interactions have also revealed key when designing effective
drug combinations to prevent or contrast diseases. Indeed, in multiple
cases, from cancer to tuberculosis, the combined action of the 
ingredients of a so-called drug cocktail, even when administered at a
low dose, has been shown to be 
more beneficial than that of single drugs in isolation. Recently, the
dose model has been developed as an efficient tool to discover
effective drug combinations based on pairwise
interactions~\cite{zimmer2016prediction}. While the dose model was
originally tested only on triplets and quadruplets of antibiotics, as
well as triplets of cancer drugs, very recently the model was shown to
successfully predict effective combinations of up to ten
drugs used for E. coli and the M. tuberculosis
pathogen~\cite{katzir2019prediction}. However, when noise in the
dataset is more important, prediction of higher-order interactions
based on pairs of drugs is less efficient~\cite{tendler2019noise}.
Notice that alternative approaches to detect effective drug cocktails
are available, such as the so-called pairs model, which was shown to
be more noise-resistant but less precise~\cite{zimmer2017prediction}.

Related to both robustness and drug resistance, the feedbacks between the
different levels of genetic information shape to a large degree the
link between genotype and phenotype
\citep{otwinowski2014inferring,crona2017inferring}.
A relevant
example is that of epistasis, which \citet{weinreich2013should} have
defined as 
``the surprise at the phenotype when mutations are combined, given the
constituent mutations’ individual effects''.  In other words,
epistasis describes the somewhat surprising observation that the
effects of multiple individual mutations appear to interact with each
other in ways that cannot be quantitatively reduced to sums of pairwise
interactions \citep{sanchez2018high}.  While the concept of higher
order epistasis is not new \citep{mackay2014epistasis}, only recently
it has become possible to start a quantitative analysis of its effects
in different contexts \citep{sanchez2019defining}.  For example,
\citet{guerrero2019proteostasis} have investigated how protein quality
control machinery influenced the epistasis in traits related to
bacterial antibiotic resistance, separating the mutations affecting an
essential bacterial enzyme from species-specific effects. 
\citet{yitbarek2019deconstructing} have explored higher-order interactions
among gut taxa and their effects on host infection risk, using a
theoretical model tailored to the type of data that might be
empirically collected in the near future. \citet{mickalide2019higher} have 
studied a controlled microbial trophic chain and detected an increased
invasion resistance of the community, stemming not from resource
allocation but from high-order interactions between its species.

Finally, dynamical processes have been recently used to extract
information on the structure of higher-order biological networks. For
instance, ~\citet{niu2019rwhmda} have considered a microbe-disease
hypergraph, where nodes are microbes and diseases are hyperedges. A
higher-order random walk (see Section \ref{sec:rw}) was shown to have a
greater accuracy in the prediction of disease-microbe associations
compared to that of traditional random walks.

\section{Outlook and conclusions}
\label{sec:conclusions}

In this review, we have discussed ways and methods to detect,  
represent, measure and model systems with higher-order interactions,
and we have illustrated models of higher-order systems 
substantially differ both structurally and dynamically
from traditional pairwise models. 

We have seen time and time again that there are crucial conceptual
differences between modeling pairwise and higher-order interactions.
For example, we have seen that higher-order interactions typically
lead to new sources of non-linearity in the systems under study, which
are not present in standard network approaches.  Further,
considering objects richer than links opens up new possibilities and
questions: for example, state variables can now be defined not only on
nodes, as in standard practices, but also on edges, triangles,
tetrahedra and so on, paving the way to concepts like group states,
but also necessitating a consideration about their meaning and
interpretation \citep{millan2019synchronization}.
Much of this new landscape is yet unexplored, but we can already
make a few important observations.

At the dynamical level, it is evident that the presence or absence of
higher-order interactions is especially important.  We reviewed
explicit examples in which higher-order interactions profoundly change
the critical behavior of dynamical processes in both simplicial
complexes
\cite{iacopini2019simplicial,torres2020simplicial,schaub2020random,muhammad2006control,skardal2019higher,
  bick2016chaos} and hypergraphs
\cite{de2020social,alvarez2020evolutionary,zhou2007learning,matheny2019exotic}.
However, even when a dynamical process does not explicitly contain
higher-order \emph{dynamical} terms, it is possible to find new
effects due to higher-order terms in the \emph{structural} patterns
underlying the dynamics.  For example, simple contagion processes are
usually considered to be largely oblivious to higher-order structures
(e.g. large groups, heterogeneous and/or hierarchical clique
structure) beyond clustering in the underlying contact patterns.
However, \citet{stonge2020school}
recently showed that membership of
nodes to cliques of heterogeneous sizes can result in unexpected
mesoscopic localization phenomena, in turn yielding possible outbreak
persistence for cases in which standard diffusion models would predict
outbreak extinction.  Along similar lines, \citet{petri2018simplicial}
showed that including group activations in a simple contagion model on
a temporally evolving contact substrate can shift the critical
infectivity, and that the shift depends on a trade-off between the
distributions of group sizes and of activity of the nodes.

The importance of higher-order interactions is naturally not limited
to their effects on dynamics.  Recent examples include applications in
which higher-order terms allowed better descriptions of group
formation in scientific collaborations
\citep{milojevic2014teams,patania2017shape,salnikov2018simplicial} and
finer classification of the local environment of node
\citep{kartun2019beyond}.
In other cases, they improved the predictions of new interactions beyond the capacity of link-based prediction models, and also significantly denoised signals in complex environments \citep{schaub2018flow,pokorny2016topological}.
Topological descriptions have even been proposed as a convenient tool to model epistemic models with distributed computing tasks~\cite{goubault2018simplicial,van2020knowledge}.\\

The study of higher-order systems, of their characteristic properties
and their effects on dynamics is a recent field, and there are still 
many open and unexplored directions. Below, we list some of them:\\

\emph{Measures for higher-order structures.}  We have described the
most common measures used in the description of HOrSs.  With the
exception of the intrinsically algebraic ones, most of these measures
however are straightforward generalizations of those used for
networks.  Temporal, multiplex and multilayer measures are still
lacking, and generally there is a large space to be filled.  An
example are measures that simply cannot be defined in pairwise
networks, the simplicial closure being one of these
\citep{patania2017shape,benson2018simplicial,kartun2019beyond}.  Other
examples touch on state variables defined on simplices or hyperedges
of arbitrary dimensions: while we have a clear understanding of what
synchronization among nodes looks like in models of oscillators, it is
much harder to have an intuitive grasp of what the state of an edge,
or of a triangle, might mean \citep{reitz2020higher}.
Defining measures able to capture these quantities would also be a
step toward an understanding of their role and quantitative insights
about their effects.  A further example is homological information
obtained from topological data analysis techniques: it is defined as
an equivalence class and therein lies its power and curse, because its
resulting non-local nature makes it a powerful descriptive tool, but
also very hard to localize on specific elements of the HOrS.  Efforts
to find a solution to this issue already exist
\citep{petri2014homological,kalivsnik2019higher,guerra2020homological},
but in many cases the problem is ill-defined and the solution specific
to the problem at hand. So, are there standard or, at
least, acceptable ways to localize homological features as to use
them in further analysis? Or should we give up on localizing shapes,
and think only about manifolds?  Finally, while hypergraph
partitioning \citep{li2018submodular,karypis1999multilevel} has a long
history, little work has focused on characterizing the mesoscopic
structure of simplicial complexes, both in terms of the definition and
detection of communities
(e.g. \citep{neubauer2009towards,billings2019simplex2vec}) and of
other types of (quasi-)local (e.g. rich club, assortative behavior,
etc) and spanning structures (e.g. cores \citep{marietti2008cores},
minimal spanning trees \citep{duval2013critical}, expander properties
\citep{steenbergen2014cheeger,parzanchevski2017mixing}).  \\

\emph{Generative models for higher-order structures.}
Models able to constrain various features of higher-order structures are crucial because they provide a principled answer to the question of what constitutes a non-trivial and topologically rich HOrS. 
As we have discussed in this review, there are currently few random models of simplicial complexes.
Some of the existing models of simplicial complexes reproduce
the local connectivity patterns\citep{courtney2016generalized,young2017construction}, but none exist that are able to reproduce or approximate more refined topological structures, like a specific target homology or mesoscopic structures.
Finally, exactly like conventional networks, HOrSs can change in time or be composed by different qualitatively different types of interactions. 
With few notable exceptions \citep{babichev2018robust,petri2018simplicial}, to date there are practically no models taking into account the temporal or multiplex structure of higher-order interactions---the vast majority model growth instead \cite{wu2015emergent}.\\

\emph{Understanding the driving mechanisms of higher-order dynamics.}
Developing new measures and generative models is also important to
identify the fundamental mechanisms that lie behind the
patterns we observe.  There is in fact clear evidence that
non-trivial higher-order topologies emerge in social
\citep{sizemore2017classification,petri2013topological} as well as in 
biological systems
\citep{petri2014homological,giusti2015clique,dabaghian2012topological},
but very little understanding of how or why they do emerge. 
Currently, only few models focus
on describing coordination, group interactions and in general growth
of HOrSs at the group level, and none reproduce higher-order
topological invariants.  This is in part due to the predominance of
network descriptions up until now, and in part to the actual
difficulties to provide an analytical description that one encounters
as soon as higher-order terms are introduced.
generalizing well-known dynamical systems.  In addition to the work on
contagion mentioned above, early efforts in this direction are already
under way \citep{kuehn2020universal}, including generalization of
Kuramoto models to higher-order interactions
\cite{torres2020simplicial,skardal2019higher,skardal2019abrupt} and
games \citep{alvarez2020evolutionary}.  A particularly important and
recent line of research focuses on extending concepts from percolation
to simplicial complexes, dubbed \emph{topological percolation}, both
in simplicial \citep{bianconi2018topological,bianconi2019percolation} and homological terms
\citep{bobrowski2020homological}.  Overall, however, we still lack a
general understanding of how higher-order terms affect dynamical
systems.\\

\emph{Inference from data.}
What is a truly genuine higher-order interaction? And how do we tease
it apart from low-order ones in data?  And if it is possible, what type of 
data do we need to tell the difference between low and higher-order
interactions?
These are hard questions in general, but for some systems it is easier
to approach them with some confidence.  Indeed, for systems where the
data already comes in the form of sets, it is straightforward to
extract higher-order interactions and measure their strengths.  This
is the case, for instance, of affiliation networks such as 
coauthorship data, where each paper
constitutes an interaction among all authors, or of data about joint
presence in locations, or different ingredients in recipes.
In many systems, however, interactions are not already identified, but
instead need to be inferred from the data. The most obvious
example is that of timeseries: brain functional networks are usually
estimated by computing correlations, or other 
measures based on information theory, between fMRI or EEG timeseries 
\citep{phinyomark2017resting};
similarly, financial networks are built starting from stock option
prices or timecourse of revenues, and so on
\citep{battiston2010structure}.  In all these cases, higher-order
interactions are seldom considered relevant, or even computed, due to
various reasons.  First, many-body correlations are often
computed as second order approximations of standard correlations, and
hence considered as perturbations.  Second, measures that can find
higher-order effects
\citep{faes2015estimating,faes2017multiscale,rosas2019quantifying}
often require long timeseries, which in many cases are not available.
Third, the scarce availability of rich data on 
dynamical models with and without higher-order interactions makes it impossible to define a
proper inference scheme for the presence, nature and strength of
higher-order interactions. 
This last point is crucial and links
back to the importance of models to understand the underlying
mechanisms.  Just like it is hardly possible to
distinguish complex from simple contagion
from prevalence and incidence data
\citep{hebert2020macroscopic}, it might well be the case that it is
not possible to tease apart the effects of complex contagion
from those of simplicial contagion \citep{iacopini2019simplicial}
in absence of microscopic mechanistic information.  However, currently
there are no inference schemes, akin to \citet{peixoto2019network},
able to test hypothesis about higher-order interactions and provide
guidance in these situations.  Developing such schemes is therefore
paramount to the advancement of the field.\\

The open directions discussed above focus on theoretical, modeling and
methodological issues.  HOrSs have already been fruitful in a
smattering of applications, but they still need to find concrete
applications to a wider range of topics.  Indeed, the real test of
their relevance will be in the breadth and depth of their impact on
specific problems.  While the paradigm of higher-order interactions is
general, we envision that problems in biology, ecology, population
dynamics, neuroscience and computational social sciences will be the
first and the foremost to benefit from these new tools and ideas.  We
hope that this review will provide a guiding path for researchers
interested in HOrSs, and we look forward to seeing how HOrSs
themselves will reshape the landscape of complex systems research.

\section*{Acknowledgments}

F. B. acknowledges partial support from the ERC Synergy Grant 810115 (DYNASNET).
G. C. and M. L. acknowledge partial support from the “European Cooperation in Science \& Technology” (COST): Action CA15109.
I. I. acknowledges partial support from the Urban Dynamics Lab under the EPSRC Grant No. EP/M023583/1.
V. L. acknowledges support from the Leverhulme Trust Research Fellowship ``CREATE: the network components of creativity and success'', RF-2019-059.
J.-G. Y. acknowledges support from the James S. McDonnell Foundation.
G. P. acknowledges partial support from Intesa Sanpaolo Innovation Center and from Compagnia San Paolo (ADnD project).  \\

The authors acknowledge valuable and stimulating discussions with many members of the network science community on the topic covered in our report, including Antoine Allard, Unai Alvarez-Rodriguez, Alex Arenas, Tomaso Aste, Paolo Bajardi, Albert-L{\'a}szl{\'o} Barab{\'a}si, Andrea Baronchelli,
Alain Barrat, Danielle Bassett, Demian Battaglia, Jaume Bertranpetit, Ginestra Bianconi, Christian Bick, Jacob C.W. Billings,
Stefano Boccaletti, Francesco Bonchi, Guido Caldarelli, Timoteo Carletti, Ciro Cattuto, 
Mario Chavez, Guilherme Ferraz de Arruda, Fabrizio De Vico Fallani, Tiziana Di Matteo, Tina Eliassi-Rad, Paul Expert,
Duccio Fanelli, Michael Farber, Mattia Frasca, Luca Gallo,
Lucia Gambuzza, Laetitia Gauvin, Fosca Giannotti, Tommaso Gili, Corrado Gioannini, Jesus G{\'o}mez-Garde{\~n}es, Heather A. Harrington, Laurent H\'ebert-Dufresne, Esther Ib{\'a}{\~n}ez-Marcelo, Gerardo I\~{n}iguez, Cliff Joslyn, M\'{a}rton Karsai, Sonia K\'{e}fi, J\'{a}nos Kert\'{e}sz, Julia Koltai, Dima Krioukov, Lucas Lacasa, Renaud Lambiotte, Bruno Lepri, Michael Lesnick, Daniele Marinazzo, Andrea Migliano, Yamir Moreno, M.E.J. Newman, Andr\'{e} Panisson, Daniela Paolotti, 
Luca Pappalardo, Dino Pedreschi, Tiago P. Peixoto, Matja{\v{z}} Perc, Nicola Perra, Angkoon Phinyomark, Mason Porter,  M\'{a}rton P\'{o}sfai, Mario Rasetti, Martin Rosvall, Manish Saggar, Enrica L. Santarcangelo, Samuel S. Scarpino, Michael Schaub, Martina Scolamiero, Ingo Scholtes, Olaf Sporns, Bosiljka
Tadi{\'c}, Stefan Thurner, Michele Tizzoni, Francesco Vaccarino, Alessandro Vespignani, Lucio Vinicius.

\bibliographystyle{unsrtnat}

\end{document}